\documentclass[a4paper,twoside,12pt]{article}
\usepackage{amssymb,amsmath}
\usepackage[running,mathlines]{lineno}
\usepackage{makeidx}
\usepackage{subeqnarray}
\usepackage{graphicx}
\usepackage[colorlinks=false,hidelinks]{hyperref}
\usepackage{fancyh1}
\usepackage{color}
\usepackage{epstopdf}
\usepackage{multirow}
\usepackage{rotating}
\usepackage{natbib}

\newcommand{\beq}{\begin{equation}}  
\newcommand{\eeq}{\end{equation}}    
\newcommand{\beqa}{\begin{eqnarray}} 
\newcommand{\eeqa}{\end{eqnarray}}   
\begin{document}

\looseness=-1

\def\today{\number\day\space\ifcase\month\or
  January\or February\or March\or April\or May\or June\or
  July\or August\or September\or October\or November\or December\fi,
  \number\year}
%
\def\etal{{\it et al.\/}}
\def\ie{{\it i.e.\/}}
\def\eg{{\it e.g.\/}}

\def\1o2{\textstyle {\frac{1}{2}}}
\def\alphab{\mbox{\boldmath $\alpha$}}
\def\bcb{\mbox{\boldmath ${\cal B}$}}
\def\betab{\mbox{\boldmath $\beta$}}
\def\d{{\rm d}}
\def\dcb{\mbox{\boldmath ${\cal D}$}}
\def\deltab{\mbox{\boldmath $\delta$}}
\def\Deltab{\mbox{\boldmath $\Delta$}}
\def\dotprod{\!\cdot\!}
\def\e{{\rm e}}
\def\ecb{\mbox{\boldmath ${\cal E}$}}
\def\epsilonb{\mbox{\boldmath $\epsilon$}}
\def\fcb{\mbox{\boldmath ${\cal F}$}}
\def\gammab{\mbox{\boldmath $\gamma$}}
\def\Gammab{\mbox{\boldmath $\Gamma$}}
\def\hcb{\mbox{\boldmath ${\cal H}$}}
\def\intd{\int\!\!\int}
\def\intt{\int\!\!\int\!\!\int}
\def\kappab{\mbox{\boldmath $\kappa$}}
\def\lambdabar{\lambda \! \! \! \!
^{\rule{0.8mm}{0mm}\rule[0.25mm]{1.6mm}{0.12mm}}\rule{0.15mm}{0mm}}
\def\mcb{\mbox{\boldmath ${\cal M}$}}
\def\me{{\rm m}_{\rm e}}
\def\nablab{\mbox{\boldmath $\nabla$}}
\def\omegab{\mbox{\boldmath $\omega$}}
\def\Omegab{\mbox{\boldmath $\Omega$}}
\def\pcb{\mbox{\boldmath ${\cal P}$}}
\def\phib{\mbox{\boldmath $\phi$}}
\def\Phib{\mbox{\boldmath $\Phi$}}
\def\pib{\mbox{\boldmath $\pi$}}
\def\Pib{\mbox{\boldmath $\Pi$}}
\def\psib{\mbox{\boldmath $\psi$}}
\def\Psib{\mbox{\boldmath $\Psi$}}
\def\sigmab{\mbox{\boldmath $\sigma$}}
\def\Sigmab{\mbox{\boldmath $\Sigma$}}
\def\taub{\mbox{\boldmath $\tau$}}
\def\thetab{\mbox{\boldmath $\theta$}}
\def\varepsb{\mbox{\boldmath $\varepsilon$}}
\def\varphib{\mbox{\boldmath $\varphi$}}
\def\vecprod{\!\times\!}
\def\xib{\mbox{\boldmath $\xi$}}
\def\ycb{\mbox{\boldmath ${\cal Y}$}}
\def\zetab{\mbox{\boldmath $\zeta$}}
\newcounter{listr}
\def\blr{\begin{list} {(\roman{listr})} {\usecounter{listr}}}
\def\elr{\end{list} \setcounter{listr}{0}}
\def\bla{\begin{list} {(\arabic{listr})} {\usecounter{listr}}}
\def\ela{\end{list} \setcounter{listr}{0}}
% ...
\def\boxeq#1{$$\mbox{\fbox{$\displaystyle{#1}$}}$$}
\def\boxeqn#1{\begin{equation}\mbox{\fbox{$\displaystyle{#1}$}}
\end{equation}}
% ...
%
\def\req#1{(\ref{#1})}
%

 \renewcommand{\theequation}{\thesection.\arabic{equation}}
 \setcounter{equation}{0} \setcounter{figure}{0} \setcounter{table}{0}


\def\title#1{\vglue0.05truein{\baselineskip=24 truept
    \pretolerance=10000
    \raggedright\noindent \LARGE\bf #1 \par}\vskip1.25cm}
%
\def\author#1{{\pretolerance=10000\raggedright\noindent
               {\large\bf #1}\par}\vskip-0.25cm}
%
\def\address#1{\bigskip\noindent {\rm #1} \par}
%
\def\addinfo#1{\bigskip\noindent {\rm #1} \par \smallskip}
%


\thispagestyle{empty}

\title{Inelastic collisions of fast charged particles with atoms.
Relativistic plane-wave Born approximation}

\author{Francesc Salvat}

\vspace{2mm}
\address{Facultat de F\'{\i}sica (FQA and ICC). Universitat de Barcelona.
\\ Diagonal 645, 08028 Barcelona, Catalonia, Spain}

\vspace{\stretch{+2.0}}

\noindent{\large\bf Abstract}

\noindent  A detailed formulation of the relativistic plane-wave Born
approximation (PWBA) for inelastic collisions of charged particles with
free atoms and positive ions is presented. The wave functions of the
target atom or ion are calculated from a central-field
in\-de\-pen\-dent-electron model with the Dirac--Hartree--Fock--Slater
self-consistent potential, and the electromagnetic field is expressed in
the Coulomb gauge. The double-differential cross section, depending on
the energy loss and the recoil energy, is given as a sum of two terms
which are products of purely kinematical factors and the generalized
oscillator strengths (GOSs). Transitions induced by the instantaneous
Coulomb interaction between the projectile and the active target
electron are described by the longitudinal GOS. Transitions caused by
the transverse interaction (exchange of virtual photons) are accounted
for by a transverse GOS. We derive closed expressions for the
longitudinal and transverse GOSs in terms of vector coupling
coefficients and radial integrals. A set of Fortran programs have been
written to compute the GOSs, the energy-loss differential cross section,
and integrals of the latter. A complete numerical database of GOSs has
been calculated for all the subshells of the ground-state configuration
of neutral atoms of the elements with atomic numbers from 1 (hydrogen)
to 99 (einsteinium). A systematic derivation of asymptotic formulas for
the total cross section, the stopping cross section and the
energy-straggling cross section is presented. The shell correction to
the asymptotic formula for the stopping cross section of protons is
obtained from the difference between computed numerical values and the
predictions of that formula.

\vspace{5mm}

\noindent {\bf Keywords}: Inelastic collisions of charged particles.
Relativistic plane wave Born approximation (PWBA). Bethe asymptotic
formulas. Stopping power. Shell corrections.

\vfill

\noindent Date: 22 September, 2021 

\rule{0mm}{10mm}

\newpage
\thispagestyle{empty}

\rule{1mm}{1mm}

  \newpage
  \setcounter{secnumdepth}{3}
  \setcounter{tocdepth}{3}
  \addtolength{\baselineskip}{-0.75mm}
  \thispagestyle{empty}
  \pagenumbering{roman}
  \pagestyle{fancyplain}
  \lhead[\fancyplain{}{\sl \thepage}]{\fancyplain{}{\sl Contents}}
  \rhead[\fancyplain{}{\sl Contents }]{\fancyplain{}{\sl \thepage}}

  \tableofcontents
  \addtolength{\baselineskip}{+0.75mm}

  \clearpage
  \newpage


\def\rtitle{Relativistic PWBA}
\def\ltitle{F. Salvat}

\lhead[\fancyplain{}{\sl \thepage}]{\fancyplain{}{\sl \rtitle }}
\rhead[\fancyplain{}{\sl \ltitle }]{\fancyplain{}{\sl \thepage}}

\pagenumbering{arabic}
\pagestyle{fancyplain}

\newpage


\vspace*{1mm}
\section{Introduction \label{sec1}}
\setcounter{equation}{0}

A fast charged particle moving in a material medium losses energy
through inelastic collisions (\ie, collisions causing electronic
excitations of the medium) and through the emission of breaking
radiation (Bremsstrahlung). The latter mechanism has an appreciable
stopping effect only for the lighter particles (\ie, electrons and
positrons) with high energies. For heavier particles, the slowing down
of the projectile is due almost exclusively to inelastic collisions.
Collisions with energy transfers larger than the ionization energies of
the active target electrons cause the ionization of the target atoms,
resulting in the emission of knock-on electrons, also know as delta
rays. When the active target electron is in an inner subshell, after the
ionizing collision the atom or the residual ion is left in a highly
excited state, which decays towards the ground state by emission of
Auger electrons and x rays. Hence, a reliable theoretical description of
inelastic collisions is needed for describing not only the slowing down
of charged particles, but also the generation of delta rays, Auger
electrons and x rays.

The present report contains a detailed study of inelastic collisions of
charged particles with free neutral atoms or positive ions using the
relativistic plane-wave first-order Born approximation (PWBA). In this
approximation, the interaction between the projectile and the atomic
electrons is treated as a perturbation to first order, and the wave
functions of the projectile before and after the interaction are
represented as Dirac plane waves. In principle, the PWBA is applicable
when the speed of the projectile is much larger than the speeds of the
atomic electrons. In the case of ionization of inner subshells by impact of
electrons and positrons, the theory yields results accurate to within
$\sim 1$ percent when the kinetic energy of the projectile is larger
than about 30 times the ionization energy of the active subshell
\citep{BoteSalvat2008}. The PWBA is the basis of the conventional theory
of the collisional stopping of fast charged particles \citep{Bethe1932,
Fano1963, Inokuti1971}. In principle, the results of the PWBA for
neutral atoms and ions are appropriate for describing the passage of
fast charged projectiles through thin monoatomic gases. The
generalization to molecules and condensed media is not trivial, mostly
because the presence of neighboring atoms causes a distortion of the
wave functions of free electrons. Moreover, for relativistic
projectiles, the dielectric polarisability of the medium screens the
electromagnetic field of the projectile, giving rise to the so-called
density-effect correction \citep{Fano1963, ICRU37}. Nevertheless, the
PWBA does give a fairly reliable description of the ionization of inner
subshells of atoms bound in molecules or solids, and provides a convenient
theoretical framework to build semiempirical models for inelastic
interactions of charged particles in condensed matter \citep[see,
\eg,][]{FernandezVarea2005}.

We give a detailed derivation of the double-differential cross section
(DDCS), differential in the energy loss $W$ and the recoil energy $Q$,
for inelastic collisions of charged particles with free neutral atoms
and positive ions. The states of the target atom or ion are described by means
of a central-field independent-electron model, \ie, as single Slater
determinants, and the electromagnetic field is expressed in the Coulomb
gauge \citep{Fano1963}. Considering the interaction between the
projectile and the atomic electrons as a perturbation to first order,
the transition matrix elements reduce to sums of one-electron integrals,
in accordance with the intuitive picture known as the
one-active-electron approximation. The resulting DDCS is expressed in
terms of the longitudinal generalized oscillator strength (GOS), which
summarizes the response of the target atom or ion to the instantaneous
Coulomb interaction, and a transverse generalized oscillator strength
(TGOS) which accounts for the so-called transverse interaction (exchange
of virtual photons). Integration of the DDCS over recoil energies gives
the energy-loss differential cross section (DCS); the moments of order
0, 1 and 2 of the energy-loss DCS are, respectively, the atomic total
cross section, the stopping cross section and energy-straggling cross
section.

The well-known asymptotic formulas for the total cross section and the
stopping cross section are derived by using a method similar to the one
adopted by \citet{Fano1963} to arrive at the stopping power formula.  We
also derive a less-well-known asymptotic formula for the
energy-straggling cross section. These formulas are valid in the limit
of high kinetic energies. Their departures from the exact cross
sections, obtained by integrating the energy-loss DCS, are known as
shell corrections \citep{Fano1963}. Existing calculations of the shell
correction to the stopping cross section are based on non-relativistic
calculations with hydrogenic wave functions or Hartree--Slater wave
functions \citep[][and references therein]{Bichsel2002}, or on the
free-electron gas theory \citep{Bonderup1967}. In the present work,
shell corrections are obtained as differences between the numerical
integrated cross sections and the corresponding asymptotic formulas.

Starting from the program of \citet{BoteSalvat2008}, we have developed
and assembled a set of Fortran programs that perform the complete
sequence of calculations leading to the energy-loss DCS, the integrated
cross sections and the shell corrections. The program {\sc gosat}
computes the GOS and the TGOS for {\it excitation} and {\it ionization}
of closed electron subshells of neutral atoms or positive ions.  The
initial and final orbitals of the active target electron are solutions
of the Dirac equation for a central potential $V(r)$, which in the
present study is set equal to the Dirac--Hartree--Fock--Slater (DHFS)
self-consistent potential \citep[see, \eg,][]{Liberman1971}. This
potential is calculated with the program {\sc dhfs}
\citep{SalvatFernandezVarea2019}; the distribution package contains the
DHFS potentials of neutral atoms of all elements from hydrogen to
einsteinium (atomic numbers $Z=1$ to 99). In the case of ions, the
self-consistent potential must be provided by the user, who can
calculate it by running the program {\sc dhfs}. The initial and final
orbitals of the active target electron are solutions of the Dirac
equation for the potential $V(r)$, which are obtained by solving the
Dirac radial equations by means of the {\sc radial} subroutine package
of \citet{SalvatFernandezVarea2019}. The expression of the TGOS involves
matrix elements that are identical to the ones occurring in the
calculation of the atomic photoelectric effect. Parts of the {\sc
gosat} code have been reused in a program, named {\sc photacs}, that
calculates cross sections for absorption of photons by atoms and ions
\citet{SabbatucciSalvat2016}. We have verified that {\sc photacs} yields
results in good agreement with previous calculations by
\citet{Scofield1973} \citep[see also][]{Cullen1997}, thus confirming the
correctness of the algorithm for the calculation of matrix elements.
The {\sc gosat} program has been used to calculate tables of the GOS and
the TGOS for all the subshells of the ground state configuration of the
elements ($Z=1$ to 99).

A second Fortran program, named {\sc pwacs}, calculates energy-loss
DCSs, for heavy charged particles, electrons and positrons, from tables
of the GOS and the TGOS pre-calculated with the {\sc gosat} program. To
ensure accuracy of the energy-loss DCS and its integrals, specific
schemes are adopted for interpolating and extrapolating the numerical
GOS tables. The program {\sc pwacs} also calculates the total cross
section, the stopping cross section, and the energy-straggling cross
section by numerical integration of the energy-loss DCS.

The goal of the present work is {\it not} to compute realistic total
cross sections and stopping powers because the independent-electron
model and the DHFS potential are too simplistic to reproduce the details
of the actual excitation spectrum of isolated atoms. Indeed, our
framework (PWBA with DHFS potential) provides reliable results only for
the ionization of inner-electron subshells by impact of high-energy
charged particles, mostly because the relevant one-electron wave
functions are practically unaffected by the existence of neighbor atoms.
Nevertheless, it allows the calculation of theoretical GOSs to very high
accuracy, from which energy-loss DCSs can be obtained by numerical
integration of the DDCS. An interesting aspect of this framework is
that, by including in the calculation all possible excitations of the
target atom (to both bound and free final states), we can analyze the
consistency of the Bethe sum rule, which states that the integral of the
GOS over $W$ equals the number of electrons in the target atom or ion.
This sum rule plays a key role in the derivation of the asymptotic
formula for the stopping power, notwithstanding the fact that it holds
only in the non-relativistic domain, \ie, for atoms with low atomic
numbers. Since both shell-corrections and departures from the Bethe sum
rule get the largest contributions from inner-electron subshells, they
are expected to be accurately described within our theoretical
framework. Motivated by this expectation, our aims are to devise a
consistent scheme to evaluate the impact of departures from the Bethe
sum rule on the asymptotic formulas, and to calculate a comprehensive
database of shell corrections to the stopping cross section for neutral
atoms with $Z=1$ to 99.

The present report is structured as follows. Section \ref{sec2} contains
a brief summary of the assumptions and simplifications underlying the
relativistic first-order PWBA. The interaction Hamiltonian
of the projectile with target electrons is considered in Section
\ref{sec3}. Section \ref{sec4} provides a detailed derivation of the
PWBA, which leads to a closed expression for the DDCS in terms of the
longitudinal and transverse GOSs, which are expressed as sums and
averages of squared transition matrix elements. In Section \ref{sec5},
the theoretical formulas for the GOS and the TGOS are reduced to a
numerically workable form. We also describe the numerical methods
employed to calculate the GOSs, as well as the interpolation tricks and
approximations used to convert the numerical tables of GOSs into smooth
continuous functions. In Section \ref{sec6} we introduce the energy-loss
DCS, and several relevant integrals of it; we also obtain simple
formulas for the contributions of close collisions with large energy
transfers. The modifications of the theory needed for describing
collisions of electrons and positrons are discussed in Section
\ref{sec7}. The Fortran codes {\sc gosat} and {\sc pwacs} are described
in Section \ref{sec8}, where illustrative sample calculation results are
presented. In Section \ref{sec9} we derive asymptotic formulas for the
total cross section, the stopping cross section, and the energy
straggling cross section of closed electron subshells for heavy charged
particles, electrons and positrons; asymptotic formulas for atoms are
obtained by adding the contributions from the various subshells. Shell
corrections are discussed in Section \ref{sec10}, where we present
numerical estimates of the corrections for projectile protons and
selected elements. In Section \ref{sec11} we introduce approximate
corrections to account for binding and Coulomb-deflection effects in
ionizing collisions of heavy particles.  Finally, Section \ref{sec12}
offers some concluding remarks.  The Appendices contain a summary of
kinematic relations relevant to inelastic collisions (\ref{secA}), a
brief description of Dirac wave functions of electrons in central fields
(\ref{secB}), a set of formulas for matrix elements of Racah tensors
that are used to evaluate angular integrals (\ref{secC}), an analytical
calculation of the DDCS for collisions of charged particles with free
electrons at rest (\ref{secD}), and a brief derivation of
angle-restricted DCSs used in electron energy-loss spectroscopy
(\ref{secE}).

In the following, all electromagnetic quantities are expressed in the
Gaussian system of units \citep[see, \eg][]{Jackson1975} and, as
mentioned above, electromagnetic potentials are in the Coulomb gauge. As
pointed out by \citet{Fano1956}, regarding electromagnetic interactions
in material media, the representations in different gauges are not
equivalent. The Coulomb gauge seems to be the most adequate since it
provides a natural decomposition of the fields into their longitudinal
and transverse parts, the latter being observable as electromagnetic
radiation under certain circumstances (\eg, Cerenkov radiation).
However, it is appropriate to mention here that \citet{Scofield1978}
calculated cross sections for the ionization of the K shell and L
subshells of atoms by impact of electrons using the relativistic PWBA
with the electromagnetic potentials in the Lorentz gauge. His
theoretical model is equivalent to ours, although it does not allow the
identification of the longitudinal and transverse GOSs, which are
essential for our purposes. In the text $e$ denotes the elementary
charge (\ie, the absolute value of the electron charge), $\me$ is the
electron mass, $\hbar = h/(2\pi)$ is the reduced Planck constant, and
$c$ is the speed of light in vacuum.


\section{The first-order Born approximation \label{sec2}}
\setcounter{equation}{0}

In this Section we present the elementary quantum theory of inelastic
collisions of fast charged particles with isolated atoms and ions. We
consider inelastic collisions of a projectile of mass $M$ and charge
$Z_0 e$ with an atom or ion of the element of atomic number $Z$ with a
nucleus of mass $M_{\rm nuc}$ and $N$ electrons in its ground state. The
theory is applicable to neutral atoms ($N=Z$) and to positive ions
($N<Z$).

The non-relativistic Hamiltonian ${\cal H}$ of the system (projectile
and target atom) can be expressed in terms of the position ${\bf R}$ of
the center of mass and the positions ${\bf r}_I$ of the electrons (indexes $I=1,
2, \ldots, N$) and the projectile (index $I=0$) relative to the nucleus
\citep[see, \eg,][Appendix 8]{BransdenJoachain1983}. Neglecting a
usually negligible mass polarization term, we have
\beqa
{\cal H}_{\rm sys} &=&
\frac{1}{2M_{\rm T}} \, {\bf P}^2 + \left[ \frac{1}{2\mu_0} {\bf
p}_0^2 + \frac{1}{2\mu_{\rm e}} \sum_{I=1}^N  {\bf p}_I^2
+ V (0, 1, \ldots, N ) \right], \rule{10mm}{0mm}
\label{2.1}\eeqa
where ${\bf P} = -{\rm i} \hbar \nabla_{\rm R}$ and ${\bf p}_I
= -{\rm i} \hbar \nabla_{{\rm r}_I}$ are the canonical momenta
corresponding to the variables ${\bf R}$ and ${\bf r}_I$, respectively,
\beq
M_{\rm T} = M_{\rm nuc} + Z \me + M
\label{2.2}\eeq
is the total mass, and
\beq
\mu_0 = \frac{M M_{\rm nuc}}{M+M_{\rm nuc}}
\qquad \mbox{and} \qquad
\mu_{\rm e} = \frac{\me M_{\rm nuc}}{\me+M_{\rm nuc}}
\label{2.3}\eeq
are the reduced masses of the projectile and the electron with the
nucleus. In a function or operator the argument 0 denotes the variables
(space coordinates and spin projection) of the projectile, and the
arguments $I=1, \ldots, N$ indicate variables of the atomic electrons.

The first term in Eq.\ \req{2.1} is the kinetic energy associated to
the motion of the center of mass, and the term $ V (0, 1, \ldots, N)$
represents the total interaction energy of all the particles in the
system. The expression in square brackets is the Hamiltonian of the
projectile and the target atom in the reference frame of the nucleus,
\ie, a frame where the nucleus is anchored to the origin of coordinates.
The theory of inelastic collisions is usually formulated in that frame,
with the additional simplification of replacing the reduced masses with
the rest masses of the projectile and the electrons. Therefore
theoretical results correspond to collisions with an atom or ion whose
nucleus has infinite mass. If the results of calculations are given
in atomic units ($\me = e = \hbar =1$), \ie, in units of the Bohr
radius
\beq
a_0 =  \frac{h^2}{\me e^2} = 5.291 \, 772 \time 10^{-9} \; {\rm cm}
\label{2.4}\eeq
and the Hartree energy
\beq
E_{\rm h} = \frac{\me e^4}{\hbar^2} = 27.211\, 386 \; {\rm eV},
\label{2.5}\eeq
the effect of the finite nuclear mass can be accounted for {\it a
posteriori} by simply redefining the atomic units of length and energy
(that is, by replacing $\me$ with $\mu_{\rm e}$).

\begin{figure}[htb]
\begin{center}
\includegraphics*[width=13.0cm]{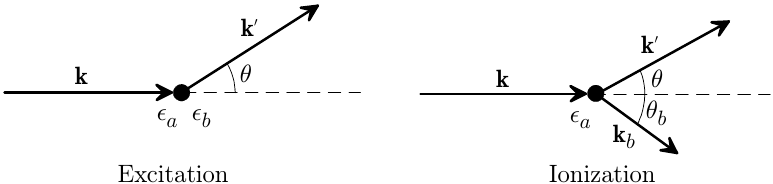}
\caption{Kinematics of inelastic collisions. The quantities $\epsilon_a$
and $\epsilon_b$ are the energies of the initial and final states of the
active target electron, respectively; $\theta$ is the polar scattering
angle of the projectile and, in the case of ionizing collisions,
$\theta_b$ is the polar angle of the direction of emission of the
knock-on electron.
\label{fig1}}
\end{center}\end{figure}

Figure \ref{fig1} displays the kinematics of the collision. Before the
interaction, the projectile moves with linear momentum ${\bf p}=\hbar
{\bf k}$, velocity ${\bf v}$, and kinetic energy $E$, the corresponding
values after the collision are ${\bf p}'=\hbar {\bf k}'$, ${\bf v}'$,
and $E'$, respectively. We assume that the speed $v=|{\bf v}|$ of the
projectile is large enough to justify the use of the first-order Born
approximation, either with plane waves or with distorted waves (see
Section \ref{sec4.3}).

The Hamiltonian of the system in the
reference frame of the nucleus can be written as
\beq
{\cal H}(0,1,\ldots, N)  = {\cal H}_{\rm T}(1, \ldots, N) + {\cal
H}_{\rm P}(0) + {\cal H}'(0,1,\ldots,N),
\label{2.6}\eeq
where ${\cal H}_{\rm T}$ and ${\cal H}_{\rm P}$ are the Hamiltonians of
the target atom and of the free projectile respectively, and the term
${\cal H}'$ describes the interaction of the projectile with
the nucleus and the $N$ electrons of the atom or ion. Since interactions
between atomic electrons and of the electrons with the nucleus are
dominated by the instantaneous Coulomb field, the Hamiltonian of the
target atom can be approximated as
\beq
{\cal H}_{\rm T} = \sum_{I=1}^N \left[
-e \varphi_{\rm nuc}(r_I) + {\cal K}(I) \right] + \sum_{I<J}
\frac{e^2}{|{\bf r}_I - {\bf r}_J|} \, ,
\label{2.7}\eeq
where ${\bf r}_I$ are the position vectors of the atomic electrons, and
$\varphi_{\rm nuc}(r)$ is the electrostatic potential of the nucleus. It
is assumed that the charge distribution of the nucleus has spherical
symmetry and is centered at the origin of coordinates. In practice, the
finite size of the nucleus only has a small effect on the atomic wave
functions and no large errors would be incurred by considering a point
nucleus, for which
\beq
\varphi_{\rm nuc}(r) = Ze/r.
\label{2.8}\eeq
${\cal K}$ is the kinetic
energy operator of the one-particle Dirac theory, \ie,
\beq
{\cal K}(I) = c \widetilde{\alphab}_I \dotprod {\bf p}_I
+ (\widetilde{\beta}_I-1) \me c^2,
\label{2.9}\eeq
where $\me$ is the electron mass and $\widetilde{\alphab}_I$ and
$\widetilde{\beta}_I$ are the Dirac matrices for the $I$-th electron.
For the sake of concreteness, we assume that the projectile is a
spin-$\1o2$ particle, and that it is described by the Dirac equation, so
that
\beq
{\cal H}_{\rm P}(0) = {\cal K}(0) = c \widetilde{\alphab}_0
\dotprod {\bf p}_0 + (\widetilde{\beta}_0-1) M c^2,
\label{2.10}\eeq
where $\widetilde{\alphab}_0$ and $\widetilde{\beta}_0$ are the Dirac
matrices for the projectile. Note that this assumption is appropriate
only for electrons and positrons. Theoretical
results for other charged particles (protons, alphas and heavier ions)
will only be accurate to the extent that the transition
probabilities do not depend too strongly on the details of the
projectile wave functions. The interaction Hamiltonian is given by
\beq
{\cal H}'(0,1,\ldots, Z)
= Z_0 e \varphi_{\rm nuc}(r_0) +\sum_{I=1}^Z {\cal H}_{\rm int}(0,I).
\label{2.11}\eeq
The first term is the Coulomb interaction of the projectile with the
atomic nucleus and the operator ${\cal H}_{\rm int}(0,I)$ represents the
effective interaction of the projectile with the $I$-th electron (see
Section \ref{sec3.2}).

In the simplest formulation of the theory, the interaction ${\cal H}'$
is treated as a first-order perturbation that induces transitions
between the eigenstates $\phi(0)\, \Psi(1, \ldots, Z)$ of the unperturbed
Hamiltonian ${\cal H}_{\rm T} + {\cal H}_{\rm P}$. Since the states
$\phi(0)$ of the free projectile are represented as plane waves, this
kind of approach is usually referred to as the (first-order) plane-wave
Born approximation (PWBA). A more effective theoretical scheme, still
based on first-order perturbation theory, is provided by the
(first-order) distorted-wave Born approximation (DWBA), which is
equivalent to the Furry (\citeyear{Furry1951}) representation of quantum
electrodynamics. To formulate the DWBA, we rewrite the Hamiltonian
\req{2.6} as
\beq
{\cal H}(0,1,\ldots, Z)  = {\cal H}_{\rm T}(1, \ldots, Z) + \left[ {\cal
H}_{\rm P}(0) + V_{\rm P}(r_0) \right] + {\cal H}''
\label{2.12}\eeq
with ${\cal H}'' = {\cal H}'-V_{\rm P}(r_0)$. That is, we have added and
subtracted an arbitrary central potential $V_{\rm P}(r_0)$ that depends
only on the coordinates of the projectile. In the DWBA, the states
$\psi(0)$ of the projectile before and after the collision are
represented by Dirac distorted plane waves, Eq.\ \req{B.22}, which
are exact solutions of the Dirac equation for the central potential
$V_{\rm P}(r_0)$,
\beq
\left[ c \widetilde{\alphab}_0 \dotprod {\bf p}_0
+ (\widetilde{\beta}_0-1) M c^2
+ V_{\rm P}(r_0) \right]  \psi(0) = E \psi(0).
\label{2.13}\eeq
We can thus consider
\beq
{\cal H}'' = {\cal H}' - V_{\rm P}(r_0)
= Z_0 e \varphi_{\rm nuc}(r_0) +\sum_{I=1}^Z {\cal H}_{\rm int}(0,I)
- V_{\rm P}(r_0)
\label{2.14}\eeq
as a perturbation that causes transitions between eigenstates
$\psi(0)\, \Psi(1, \ldots, Z)$ of the ``unperturbed'' Hamiltonian ${\cal
H}_{\rm T} + [{\cal H}_{\rm P} + V_{\rm P}]$. The effectiveness of the
DWBA lies in the fact that ${\cal H}''$ can be made much ``weaker'' than the
original interaction ${\cal H}'$. Unfortunately, since the interaction
${\cal H}'$ does depend on the coordinates of the atomic electrons,
${\cal H}''$ cannot be reduced to zero. Nonetheless, it is assumed that,
with a proper choice of the distorting potential $V_{\rm P}$, ${\cal
H}''$ can be made small enough to be treated as a perturbation to first
order.


\subsection{The one-active-electron approximation \label{sec2.1}}

For the sake of concreteness, and also to facilitate numerical
computations, the states of the target atom or ion will be described by
using a central-field independent-electron approximation (iea), \ie, atomic
electrons will be considered to move independently in a common central
potential $V_{\rm T}(r)$ (\eg, the DHFS potential).
This approximation amounts to replacing the atomic Hamiltonian \req{2.7}
with
\beq
{\cal H}_{\rm T}^{\rm iea} =
\sum_{I=1}^N \left[ {\cal K}(I) + V_{\rm T}(r_I) \right]\, .
\label{2.15}\eeq
We assume that the potential $V_{\rm T}(r)$ describes both the initial
and final atomic states $\Psi_n$. These can  be represented as single
Slater determinants, build with $N$ one-electron orbitals,
$\psi_{n\kappa m}({\bf r})$, which are solutions of the Dirac equation,
\beq
\left[ c \widetilde{\alphab} \dotprod {\bf p} + (\widetilde{\beta}-1)
\me c^2 + V_{\rm T}(r) \right] \psi_{n\kappa m}({\bf r}) =
\epsilon_{n\kappa} \psi_{n\kappa m}({\bf r}),
\label{2.16}\eeq
where $\epsilon_{n\kappa}$ is the energy of the electron, exclusive of
the rest energy. The Slater determinants constitute a complete basis of
antisymmetric eigenfunctions of ${\cal H}_{\rm T}^{\rm iea}$.
Under this independent-electron approximation, and by virtue of the
Slater--Condon rules \citep{Slater1929, Condon1930}, when the
interaction with the projectile is treated as a first-order
perturbation, the only allowed transitions of the target atom are
single-electron excitations. That is, the interaction causes the
excitation of the target atom from the initial state $\Psi_a$ (usually
the ground state) to a final state $\Psi_b$, which differs from $\Psi_a$
by a single orbital. This is equivalent to the so-called
one-active-electron approximation, which consists of considering only
the excitations of a single electron from a bound orbital $\psi_a$ to an
unoccupied (bound or free) orbital $\psi_b$, whereas the other atomic
electrons behave as mere spectators and their orbitals remain frozen in
the course of the interaction. Notice that the orbitals $\psi_a$ and
$\psi_b$ of the active electron, as well as those of the spectator
electrons, are solutions of the Dirac equation for the same potential
and, therefore, they are mutually orthogonal. This fact allows the
application of the Slater--Condon rules which lead to substantial
simplifications in both the theory and the numerical calculations; it
also permits a consistent description of exchange effects in the case of
electron collisions.

Formally, the theory can be freed from the one-active-electron
approximation by replacing the active electron orbitals with generic
atomic wave functions and redefining the interaction Hamiltonian by
adding the contributions of all the electrons in the atom. It should be
noted, however, that the calculation of accurate atomic wave functions
(such that, \eg, they reproduce the observed excitation energy spectrum
in the optical range) is extremely difficult. These wave functions
should account for correlations between the spatial and spin variables
of the bound electrons, which cannot be fully described by a single
Slater determinant. In practice, the one-active-electron approximation
yields a reasonably accurate description of the excitations and
ionizations of inner subshells, because the effective potential felt by the
electrons in these subshells is dominated by the screened electrostatic
interaction and correlation effects represent only a slight
perturbation. However, the approximation fails to describe excitation
and ionization of weakly bound electrons, which are more
sensitive to correlation effects.

Following \citet{Segui2003} and \citet{BoteSalvat2008}, the potential
$V_{\rm T}(r)$ of atomic electrons is set equal to the self-consistent,
spherically averaged DHFS potential, $V_{\rm DHFS}(r)$. This potential
is completely determined by the density $\rho(r)$ of the atomic
electrons,
\beq
\rho(r) = \sum_a \psi_a^\dagger({\bf r}) \,
\psi_a({\bf r}),
\label{2.17}\eeq
where the sum is over the occupied orbitals of the ground-state
configuration of the target atom or ion. In the case of configurations
with partially filled subshells, an average over the orbitals of the open
subshells is implied. The DHFS potential is given by
\beq
V_{\rm DHFS} (r) = -e \varphi_{\rm nuc}(r) - e \varphi_{\rm el}(r)
+ V_{\rm ex}^{\rm Slater}(r),
\label{2.18}\eeq
where $\varphi_{\rm nuc}(r)$ is the electrostatic potential of the
nucleus,
\beq
\varphi_{\rm el}(r) = \frac{e}{r} \int_0^r \rho(r') \, 4\pi r'^2
\, \d r' + e \int_r^\infty \rho(r') \, 4\pi r' \, \d r'
\label{2.19}\eeq
is the electrostatic potential of the atomic electron cloud, and
\beq
  V_{\rm ex}^{\rm Slater}(r) = -e^2\,(3/\pi)^{1/3}\,[\rho(r)]^{1/3}
\label{2.20}\eeq
is the \citet{Slater1951} local approximation to the exchange
interaction.  To reproduce the correct large-$r$ behavior of the
potential, $-(Z-N+1)e^2/r$, we adopt Latter's (\citeyear{Latter1955}) tail
correction and define
\begin{subequations}
\label{2.21}
\beq
  V_{\rm DHFS} (r) \equiv \left\{
    \begin{array}{ll}
      -e \varphi_{\rm nuc}(r) - e \varphi_{\rm el}(r)
      + V_{\rm ex}^{\rm Slater}(r)
      \rule{10mm}{0mm}
      & \mbox{if $r < r_{\rm Latter}$,}
      \\ [2mm]
      - (Z-N+1) e^2/r
      & \mbox{if $r>r_{\rm Latter}$,}
    \end{array} \right.
\label{2.21a}\eeq
where the cutoff radius $r_{\rm Latter}$ is the
outer root of the equation
\beq
  - e \varphi_{\rm nuc}(r) - e \varphi_{\rm el}(r)
  + V_{\rm ex}^{\rm Slater}(r) = - (Z-N+1) e^2/r.
\label{2.21b}\eeq
\end{subequations}
A practical reason for choosing the DHFS potential is that, for inner
subshells with ionization energies higher than about 200 eV, the negative
eigenvalues $-\epsilon_{n\kappa}$ of the one-electron Dirac equation
with the DHFS potential are very close to the experimental subshell
ionization energies \citep[see the complementary file
of][]{SalvatFernandezVarea2019}. In the following, we will assume that
the ionization energy $E_{n\kappa}$ of a subshell $n\kappa$ coincides with
the negative of the DHFS eigenvalue, \ie, $E_{n\kappa} =
-\epsilon_{n\kappa}$.

For inelastic collisions of electrons, \citet{Segui2003} and
\citet{BoteSalvat2008} took $V_{\rm P}(r) = V_{\rm T}(r) = V_{\rm DHFS}(r)$ so
that the orbitals of the projectile and the active target electron are
mutually orthogonal (because they are solutions of the Dirac equation
with the same potential). This choice of potential amounts to assuming
that the projectile and target electrons interact with the inactive
(spectator) atomic electrons in the same way, which is a plausible
assumption. Notice, however, that we are disregarding the dependence of
the exchange interaction on the electron speed \citep[see, \eg][and
references therein]{RileyTruhlar1975}. For collisions of
positrons and other charged particles, it seems natural to take $V_{\rm
P}(r)$ equal to the electrostatic interaction energy with the atomic
charge distribution,
\beq
V_{\rm static} (r) =
Z_0 e \varphi_{\rm nuc}(r) + Z_0 e \varphi_{\rm el}(r).
\label{2.22}\eeq
Unfortunately, this potential reaches its asymptotic Coulomb form at
distances that are much larger than $r_{\rm Latter}$ and this makes the
numerical computation of the ionization cross section much more
difficult and lengthier than for electrons. To circumvent this
difficulty, \citet{Segui2003} took $V_{\rm P}(r) = - Z_0 V_{\rm
DHFS}(r)$, \ie, the distorting potential was obtained by rescaling
the DHFS potential according to the charge of the projectile. This
potential includes exchange contributions, which may seem inappropriate
for particles that are different from the electron. Nonetheless, at
large radial distances the potential $ - Z_0 V_{\rm DHFS}(r)$ does
represent the interaction of the projectile with the nucleus and the
spectator atomic electrons. Notice that a part of the local exchange
potential serves to eliminate the self-interaction of the atomic
electrons (\ie, the electrostatic interaction energy of each electron
with its own charge distribution); a similar term must be subtracted
from the electrostatic potential \req{2.22} to give the effective
interaction of the projectile with the inactive atomic charges.


\subsection{Differential cross sections \label{sec2.2}}

The DCS for excitation of the active electron from the orbital $\psi_a$
to an unoccupied bound orbital $\psi_b$ is given by \citep[see,
\eg,][]{Bethe1932, Fano1963, Joachain1975}
\beq
\d \sigma^{\rm exc} = \frac{(2\pi)^4}{\hbar v} \,
|T_{fi}|^2\, \delta(E - E'-\epsilon_b+\epsilon_a)
\, \d {\bf k}',
\label{2.23}\eeq
where $T_{fi}$ is the transition matrix element (see below). Using the
relation
$$
(c\hbar k')^2 = E'(E'+2M c^2),
$$
we have
\beq
\d {\bf k}' = {k'}^2 \, \d k' \, \d \hat{\bf k}'
= {k'}^2 \frac{\d k'}{\d E'} \, \d E'\,\d\hat{\bf k}'
= k'\, \frac{E'+Mc^2}{c^2\hbar^2} \, \d E'\,\d\hat{\bf k}',
\label{2.24}\eeq
and, therefore,
\beq
\d \sigma^{\rm exc} = \frac{(2\pi)^4}{\hbar v} \,
k'\, \frac{E'+Mc^2}{c^2\hbar^2} \,
|T_{fi}|^2\, \delta(W-\epsilon_b+\epsilon_a)
\, \d E' \,\d\hat{\bf k}'.
\label{2.25}\eeq
Although the delta function, which forces energy conservation, can be
removed by integration over $E'$, it is more convenient to keep it
and consider the excitation DCS as a function of the angular deflection
$\hat{\bf k}'$ and the energy loss $W\equiv E - E' = \epsilon_b -
\epsilon_a$ of the projectile,
\beq
\frac{\d^2 \sigma^{\rm exc}}{\d W \, \d\hat{\bf k}'}
 = \frac{(2\pi)^4}{\hbar v}
\, k'\, \frac{E-W+Mc^2}{c^2\hbar^2} \, \delta(W - \epsilon_b +
\epsilon_a) \, |T_{fi}|^2,
\label{2.26}\eeq
where the $T$-matrix element is ``on the energy shell'', \ie, the
initial and final states (of the projectile {\it and} the active
electron) have the same total energy, $E + \epsilon_a = E' +\epsilon_b$.

The DCS for ionization (\ie, for transitions where $\psi_b$ is a
free orbital) is given by \citep[see, \eg,][]{Bethe1932, Joachain1975}
\beq
\d \sigma^{\rm ion} = \frac{(2\pi)^4}{\hbar v} \,
|T_{fi}|^2\, \delta(E - E' -\epsilon_b+\epsilon_a)
\, \d {\bf k}'\, \d {\bf k}_b \, ,
\label{2.27}\eeq
where $\hbar {\bf k}_b$ is the linear momentum of the ejected electron.
Using the relation \req{2.24} and the analogous one for ${\bf k}_b$,
\beq
\d {\bf k}_b = k_b \, \frac{\epsilon_b +\me c^2}{c^2\hbar^2} \, \d
\epsilon_b \, \d \hat{\bf k}_b,
\label{2.28}\eeq
and performing the integration over $E'$, we get
\beq
\d \sigma^{\rm ion} = \frac{(2\pi)^4}{\hbar v} \,
\, k' k_b \, \frac{E-W+Mc^2}{c^2\hbar^2} \,
\frac{\epsilon_b +\me c^2}{c^2\hbar^2} \,
|T_{fi}|^2\, \d\hat{\bf k}' \, \d \epsilon_b \, \d\hat{\bf k}_b\, .
\label{2.29}\eeq
That is,
\beq
\frac{\d^2 \sigma^{\rm ion}}{\d W \, \d \hat{\bf k}' \, \d \hat{\bf
k}_b} = \frac{(2\pi)^4}{\hbar v} \,
\, k' k_b \, \frac{E-W+Mc^2}{c^2\hbar^2} \,
\frac{\epsilon_b +\me c^2}{c^2\hbar^2} \,
|T_{fi}|^2,
\label{2.30}\eeq
where use has been made of the fact that $\epsilon_b = \epsilon_a + W$.
Hereafter, $T$-matrix elements are on the energy shell.
In many cases (\eg, in calculations of the stopping power of gases)
only the effect of the interactions on the projectile is of interest.
Then, the relevant DCS is obtained by integrating Eq.\ \req{2.30}
over the direction $\hat{\bf k}_b$ of the emitted electron,
\beq
\frac{\d^2 \sigma^{\rm ion}}{\d W \, \d \hat{\bf k}'}  =
\frac{(2\pi)^4}{\hbar v} \,
\, k' k_b \, \frac{E-W+Mc^2}{c^2\hbar^2} \,
\frac{\epsilon_b +\me c^2}{c^2\hbar^2} \, \int
|T_{fi}|^2 \, \d \hat{\bf k}_b \, .
\label{2.31}\eeq

The transition matrix element is given by
\beq
T_{fi} =
\left< \psi^{(-)}_{{\bf k}', m'_{\rm S}} (0) \, \psi_b(1)
\left| \rule{0mm}{4mm}{\cal H}_{\rm int}(0, 1)
\right| \psi^{(+)}_{{\bf k}, m_{\rm
S}} (0) \, \psi_a(1) \right>\, ,
\label{2.32}\eeq
where the indexes 0 and 1 denote the projectile and the active
target electron, respectively. The Hamiltonian ${\cal H}_{\rm int}(0,
1)$ describes the interaction between these two particles.
$\psi^{(+)}_{{\bf k}, m_{\rm S}}$ and $\psi^{(-)}_{{\bf k}' m'_{\rm S}}$
are distorted plane waves corresponding to the initial and final state
of the projectile, respectively (see Section \ref{secB.2} of Appendix
\ref{secB}). Notice
that distorted plane waves of initial (final) states have outgoing
(incoming) spherical distortions \citep[see, \eg,][]{BreitBethe1954}. The
above expressions for the DCSs apply when bound orbitals are normalized
to unity and distorted plane waves are normalized in wave-number space,
\ie,
\beq
\int \left[ \psi^{(\pm)}_{{\bf k}', m'_{\rm S}}({\bf r})
\right]^\dagger
\psi^{(\pm)}_{{\bf k}, m_{\rm S}}({\bf r}) \, \d {\bf r}
= \delta({\bf k} - {\bf k}')
\, \delta_{m_{\rm S},m'_{\rm S}}.
\label{2.33}\eeq
With this normalization, the density of states per unit volume in ${\bf
k}$-space is unity, and Eqs.\ \req{2.23} and \req{2.27} follow directly
from Fermi's golden rule \citep[see, \eg,][]{Baym1974}.

From now on, we shall limit to consider the DDCSs given by Eqs.\
\req{2.26} and \req{2.31}. In the derivation of these formulas, we
have assumed transitions from a given initial state $i=$$\{\psi_{{\bf
k}, m_{\rm S}}^{(+)}(0)$, $\psi_{n_a \kappa_a m_a}(1)\}$ to a well defined
final state $f=$$\{\psi_{{\bf k}', m'_{\rm S}}^{(-)}(0)$, $\psi_{b} (1)\}$. In
most practical cases, the target atoms are randomly oriented, the
incident beam is unpolarized and final magnetic and spin states are not
distinguished. Under these circumstances, the observed DCS is obtained
by averaging over initial degenerate magnetic and spin states and
summing over final degenerate states. Thus, the observed DDCS for
excitation to bound states is given by
\beq
\frac{\d^2 \sigma^{\rm exc}}{\d W \, \d\hat{\bf k}'}
 = \frac{(2\pi)^4}{\hbar v}
\, k'\, \frac{E-W+Mc^2}{c^2\hbar^2} \,
{\cal I}_{fi} \, ,
\label{2.34}\eeq
where
\beqa
{\cal I}_{fi}
&\equiv& \delta(W - \epsilon_b + \epsilon_a) \,
\frac{1}{2(2j_a+1)}
\sum_{m_a, m_{\rm S}} \sum_{m_b, m'_{\rm S}} \, \left| T_{fi} \right|^2
\nonumber \\ [2mm]
&=& \frac{\delta(W - \epsilon_b + \epsilon_a)}{2(2j_a+1)} \sum_{m_a,
m_{\rm S}} \sum_{m_b, m'_{\rm S}}
\left| \left< \psi^{(-)}_{{\bf k}', m'_{\rm S}} \, \psi_{n_b \kappa_b
m_b} \left| \rule{0mm}{4mm}{\cal H}_{\rm int}
\right| \psi^{(+)}_{{\bf k}, m_{\rm
S}} \, \psi_{n_a \kappa_a m_a} \right> \right|^2. \rule{10mm}{0mm}
\label{2.35}\eeqa
The DCS for ionization can also be written in the form \req{2.34},
\beq
\frac{\d^2 \sigma^{\rm ion}}{\d W \,\d\hat{\bf k}'}
 = \frac{(2\pi)^4}{\hbar v}
\, k'\, \frac{E-W+Mc^2}{c^2\hbar^2} \,
{\cal J}_{fi},
\label{2.36}\eeq
with
\beqa
{\cal J}_{fi} &\equiv&
k_b \, \frac{\epsilon_b +\me c^2}{c^2\hbar^2} \,
\frac{1}{2(2j_a+1)} \sum_{m_a,
m_{\rm S}} \sum_{m_{{\rm S}b},m'_{\rm S}} \, \int \d \hat{\bf k}_b
\left| T_{fi} \right|^2
\label{2.37}\\ [2mm]
&=&
k_b \, \frac{\epsilon_b +\me c^2}{c^2\hbar^2} \,
\frac{1}{2(2j_a+1)} \sum_{m_a,
m_{\rm S}} \sum_{m_{{\rm S}b},m'_{\rm S}} \,
\nonumber \\ [2mm]
&& \mbox{} \times
\int \d \hat{\bf k}_b \left|
\left< \psi^{(-)}_{{\bf k}', m'_{\rm S}} \, \psi^{(-)}_{{\bf k}_b,
m_{{\rm S}b}}
\left| \rule{0mm}{4mm}{\cal H}_{\rm int}
\right| \psi^{(+)}_{{\bf k}, m_{\rm
S}} \, \psi_{n_a \kappa_a m_a} \right> \right|^2.
\label{2.38}\eeqa
Note that the final orbital of the active target electron is represented
here as a distorted plane wave, \ie, an exact solution of the Dirac
equation for the atomic potential.

Introducing the recoil energy $Q$, defined by [cf.\ Eq.\ \req{A.18}],
\beq
Q(Q+2\me c^2) = c^2 \hbar^2 ({\bf k} - {\bf k}')^2 =
c^2 \hbar^2 \left( k^2 + k'^2 - 2 k k' \cos\theta \right).
\label{2.39}\eeq
and noting that
\beq
\d \hat{\bf k}' = 2\pi \, \d (\cos\theta) = \frac{2 \pi (Q + \me c^2)}{
c^2 \hbar^2 k k'} \, \d Q,
\label{2.40}\eeq
the DDCSs are written as
\beq
\frac{\d^2 \sigma^{\rm exc}}{ \d W \, \d Q } =
\frac{(2\pi)^5}{c^2 \hbar^4 v^2} \, \frac{E-W+M c^2}{E+Mc^2} \,
(Q+\me c^2) \, {\cal I}_{fi}
\label{2.41}\eeq
and
\beq
\frac{\d^2 \sigma^{\rm ion}}{\d W \, \d Q } =
\frac{(2 \pi)^5}{c^2 \hbar^4 v^2} \, \frac{E-W+M c^2}{E+Mc^2} \,
(Q+\me c^2) \, {\cal J}_{fi}.
\label{2.42}\eeq

To evaluate the DDCS for ionization, we can expand the distorted
plane wave $\psi^{(-)}_{{\bf k}_b, m_{{\rm S}b}}$ as in Eq.\ \req{B.22},
\beq
\psi_{{\bf k}_b m_{{\rm S}b}}^{(-)} =
\frac{1}{k_b} \,
\sqrt{\frac{\epsilon_b+2\me c^2}{\pi(\epsilon_b + \me c^2)}}
\sum_{\kappa_b,m_b} {\rm i}^{\ell_{b}} \,
\exp \left( - {\rm i} \delta_{\kappa_b} \right) \,
\left\{ \Omega_{\kappa_b m_b}^\dagger (\hat{\bf k}_b)
\chi_{m_{{\rm S}b}} \right\} \psi_{\epsilon_b\kappa_b m_b}
\label{2.43}\eeq
and write
\beqa
{\cal J}_{fi} &=&
k_b \, \frac{\epsilon_b +\me c^2}{c^2\hbar^2} \,
\frac{1}{2(2j_a+1)} \sum_{m_a,
m_{\rm S}} \sum_{m_{{\rm S}b},m'_{\rm S}} \, \int \d \hat{\bf k}_b
\frac{\epsilon_b+2\me c^2}{k_b^2 \pi(\epsilon_b + \me c^2)}
\nonumber \\ [2mm]
&& \mbox{} \times
\sum_{\kappa_b,m_b}
\sum_{\kappa'_b,m'_b} {\rm i}^{\ell'_b - \ell_b} \, \exp \left[ {\rm i}
\left( \delta_{\kappa_b} - \delta_{\kappa'_b}
\right) \right] \,
\left\{ \Omega_{\kappa'_b m'_b}^\dagger (\hat{\bf k}_b)
\chi_{m_{{\rm S}b}} \right\}
\left\{
\chi_{m_{{\rm S}b}}^\dagger
\Omega_{\kappa_b m_b} (\hat{\bf k}_b)
\right\}
\nonumber \\ [2mm]
&& \times
\left< \psi^{(-)}_{{\bf k}', m'_{\rm S}} \, \psi_{\epsilon_b\kappa_b m_b}
\left| \rule{0mm}{4mm}{\cal H}_{\rm int}
\right| \psi^{(+)}_{{\bf k}, m_{\rm
S}} \, \psi_{n_a \kappa_a m_a} \right> \,
\left<
\psi^{(+)}_{{\bf k}, m_{\rm S}} \, \psi_{n_a \kappa_a m_a}
\left| \rule{0mm}{4mm}{\cal H}_{\rm int} \right|
\psi^{(-)}_{{\bf k}', m'_{\rm S}} \, \psi_{\epsilon_b\kappa'_b m'_b}
\right> \, .
\nonumber \eeqa
Using the completeness property of the spherical spinors \req{B.12},
\beq
\sum_{m_{{\rm S}b}} \int \d \hat{\bf k}_b \,
\Omega_{\kappa_b m_b}^\dagger (\hat{\bf k}_b)
\chi_{m_{{\rm S}b}} \,
\chi_{m_{{\rm S}b}}^\dagger
\Omega_{\kappa'_b m'_b} (\hat{\bf k}_b) = \delta_{\kappa'_b, \kappa_b}
\, \delta_{m'_b,m_b},
\label{2.44}\eeq
we obtain
\beqa
{\cal J}_{fi} &=&
\frac{k_b}{\epsilon_b \pi} \,
\frac{1}{2(2j_a+1)} \sum_{m_a,
m_{\rm S}} \sum_{m'_{\rm S}}
\sum_{\kappa_b, m_b}
\left|
\left< \psi^{(-)}_{{\bf k}', m'_{\rm S}} \, \psi_{\epsilon_b
\kappa_b m_b}
\left| \rule{0mm}{4mm}{\cal H}_{\rm int}
\right| \psi^{(+)}_{{\bf k}, m_{\rm
S}} \, \psi_{n_a \kappa_a m_a} \right> \right|^2.
\label{2.45}\eeqa
We see that, apart from a global factor and different summation indices,
${\cal J}_{fi}$ has the same form as ${\cal I}_{fi}$, Eq.\ \req{2.35}.
Thus, when the direction of the ejected electron is not observed, its
final state can be described either as a distorted plane wave or as a
spherical wave. Of course, the numerical calculation of ${\cal J}_{fi}$
is simpler with spherical waves. Moreover, spherical waves have defined
parity under space inversion, a property that will be of great help in
the following.


\subsubsection{Collisions of electrons with atoms \label{sec2.2.1}}

The expressions for the DDCSs obtained above are appropriate for
describing collisions of spin-$\1o2$ projectiles that are different from
the electron. When the projectile is an electron, it is
indistinguishable from the active target electron and, therefore, they can
undergo re-arrangement collisions (\ie, the projectile and the target
electrons can ``exchange places''). The effect of exchange is described
by antisymmetrizing the initial and final states in the transition
matrix elements. That is, the transition matrix elements \req{2.32}
are to be replaced with
\beq
T_{fi}^{\rm el} =
\left< \sqrt{2} \, {\cal A} \left[ \psi^{(-)}_{{\bf k}', m'_{\rm S}} (0) \,
\psi_b(1) \right]
\left| \rule{0mm}{4mm}{\cal H}_{\rm int}(0, 1)
\right| \sqrt{2} \, {\cal A} \left[ \psi^{(+)}_{{\bf k}, m_{\rm
S}} (0) \, \psi_a(1) \right] \right> \, ,
\label{2.46}\eeq
where the operator ${\cal A}$ is the 2-particle antisymmetrizer,
\beq
{\cal A} \psi_a(0) \psi_b(1) \equiv \frac{1}{2!} \left[
 \psi_a(0) \psi_b(1) - \psi_a(1) \psi_b(0)\right].
\label{2.47}\eeq
Note that this operator is Hermitian and ${\cal A}^2 = {\cal A}$. As the
interaction is symmetrical, and the four orbitals are mutually
orthogonal, we have
\beqa
T_{fi}^{\rm el} &=& 2
\left< {\cal A} \left[ \psi^{(-)}_{{\bf k}', m'_{\rm S}} (0) \,
\psi_b(1) \right]
\left| \rule{0mm}{4mm}{\cal H}_{\rm int}(0, 1)  \right|
\psi^{(+)}_{{\bf k}, m_{\rm S}} (0) \, \psi_a(1) \right>
\nonumber \\ [2mm]
&=&
\left< \psi^{(-)}_{{\bf k}', m'_{\rm S}} (0) \, \psi_b(1) \left|
\rule{0mm}{4mm}{\cal H}_{\rm int}(0, 1) \right| \psi^{(+)}_{{\bf k},
m_{\rm S}} (0) \, \psi_a(1) \right>
\nonumber \\ [2mm]
&& \mbox{} -
\left< \psi^{(-)}_{{\bf k}', m'_{\rm S}} (1) \, \psi_b(0) \left|
\rule{0mm}{4mm}{\cal H}_{\rm int}(0, 1) \right| \psi^{(+)}_{{\bf k},
m_{\rm S}} (0) \, \psi_a(1) \right>.
\label{2.48}\eeqa
The first and second terms in this expression describe direct and
re-arrangement collisions, respectively.

It is worth noting that in the derivation of the electron matrix
elements \req{2.48} we have ignored the presence of the inactive
electrons of the target atom. The result is correct whenever the
orbitals of these electrons and the initial and final orbitals of the
target {\it and} the projectile are mutually orthogonal. In this case,
matrix elements of the operator ${\cal H}''$ between initial and final
states (Slater determinants) of the $(N+1)$-electron system (\ie, the
projectile and the atomic electrons) reduce to the expression
\req{2.48} \citep{Condon1930}. Indeed, this is the reason why
\citet{Segui2003} assumed that the projectile feels the same potential as the
active electron, $V_{\rm P} \equiv V_{\rm T}$.


\section{Effective interaction with electrons \label{sec3}}
\setcounter{equation}{0}

To complete the formulation of the theory, we will determine the
Hamiltonian ${\cal H}_{\rm int}(0,1)$ that describes the interaction of the
projectile and the active atomic electron. Because the charged projectile is
moving fast with respect to the target atom, the latter is subject to a
rapidly varying electromagnetic field. In the Coulomb gauge, this field
is described by the instantaneous Coulomb potential and a transverse
vector potential that vanishes when the speed of the projectile is
small.  As a first approximation, we can adopt a semi-classical picture
in which the atomic electrons interact with the classical
electromagnetic field of the projectile. Moreover, we may assume that
the interaction is ``weak'' enough to be considered as a first order
perturbation. This scheme gives transition matrix elements that are
proportional to the field strength, \ie, we get a ``linear response''
approximation analogous to the one adopted in the classical dielectric
theory of stopping \citep{Lindhard1954, Jackson1975}.


\subsection{Semi-classical interaction \label{sec3.1}}

For a moment, let us assume that the projectile follows a straight
trajectory ${\bf r}_0 = {\bf v}t$, such that it passes by the
origin of coordinates at the time $t=0$. The corresponding charge and
current distributions are,
\beq
\rho_0({\bf r},t) = Z_0 e \, \delta ({\bf r} - {\bf v}t)\, ,
\qquad
{\bf j}_0({\bf r},t) = Z_0 e \, {\bf v} \, \delta ({\bf r} -
{\bf v}t)\, ,
\label{3.1}\eeq
respectively. We wish to determine the scalar potential $\varphi({\bf r},t)$
and the vector potential ${\bf A}({\bf r},t)$ of the electromagnetic
field created by
these distributions. For this purpose, it is convenient
to consider the Fourier transforms of the charge distributions and the
potentials. To simplify the notation, we shall use the same
symbol for the original function and its transform, which are
distinguished by their arguments. Thus, for instance,
\beq
\varphi({\bf q},\omega) \equiv (2\pi)^{-2} \int \d {\bf r} \int \d t \,
\exp\left[ - {\rm i} \left( {\bf q} \dotprod {\bf r} - \omega t \right)
\right] \, \varphi({\bf r},t)\, .
\label{3.2}\eeq
The inverse transform is
\beq
\varphi({\bf r},t) \equiv (2\pi)^{-2} \int \d {\bf q} \int \d \omega \,
\exp\left[ {\rm i} \left( {\bf q} \dotprod {\bf r} - \omega t \right)
\right] \, \varphi({\bf q},\omega)\, .
\label{3.3}\eeq
The electromagnetic potentials in the Coulomb gauge can be readily
obtained by noting that their Fourier transforms satisfy the equations
\citep[see, \eg,][]{Jackson1975}
\beq
q^2 \varphi({\bf q},\omega) = 4 \pi \rho_0({\bf q},\omega)
\label{3.4}\eeq
and
\beq
\left[ q^2 -
\frac{\omega^2}{c^2} \right] {\bf A}({\bf q},\omega) =
\frac{4\pi}{c}\,
{\bf j}^{\rm T}_0({\bf k},\omega)\, .
\label{3.5}\eeq
Here ${\bf j}^{\rm T}_0({\bf q},\omega)$ denotes the ``transverse''
component (perpendicular to $\hat{\bf q}$) of the current-density vector.
The Fourier transform of the charge density \req{3.1} is
\beqa
\rho_0({\bf q},\omega) &=&
(2\pi)^{-2} \int \d {\bf r} \int \d t \,
\exp\left[ - {\rm i} \left( {\bf q} \dotprod {\bf r} - \omega t \right)
\right] \, Z_0 e \, \delta({\bf r} - {\bf v} t)
\nonumber \\ [2mm]
&=& \frac{Z_0 e }{(2\pi)^2} \int \d t \,
\exp\left[ - {\rm i} \left( {\bf q} \dotprod {\bf v} - \omega \right) t
\right] = \frac{Z_0 e }{2\pi} \, \delta( {\bf q} \dotprod {\bf v} -
\omega)\, .
\label{3.6}\eeqa
Similarly, the Fourier transform of the current density \req{3.1} is
\beq
{\bf j}_0({\bf q},\omega) = \frac{Z_0 e }{2\pi} \,
{\bf v} \, \delta( {\bf q} \dotprod
{\bf v} - \omega)\, .
\label{3.7}\eeq
Hence, the longitudinal and transverse components of ${\bf j}_0$ are
\beqa
&&
{\bf j}_0^{\rm L}({\bf q},\omega) = \frac{Z_0 e }{2\pi} \,
(\hat{\bf q} \dotprod {\bf v}) \hat{\bf q}\, \delta( {\bf q} \dotprod
{\bf v} - \omega)\, ,
\nonumber \\ [2mm]
&&
{\bf j}_0^{\rm T}({\bf q},\omega) = \frac{Z_0 e }{2\pi} \, [
{\bf v} - (\hat{\bf q} \dotprod {\bf v}) \hat{\bf q} ]\, \delta( {\bf q}
\dotprod {\bf v} - \omega)\, .
\label{3.8}\eeqa
Now, using Eqs.\ \req{3.4} and \req{3.5}, we have
\beq
\varphi ({\bf q}, \omega) = \frac{4\pi}{q^2} \, \rho_0 ({\bf
q},
\omega) = \frac{2Z_0 e}{q^2} \, \delta({\bf q} \dotprod {\bf v} -
\omega)\, ,
\label{3.9}\eeq
and
\beq
{\bf A} ({\bf q}, \omega) = \frac{4\pi}{c} \, \frac{ {\bf
j}^{\rm T}_0({\bf q}, \omega)}{q^2 - (\omega/c)^2}
= \frac{2Z_0 e}{q^2-(\omega/c)^2} \, \left[ \betab - (\hat{\bf q}
\dotprod \betab ) \hat{\bf q} \right] \,
\delta({\bf q} \dotprod {\bf v} - \omega)\, ,
\label{3.10}\eeq
with $\betab \equiv {\bf v}/c$.
Therefore
\beqa
\varphi({\bf r},t) &=& (2\pi)^{-2} \int \d {\bf q} \int \d \omega \,
\exp\left[ {\rm i} \left( {\bf q} \dotprod {\bf r} - \omega t \right)
\right] \, \frac{2Z_0 e}{q^2} \delta({\bf q} \dotprod {\bf v} - \omega)
\nonumber \\ [2mm]
&=&
\frac{Z_0 e}{2\pi^2}
\int \d {\bf q} \, \frac{1}{q^2} \,
\exp\left[ {\rm i} {\bf q} \dotprod \left({\bf r} -
{\bf r}_0 \right) \right]\, .
\label{3.11}\eeqa
In the same manner,
\beqa
{\bf A} ({\bf r},t)
&=& (2\pi)^{-2} \int \d {\bf q} \int \d \omega \,
\exp\left[ {\rm i} \left( {\bf q} \dotprod {\bf r} - \omega t \right)
\right] \,
\frac{2Z_0 e}{q^2-(\omega/c)^2} \, \left[ \betab - (\hat{\bf q}
\dotprod \betab ) \hat{\bf q} \right] \,
\delta({\bf q} \dotprod {\bf v} - \omega)
\nonumber \\ [2mm]
&=& \frac{Z_0 e}{2\pi^2} \int \d {\bf q} \, \frac{
\betab - (\hat{\bf q} \dotprod \betab ) \hat{\bf q}}{q^2-(
{\bf q} \dotprod \betab)^2}
\, \exp\left[ {\rm i} {\bf q} \dotprod \left({\bf r} - {\bf r}_0 \right)
\right]\, .
\label{3.12}\eeqa

To reduce the scalar potential to a more familiar form, we note that
\beqa
\int \frac{\exp({\rm i}{\bf k}\dotprod{\bf r})}{|{\bf r}-{\bf
r}_{0}|} \, \d{\bf r} &=&
-\frac{1}{k^2} \int \frac{1}{|{\bf r}-{\bf r}_{0}|}
\nabla^{2} \exp({\rm i}{\bf k}\dotprod{\bf r}) \, \d{\bf r}
\nonumber \\[2mm]
&=& - \frac{1}{k^2}
\int \exp({\rm i}{\bf k}\dotprod{\bf r}) \nabla^{2}
\left( \frac{1}{|{\bf r}-{\bf r}_{0}|} \right)
\d{\bf r}
\nonumber \\[2mm]
&=& -\frac{1}{k^2}
\int \exp({\rm i}{\bf k}\dotprod{\bf r}/\hbar)
\left[ -4\pi\delta({\bf r}-{\bf r}_{0}) \right] \d{\bf r} \nonumber
\\[2mm]
&=& \frac{4\pi}{k^2} \exp({\rm i}{\bf k}\dotprod{\bf r}_{0}).
\label{3.13}\eeqa
This result, which was first obtained by \citet{Bethe1930}, is known as the
Bethe integral. It allows us to write
\beqa
\varphi({\bf r},t) &=& \frac{Z_0 e}{2\pi^2} \int \d {\bf q} \, \exp (
{\rm i} {\bf q} \dotprod {\bf r}) \frac{1}{q^2} \, \exp(- {\rm i} {\bf
q} \dotprod {\bf r}_0)
\nonumber \\ [2mm]
&=& \frac{Z_0 e}{(2\pi)^3} \int \d {\bf r}' \, \frac{1}
{|{\bf r'}-{\bf r}_{0}|}
\int \exp\left[ {\rm i}{\bf q}\dotprod \left( {\bf r}-{\bf
r}' \right) \right] \ \d {\bf q}
\nonumber \\ [2mm]
&=& Z_0 e \int \frac{1}{|{\bf r}'-{\bf r}_{0}|}
\delta({\bf r}-{\bf r}') \, \d {\bf r}' =
\frac{Z_0e}{|{\bf r}-{\bf r}_{0}|}\, .
\label{3.14}\eeqa
We thus see that $\varphi({\bf r},t)$ is simply the unretarded
Coulomb field of the projectile, a peculiarity of the Coulomb gauge.

The semi-classical Hamiltonian for the interaction of the target
electron with the electromagnetic field of the projectile is \citep[see,
\eg,][]{Sakurai1967}
\beqa
{\cal H}_{\rm int,cl} (0, 1) &=&
- e \varphi({\bf r}_1,t)
+ e \widetilde{\alphab}_1 \dotprod {\bf A}({\bf r}_1,t)
\nonumber \\ [2mm]
&=& - \frac{Z_0 e^2}{2\pi^2} \int \d {\bf q} \left( \frac{1}{q^2} -
\frac{ \widetilde{\alphab}_1 \dotprod
\left[ \betab - (\hat{\bf q} \dotprod \betab ) \hat{\bf
q}\right] }{q^2-({\bf q} \dotprod \betab)^2} \right)
\, \exp\left[ {\rm i} {\bf q} \dotprod \left({\bf r}_1 - {\bf r}_0
\right) \right]\, , \rule{10mm}{0mm}
\label{3.15}\eeqa
where $c\widetilde{\alphab}_1$ is the Dirac velocity operator of the
active electron. The first term represents the longitudinal interaction
through the instantaneous Coulomb field, whereas the second term
corresponds to the transverse interaction that arises from the exchange
of virtual photons between the projectile and the target electron (see
below). The semi-classical Hamiltonian \req{3.15} contains most of the
physics of inelastic interactions.

\subsection{Quantized interaction \label{sec3.2}}

We shall now derive a more rigorous form of the interaction Hamiltonian
using elementary quantum electrodynamics in the Coulomb gauge, as done
by \citet{Fano1956, Fano1963}. In this gauge, the interaction is
split into its longitudinal and transverse parts and only the transverse
field needs to be quantized. For a thorough discussion of the
consistency of this procedure see, \eg, \citet{Sakurai1967}. Thus, we
have
\beq
{\cal H}_{\rm int} (0, 1) =
{\cal H}^{\rm L} (0, 1) +
{\cal H}^{\rm T} (0, 1)\, .
\label{3.16}\eeq
The longitudinal term is simply the unretarded Coulomb interaction [see
Eq.\ \req{3.30} below],
\beq
{\cal H}^{\rm L} (0, 1) =
- \frac{Z_0 e^2}{ |{\bf r}_1 - {\bf r}_0|} =
- \frac{Z_0 e^2}{2\pi^2}
\int \d {\bf q} \, \frac{1}{q^2} \,
\exp\left[ {\rm i} {\bf q} \dotprod \left({\bf r}_1 -
{\bf r}_0 \right) \right]\, .
\label{3.17}\eeq
To determine the effective Hamiltonian for the transverse interaction,
we consider the interaction Hamiltonian of the active electron and the
projectile with the quantized transverse field, which is given by
\beqa
{\cal H}^{\rm T}(t) &=&
\sum_{\nu=1,2} \int \d {\bf q}
\sqrt{\frac{\hbar c^2}{(2\pi)^2 \omega}}
\left[ a_{{\bf q},\nu} \left( \rule{0mm}{4mm} e M_{{\bf
q},\nu}(1)
- Z_0 e M_{{\bf q},\nu}(0) \right) \exp(-{\rm i} \omega t)
\right.
\nonumber \\ [2mm]
&& \rule{4mm}{0mm} \left. \mbox{}
+ a^\dagger_{{\bf q},\nu}
\left( e M^\dagger_{{\bf q},\nu}(1) - Z_0 e  M^\dagger_{{\bf
q},\nu}(0) \right)
\exp({\rm i} \omega t) \right],
\label{3.18}\eeqa
where $a_{{\bf q}\nu}$ and $a^\dagger_{{\bf q}\nu}$ are, respectively, the
annihilation and creation operators of photons with wave-vector ${\bf
q}$ and polarization $\nu$. Photon polarization states are expressed as
linear combinations of the base states of linear polarization in the directions defined by the unit vectors
$\hat{\mbox{\boldmath $\epsilon$}}_{{\bf q},1}$ and
$\hat{\mbox{\boldmath $\epsilon$}}_{{\bf q},2}$, which are real, orthogonal to
each other and to the propagation vector ${\bf q}$. The frequency
$\omega$ is that of a photon with wave number $q$ in vacuum, \ie,
$\omega=cq$. The time-independent operator
\beq
M_{{\bf q},\nu}(j) \equiv \hat{\mbox{\boldmath
$\epsilon$}}_{{\bf q},\nu} \dotprod \widetilde{\alphab}_j
\, \exp ({\rm i} {\bf q}\dotprod {\bf r}_j )
\label{3.19}\eeq
describes the absorption (annihilation) by a Dirac particle (the
projectile or the target electron) of a radiation quantum of momentum
$\hbar {\bf q}$ and polarization $\nu$. Its Hermitian conjugate
\beq
M^\dagger_{{\bf q},\nu}(j) = \hat{\mbox{\boldmath
$\epsilon$}}_{{\bf q},\nu} \dotprod \widetilde{\alphab}_j
\, \exp (-{\rm i} {\bf q}\dotprod {\bf r}_j )
\label{3.20}\eeq
describes the emission of photons by the particle.

{\allowdisplaybreaks
The transitions to be considered go from the initial state $i$$=$$\{
\psi_{{\bf k}, m_{\rm S}}^{(+)}(0)$, $\psi_{a}(1)$, zero
photons$\}$ to the final state $f$$=$$\{ \psi_{{\bf k}', m'_{\rm
S}}^{(-)} (0)$, $\psi_{b}(1)$, zero photons$\}$. In a
second-order perturbation treatment, the transition-matrix element is
given by \citep[][p. 185]{Sakurai1967}
\beq
T^{\rm T}_{fi} = {\cal H}^{\rm T}_{fi} +
\lim_{\eta \rightarrow 0^+}
\sum_n \frac{ {\cal H}^{\rm T}_{fn}
{\cal H}^{\rm T}_{ni}}{E_i - E_n + {\rm i} \eta}\, ,
\label{3.21}\eeq
where $n$ indicates the complete set of intermediate states with one
photon and $E_n$ denotes the total energy of the system, \eg, $E_i= E +
\epsilon_a$, $E_f = E-W + \epsilon_b$. States $n$ with more than one
radiation quanta do not contribute to second order, since ${\cal H}^{\rm
T}$ can only create or annihilate one photon. As the number of real
photons remains unaltered in the course of the interaction, the
first-order term vanishes and we must extend the calculation to second
order. The only contributions to second order correspond to the emission
of one photon by the projectile or the target electron, followed by its
absorption by the other particle (we disregard higher order processes,
in which a single particle emits and absorbs virtual quanta). Then, the
sum over intermediate states reduces to a sum over photon wave vectors
${\bf q}$ and polarizations $\nu$ of two terms, which correspond to the
following processes: \\
1) A photon with momentum $\hbar {\bf q}$ and polarization $\nu$ is
emitted by the projectile. The energy of the intermediate state is $E_n
= (E-W) + \epsilon_a + \hbar c q = E_i - W + \hbar c q$. \\
2) The target electron emits a photon with momentum $-\hbar {\bf q}$ and
polarization $\nu$. The energy of the intermediate state is $E_n =
E + \epsilon_b + \hbar c q = E_i + W + \hbar c q$. \\
Hence, the transition matrix element takes the form
\beqa
T^{\rm T}_{fi} &=& - \frac{Z_0 e^2}{(2\pi)^2}
\lim_{\eta \rightarrow 0^+}
\sum_{\nu} \int \d {\bf q} \;
\frac{\hbar c^2}{\omega}
\left[
\frac{
\left< \psi_{b} \left| M_{{\bf
q},\nu}(1) \rule{0mm}{4mm}\right| \psi_{a}
\right>
\left< \psi_{{\bf k}', m'_{\rm S}}^{(-)} \left| M^\dagger_{{\bf
q},\nu}(0) \right| \psi_{{\bf k}, m_{\rm S}}^{(+)}
\right>}{-\hbar c q +W + {\rm i}\eta } \right.
\nonumber \\ [2mm]
&& \mbox{} + \left.
\frac{
\left< \psi_{{\bf k}', m'_{\rm S}}^{(-)} \left| M_{-{\bf
q},\nu}(0) \right| \psi_{{\bf k}, m_{\rm S}}^{(+)}
\right>
\left< \psi_{b} \left| M^\dagger_{-{\bf
q},\nu}(1) \rule{0mm}{4mm}\right| \psi_{a}
\right>}{-\hbar c q -W + {\rm i}\eta} \right]
\nonumber \\ [2mm]
&=& \frac{Z_0 e^2}{(2\pi)^2}
\lim_{\eta \rightarrow 0^+}
\sum_{\nu} \int \d {\bf q} \;
\frac{\hbar c}{q}
\left( \frac{1}{\hbar c q -W -{\rm i}\eta}+ \frac{1}{\hbar c q +W -
{\rm i} \eta} \right)
\nonumber \\ [2mm]
&& \mbox{} \times
\left< \psi_{{\bf k}', m'_{\rm S}}^{(-)} \left| M^\dagger_{{\bf
q},\nu}(0) \right| \psi_{{\bf k}, m_{\rm S}}^{(+)}
\right>
\left< \psi_{b} \left| M_{{\bf
q},\nu}(1) \rule{0mm}{4mm}\right| \psi_{a}
\right>\, ,
\label{3.22}\eeqa
where, to write the last expression, we have made a change of variable
$-{\bf q} \rightarrow {\bf q}$ in the second term and set $\omega/c=q$.
Replacing the $M$ operators by their expressions \req{3.19} and
\req{3.20}, we have
\beqa
T^{\rm T}_{fi} &=&
\frac{Z_0 e^2}{(2\pi)^2}
\lim_{\eta \rightarrow 0^+}
\sum_{\nu} \int \d {\bf q} \;
\frac{\hbar c}{q}
\left( \frac{1}{\hbar c q -W -{\rm i}\eta}+ \frac{1}{\hbar c q +W -
{\rm i} \eta } \right)
\nonumber \\ [2mm]
&& \mbox{} \times
\left< \psi_{{\bf k}', m'_{\rm S}}^{(-)} (0)\left|
\hat{\mbox{\boldmath
$\epsilon$}}_{{\bf q},\nu} \dotprod \widetilde{\alphab}_0
\, \exp (-{\rm i} {\bf q}\dotprod {\bf r}_0 )
\rule{0mm}{4mm}\right| \psi_{{\bf k}, m_{\rm S}}^{(+)}(0)
\right>
\nonumber \\ [2mm]
&& \mbox{} \times
\left< \psi_{b} (1) \left|
\hat{\mbox{\boldmath
$\epsilon$}}_{{\bf q},\nu} \dotprod \widetilde{\alphab}_1
\, \exp ({\rm i} {\bf q}\dotprod {\bf r}_1 )
\rule{0mm}{4mm}\right| \psi_{a} (1)
\right>.
\label{3.23}\eeqa

Now, the effective transverse Hamiltonian can be defined by requiring
that
\beq
T^{\rm T}_{fi} = \left< \psi^{(-)}_{{\bf k}', m'_{\rm S}} (0)
\, \psi_b(1) \left| \rule{0mm}{4mm}{\cal H}^{\rm T}(0,
1) \right| \psi^{(+)}_{{\bf k}, m_{\rm S}} (0) \,
\psi_a(1) \right>\, .
\label{3.24}\eeq
That is,
\beqa
{\cal H}^{\rm T} (0, 1) &=&
\frac{Z_0 e^2}{(2\pi)^2} \lim_{\eta \rightarrow 0^+}
\int \d {\bf q} \;
\frac{1}{q}
\left( \frac{1}{q -(W/\hbar c) -{\rm i}\eta}+ \frac{1}{q
+(W/\hbar c) - {\rm i} \eta } \right)
\nonumber \\ [2mm]
&& \rule{20mm}{0mm} \times
(\hat{\mbox{\boldmath
$\epsilon$}}_{{\bf q},\nu} \dotprod \widetilde{\alphab}_0)
(\hat{\mbox{\boldmath
$\epsilon$}}_{{\bf q},\nu} \dotprod \widetilde{\alphab}_1)
\, \exp [{\rm i} {\bf q}\dotprod ({\bf r}_1-{\bf r}_0 )].
\label{3.25}\eeqa
The first term in the parenthesis has a pole at the point
$q=W/\hbar c$, which arises from the possibility of a spontaneous
radiative transition of one of the particles with emission of a real
photon of energy $W$. When integrated over $q$, the contribution of this
term can be evaluated by using the formula \citep[see][p.\
85]{Merzbacher1970}
\beq
\lim_{\eta \rightarrow 0^+}
\frac{1}{q -(W/\hbar c) -{\rm i}\epsilon} =
{\sf P} \frac{1}{q -(W/\hbar c)} + {\rm i} \pi \, \delta(q -W/\hbar c),
\label{3.26}\eeq
where ${\sf P}$ denotes the principal value of the integral. Using the
identity
\beq
\sum_{\nu}
(\hat{\mbox{\boldmath
$\epsilon$}}_{{\bf q},\nu} \dotprod \widetilde{\alphab}_0)
(\hat{\mbox{\boldmath
$\epsilon$}}_{{\bf q},\nu} \dotprod \widetilde{\alphab}_1) =
\widetilde{\alphab}_0 \dotprod \widetilde{\alphab}_1 -
(\widetilde{\alphab}_0 \dotprod \hat{\bf q})
(\widetilde{\alphab}_1 \dotprod \hat{\bf q}) \,
\label{3.27}\eeq
and recalling that $q +(W/\hbar c) > 0$, we can write,
with a principal-value integration understood,
\beqa
{\cal H}^{\rm T} (0,1) &=&
\frac{Z_0 e^2}{(2\pi)^2} \int \d {\bf q} \;
\left( \frac{{\rm i} \pi}{q} \, \delta(q -W/\hbar c) +
\frac{2}{q^2 - (W/\hbar c)^2} \right)
\nonumber \\ [2mm]
&& \rule{20mm}{0mm} \times
\left[
\widetilde{\alphab}_0 \dotprod \widetilde{\alphab}_1 -
(\widetilde{\alphab}_0 \dotprod \hat{\bf q})
(\widetilde{\alphab}_1 \dotprod \hat{\bf q}) \right]
\, \exp [{\rm i} {\bf q}\dotprod ({\bf r}_1-{\bf r}_0)] \, .
\rule{10mm}{0mm}
\label{3.28}\eeqa
We see that ${\cal H}^{\rm T} (0,1)$ remains unaltered when the
positions of the two particles are exchanged, as required. Notice that
the transverse Hamiltonian defined by Eq.\ \req{3.28} is correct only to
first order in $e^2$.
}

Incidentally, we note that when the energy transfer $W$ is negligible in
front of $c\hbar q$, the delta function vanishes and the term
$(W/c\hbar)^2$ in the denominator can be deleted, to give the
approximation
\beq
{\cal H}^{\rm T} (0,1) \simeq
\frac{Z_0 e^2}{(2\pi)^2} \int \d {\bf q} \;
\frac{2}{q^2} \left[
\widetilde{\alphab}_0 \dotprod \widetilde{\alphab}_1 -
(\widetilde{\alphab}_0 \dotprod \hat{\bf q})
(\widetilde{\alphab}_1 \dotprod \hat{\bf q}) \right]
\, \exp \left( {\rm i} {\bf q} \dotprod {\bf R} \right)
\label{3.29}\eeq
with ${\bf R} \equiv {\bf r}_1 - {\bf r}_0$. We have
\beqa
\lefteqn{
\int \d {\bf q} \; \frac{1}{q^2}
\, \exp \left( {\rm i} {\bf q} \dotprod {\bf R} \right) =
\int_0^{2\pi} \d \phi \int_{-1}^{1} \d(\cos\theta) \int_0^\infty \d q
\;  \exp \left( {\rm i} q R \, \cos\theta \right) }
\nonumber \\ [2mm]
&=& 2\pi \int_0^\infty \d q \, \left[ \frac{\exp({\rm i} qR \,
\cos\theta)}{qR} \right]_{-1}^1 =
\frac{4\pi}{R} \int_0^\infty \d q \, \frac{\sin(qR)}{q}=
\frac{2\pi^2}{R}.
\label{3.30}\eeqa
Using the equalities
\beq
\nablab_{\bf R} \, \exp \left( {\rm i} {\bf q} \dotprod {\bf R} \right)
= {\rm i} {\bf q} \, \exp \left( {\rm i} {\bf q} \dotprod {\bf R} \right)
\quad \mbox{and} \quad
\nablab_{\bf q} \, \frac{1}{q^2} = - \, \frac{2{\bf q}}{q^4},
\nonumber \eeq
the second part of the integral in \req{3.29} can be evaluated as
\beqa
&& \! \! \! \! \! \! \! \! \! \! \! \! \! \! \! \! \! \! \! \! \! \!
\int \d {\bf q} \; \frac{1}{q^2} \,
\left( \widetilde{\alphab}_0 \dotprod \hat{\bf q} \right)
\left(\widetilde{\alphab}_1 \dotprod \hat{\bf q} \right)
\exp \left( {\rm i} {\bf q} \dotprod {\bf R} \right)
\nonumber \\ [2mm]
&=& \int \d {\bf q} \; \frac{1}{q^2} \,
\left[ \widetilde{\alphab}_0 \dotprod \left(
\frac{-{\rm i}}{q} \nablab_{\bf R}
\exp \left( {\rm i} {\bf q} \dotprod {\bf R} \right) \right)
\right]
\left[\widetilde{\alphab}_1 \dotprod
\left( - \frac{q^3}{2} \, \nablab_{\bf q} \, \frac{1}{q^2} \right)\right]
\nonumber \\ [2mm]
&=& \frac{\rm i}{2}
\left( \widetilde{\alphab}_0 \dotprod \nablab_{\bf R} \right)
\left( \widetilde{\alphab}_1 \dotprod
\int \d {\bf q} \;
\exp \left( {\rm i} {\bf q} \dotprod {\bf R} \right)
\nablab_{\bf q} \, \frac{1}{q^2} \right)
\nonumber \\ [2mm]
&=& \frac{-{\rm i}}{2}
\left( \widetilde{\alphab}_0 \dotprod \nablab_{\bf R} \right)
\left( \widetilde{\alphab}_1 \dotprod
\int \d {\bf q} \; \left[ \nablab_{\bf q} \,
\exp \left( {\rm i} {\bf q} \dotprod {\bf R} \right) \right]
\frac{1}{q^2} \right)
\nonumber \\ [2mm]
&=& \frac{1}{2}
\left( \widetilde{\alphab}_0 \dotprod \nablab_{\bf R} \right)
\left( \widetilde{\alphab}_1 \dotprod {\bf R}
\int \d {\bf q} \;
\exp \left( {\rm i} {\bf q} \dotprod {\bf R} \right)
\frac{1}{q^2} \right)  \qquad \mbox{[using \req{3.30}]}
\nonumber \\ [2mm]
&=& \frac{1}{2}
\left( \widetilde{\alphab}_0 \dotprod \nablab_{\bf R} \right)
\left( \widetilde{\alphab}_1 \dotprod {\bf R} \; \frac{2\pi^2}{R}
\right) = \frac{\pi^2}{R} \left[
\widetilde{\alphab}_0 \dotprod \widetilde{\alphab}_1-
\left( \widetilde{\alphab}_0 \dotprod \hat{\bf R} \right)
\left( \widetilde{\alphab}_1 \dotprod \hat{\bf R} \right)
\rule{0mm}{3mm}\right].
\label{3.31}\eeqa
Inserting the results \req{3.30} and \req{3.31} into Eq.\ \req{3.29},
the transverse part of the effective Hamiltonian reduces to the
Breit operator,
\beq
{\cal H}^{\rm (Breit)} (0,1) = \frac{Z_0 e^2}{2 R}
\left[ \widetilde\alphab_0 \dotprod \widetilde\alphab_1 +
(\widetilde\alphab_0 \dotprod \hat{\bf R})
(\widetilde\alphab_1 \dotprod \hat{\bf R}) \right].
\label{3.32}\eeq
Removing the energy-transfer dependence in this way implies the neglect
of retardation, \ie, of the fact that electromagnetic signals propagate
with the speed of light. As pointed out by \citet{Rose1961}, in an atom
of atomic number $Z$ the relevant values of $q$ are of the order
$\langle r^{-1} \rangle \sim \alpha Z$ atomic units, where $\alpha$ is
the fine-structure constant, while $W \sim (\alpha Z)^2$ atomic units.
Hence the use of the Breit Hamiltonian in atomic structure calculations
is justified only if $\alpha Z \ll 1$. In addition, the Breit
interaction must be treated as a first-order perturbation to the
self-consistent orbitals because otherwise we would introduce terms that
were neglected in the derivation of the effective Hamiltonian
\req{3.28}.

When inelastic collisions are treated within the PWBA, $q$ is always
larger than $W/\hbar c$ (see Appendix \ref{secA}) and the delta function
in Eq.\ \req{3.28} does not contribute to the $T$-matrix elements. Thus,
the effective Hamiltonian to be used in PWBA calculations of inelastic
collisions of charged particles is the sum of the longitudinal and
transverse interactions, given by Eqs.\ \req{3.17} and \req{3.28}:
\beqa
{\cal H}_{\rm int} (0,1) &=&
{\cal H}^{\rm L} (0,1) +
{\cal H}^{\rm T} (0,1)
\nonumber \\ [2mm]
&=& - \frac{Z_0 e^2}{2\pi^2}
\int \d {\bf q} \, \left( \frac{1}{q^2} - \frac{
\widetilde{\alphab}_0 \dotprod \widetilde{\alphab}_1 -
(\widetilde{\alphab}_0 \dotprod \hat{\bf q})
(\widetilde{\alphab}_1 \dotprod \hat{\bf q})}{q^2 - (W/\hbar c)^2} \right)
\exp\left[ {\rm i} {\bf q} \dotprod \left({\bf r}_1 -
{\bf r}_0 \right) \right].
 \rule{10mm}{0mm}
\label{3.33}\eeqa
Indeed, this is the Hamiltonian used by \citet{Fano1963} in his study of
the stopping power of heavy charged particles.

It is worth recalling that, for a Dirac free particle, $\left<
\widetilde{\alphab}_0 \right> = \betab$ \citep[see,
\eg,][]{Sakurai1967}. On the other hand, in a semi-classical formulation
the quantity $\hbar q$ is interpreted as the momentum transfer
\citep{Lindhard1954, Salvat2022c} and, when the energy loss $W$ is much
less than $E$, Eq.\ \req{A.17} holds and it implies that ${\bf q} \cdot
\betab \simeq W/(\hbar c)$. Under that circumstance, the semi-classical
Hamiltonian \req{3.15} is seen to be consistent with the quantum
interaction \req{3.33}.

\subsection{Generalized Breit interaction \label{sec3.3}}

The expression \req{3.33} of the effective Hamiltonian is useful for
calculations within the PWBA because the exponential function simplifies
the handling of the projectile plane waves (see Section \ref{sec4}). For
other calculations, it is more convenient to cast it in a different form
by means of the following argument, which parallels the one used by
\citet{BetheSalpeter1957} to derive the electron-electron interaction.
We introduce the operator
\beq
{\cal B} = \left[
(\widetilde{\alphab}_0 \dotprod {\bf q})
(\widetilde{\alphab}_1 \dotprod {\bf q}) - (W/\hbar c)^2 \right]
\exp [{\rm i} {\bf q}\dotprod ({\bf r}_1-{\bf r}_0 )],
\label{3.34}\eeq
and write the transverse Hamiltonian \req{3.28} as
\beqa
{\cal H}^{\rm (T)} (0,1) &=&
\frac{Z_0 e^2}{2\pi^2} \int \d {\bf q} \;
\left( \frac{1}{2q} {\rm i} \pi \, \delta(q -W/\hbar c) +
\frac{1}{q^2 - (W/\hbar c)^2} \right)
\nonumber \\ [2mm]
&& \rule{10mm}{0mm} \times
\left[
\left( \widetilde\alphab_0 \dotprod
\widetilde\alphab_1 - \frac{(W/\hbar c )^2}{q^2} \right)
\, \exp[{\rm i} {\bf q}\dotprod ({\bf r}_1-{\bf r}_0 )]
- \frac{{\cal B}}{q^2} \right]
\label{3.35}\eeqa
Using the fact that $W=\epsilon_b - \epsilon_a = E - E'$
and the property \req{B.3} of the Dirac Hamiltonian, which implies that
\beq
[{\cal H}_{\rm D}, \exp\left({\rm i} {\bf q} \dotprod
{\bf r} \right) ] = c\hbar (\widetilde{\alphab} \dotprod {\bf q})
\exp\left({\rm i} {\bf q} \dotprod {\bf r} \right),
\label{3.36}\eeq
we have
\beqa
W \, \left< \, \psi_b({\bf r}_1) \left|
\rule{0mm}{4mm}
\exp ({\rm i} {\bf q}\dotprod {\bf r}_1 )
\right|
\psi_a({\bf r}_1) \right> &=&
\left< \, \psi_b \left|
\rule{0mm}{4mm} \left[ {\cal H}_{\rm D}(1),
\exp ({\rm i} {\bf q}\dotprod {\bf r}_1 ) \right]
\right| \psi_a \right>
\nonumber \\ [2mm]
&=& c \hbar
\left< \, \psi_b \left|
\rule{0mm}{4mm} \widetilde{\alphab}_1 \dotprod {\bf q} \,
\exp ({\rm i} {\bf q}\dotprod {\bf r}_1 )
\right| \psi_a \right>
\label{3.37}\eeqa
and
\beqa
W \left< \psi_{{\bf k}', m'_{\rm S}}^{(-)} ({\bf r}_0)\left|
\exp (-{\rm i} {\bf q}\dotprod {\bf r}_0 )
\rule{0mm}{4mm}\right| \psi_{{\bf k}, m_{\rm S}}^{(+)}({\bf r}_0)
\right> &=& - \left< \psi_{{\bf k}', m'_{\rm S}}^{(-)} \left|
\left[ {\cal H}_{\rm D}(0), \exp (-{\rm i} {\bf q}\dotprod {\bf r}_0 )
\right] \rule{0mm}{4mm}\right| \psi_{{\bf k}, m_{\rm S}}^{(+)} \right>
\nonumber \\ [2mm]
&& \! \! \! \! \! \! \! \! \! \! \! \! \! \! \! \! \! \! \! \! \! \!
\! \! \!
=  c \hbar
\left< \psi_{{\bf k}', m'_{\rm S}}^{(-)} \left|
\widetilde{\alphab}_0 \dotprod {\bf q} \,
\exp (-{\rm i} {\bf q}\dotprod {\bf r}_0 )
\rule{0mm}{4mm}\right| \psi_{{\bf k}, m_{\rm S}}^{(+)}
\right>.\rule{10mm}{0mm}
\label{3.38}\eeqa
Therefore, the matrix element of the operator ${\cal B}$ between the
initial and final states vanishes, and the effective transverse
Hamiltonian \req{3.35} can be replaced with
\beqa
{\cal H}^{\rm (T)} (0,1) &=&
\frac{Z_0 e^2}{2\pi^2} \int \d {\bf q} \;
\left( \frac{1}{2q} {\rm i} \pi \, \delta(q -W/\hbar c) +
\frac{1}{q^2 - (W/\hbar c)^2} \right)
\nonumber \\ [2mm]
&& \rule{20mm}{0mm} \times
\left( \widetilde\alphab_0 \dotprod \widetilde\alphab_1
- \frac{(W/\hbar c )^2}{q^2} \right)
\, \exp ({\rm i} {\bf q}\dotprod {\bf R}),
\label{3.39}\eeqa
where ${\bf R} = {\bf r}_1-{\bf r}_0$.
The integral over the direction of ${\bf q}$ is
\beq
\int \, \exp ({\rm i} {\bf q}\dotprod {\bf R}) \, \d \hat{\bf q} =
\frac{2\pi}{{\rm i} q R}
\left[ \exp ({\rm i} q R) - \exp (-{\rm i} q R) \right],
\label{3.40}\eeq
so that
\beqa
{\cal H}^{\rm (T)} (0,1) &=&
\frac{Z_0 e^2}{\pi {\rm i} R } \int_0^\infty \d q \;
\left( {\rm i} \frac{\pi}{2} \, \delta(q -W/\hbar c) +
\frac{q }{q^2 - (W/\hbar c)^2} \right)
\nonumber \\ [2mm]
&& \times
\left( \widetilde\alphab_0 \dotprod \widetilde\alphab_1
- \frac{(W/\hbar c )^2}{q^2} \right)
\, \left[ \exp ({\rm i} q R) - \exp (-{\rm i} q R) \right]
\nonumber \\ [2mm]
&=&
\frac{Z_0 e^2}{R} \,
\left( \widetilde\alphab_0 \dotprod \widetilde\alphab_1 - 1 \right)
{\rm i} \sin(WR/\hbar c) +
\frac{Z_0 e^2}{\pi {\rm i} R } \int_0^\infty \d q \;
\frac{q }{q^2 - (W/\hbar c)^2}
\nonumber \\ [2mm]
&& \times
\left( \widetilde\alphab_0 \dotprod \widetilde\alphab_1
+ \frac{q^2 - (W/\hbar c )^2}{q^2} -
1 \right) \, \left[ \exp ({\rm i} q R) - \exp (-{\rm i} q R) \right]
\nonumber \\ [2mm]
&=& \frac{Z_0 e^2}{R} \,
\left( \widetilde\alphab_0 \dotprod \widetilde\alphab_1 - 1 \right)
{\rm i} \sin(WR/\hbar c)
\nonumber \\ [2mm]
&+& \frac{Z_0 e^2}{\pi {\rm i} R }
\left( \widetilde\alphab_0 \dotprod \widetilde\alphab_1 - 1 \right)
\int_0^\infty \d q \;
\frac{q }{q^2 - (W/\hbar c)^2}
\, \left[ \exp ({\rm i} q R) - \exp (-{\rm i} q R) \right]
\nonumber \\ [2mm]
&+&
\frac{Z_0 e^2}{\pi {\rm i} R } \int_0^\infty \d q \;
\frac{1}{q}
\, \left[ \exp ({\rm i} q R) - \exp (-{\rm i} q R) \right].
\label{3.41}\eeqa
Now,
\beq
\int_0^\infty \d q \;
\frac{1}{q} \, \left[ \exp ({\rm i} q R) - \exp (-{\rm i} q R) \right] =
2 {\rm i} \int_0^\infty \d q \; \frac{\sin (qR)}{q} = {\rm i} \pi
\label{3.42}\eeq
The remaining integral can be easily evaluated as a path integral in
the complex plane \citep[see, \eg,][pp 409-411]{Arfken1985},
\beqa
&&
\! \! \! \! \! \! \! \! \! \! \! \! \! \! \! \! \! \!
\int_0^\infty \frac{q}{q^2-(W/\hbar c)^2}
\left[ \exp ({\rm i} q R) - \exp (-{\rm i} q R) \right] \, \d q
\nonumber \\ [2mm]
&=&
\int_{-\infty}^\infty \frac{q \exp ({\rm i} q R)
}{q^2-(W/\hbar c)^2}
\, \d q = \pi {\rm i} \cos(WR/\hbar c).
\label{3.43}\eeqa
We thus obtain
\beqa
{\cal H}^{\rm (T)} (0,1) &=& \frac{Z_0 e^2}{R} \,
\left( \widetilde\alphab_0 \dotprod \widetilde\alphab_1 - 1 \right)
\left[ \cos(WR/\hbar c)
+ {\rm i} \sin(WR/\hbar c) \right] + \frac{Z_0 e^2}{R}
\nonumber \\ [2mm]
&=&
\frac{Z_0 e^2}{R} \,
\left( \widetilde\alphab_0 \dotprod \widetilde\alphab_1 - 1 \right)
\exp({\rm i} WR/\hbar c) + \frac{Z_0 e^2}{R}.
\label{3.44}\eeqa
This Hamiltonian is known as the {\it generalized Breit interaction}.
Recalling the identity \req{3.17}, the effective Hamiltonian
then becomes
\beqa
{\cal H}_{\rm int} (0,1) &=& {\cal H}^{\rm L} (0,1) +
{\cal H}^{\rm T} (0,1)
\nonumber \\ [2mm]
&=& \frac{Z_0 e^2}{R} \, \left(
\widetilde\alphab_0 \dotprod \widetilde\alphab_1 - 1 \right)
\exp({\rm i} WR/\hbar c).
\label{3.45}\eeqa
With $Z_0=-1$, it reduces to the well-known M\o ller Hamiltonian,
\beq
{\cal H}^{\rm (Moller)} (0,1) = - \frac{e^2}{R} \,
\left( \widetilde\alphab_0 \dotprod \widetilde\alphab_1 - 1 \right)
\exp({\rm i} WR/\hbar c),
\label{3.46}\eeq
which describes the retarded interaction between two electrons. The
effect of retardation is expressed through the factor $\exp({\rm i}
WR/c\hbar)$, which depends explicitly on the energy transfer. The usual
derivation of the M\o ller Hamiltonian \citep[see, \eg,][]{Rose1961}
proceeds through an argument similar to the one presented above, but
using the Lorentz gauge instead of the Coulomb gauge. The present
derivation explicitly shows that, for orbitals from a local potential
such as the DHFS self-consistent potential, the effective interaction,
given by expression \req{3.32} with $Z_0=-1$, and the M\o ller
Hamiltonian \req{3.46} are equivalent, \ie, their matrix elements
between spherical orbitals or DPWs are equal. Notice also that the form
\req{3.46} is valid only when the matrix elements of the operator ${\cal
B}$, eq.\ \req{3.34}, between the initial and final states vanishes;
otherwise, the original expression \req{3.33} must be used.


\section{The plane-wave Born approximation \label{sec4}}
\setcounter{equation}{0}

Introducing the Hamiltonian \req{3.33}, the transition
matrix elements \req{2.32} take the form
\beqa
T_{fi} &=&
\left< \psi^{(-)}_{{\bf k}', m'_{\rm S}} (0) \, \psi_b(1)
\left| \rule{0mm}{4mm}{\cal H}_{\rm int}(0, 1)
\right| \psi^{(+)}_{{\bf k}, m_{\rm
S}} (0) \, \psi_a(1) \right>
\nonumber \\ [2mm]
&=& - \frac{Z_0 e^2}{2\pi^2}
\int \d {\bf q} \,
\int \d {\bf r}_0 \,
\int \d {\bf r}_1 \,
\left[ \psi^{(-)}_{{\bf k}', m'_{\rm S}} ({\bf r}_0)\right]^\dagger \,
\psi_b^\dagger({\bf r}_1)
\nonumber \\ [2mm]
&& \mbox{} \times
\left( \frac{1}{q^2} -
\frac{\widetilde{\alphab}_0
\dotprod \widetilde{\alphab}_1 - (\widetilde{\alphab}_0 \dotprod \hat{\bf q}) (\widetilde{\alphab}_1
\dotprod \hat{\bf q})}{q^2 - (W/\hbar c)^2} \right)
\nonumber \\ [2mm]
&& \mbox{} \times
\exp\left[ {\rm i} {\bf q} \dotprod \left({\bf r}_1 -
{\bf r}_0 \right) \right] \,
\psi^{(+)}_{{\bf k}, m_{\rm
S}} ({\bf r}_0) \, \psi_a({\bf r}_1).
\label{4.1}\eeqa
These matrix elements correspond to the distorted-wave Born
approximation (DWBA). Note that the final state of the active target
electron, $\psi_{b}$, is described by a spherical orbital:
$\psi_{\epsilon_b \kappa_b m_b}$ in the case of ionization ($\epsilon_b
>0$) and $\psi_{n_b \kappa_b m_b}$ in the case of excitation to discrete
bound levels ($\epsilon_b < 0$).

\allowdisplaybreaks{
The theory described up to this point is the basis of calculations by
\citet{Segui2003}, \citet{Colgan2006}, \citet{BoteSalvat2008}, and
others. The difficulty of these DWBA calculations increases rapidly with
the kinetic energy of the projectile, because of the slow convergence of
the partial-wave series. In the case of inelastic collisions of
electrons and positrons with free atoms, calculations are feasible only
when the kinetic energy of the projectile is less than about 30 times
the ionization energy of the active electron subshell. For electrons and
positrons with higher energies, and for projectiles with masses larger
than $\me$, we have to rely on the plane-wave Born approximation (PWBA),
which allows the calculations to be extended up to arbitrarily high
energies. The formulation of the PWBA presented below is based on the
work of \citet{Fano1963}, which we follow closely. However, it is
slightly more general, and is tailored for practical numerical
computations.

The PWBA consists of replacing the projectile distorted waves
$\psi^{(\pm)}_{{\bf k}, m_{\rm S}} ({\bf r})$ in the $T$-matrix element
\req{4.1} with the corresponding positive-energy plane waves [Eqs.\
\req{B.5} and \req{B.6a}],
\beq
\phi_{{\bf k}, m_{{\rm S}},+1}({\bf r}) =
\frac{{\rm e}^{{\rm i} {\bf k} \cdot {\bf r} }}{(2\pi)^{3/2}}
U_{{\bf k},m_{{\rm S}},+1}
\label{4.2}\eeq
with
\beq
U_{{\bf k}, m_{{\rm S}},+1} \equiv
\left[ \frac{E+2M c^2}{2E+2 M c^2} \right]^{1/2}
\left( \begin{array}{c}  I_2 \\ [2mm]
\displaystyle{\frac{c \hbar (\sigmab \dotprod {\bf k})
}{E+2M c^2}}
\end{array} \right) \chi_{m_{\rm S}}.
\label{4.3}\eeq
In principle, this replacement is permissible only when the energies of
the projectile before and after the collision, $E$ and $E'=E-W$, are
both sufficiently large so that the distortion of the projectile wave
functions caused by the electrostatic field of the target atom is
negligible. In practice, however, the PWBA does work under much less
restrictive conditions, namely, when the initial speed $v$ of the
projectile is larger than a few times the average speed of the active
electron in its initial bound state (see Section \ref{sec4.3}).
\citet{BoteSalvat2008} have shown that, in the case of
projectile electrons and positrons, total cross sections for ionization
of inner subshells of atoms calculated from the DWBA and from the PWBA
differ by less than about 1 percent when the kinetic energy of the
projectile is higher than about 30 times the ionization energy of the
active electron. In most radiation transport studies, however, interest
is on the stopping of the projectile and contributions from excitations
of different electron subshells are not distinguished. As the contributions
of tightly bound electrons to the atomic cross section are relatively
small (see below), the PWBA yields fairly accurate results even for slow
projectiles \citep{Salvat2022a}.

Introducing the plane waves \req{4.2}, the PWBA $T$-matrix elements take
the form
\beqa
T_{fi}^{\rm PW}=&-& \frac{Z_0 e^2}{2\pi^2}
\int \d {\bf q} \,
\int \d {\bf r}_0 \,
\int \d {\bf r} \,
\phi^\dagger_{{\bf k}', m'_{\rm S},+1} ({\bf r}_0) \,
\psi_{b}^\dagger({\bf r})
\nonumber \\ [2mm]
&\times&
\left( \frac{1}{q^2} - \frac{\widetilde{\alphab}_0
\dotprod \widetilde{\alphab} -
(\widetilde{\alphab}_0 \dotprod \hat{\bf q}) (\widetilde{\alphab}
\dotprod \hat{\bf q})
}{q^2 - (W/\hbar c)^2} \right)
\nonumber \\ [2mm]
 &\times&
\exp\left[ {\rm i} {\bf q} \dotprod \left({\bf r} -
{\bf r}_0 \right) \right] \,
\phi_{{\bf k}, m_{\rm S},+1} ({\bf r}_0)
\, \psi_{n_a \kappa_a m_a}({\bf r}).
\label{4.4}\eeqa
The advantage of the PWBA is that integrals over the space coordinates
${\bf r}_0$ of the projectile can be evaluated analytically. The one
corresponding to the first term within the parentheses is elementary,
\beqa
&& \rule{-2cm}{0cm}
\left< \phi_{{\bf k}', m'_{\rm S},+1} ({\bf r}_0)\left|
\exp
(-{\rm i} {\bf q}\dotprod {\bf r}_0 ) \rule{0mm}{4mm}\right| \phi_{{\bf
k}, m_{\rm S},+1}({\bf r}_0) \right>
\nonumber \\ [2mm]
&=&
U_{{\bf k}', m'_{\rm S},+1}^\dagger \, U_{{\bf k}, m_{\rm S},+1}\;
\frac{1}{(2\pi)^3}
\int \exp\left[ - {\rm i} ({\bf k}'+ {\bf q} - {\bf k}) \dotprod
{\bf r}_0 \right] \, \d {\bf r}_0
\nonumber \\ [2mm]
&=&
U_{{\bf k}', m'_{\rm S},+1}^\dagger \, U_{{\bf k}, m_{\rm S},+1} \;
\delta ({\bf q} - {\bf k} + {\bf k}').
\label{4.5}\eeqa
The second term in \req{4.4} involves the matrix elements
\beqa
&& \rule{-2cm}{0cm}
\left< \phi_{{\bf k}', m'_{\rm S},+1} ({\bf r}_0)\left|
\widetilde{\alphab}_0 \, \exp
(-{\rm i} {\bf q}\dotprod {\bf r}_0 ) \rule{0mm}{4mm}\right| \phi_{{\bf
k}, m_{\rm S},+1}({\bf r}_0) \right>
\nonumber \\ [2mm]
&=&
U_{{\bf k}', m'_{\rm S},+1}^\dagger \,
\widetilde{\alphab}_0 \, U_{{\bf k}, m_{\rm S},+1} \;
\frac{1}{(2\pi)^3}
\int \exp\left[ - {\rm i} ({\bf k}'+ {\bf q} - {\bf k})
\dotprod {\bf r}_0 \right] \, \d {\bf r}_0
\nonumber \\ [2mm]
&=&
U_{{\bf k}', m'_{\rm S},+1}^\dagger \,
\widetilde{\alphab}_0 \, U_{{\bf k}, m_{\rm S},+1} \;
\delta ({\bf q} - {\bf k} + {\bf k}').
\label{4.6}\eeqa
and
\beq
\left< \phi_{{\bf k}', m'_{\rm S},+1} ({\bf r}_0)\left|
\widetilde{\alphab}_0 \dotprod \hat{\bf q} \, \exp
(-{\rm i} {\bf q}\dotprod {\bf r}_0 ) \rule{0mm}{4mm}\right| \phi_{{\bf
k}, m_{\rm S},+1}({\bf r}_0) \right>.
\label{4.7}\eeq
The latter can be calculated easily by noting that the Dirac Hamiltonian
${\cal H}_{\rm D}$ for a central potential $V(r)$, Eq.\  \req{B.1},
satisfies the commutation relation
\beq
[{\cal H}_{\rm D}, \exp\left({\rm i} {\bf q} \dotprod
{\bf r} \right) ] = c\hbar (\widetilde{\alphab} \dotprod {\bf q})
\exp\left({\rm i} {\bf q} \dotprod {\bf r} \right),
\label{4.8}\eeq
which follows from Eq.\ \req{B.3} and it is valid also for a free
particle ($V=0$). Hence, we can write
\beqa
&& \rule{-2cm}{0cm}
\left< \phi_{{\bf k}', m'_{\rm S},+1} ({\bf r}_0)\left|
\widetilde{\alphab}_0 \dotprod \hat{\bf q} \, \exp
(-{\rm i} {\bf q}\dotprod {\bf r}_0 ) \rule{0mm}{4mm}\right| \phi_{{\bf
k}, m_{\rm S},+1}({\bf r}_0) \right>
\nonumber \\ [2mm]
&=& \frac{-1}{c\hbar q}
\left< \phi_{{\bf k}', m'_{\rm S},+1} ({\bf r}_0)\left|
 {\cal H}_{\rm D} \,
\exp (-{\rm i} {\bf q}\dotprod {\bf r}_0) -
\exp (-{\rm i} {\bf q}\dotprod {\bf r}_0) {\cal H}_{\rm D}
\rule{0mm}{4mm}\right| \phi_{{\bf
k}, m_{\rm S},+1}({\bf r}_0) \right>
\nonumber \\ [2mm]
&=& \frac{E-E'}{c\hbar q}
\left< \phi_{{\bf k}', m'_{\rm S},+1} ({\bf r}_0)\left|
 \exp
(-{\rm i} {\bf q}\dotprod {\bf r}_0 ) \rule{0mm}{4mm}\right|
\phi_{{\bf k}, m_{\rm S},+1}({\bf r}_0) \right>
\nonumber \\ [2mm]
&=& \frac{W}{c\hbar q} \,
U_{{\bf k}', m'_{\rm S},+1}^\dagger \, U_{{\bf k}, m_{\rm S},+1}\;
\frac{1}{(2\pi)^3}
\int \exp\left[ - {\rm i} ({\bf k}'+ {\bf q} - {\bf k}) \dotprod
{\bf r}_0 \right] \, \d {\bf r}_0
\nonumber \\ [2mm]
&=&
\frac{W}{c\hbar q} \,
U_{{\bf k}', m'_{\rm S},+1}^\dagger \, U_{{\bf k}, m_{\rm S},+1} \;
\delta ({\bf q} - {\bf k} + {\bf k}').
\label{4.9}\eeqa
Inserting these expressions, the matrix element \req{4.4} becomes
\beqa
T_{fi}^{\rm PW} &=&
- \frac{Z_0 e^2}{2\pi^2} \,
\left\{ \rule{0mm}{8mm}
U_{{\bf k}', m'_{\rm S},+1}^\dagger \, U_{{\bf k}, m_{\rm S},+1} \;
\frac{1}{q^2}  \left< \psi_{_b} \left|
\exp ({\rm i} {\bf q}\dotprod {\bf r}) \rule{0mm}{4mm}
\right| \psi_{n_a \kappa_a m_a} \right> \right.
\nonumber \\ [2mm]
&& \mbox{} - U_{{\bf k}', m'_{\rm S},+1}^\dagger \,
\left( \widetilde{\alphab}_0 - \frac{W}{c\hbar q^2} \, {\bf q} \right)
U_{{\bf k}, m_{\rm S},+1} \dotprod  \,
\left. \frac{\left< \psi_{b} \left| \widetilde{\alphab} \,
\exp ({\rm i} {\bf q}\dotprod {\bf r}) \rule{0mm}{4mm}
\right| \psi_{n_a \kappa_a m_a} \right>}{q^2-(W/c\hbar)^2} \right\},
\rule{10mm}{0mm}
\label{4.10}\eeqa
where ${\bf q} = {\bf k} - {\bf k}'$ is the momentum transfer in units
of $\hbar$.

As indicated by \citet{BoteSalvat2008}, the present treatment is
strictly equivalent to that of \citet{Bethe1932, Bethe1933}.
Unfortunately, the derivation given by Bote and Salvat is incomplete.
The equivalence of the two formulations can be proved readily by noting
that the commutation relation \req{4.8} implies that
\beqa
{\bf q} \dotprod \left< \psi_{b} \left| \widetilde{\alphab} \,
\exp ({\rm i} {\bf q}\dotprod {\bf r}) \rule{0mm}{4mm}
\right| \psi_{n_a \kappa_a m_a} \right>
&=& \frac{1}{c\hbar} \left< \psi_{b} \left| {\cal H}_{\rm D} \,
\exp ({\rm i} {\bf q}\dotprod {\bf r}) -
\exp ({\rm i} {\bf q}\dotprod {\bf r}) {\cal H}_{\rm D}
\rule{0mm}{4mm}
\right| \psi_{n_a \kappa_a m_a} \right>
\nonumber \\ [2mm]
&=& \frac{\epsilon_b - \epsilon_a}{c\hbar} \left< \psi_{b} \left|
\exp ({\rm i} {\bf q}\dotprod {\bf r})
\rule{0mm}{4mm}
\right| \psi_{n_a \kappa_a m_a} \right>.
\label{4.11}\eeqa
Now, the $T$-matrix element \req{4.10} can be expressed as
\beqa
T_{fi}^{\rm PW} &=&
- \frac{Z_0 e^2}{2\pi^2} \,
\left\{ \rule{0mm}{8mm}
U_{{\bf k}', m'_{\rm S},+1}^\dagger \, U_{{\bf k}, m_{\rm S},+1} \;
\left( \frac{1}{q^2} + \frac{(W/c\hbar q)^2 }{q^2 -(W/c\hbar)^2} \right)
\left< \psi_{_b} \left|
\exp ({\rm i} {\bf q}\dotprod {\bf r}) \rule{0mm}{4mm}
\right| \psi_{n_a \kappa_a m_a} \right> \right.
\nonumber \\ [2mm]
&& \mbox{} - U_{{\bf k}', m'_{\rm S},+1}^\dagger \,
\widetilde{\alphab}_0 \,
U_{{\bf k}, m_{\rm S},+1} \dotprod  \,
\left. \frac{\left< \psi_{b} \left| \widetilde{\alphab} \,
\exp ({\rm i} {\bf q}\dotprod {\bf r}) \rule{0mm}{4mm}
\right| \psi_{n_a \kappa_a m_a} \right>}{q^2-(W/c\hbar)^2} \right\}
\label{4.12}\eeqa
or, more succinctly,
\beqa
T_{fi}^{\rm PW}
&=& - \frac{Z_0 e^2}{2\pi^2}
\frac{1}{q^2-(W/\hbar c)^2}
\left< \psi_{_b} \left| \left(A_0+{\bf A} \dotprod\widetilde{\alphab} \right)
\exp ({\rm i} {\bf q}\dotprod {\bf r}) \rule{0mm}{4mm}
\right| \psi_{n_a \kappa_a m_a} \right>
\label{4.13}\eeqa
with
\begin{subequations}
\label{4.14}
\beq
A_0=U_{{\bf k}', m_{{\rm S}'},+1}^\dagger \, U_{{\bf k},
m_{{\rm S}},+1},
\label{4.14a}
\eeq
\beq
{\bf A}= - U_{{\bf k}', m_{{\rm S}'},+1}^\dagger \, \widetilde{\alphab}_0 \; U_{{\bf k}
, m_{{\rm S}},+1}.
\label{4.14b}
\eeq
\end{subequations}
Expression \req{4.13} agrees exactly with the matrix elements in
Bethe's formulation, Eq.\ (5) of \citet{Bethe1932} and Eq.\ (50.1) of
\citet{Bethe1933}. We prefer using the form \req{4.10} because it keeps the
contributions from longitudinal and transverse interactions separated.

In what follows, we will limit our considerations to the case of
excitation or ionization of a closed subshell, $n_a\kappa_a$, with
$2|\kappa_a|=2j_a+1$ electrons. The DDCS for an open subshell, with $q_a <
2j_a+1$ electrons, will be approximated as the product of the DDCS of the
closed subshell and the fractional occupancy, $q_a /(2j_a+1)$.
The DDCSs for excitation and ionization of the closed subshell
are given by [see Eqs.\ \req{2.34} and \req{2.36}]
\beqa
\frac{\d^2 \sigma^{\rm exc}}{\d W \, \d Q}
=  \frac{(2 \pi)^5}{c^2 \hbar^4 v^2} \, \frac{E-W+M c^2}{E+Mc^2} \,
(Q+\me c^2) \,
 {\cal I}_{fi}^{\rm PW}
\label{4.15}\eeqa
and
\beqa
\frac{\d^2 \sigma^{\rm ion}}{\d W \, \d Q}
=  \frac{(2 \pi)^5}{c^2 \hbar^4 v^2} \, \frac{E-W+M c^2}{E+Mc^2} \,
(Q+\me c^2)  \,
{\cal J}_{fi}^{\rm PW},
\label{4.16}\eeqa
with
\beq
{\cal I}_{fi}^{\rm PW}
= \delta(W-\epsilon_b+\epsilon_{n_a \kappa_a})
\, {\cal T}^{\rm eff}_{fi}
\label{4.17}\eeq
and
\beq
{\cal J}_{fi}^{\rm PW}
= \frac{k_b}{\epsilon_b \pi}
\sum_{\kappa_b} \, {\cal T}^{\rm eff}_{fi}
\label{4.18}\eeq
where
\beq
{\cal T}^{\rm eff}_{fi} \equiv \sum_{m_b}\sum_{m_{a}}
\frac{1}{2} \sum_{m'_{{\rm S}},m_{{\rm S}}}
|T_{fi}^{\rm PW}|^2.
\label{4.19}\eeq

}

\subsection{Transition-matrix elements \label{sec4.1}}

\allowdisplaybreaks{

The calculation of the squared $T$-matrix elements, $|T_{fi}^{\rm
PW}|^2$, is simplified by noting that the operators in the longitudinal
and transverse terms of expression \req{4.10} have different parities
under reflection on any plane that contains ${\bf q}$ (the
$\widetilde{\alphab}$ matrices are odd under space inversion; see
\citet{SalvatDE2021}). As the
spherical waves also have definite parity, it follows that for a given
transition, the transverse and longitudinal terms cannot be different
from zero simultaneously. That is, the longitudinal and transverse
interactions excite transitions of the active electron from its initial
bound orbital to final orbitals of different parities. Therefore,
\beqa
\left| T_{fi}^{\rm PW} \right|^2 &=&
\frac{Z_0^2 e^4}{4\pi^4} \, \frac{1}{q^4}
\left| U_{{\bf k}', m'_{\rm S},+1}^\dagger \, U_{{\bf k}, m_{\rm
S},+1}\right|^2  \;
\left| \left< \psi_{b} \left|
\exp ({\rm i} {\bf q}\dotprod {\bf r}) \rule{0mm}{4mm}
\right| \psi_{n_a \kappa_a m_a} \right> \right|^2
\nonumber \\ [2mm]
&+& \frac{Z_0^2 e^4}{4\pi^4} \, \frac{1}{[q^2-(W/c\hbar)^2]^2}
\left| U_{{\bf k}', m'_{\rm S},+1}^\dagger \,
\left( \widetilde{\alphab}_0 - \frac{W}{c\hbar q^2} \, {\bf q} \right)
\dotprod {\bf G} \;
U_{{\bf k}, m_{\rm S},+1} \right|^2, \rule{10mm}{0mm}
\label{4.20}\eeqa
with
\beq
{\bf G} \equiv \left<
\psi_{b}
\left| \widetilde{\alphab} \,
\exp ({\rm i} {\bf q}\dotprod {\bf r}) \rule{0mm}{4mm}
\right| \psi_{n_a \kappa_a m_a} \right>.
\label{4.21}\eeq
This clean separation of longitudinal and transverse contributions
occurs only when the active target electron moves in a spherical
potential. Any departure from spherical symmetry would induce
interference between the longitudinal and transverse terms
\citep{Schattschneider2005}.

To evaluate the expression \req{4.19}, it is convenient to perform the
various summations in the order in which they appear. The spin-averaged
squared matrix element
\beq
{\cal T}_1 \equiv
\frac{1}{2} \sum_{m_{{\rm S}'}, m_{{\rm S}}}
\left|T_{fi}^{\rm PW}\right|^2
\label{4.22}\eeq
can be expressed in the form
\beq
{\cal T}_1 =
\frac{Z_0^2 e^4}{4\pi^4} \, \frac{1}{q^4} \left|
\left< \psi_{b} \left|
\exp ({\rm i} {\bf q}\dotprod {\bf r}) \rule{0mm}{4mm}
\right| \psi_{n_a \kappa_a m_a} \right> \right|^2 {\cal S}_{\rm L}
+ \frac{Z_0^2 e^4}{4\pi^4} \, \frac{1}{[q^2-(W/c\hbar)^2]^2}
\, {\cal S}_{\rm T},
\label{4.23}\eeq
where
\beqa
{\cal S}_{\rm L} &\equiv&
\frac{1}{2} \sum_{m_{{\rm S}'}, m_{{\rm
S}}} \left| U_{{\bf k}', m_{{\rm S}'},+1}^\dagger \, U_{{\bf k},
m_{{\rm S}},+1}\right|^2
\nonumber \\ [2mm]
&=& \frac{1}{2} \sum_{m'_{\rm S}} U_{{\bf k}', m'_{\rm
S},+1}^\dagger \, \left( \sum_{m_{\rm S}} U_{{\bf k}, m_{\rm S},+1}  \,
U_{{\bf k}, m_{\rm S},+1}^\dagger \right) U_{{\bf k}', m'_{\rm S},+1}\, ,
\label{4.24}\eeqa
and
\beqa
{\cal S}_{\rm T} &\equiv&
\frac{1}{2} \sum_{m_{{\rm S}'}, m_{{\rm S}}}
\left| U_{{\bf k}', m_{{\rm S}'},+1}^\dagger \, \left( \widetilde{\alphab}_0 -
\frac{W}{c\hbar q^2} \, {\bf q} \right) \, \dotprod \, {\bf G} \; U_{{\bf k}
, m_{{\rm S}},+1} \right|^2
\nonumber \\ [2mm]
&=&
\frac{1}{2} \sum_{m_{\rm S}, m'_{\rm S}} \left[ U_{{\bf k}', m'_{\rm
S},+1}^\dagger \, \left( \widetilde{\alphab}_0 - \frac{W}{c\hbar q^2} \,
{\bf q}\right) \dotprod {\bf G} \; U_{{\bf k}, m_{\rm S},+1} \right]
\nonumber \\ [2mm]
&& \times
\left[ U_{{\bf k}, m_{\rm S},+1}^\dagger \, \left(
\widetilde{\alphab}_0 - \frac{W}{c\hbar q^2} \,
{\bf q}\right) \dotprod {\bf G}^\ast \; U_{{\bf k}', m'_{\rm S},+1}
\right].
\label{4.25}\eeqa
The summations in Eqs. \req{4.24} and \req{4.25} can be evaluated by
using conventional projection tricks \citep[see, \eg,][]{Heitler1954}. The
calculation is easier for the longitudinal term,
\beq
{\cal S}_{\rm L} =
\frac{1}{2} \sum_{m'_{\rm S}} U_{{\bf k}', m'_{\rm
S},+1}^\dagger \, \Pi_{{\bf k},+1} U_{{\bf k}', m'_{\rm S},+1}\, ,
\label{4.26}\eeq
where we have introduced the operator
\beqa
\Pi_{{\bf k},+1} &\equiv& \sum_{\mu=\pm 1/2} U_{{\bf k}, \mu, +1} (\hat{\bf z})
U_{{\bf k}, \mu, +1}^\dagger (\hat{\bf z})
\nonumber \\ [2mm]
&=& \frac{1}{2(E+M c^2)} \left( E + M c^2 + c\hbar\,
\widetilde{\alphab}_0 \dotprod {\bf k} + \widetilde{\beta}_0 M c^2
\right),
\label{4.27}\eeqa
which is the projector on the positive-energy subspace, Eq.\ \req{B.9}.
To evaluate this expression, we replace the spinors
$ U_{{\bf k}', m'_{\rm S},+1}$ with $\Pi_{{\bf k}',+1} \, U_{{\bf k}',
m'_{\rm S},\tau}$ and extend the summation to positive- and
negative-energy spinors (negative-energy spinors do not contribute
because their projections vanish). We thus have
\beq
{\cal S}_{\rm L} =
\frac{1}{2} \sum_{m'_{\rm S},\tau}
 U_{{\bf k}', m'_{\rm S},\tau}^\dagger \,
\Pi_{{\bf k}',+1}
\, \Pi_{{\bf k},+1} \Pi_{{\bf k}',+1} \, U_{{\bf k}', m'_{\rm S},\tau}
= \frac{1}{2} {\rm Tr} \left\{ \Pi_{{\bf k}',+1}
\, \Pi_{{\bf k},+1} \Pi_{{\bf k}',+1}
\right\},
\label{4.28}\eeq
where the symbol ${\rm Tr}$ stands for the trace (the sum of diagonal
elements) of the matrix. Noting that the trace of a product of
matrices is not altered by a cyclic permutation of the factors, and
recalling that $\Pi_{{\bf k}',+1}$ is a projector ($\Pi_{{\bf
k}',+1}=\Pi_{{\bf k}',+1}^\dagger= \Pi_{{\bf k}',+1}^2$), we can write
\beqa
{\cal S}_{\rm L} &=&
\frac{1}{2} {\rm Tr} \left\{ \Pi_{{\bf k},+1} \Pi_{{\bf k}',+1} \right\}
\nonumber \\ [2mm]
&=&
\frac{1}{8\, (E+M c^2)(E'+M c^2)} {\rm Tr} \left\{
\left( E + M c^2 + c\hbar\,
\widetilde{\alphab}_0 \dotprod {\bf k} + \widetilde{\beta}_0 M c^2
\right) \right.
\nonumber \\ [2mm]
&& \times \left.
\left( E - W + M c^2 + c\hbar\,
\widetilde{\alphab}_0 \dotprod {\bf k}' + \widetilde{\beta}_0 M c^2
\right) \right\}.
\label{4.29}\eeqa
Using the formulas of the traces of products of Dirac matrices
\citep[given, \eg, by][]{Heitler1954} this expression can be evaluated
analytically.  The result is
\beq
{\cal S}_{\rm L} = \frac{(2E - W + 2 M c^2)^2 - (c\hbar q)^2 }{4\,
(E+M c^2)(E-W+M c^2)}.
\label{4.30}\eeq

The summation in the transverse term, ${\cal S}_{\rm T}$, can be evaluated
similarly, although the calculation is more laborious. We replace the
spinors $ U_{{\bf k}', m'_{\rm S},+1}$ with $\Pi_{{\bf k}',+1} \, U_{{\bf
k}', m'_{\rm S},\tau}$, and extend the summation to positive- and
negative-energy spinors,
\beqa
{\cal S}_{\rm T} &=&
\frac{1}{2} \sum_{m_{{\rm S}'},\tau} U_{{\bf k}', m'_{\rm S},\tau}^\dagger
\, \Pi_{{\bf k}',+1} \left( \widetilde{\alphab}_0 - \frac{W}{c\hbar q^2}
\, {\bf q} \right) \, \dotprod \, {\bf G} \, \Pi_{{\bf k},+1} \, \left(
\widetilde{\alphab}_0 - \frac{W}{c\hbar q^2} \, {\bf q} \right) \,
\dotprod \, {\bf G}^\ast \; \Pi_{{\bf k}',+1} \, U_{{\bf k}', m'_{\rm
S},\tau}
\nonumber \\ [2mm]
&=& \frac{1}{2} {\rm Tr} \left\{
\Pi_{{\bf k}',+1}
\left( \widetilde{\alphab}_0\dotprod {\bf G}
- \frac{W}{c\hbar q^2} \, {\bf q}\dotprod {\bf G} \right) \,
\Pi_{{\bf k},+1} \, \left( \widetilde{\alphab}_0\dotprod {\bf G}^\ast -
\frac{W}{c\hbar q^2} \,
{\bf q}\dotprod {\bf G}^\ast \right) \,
\Pi_{{\bf k}',+1} \right\} ,
\nonumber \\ [2mm]
&=&
\frac{1}{2} {\rm Tr} \left\{ \left( \widetilde{\alphab}_0\dotprod {\bf G} -
\frac{W}{c\hbar q^2} \, {\bf q}\dotprod {\bf G} \right) \, \Pi_{{\bf
k},+1} \, \left( \widetilde{\alphab}_0\dotprod {\bf G}^\ast - \frac{W}{c\hbar q^2} \,
{\bf q}\dotprod {\bf G}^\ast \right) \Pi_{{\bf k}',+1} \right\}.
\label{4.31}\eeqa
Using Eq.\ \req{4.27}, we have
\beqa
{\cal S}_{\rm T} &=&
\frac{1}{8(E'+M c^2)\, (E+M c^2)} {\rm Tr} \left\{
\left( \widetilde{\alphab}_0\dotprod {\bf G}
- \frac{W}{c\hbar q^2} \, {\bf q}\dotprod {\bf G} \right)
\left( E + M c^2 + c\hbar\,
\widetilde{\alphab}_0 \dotprod {\bf k} + \widetilde{\beta}_0 M c^2
\right)
\right.
\nonumber \\ [2mm]
&& \times
\left.
\, \left( \widetilde{\alphab}_0\dotprod {\bf G}^\ast - \frac{W}{c\hbar q^2} \,
{\bf q}\dotprod {\bf G}^\ast \right) \,
\left( E' + M c^2 + c\hbar\,
\widetilde{\alphab}_0 \dotprod {\bf k}' + \widetilde{\beta}_0 M c^2
\right) \right\}.
\label{4.32}\eeqa
After a straightforward, but tedious, calculation, we obtain
\beqa
{\cal S}_{\rm T} &=&
\frac{1}{4(E+M c^2)\, (E-W+M c^2)}
\nonumber \\ [2mm]
&& \times
\left\{ \left[ \left(
c\hbar (2 {\bf k} - {\bf q} )
 - \frac{W}{c\hbar q^2}  (2E-W+2M c^2)
{\bf q} \right) \dotprod {\bf G} \right] \right.
\nonumber \\ [2mm]
&& \rule{0.5em}{0mm} \times
\left[ \left( c\hbar (2 {\bf k} -{\bf q})
 - \frac{W}{c\hbar q^2}  (2E-W+2M c^2)
{\bf q} \right) \dotprod {\bf G} \right]^\ast
\nonumber \\ [2mm]
&& \left. +
\left[ (c\hbar q)^2 - W^2 \right] \left[ ({\bf G}\dotprod {\bf G}^\ast)
- \frac{1}{q^2} ({\bf q}\dotprod {\bf G}^\ast) ({\bf q}\dotprod {\bf G})
\right] \right\}.
\label{4.33}\eeqa
We note that
\beqa
&& \! \! \! \! \! \! \! \! \! \! \! \! \! \! \! \! \! \! \! \! \! \!
\frac{1}{2(E+M c^2)} \left(
c\hbar (2 {\bf k} - {\bf q} )
 - \frac{W}{c\hbar q^2}  (2E-W+2M c^2)
{\bf q} \right)
\nonumber \\ [2mm]
&=& \frac{2 c\hbar {\bf k}}{2(E+M c^2)}
- \frac{ -W^2 + 2W (E+M c^2) + (c\hbar q)^2}{2 (c \hbar q^2)  (E+M c^2)}
\, {\bf q}
\nonumber \\ [2mm]
&=& \frac{2 c\hbar k}{2(E+M c^2)} \, \hat{\bf k}
- \frac{W}{(c\hbar q)} \left( 1 + \frac{(c\hbar q)^2-W^2}{2 W (E+M
c^2)}\right) \, \hat{\bf q}
\nonumber \\ [2mm]
&=& \beta \, (\hat{\bf k}
- \cos\theta_{\rm r} \, \hat{\bf q} )
= \beta \, [\hat{\bf k}
- (\hat{\bf k} \dotprod \hat{\bf q}) \, \hat{\bf q} ] = \beta \hat{\bf
k}_\perp,
\label{4.34}\eeqa
where $\beta=v/c$ and we have introduced the ``recoil
angle'' $\theta_{\rm r}$, the angle between the vectors $\hat {\bf k}$
and $\hat {\bf q}$ (see Fig.\ \ref{fig2}), which is given by Eq.\
\req{A.16},
\beqa
\cos\theta_{\rm r} = \hat{\bf k} \cdot \hat{\bf q} =
\frac{W}{\beta (c\hbar q)} \left( 1 + \frac{ (c\hbar q)^2-W^2}
{2 W (E+M c^2)} \right).
\label{4.35}\eeqa
The vector ${\bf k}_\perp =  \hat{\bf k} - (\hat{\bf k} \dotprod
\hat{\bf q}) \, \hat{\bf q} $ is the ``transverse'' component
(perpendicular to $\hat{\bf q}$) of $\hat{\bf k}$. The
magnitude of this vector is given by
\beq
{\bf k}_\perp^2 = 1-\cos^2\theta_{\rm r} = 1 -
\frac{W^2}{\beta^2 (c\hbar q)^2} \left( 1 + \frac{ (c\hbar q)^2-W^2}
{2 W (E+M c^2)} \right)^2.
\label{4.36}\eeq
With all this, we can write
\beqa
{\cal S}_{\rm T} &=&
\frac{E+M c^2}{E-W+M c^2}
\left\{ \beta^2 \left| {\bf k}_\perp \dotprod {\bf G} \right|^2
+ \frac{(c\hbar q)^2 - W^2}{4(E+M c^2)^2}\,
\left[ ({\bf G}\dotprod {\bf G}^\ast) - (\hat{\bf q}\dotprod
{\bf G}^\ast) (\hat{\bf q}\dotprod {\bf G}) \right] \right\}
\nonumber \\ [2mm]
&=&
\frac{E+M c^2}{E-W+M c^2}
\left\{ \beta^2 \left| \hat{\bf k}_\perp \dotprod {\bf G}_\perp \right|^2
+ \frac{(c\hbar q)^2 - W^2 }{4(E+M c^2)^2}\,
\left| {\bf G}_\perp \right|^2 \right\},
\label{4.37}\eeqa
where ${\bf G}_\perp = {\bf G} - ({\bf G} \dotprod \hat{\bf q}) \hat{\bf
q}$ is the ``transverse'' component of ${\bf G}$.  Note that
${\bf k}_\perp \dotprod {\bf G} = {\bf k}_\perp \dotprod {\bf G}_\perp$.

}

\begin{figure}[htb]
\begin{center}
\includegraphics*[width=5.5cm]{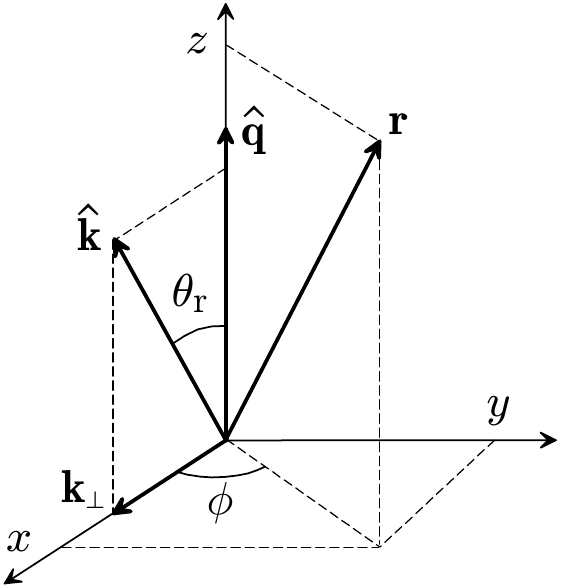}
\caption{
Reference frame used in the evaluation of the PWBA transition-matrix
elements.
\label{fig2}}
\end{center}\end{figure}

Now, the spin-averaged squared transition-matrix elements \req{4.22} can
be expressed as
\beqa
{\cal T}_1 &=&
\frac{Z_0^2 e^4}{4\pi^4} \, \frac{1}{q^4} \,
\frac{(2E -W + 2 M c^2)^2 - (c\hbar q)^2 }{4\,
(E+M c^2)(E-W+M c^2)}
\left|
\left< \psi_{b} \left|
\exp ({\rm i} {\bf q}\dotprod {\bf r}) \rule{0mm}{4mm}
\right| \psi_{n_a \kappa_a m_a} \right> \right|^2
\nonumber \\ [2mm]
&+& \frac{Z_0^2 e^4}{4\pi^4} \, \frac{1}{[q^2-(W/c\hbar)^2]^2} \,
\frac{E+M c^2}{E-W+M c^2}
\nonumber \\ [2mm]
&& \rule{5mm}{0mm} \times  \,
\left( \beta^2 \left| {\bf k}_\perp \dotprod {\bf G}_\perp \right|^2
+ \frac{(c\hbar q)^2 - W^2 }{4(E+ M c^2)^2}\,
\left| {\bf G}_\perp \right|^2 \right). \rule{14mm}{0mm}
\label{4.38}\eeqa

Continuing with the evaluation of the quantity \req{4.19}, we can now
perform the summations over $m_a$ and $m_b$,
\beqa
{\cal T}^{\rm eff}_{fi}
&=& \sum_{m_b}\sum_{m_{a}}
\frac{1}{2} \sum_{m_{{\rm S}'}, m_{{\rm S}}}
|T_{fi}^{\rm PW}|^2
= \sum_{m_b} \sum_{m_{a}} {\cal T}_1
\nonumber \\ [2mm]
&=&
\frac{Z_0^2 e^4}{4\pi^4} \, \frac{1}{q^4} \,
\frac{(2E -W + 2 M c^2)^2 - (c\hbar q)^2 }{4\,
(E+M c^2)(E-W+M c^2)}
\sum_{m_b}\sum_{m_{a}}  \left|
\left< \psi_{b} \left|
\exp ({\rm i} {\bf q}\dotprod {\bf r}) \rule{0mm}{4mm}
\right| \psi_{n_a \kappa_a m_a} \right> \right|^2
\nonumber \\ [2mm]
&+& \frac{Z_0^2 e^4}{4\pi^4} \, \frac{1}{[q^2-(W/c\hbar)^2]^2} \,
\frac{E+M c^2}{E-W+M c^2}
\nonumber \\ [2mm]
&& \rule{5mm}{0mm} \times
\sum_{m_b}\sum_{m_{a}}  \,
\left( \beta^2 \left| {\bf k}_\perp \dotprod {\bf G}_\perp \right|^2
+ \frac{(c\hbar q)^2 - W^2 }{4(E+ M c^2)^2}\,
\left| {\bf G}_\perp \right|^2 \right). \rule{5mm}{0mm}
\label{4.39}\eeqa
For the evaluation of the matrix elements, it is convenient to select a
reference frame with the $z$ axis in the direction of ${\bf q}$ and the
$x$ axis parallel to ${\bf k}_\perp$ (see Fig.\ \ref{fig2}). Noting the axial
symmetry of the system about this $z$ axis, we can write
\beqa
{\cal T}^{\rm eff}_{fi}
&=&
\frac{Z_0^2 e^4}{4\pi^4} \, \frac{1}{q^4} \,
\frac{(2E-W + 2 M c^2)^2 - (c\hbar q)^2 }{4\,
(E+M c^2)(E-W+M c^2)}
\sum_{m_b}\sum_{m_{a}}  \left|
\left< \psi_{b} \left|
\exp ({\rm i} {\bf q}\dotprod {\bf r}) \rule{0mm}{4mm}
\right| \psi_{n_a \kappa_a m_a} \right> \right|^2
\nonumber \\ [2mm]
&+& \frac{Z_0^2 e^4}{4\pi^4} \, \frac{1}{[q^2-(W/c\hbar)^2]^2} \,
\frac{E+M c^2}{E-W+M c^2}
\nonumber \\ [2mm]
&& \rule{5mm}{0mm}\times
\left( \beta^2 \sin^2\theta_{\rm r}
+ \frac{(c\hbar q)^2 - W^2 }{2(E+ M c^2)^2}\,
\right) \sum_{m_b}\sum_{m_{a}}  \,
\left| G_x \right|^2. \rule{10mm}{0mm}
\label{4.40}\eeqa

\subsection{Differential cross sections \label{sec4.2}}

We can now evaluate the DDCS for inelastic collisions with the $2j_a+1$
electrons in a closed subshell. As indicated above, for open subshells with
$q_a < 2j_a+1$ electrons, the DDCS is obtained by multiplying the DDCS
of the closed subshell by the fractional occupancy, $q_a /(2j_a+1)$. In
general, collisions may induce transitions of the active electron to
bound orbitals and to free orbitals.

\noindent $\bullet$ {\bf Excitation.} \\ [2mm]
The DDCS for excitations of electrons in the closed subshell
$n_a\kappa_a$ to states of the bound level $n_b\kappa_b$ is given by
Eq.\ \req{2.34},
\beqa
\frac{\d^2 \sigma^{\rm exc}_a}{\d W \, \d Q}
=  \frac{(2 \pi)^5}{c^2 \hbar^4 v^2} \, \frac{E-W+M c^2}{E+Mc^2} \,
(Q+\me c^2) \,
 {\cal I}_{fi}^{\rm PW},
\label{4.41}\eeqa
with
\beqa
{\cal I}_{fi}^{\rm PW}
&=& \delta(W-\epsilon_b+\epsilon_{n_a \kappa_a})
\, {\cal T}^{\rm eff,exc}_{fi}.
\label{4.42}\eeqa
The quantity ${\cal T}^{\rm eff,exc}_{fi}$ is the average squared
matrix element \req{4.40} with the final orbital of the active
electron, $\psi_b$ represented by a bound spherical orbital,
$\psi_{n_b\kappa_bm_b}$.

{\allowdisplaybreaks
\noindent $\bullet$ {\bf Ionization.} \\ [2mm]
The DDCS for ionization of the subshell $n_a\kappa_a$ is
given by Eq.\ \req{2.36},
\beqa
\frac{\d^2 \sigma^{\rm ion}_a}{\d W \, \d Q}
= \frac{(2 \pi)^5}{c^2 \hbar^4 v^2} \, \frac{E-W+M c^2}{E+Mc^2} \,
(Q+\me c^2)  \,
{\cal J}_{fi}^{\rm PW},
\label{4.43}\eeqa
where [see Eq.\ \req{2.45}]
\beqa
{\cal J}_{fi}^{\rm PW}
&=&
\frac{k_b}{\epsilon_b \pi}
\sum_{\kappa_b} \, {\cal T}^{\rm eff,ion}_{fi}.
\label{4.44}\eeqa
Here ${\cal T}^{\rm eff,ion}_{fi}$ is the average squared
matrix element \req{4.40} with $\psi_b$ replaced with free
spherical waves,
$\psi_{\epsilon_b\kappa_b m_b}$ ($\epsilon_b = \epsilon_a + W$).

The DDCS, for both excitation and ionization, is
\beqa
\frac{\d^2 \sigma_a}{\d W \, \d Q}
&=&  \frac{(2 \pi)^5}{c^2 \hbar^4 v^2} \, \frac{E-W+M c^2}{E+Mc^2} \,
(Q+\me c^2)
\nonumber \\ [2mm]
&& \times \left(
 \delta(W -\epsilon_{n_b \kappa_b}+\epsilon_{n_a \kappa_a})
\, {\cal T}^{\rm eff, exc}_{fi}
+ \frac{k_b}{\epsilon_b \pi}
\sum_{\kappa_b} \, {\cal T}^{\rm eff,ion}_{fi}
\right).
\label{4.45}\eeqa
After simple algebraic manipulations, we can express it in the form
{\allowdisplaybreaks
\beqa
\frac{\d^2 \sigma_a}{\d W \, \d Q}
&=& \frac{2\pi Z_0^2 e^4}{\me v^2} \,
\left[ \rule{0mm}{8mm}
\frac{2\me c^2}{WQ(Q+2\me c^2)} \right.
\nonumber \\ [2mm]
&& \rule{5mm}{0mm} \times
\left\{ \frac{(2 E -W  + 2 M c^2)^2 - Q(Q+2\me c^2)}{4\,
(E+M c^2)^2} \right\}
\frac{\d f_a(Q,W)}{\d W}
\nonumber \\ [2mm]
&+& \, \frac{2\me c^2 W}{[Q(Q+2\me c^2)-W^2]^2} \,
\nonumber \\ [2mm]
&& \rule{5mm}{0mm} \times
\left. \left( \beta^2 \sin^2\theta_{\rm r}
+ \left\{ \frac{Q(Q+2\me c^2)- W^2 }{2(E+ M c^2)^2}\,
\right\} \right) \frac{\d g_a(Q,W)}{\d W} \rule{0mm}{8mm} \right],
\label{4.46}\eeqa
}
with [see Eq.\ \req{4.35}]
\beq
\beta^2 \sin^2\theta_{\rm r} = \beta^2
- \frac{W^2}{Q(Q+2\me c^2)} \left( 1 + \frac{ Q(Q+2\me c^2)-W^2}
{2 W (E+M c^2)} \right)^2 \, .
\label{4.47}\eeq
In Eq.\ \req{4.46} we have introduced the longitudinal generalized
oscillator strength (GOS), defined by
\beqa
\frac{\d f_a(Q,W)}{\d W} &\equiv& \frac{W 2(Q+\me
c^2)}{Q(Q+2\me c^2)}\, \frac{k_b}{\epsilon_b \pi}
\nonumber \\ [2mm]
&& \rule{5mm}{0mm}\times
\sum_{\kappa_b,m_b} \sum_{m_a}
\left| \left<
\psi_{\epsilon_b \kappa_b m_b}
\left| \rule{0mm}{4mm}\exp\left( {\rm i}
{\bf q} \dotprod {\bf r} \right) \right| \psi_{n_a \kappa_a m_a}
\right> \right|^2
\nonumber \\ [2mm]
&+& \frac{W 2(Q+\me
c^2)}{Q(Q+2\me c^2)}\,
\sum_{n_b,\kappa_b}
\delta(W -\epsilon_{n_b \kappa_b}+\epsilon_{n_a \kappa_a})
\nonumber \\ [2mm]
&& \rule{5mm}{0mm}\times
\sum_{m_a,m_b} \left| \left< \psi_{n_b \kappa_b m_b}
\left| \rule{0mm}{4mm}\exp\left( {\rm i}
{\bf q} \dotprod {\bf r} \right) \right| \psi_{n_a \kappa_a m_a}
\right> \right|^2,
\label{4.48}\eeqa
and the transverse generalized oscillator strength (TGOS), defined as
follows
\beqa
\frac{\d g_{a}(Q,W)}{\d W} &\equiv& \frac{2(Q+\me
c^2)}{W}\, \frac{k_b}{\epsilon_b \pi}
\nonumber \\ [2mm]
&& \rule{5mm}{0mm}\times
\sum_{m_a} \sum_{\kappa_b,m_b}
 \left| \left<
\psi_{\epsilon_b \kappa_b m_b}
\left| \rule{0mm}{4mm} \widetilde {\alpha}_x \exp\left( {\rm i}
{\bf q} \dotprod {\bf r} \right) \right| \psi_{n_a \kappa_a m_a}
\right> \right|^2
\nonumber \\ [2mm]
&+&  \frac{2(Q+\me
c^2)}{W} \sum_{n_b,\kappa_b}  \, \delta(W -\epsilon_{n_b \kappa_b}+\epsilon_{n_a \kappa_a})
\nonumber \\ [2mm]
&& \rule{5mm}{0mm}\times
\sum_{m_a,m_b}  \left| \left< \psi_{n_b \kappa_b m_b}
\left| \rule{0mm}{4mm} \widetilde {\alpha}_x \exp\left( {\rm i}
{\bf q} \dotprod {\bf r} \right) \right| \psi_{n_a \kappa_a m_a}
\right> \right|^2.
\label{4.49}\eeqa
Because of the spherical symmetry of closed subshells, both the GOS and the
TGOS are functions of only the energy loss $W$ and the recoil energy
$Q$ (\ie, they depend only on the magnitude of the vector ${\bf q}$).
}

The numerical and kinematical factors in the definitions \req{4.48} and
\req{4.49} are such that, in the limit $Q\rightarrow 0$, both the
longitudinal and transverse GOSs reduce to the dipole optical
oscillator strength (OOS),
\beq
\frac{\d f_a(W)}{\d W} \equiv \lim_{Q\rightarrow 0}
\frac{\d f_a(Q,W)}{\d W}\, ,
\label{4.50}\eeq
which plays a fundamental role in the theory of stopping of charged
particles \citep{Fano1963, Ahlen1980}.
Expanding the exponential as a power series in ${\bf q} \cdot {\bf
r}$, recalling that the initial and final orbitals are orthogonal, and
keeping only the lowest-order non-vanishing term, we have
\beqa
\frac{\d f_a(W)}{\d W} &=& \frac{W \, 2\me}{\hbar^2}\,
\frac{k_b}{\epsilon_b \pi}
\sum_{\kappa_b} \sum_{m_a,m_b}
\left| \hat{\bf q} \dotprod \left<
\psi_{\epsilon_b \kappa_b m_b}
\left| \rule{0mm}{4mm} {\bf r} \right| \psi_{n_a \kappa_a m_a}
\right> \right|^2
\nonumber \\ [2mm]
&+& \frac{W \, 2\me}{\hbar^2}\,
\sum_{n_b,\kappa_b}
\delta(W -\epsilon_{n_b \kappa_b}+\epsilon_{n_a \kappa_a})
\sum_{m_a,m_b} \left| \hat{\bf q} \dotprod \left< \psi_{n_b \kappa_b m_b}
\left| \rule{0mm}{4mm}{\bf r} \right| \psi_{n_a \kappa_a m_a}
\right> \right|^2\, . \rule{10mm}{0mm}
\label{4.51}\eeqa
Because $\hat{\bf q}\dotprod {\bf r}$ is a scalar quantity (\ie, it is
invariant under rotations), the summation
\beqa
&& \sum_{m_a,m_b} \left| \hat{\bf q} \dotprod \left< \psi_{n_b \kappa_b m_b}
\left| \rule{0mm}{4mm}{\bf r} \right| \psi_{n_a \kappa_a m_a}
\right> \right|^2
\label{4.52}\eeqa
is independent of the direction of the unit vector $\hat{\bf q}$, as it
should be.

The equality of the GOS and the TGOS at $Q=0$ can be easily proved as
follows. From Eq.\ \req{4.49}, we can write
\beqa
\frac{\d g_{a}(0,W)}{\d W} &\equiv& \frac{2 \me
c^2}{W}\, \frac{k_b}{\epsilon_b \pi}
\sum_{m_a} \sum_{\kappa_b,m_b}
 \left| \left<
\psi_{\epsilon_b \kappa_b m_b}
\left| \rule{0mm}{4mm} \widetilde {\alpha}_x
\right| \psi_{n_a \kappa_a m_a}
\right> \right|^2
\nonumber \\ [2mm]
&+&  \frac{2 \me c^2}{W}
\sum_{n_b,\kappa_b}  \, \delta(W -\epsilon_{n_b \kappa_b}+\epsilon_{n_a \kappa_a})
\sum_{m_a,m_b}  \left| \left< \psi_{n_b \kappa_b m_b}
\left| \rule{0mm}{4mm} \widetilde {\alpha}_x
\right| \psi_{n_a \kappa_a m_a}
\right> \right|^2.
\nonumber\eeqa
Now, using the commutation relation $\widetilde{\alphab} = {\rm i}
(c\hbar)^{-1} \left[ {\cal H}_{\rm D}, {\bf r} \right]$, where ${\cal
H}_{\rm D}$ is the Dirac Hamiltonian \req{B.1}, we obtain
\beqa
\frac{\d g_{a}(0,W)}{\d W} &\equiv& \frac{W 2 \me}
{\hbar^2}\, \frac{k_b}{\epsilon_b \pi}
\sum_{m_a} \sum_{\kappa_b,m_b}
 \left| \left<
\psi_{\epsilon_b \kappa_b m_b}
\left| \rule{0mm}{4mm} x
\right| \psi_{n_a \kappa_a m_a}
\right> \right|^2
\nonumber \\ [2mm]
&+&  \frac{W 2 \me}{\hbar^2}
\sum_{n_b,\kappa_b}  \,
\delta(W -\epsilon_{n_b \kappa_b}+\epsilon_{n_a \kappa_a})
\sum_{m_a,m_b}  \left| \left< \psi_{n_b \kappa_b m_b}
\left| \rule{0mm}{4mm} x
\right| \psi_{n_a \kappa_a m_a}
\right> \right|^2\, , \rule{10mm}{0mm}
\label{4.53}\eeqa
which is equivalent to expression \req{4.51}.

For recoil energies $Q$ much larger than
the ionization energy $E_a = -\epsilon_{n_a\kappa_a}$, both the GOS and
the TGOS differ from zero only in the vicinity of the line $Q\sim W$,
the Bethe ridge, because the target electrons react as if they were
essentially free and at rest. The GOS and the TGOS for a free stationary
target electron are (see Appendix \ref{secD}),
\beq
\frac{\d f^{\rm free}(Q,W)}{\d W} = \delta(Q-W) \quad \mbox{and} \quad
\frac{\d g^{\rm free}(Q,W)}{\d W} = \delta(Q-W),
\label{4.54}\eeq
respectively. It is thus clear that, with the definitions
\req{4.48} and \req{4.49}, in the high-$Q$ limit the GOS and
the TGOS satisfy approximately the Bethe sum rule,
\beq
\int_0^\infty \frac{\d f_a(Q,W)}{\d W} \, \d W \simeq
2|\kappa_a|, \qquad
\int_0^\infty \frac{\d g_a(Q,W)}{\d W} \, \d W \simeq 2|\kappa_a|,
\label{4.55}\eeq
where $2|\kappa_a|=2j_a+1$ is the number of electrons in the active {\it
closed} subshell. Within the non-relativistic PWBA, \citet{Bethe1930}
proved that the longitudinal GOS satisfies the sum rule \req{4.55} for
all $Q$. The proof of this important sum rule can be found, \eg, in the
book of \citet{BetheJackiw1997}. The validity of the Bethe sum rule in
the relativistic theory will be analyzed in Section \ref{sec5.5}.

As mentioned above, our derivation of the DDCS \req{4.46} is similar to
that of \citet{Fano1963}, who explicitly assumed that the mass of the
projectile is much larger than the electron mass or, equivalently, that
the momentum transfer is much smaller than the momentum of the
projectile ($q \ll k$). The effect of this assumption is simply to
remove the quantities in braces on the right-hand side of Eq.\
\req{4.46}.

\subsection{Validity of the PWBA \label{sec4.3}}

The PWBA is expected to be valid for projectiles with sufficiently high
energies, whose wave functions are only slightly distorted by the
atomic potential. Qualitative arguments indicate that the approximation
is applicable when the speed $v$
of the projectile is much larger than the velocity $u$ of atomic (bound)
electrons \citep{MottMassey1965, Schiff1968}. In the least favorable
case of the K shell, the unscreened hydrogenic model gives
\beq
u^2 = \frac{2 \left< {\cal K} \right>}{\me} =
\frac{2}{\me} \left( \frac{Z^2}{2} \, \frac{\me e^4}{\hbar^2}
\right)
\label{4.56}\eeq
and the validity condition for the PWBA reads
\beq
\frac{u^2}{v^2} = \frac{Z}{\beta^2} \, \frac{e^4}{\hbar^2 c^2} =
\frac{Z^2 \alpha^2}{\beta^2} \ll 1,
\label{4.57}\eeq
where $\alpha = e^2/(\hbar c)\simeq 1/137$ is the fine-structure constant.
Hence, the PWBA should be applicable when
\beq
Z \alpha \ll \beta.
\label{4.58}\eeq
\citet{Inokuti1971} pointed out that the equivalence of the PWBA and the
impact-parameter approximation \citep{BetheJackiw1997}
implies that the PWBA should be valid when the impact-parameter
approximation is applicable, that is, when
\beq
\frac{\me}{M} \, Z \alpha \ll \beta,
\label{4.59}\eeq
where $M$ is the mass of the projectile. More quantitative indications
on the limits of validity of the PWBA can be obtained by comparing its
results with those from the more elaborate DWBA (see Section
\ref{sec2.2}). \citet{BoteSalvat2008} have performed PWBA and DWBA
calculations of ionization by electron impact for all the subshells of
atoms with $Z=1$ to 99 and concluded that the PWBA is valid (to within
1~\% or so) for electrons with kinetic energies larger than about 30 times
the ionization energy or, equivalently (invoking again the hydrogenic
formula for the binding energy)
\beq
v^2 \gtrsim 30 u^2 = 30 \, \frac{Ze^2}{n^2 \hbar^2}.
\nonumber \eeq
With evident rearrangements, we can write
\beq
\beta \gtrsim 6 \, \frac{Z\alpha}{n}
\label{4.60}\eeq
which agrees with the conditions stated above. Although the latter
result has been numerically verified only for electrons, we may conclude
that it is also valid for heavier projectiles, because their wave
functions are less distorted by the atomic field than those of electrons
(projectiles with larger masses accelerate less than electrons).


\section{Calculation of the GOS and the TGOS \label{sec5}}
\setcounter{equation}{0}

The longitudinal and transverse GOSs of a closed subshell are proportional
to the quantities
\beq
{\cal F}_{ba} \equiv \sum_{m_b,m_a} \left|
\left< \psi_{\epsilon_b \kappa_b m_b} \left|
\rule{0mm}{4mm}\exp\left( {\rm i} {\bf q} \dotprod {\bf r} \right)
\right| \psi_{n_a \kappa_a m_a} \right> \right|^2
\label{5.1}\eeq
and
\beq
{\cal G}_{ba} \equiv
\sum_{m_b,m_a} \left| \left< \psi_{\epsilon_b \kappa_b
m_b} \left| \rule{0mm}{4mm} \zetab \dotprod \widetilde {\alphab}
\exp\left( {\rm i} {\bf q} \dotprod {\bf r} \right) \right| \psi_{n_a
\kappa_a m_a} \right> \right|^2 \, ,
\label{5.2}\eeq
respectively. Here, $\zetab$ stands for an arbitrary unit
vector
perpendicular to ${\bf q}$. The initial and final orbitals of the active
electron are spherical waves \req{B.10} of the form
\beq
\psi_{\epsilon\kappa m}({\bf r}) = \frac{1}{r}
\left( \begin{array}{c}
P_{\epsilon\kappa}(r) \, \Omega_{\kappa m} (\hat{\bf r}) \\ [1mm]
{\rm i} Q_{\epsilon\kappa}(r) \,
\Omega_{-\kappa m} (\hat{\bf r}) \end{array} \right).
\label{5.3}\eeq
We recall that, in the case of inelastic collisions of charged particles
with atoms and ions, $\psi_{\epsilon_a \kappa_a m_a}$ are bound
orbitals; for excitation (ionization) $\psi_{\epsilon_b \kappa_b m_b}$
are bound (free) spherical waves.

In the non-relativistic theory, the transverse interaction is not
considered and the matrix elements of the longitudinal interaction have
the form \req{5.1} with Schr\"{o}dinger orbitals. To clarify the
relationship between the relativistic and non-relativistic theories, it is
convenient to consider the Schr\"{o}dinger orbitals in the ``coupled
representation'', $\psi_{\epsilon \kappa m}({\bf r}) = r^{-1}
P_{\epsilon \kappa}(r) \, \, \Omega_{\kappa m} (\hat{\bf r})$, in which
case the non-relativistic transition matrix elements are obtained from
the corresponding relativistic expressions by setting the small radial
function $Q_{\epsilon\kappa}(r)=0$.

\subsection{One-electron transition-matrix elements \label{sec5.1}}

Let us consider first the calculation of matrix elements of the operators
$\exp({\rm i} {\bf q} \, \dotprod \, {\bf r})$ and $\zetab \, \dotprod
\, \widetilde{\alphab} \; \exp({\rm i} {\bf q} \, \dotprod \, {\bf r})$
for one-electron transitions in the basis of spherical waves. These
matrix elements are reduced to sums of radial integrals by introducing
the Rayleigh expansion of a plane wave \citep{AbramowitzStegun1974},
\beq
\exp ({\rm i} {\bf q} \cdot {\bf r})
= \sum_{\lambda=0}^\infty \sum_{\mu=-\lambda}^\lambda
{\rm i}^{\lambda} (2 \lambda +1) \, j_{\lambda}(qr)
\, C_{\lambda\mu}^{\ast}(\hat{\bf q})
\, C_{\lambda\mu}(\hat{\bf r}) \, ,
\label{5.4}\eeq
where $j_\lambda (qr)$ are spherical Bessel functions, and $C_{\lambda
\mu}(\hat{\bf r})\equiv [4\pi/(2\lambda+1)]^{1/2} Y_{\lambda
\mu}(\hat{\bf r})$ are the components of Racah's spherical tensors. The
spin sums and angular integrals of the matrix elements can then be
calculated by means of the Wigner-Eckart theorem \citep[see,
\eg,][]{Edmonds1960}.

\vspace{3mm}

\noindent $\bullet$ {\bf Longitudinal interaction.} \\ [2mm]
The matrix elements
{\allowdisplaybreaks
\beq
F_{ba} \equiv \left<
\psi_{\epsilon_b \kappa_b m_b}
\left| \rule{0mm}{4mm}\exp\left( {\rm i}
{\bf q} \dotprod {\bf r} \right) \right| \psi_{\epsilon_a \kappa_a m_a}
\right>,
\label{5.5}\eeq
take the form
\beqa
F_{ba} &=& \sum_{\lambda\mu} {\rm i}^{\lambda} (2\lambda+1)
\, C_{\lambda \mu}^\ast (\hat{\bf q})
\nonumber \\ [2mm]
&& \mbox{} \times
\int \frac{\d {\bf r}}{r^2}
\left( \begin{array}{c}
P_{\epsilon_b \kappa_b}(r) \, \Omega_{\kappa_b m_b} (\hat{\bf r}) \\ [1mm]
{\rm i} Q_{\epsilon_b \kappa_b}(r) \,
\Omega_{-\kappa_b, m_b} (\hat{\bf r}) \end{array} \right)^\dagger \!
j_{\lambda}(qr)
\,C_{\lambda \mu}(\hat{\bf r})
\left( \begin{array}{c}
P_{\epsilon_a \kappa_a}(r) \, \Omega_{\kappa_a m_a} (\hat{\bf r})  \\ [1mm]
{\rm i} Q_{\epsilon_a \kappa_a}(r) \,
\Omega_{-\kappa_a, m_a} (\hat{\bf r})
\end{array} \right)
\nonumber \\ [2mm]
&=& \sum_{\lambda \mu} {\rm i}^{\lambda} (2 \lambda+1)
\, C_{\lambda \mu}^\ast (\hat{\bf q})
\int \frac{\d {\bf r}}{r^2}
\left\{ \rule{0mm}{5mm}
P_{\epsilon_b \kappa_b}(r) \, P_{\epsilon_a \kappa_a}(r) \left[
\Omega_{\kappa_b m_b} (\hat{\bf r})^\dagger
\,C_{\lambda \mu}(\hat{\bf r}) \,
\Omega_{\kappa_a m_a} (\hat{\bf r}) \right] \right.
\nonumber \\ [2mm]
&& \mbox{} \left. \rule{10mm}{0mm}
+  Q_{\epsilon_b \kappa_b}(r) \, Q_{\epsilon_a \kappa_a}(r) \left[
\Omega_{-\kappa_b, m_b} (\hat{\bf r})^\dagger
\,C_{\lambda \mu}(\hat{\bf r}) \,
\Omega_{-\kappa_a, m_a} (\hat{\bf r}) \right]
\rule{0mm}{5mm} \right\}
j_{\lambda}(qr).
\label{5.6}\eeqa
The angular integrals are matrix elements of Racah tensors, which we
replace with the coefficients $d^L (\kappa_1, m_1; \kappa_2, m_2) $, Eq.\
\req{C.9}. Thus, with the aid of the symmetry
relations \req{C.10} of these coefficients, we can write
\beqa
F_{ba} &=&
\sum_{\lambda} {\rm i}^{\lambda} (2\lambda+1) \, \left[
\sum_{\mu}  d^\lambda_\mu (\kappa_b, m_b; \kappa_a, m_a)
C_{\lambda \mu}^\ast (\hat{\bf q}) \right]
\, R_{\epsilon_b \kappa_b; \epsilon_a,\kappa_a}^{\lambda}(q),
\label{5.7}\eeqa
with the radial integrals
\beq
R_{\epsilon_b \kappa_b; \epsilon_a,\kappa_a}^{\lambda}(q) =
\int_0^\infty
\left[ \rule{0mm}{3.5mm}
P_{\epsilon_b\kappa_b}(r) P_{\epsilon_a\kappa_a}(r)
+ Q_{\epsilon_b\kappa_b}(r) Q_{\epsilon_a\kappa_a}(r)
\rule{0mm}{3.5mm} \right] \,
j_{\lambda}(qr) \, \d r,
\label{5.8}\eeq
which are independent of the magnetic quantum numbers. Note
that these integrals are real.

To simplify the formulas, we introduce the quantities
\beqa
{\cal X}_{\epsilon_b \kappa_b;\epsilon_a \kappa_a}^{\lambda} (q)
&\equiv & \sqrt{2\lambda+1} \,
\left< \ell_b \1o2 j_b||
{\bf C}^{(\lambda)} || \ell_a \1o2 j_a\right> \,
R_{\epsilon_b \kappa_b; \epsilon_a,\kappa_a}^{\lambda}(q)
\label{5.9}\eeqa
Considering the symmetry properties of the Clebsch--Gordan coefficients
\citep{Rose1995},
we can write
\beqa
F_{ba}(q) &=&
\sum_{\lambda} {\rm i}^{\lambda} (2\lambda+1) \,
\left[ \sum_{\mu}
\frac{1}{\sqrt{2j_b+1}}
\langle j_a, \lambda, m_a, \mu | j_b, m_b \rangle \;
C_{\lambda \mu}^\ast (\hat{\bf q}) \right]
\frac{1}{\sqrt{2\lambda+1}}
{\cal X}_{\epsilon_b \kappa_b;\epsilon_a \kappa_a}^{\lambda} (q)
\nonumber \\ [2mm]
&=&
(-1)^{j_a-m_a}
\sum_{\lambda} {\rm i}^{\lambda}
\left[ \sum_{\mu}
\langle j_a, j_b, m_a, -m_b | \lambda, -\mu \rangle \;
C_{\lambda \mu}^\ast (\hat{\bf q}) \right]
{\cal X}_{\epsilon_b \kappa_b;\epsilon_a \kappa_a}^{\lambda} (q).
\label{5.10}\eeqa
}

\noindent $\bullet$ {\bf Transverse interaction.} \\ [2mm]
The one-electron transition-matrix elements of the transverse
interaction, Eq.\ \req{4.4}, involve the operator $M \equiv \zetab \,
\dotprod \, \widetilde{\alphab} \exp({\rm i} {\bf q} \, \dotprod \, {\bf
r})$. This operator and its Hermitian conjugate describe, respectively,
the absorption and the emission of a photon with wave vector ${\bf q}$
and unit polarization vector $\zetab$ (cf. Section \ref{sec3.2}). Therefore,
matrix elements of this operator occur in many calculations of quantum
electrodynamics (\eg, photoelectric absorption, bremsstrahlung emission
by electrons, electron-positron pair production by photons, positron
annihilation). The relevant matrix elements are
\beq
M_{ba} = \zetab \, \dotprod \,
\left< \psi_{\epsilon_b \kappa_b m_b}
\left| \widetilde{\alphab} \exp({\rm i} {\bf q} \dotprod {\bf r}) \right|
\psi_{\epsilon_a \kappa_a m_a}
\right> \, ,
\label{5.11}\eeq
where, in general, the polarization vector $\zetab$ may be complex. For
the sake of generality, we assume that $\zetab$ is arbitrary, although
for vector potentials in the Coulomb gauge $\zetab$ is perpendicular to
the wave vector. It is convenient to introduce the vector [see Eq.\
\req{4.21}]
\beq
{\bf G} = \left< \psi_{\epsilon_b \kappa_b m_b}
\left| \widetilde{\alphab} \exp({\rm i} {\bf q} \dotprod {\bf r}) \right|
\psi_{\epsilon_a \kappa_a m_a}
 \right>,
\label{5.12}\eeq
and write
\beq
M_{ba} = \zetab \dotprod {\bf G}.
\label{5.13}\eeq
Calculations are easier if all vectors are expressed in terms of the
spherical unit vectors $\xib_\nu$ ($\nu=+1, 0, -1$) defined by Eq.\
\req{C.13}, whose properties are described, \eg, in \citet{Rose1995} and
\citet{Edmonds1960}. The expansion of the vector ${\bf G}$ in the
spherical basis has the form
\beq
{\bf G} = \sum_\nu (-1)^\nu G_\nu \xib_{-\nu} \qquad
\mbox{with} \qquad G_\nu = \xib_\nu \dotprod {\bf G},
\label{5.14}\eeq
where $G_\mu$ are the spherical components of ${\bf G}$.
The dot product of two vector quantities, $\zetab$ and ${\bf G}$, is
\beq
\zetab \dotprod {\bf G} = \sum_i \zeta_i G_i = \sum_\nu
(-1)^\nu \zeta_\nu G_{-\nu}\, ,
\label{5.15}\eeq
where $\zeta_i, G_i$ and  $\zeta_\nu, G_\nu$ are, respectively, the
Cartesian and spherical components of the two vectors.

To evaluate the vector ${\bf G}$, Eq.\ \req{5.12}, we introduce the
Rayleigh expansion of a plane wave, Eq.\ \req{5.4}, and write
\beqa
{\bf G} &=& \sum_{\lambda}
{\rm i}^{\lambda} (2\lambda+1) \sum_\mu
\left< \psi_{\epsilon_b \kappa_b m_b} \left|
\widetilde{\alphab} \, j_{\lambda}(qr)
\, C_{\lambda\mu}(\hat{\bf r})
\right| \psi_{\epsilon_a \kappa_a m_a}  \right>
\, C_{\lambda \mu}^\ast(\hat{\bf q})
\nonumber\\ [2mm]
&=& \sum_{\lambda}
{\rm i}^{\lambda} (2\lambda+1) \sum_\mu
\Gammab_{\lambda \mu} \, C_{\lambda \mu}^\ast(\hat{\bf q}) \, .
\label{5.16}\eeqa
The matrix elements
\beq
\Gammab_{\lambda \mu} \equiv
\left<
\psi_{\epsilon_b \kappa_b m_b}
\left| \widetilde{\alphab} \, j_{\lambda}(qr)
\,C_{\lambda \mu}(\hat{\bf r}) \rule{0mm}{4mm}
\right| \psi_{\epsilon_a \kappa_a m_a} \right>,
\label{5.17}\eeq
are calculated by a method similar to the one adopted by
\citet{MannJohnson1971} for a related problem. Inserting the spinor
representation of the $\widetilde{\alphab}$ matrices, we have
\beqa
\Gammab_{\lambda \mu} &=& \int \frac{\d {\bf r}}{r^2}
\left( \begin{array}{c}
P_{\epsilon_b \kappa_b}(r) \, \Omega_{\kappa_b m_b} (\hat{\bf r}) \\ [1mm]
{\rm i} Q_{\epsilon_b \kappa_b}(r) \,
\Omega_{-\kappa_b m_b} (\hat{\bf r}) \end{array} \right)^\dagger \!
j_{\lambda}(qr)
\,C_{\lambda \mu}(\hat{\bf r})
\left( \begin{array}{c}
{\rm i} Q_{\epsilon_a \kappa_a}(r) \,
\sigmab \Omega_{-\kappa_a m_a} (\hat{\bf r}) \\ [1mm]
P_{\epsilon_a \kappa_a}(r) \, \sigmab \Omega_{\kappa_a m_a} (\hat{\bf r})
\end{array} \right)
\nonumber\\ [2mm]
&=& {\rm i} \int C_{\lambda \mu}(\hat{\bf r})
\left\{\rule{0mm}{5mm}
F_{\epsilon_b \kappa_b;\epsilon_a \kappa_a}^\lambda \;
\left[ \Omega_{\kappa_b m_b}^\dagger (\hat{\bf r}) \, \sigmab \,
\Omega_{-\kappa_a m_a} (\hat{\bf r}) \right] \right.
\nonumber \\ [2mm]
&& \mbox{} \left. \rule{10mm}{0mm}
- G_{\epsilon_b \kappa_b;\epsilon_a \kappa_a}^\lambda
\left[ \Omega_{-\kappa_b m_b}^\dagger (\hat{\bf r}) \, \sigmab
\, \Omega_{\kappa_a m_a} (\hat{\bf r}) \right]
\rule{0mm}{5mm} \right\} \d \hat{\bf r}
\nonumber \\ [2mm]
&=& {\rm i} \sum_\nu (-1)^\nu
\left[
F_{\epsilon_b \kappa_b;\epsilon_a \kappa_a}^\lambda \;
c_{\lambda \mu}^\nu(\kappa_b, m_b; -\kappa_a, m_a) \right.
\nonumber \\ [2mm]
&& \mbox{} \left. \rule{10mm}{0mm}
- G_{\epsilon_b \kappa_b;\epsilon_a \kappa_a}^\lambda \;
c_{\lambda \mu}^\nu(-\kappa_b, m_b; \kappa_a, m_a)
\right] \xib_{-\nu}
\label{5.18}\eeqa
where we have introduced the expansion \req{C.18}, with coefficients
[see Eq.\ \req{C.19}]
\beqa
c_{\lambda \mu}^\nu(\kappa_b m_b; \kappa_a m_a)
&=& \sqrt{\frac{1}{2\lambda+1}} \sum_{J}
\left< \lambda, 1, \mu, \nu \left| J,\mu+\nu \right>  \right.
\nonumber \\ [2mm]
\nonumber \\ [2mm]
&& \! \! \! \! \! \! \! \! \! \! \! \! \! \! \! \! \! \! \!
\mbox{} \times \left[
\sqrt{J+1} \, \left( \frac{\kappa_b+\kappa_a}{J+1} + 1 \right)
d^J_{\nu+\mu}(\kappa_b,m_b;-\kappa_a, m_a) \,
\delta_{J,\lambda-1}
\right.
\nonumber \\ [2mm]
&& \! \! \! \! \! \! \! \! \! \! \! \! \! \! \! \! \! \! \!
\mbox{} - \sqrt{\frac{2J+1}{J(J+1)}}\;
(\kappa_b-\kappa_a)
\, d^J_{\nu+\mu}(\kappa_b,m_b;\kappa_a, m_a) \,
\delta_{J,\lambda}
\nonumber \\ [2mm]
&& \! \! \! \! \! \! \! \! \! \! \! \! \! \! \! \! \! \! \!
\mbox{} + \left.
\sqrt{J} \, \left( \frac{\kappa_b+\kappa_a}{J}
- 1 \right) d^J_{\nu+\mu} \left( \kappa_b,m_b;-\kappa_a, m_a \right) \,
\delta_{J,\lambda+1}
\right],
\label{5.19}\eeqa
and the radial integrals
\beqa
F_{\epsilon_b \kappa_b;\epsilon_a \kappa_a}^\lambda &\equiv& \int_0^\infty
P_{\epsilon_b \kappa_b}(r) \,  Q_{\epsilon_a \kappa_a}(r) \,
j_{\lambda}(qr) \, \d r,
\nonumber \\ [2mm]
G_{\epsilon_b \kappa_b;\epsilon_a \kappa_a}^\lambda &\equiv& \int_0^\infty
Q_{\epsilon_b \kappa_b}(r) \,  P_{\epsilon_a \kappa_a}(r) \,
j_{\lambda}(qr) \, \d r,
\label{5.20}\eeqa
which are both real and independent of the magnetic quantum numbers.

We can thus write the vector ${\bf G}$ as
\beq
{\bf G} = \sum_{\nu} (-1)^\nu \,
G_\nu (\epsilon_b \kappa_b m_b; \epsilon_a \kappa_a m_a; {\bf q})
 \, \xib_{-\nu}
 \label{5.21}\eeq
with
\beqa
G_\nu(\epsilon_b \kappa_b m_b; \epsilon_a \kappa_a m_a; {\bf q})
&=& \sum_{\lambda,\mu}
{\rm i}^{\lambda+1} \,
(2\lambda+1)
\left[
F_{\epsilon_b \kappa_b;\epsilon_a \kappa_a}^\lambda \;
c_{\lambda \mu}^\nu(\kappa_b, m_b; -\kappa_a, m_a) \right.
\nonumber \\ [2mm]
&& \mbox{} \left.
- G_{\epsilon_b \kappa_b;\epsilon_a \kappa_a}^\lambda \;
c_{\lambda \mu}^\nu(-\kappa_b, m_b; \kappa_a, m_a)
\right] C_{\lambda \mu}^\ast(\hat{\bf q}).
\label{5.22}\eeqa
Introducing the expression \req{5.19}, using the symmetry property
\req{C.10c} of the $d^J_M$ coefficients, and reorganizing
terms, we have
\beqa
&& \! \! \! \! \! \! \! \! \! \! \! \!
G_\nu(\epsilon_b \kappa_b m_b; \epsilon_a \kappa_a m_a; {\bf q})
= \sum_{\lambda,\mu} {\rm i}^{\lambda+1} \,
\sqrt{2\lambda+1} \, C_{\lambda \mu}^\ast(\hat{\bf q}) \sum_J
\left< \lambda, 1, \mu, \nu \left| J,\mu+\nu \right>  \right.
\nonumber \\ [0mm]
&& \mbox{} \times \left\{
F_{\epsilon_b \kappa_b;\epsilon_a \kappa_a}^\lambda
\sqrt{J} \, \left( \frac{\kappa_b-\kappa_a}{J}
- 1 \right) d^J_{\nu+\mu} \left( \kappa_b,m_b;\kappa_a, m_a \right) \,
\delta_{J,\lambda+1} \right.
\nonumber \\ [0mm]
&& \mbox{} \rule{7mm}{0mm} +
G_{\epsilon_b \kappa_b;\epsilon_a \kappa_a}^\lambda
\sqrt{J} \, \left( \frac{\kappa_b-\kappa_a}{J}
+ 1 \right) d^J_{\nu+\mu} \left( \kappa_b,m_b;\kappa_a, m_a \right) \,
\delta_{J,\lambda+1}
\nonumber \\ [0mm]
&& \mbox{} \rule{7mm}{0mm} -
F_{\epsilon_b \kappa_b;\epsilon_a \kappa_a}^\lambda
\sqrt{\frac{2J+1}{J(J+1)}}\;
(\kappa_b+\kappa_a)
\, d^J_{\nu+\mu}(\kappa_b,m_b;-\kappa_a, m_a) \,
\delta_{J,\lambda}
\nonumber \\ [0mm]
&& \mbox{} \rule{7mm}{0mm} -
G_{\epsilon_b \kappa_b;\epsilon_a \kappa_a}^\lambda
\sqrt{\frac{2J+1}{J(J+1)}}\;
(\kappa_b+\kappa_a)
\, d^J_{\nu+\mu}(\kappa_b,m_b;-\kappa_a, m_a) \,
\delta_{J,\lambda}
\nonumber \\ [0mm]
&& \mbox{} \rule{7mm}{0mm} +
F_{\epsilon_b \kappa_b;\epsilon_a \kappa_a}^\lambda
\sqrt{J+1} \, \left(
\frac{\kappa_b -\kappa_a}{J+1} + 1 \right)
d^J_{\nu+\mu}(\kappa_b,m_b;\kappa_a, m_a) \,
\delta_{J,\lambda-1}
\nonumber \\ [0mm]
&& \mbox{} \rule{7mm}{0mm} + \left.
G_{\epsilon_b \kappa_b;\epsilon_a \kappa_a}^\lambda
\sqrt{J+1} \, \left(
\frac{\kappa_b-\kappa_a}{J+1} - 1 \right)
d^J_{\nu+\mu}(\kappa_b,m_b;\kappa_a, m_a) \,
\delta_{J,\lambda-1}  \right\}.
\label{5.23}\eeqa

The formulas become simpler if we consider a reference frame with the
$z$ axis parallel to the direction $\hat{\bf q}$ of the photon. In such
a frame, $C_{\lambda \mu}^{\ast} (\hat{\bf q})=\delta_{\mu 0}$, and
writing explicitly the terms corresponding to the various values of
$\lambda$ ($\lambda = J, J\pm 1$), we have
\beqa
G_\nu &\equiv&
G_\nu(\epsilon_b \kappa_b m_b; \epsilon_a \kappa_a m_a; {\bf z})
\nonumber \\ [0mm]
&=& \sum_J \left\{
{\rm i}^{J} \sqrt{2J-1}
\left< J-1, 1, 0, \nu \left| J,\nu \right>  \right.
d^J_{\nu} \left( \kappa_b,m_b;\kappa_a, m_a \right) \, \sqrt{J} \right.
\nonumber \\ [0mm]
&& \mbox{} \rule{7mm}{0mm} \times \left[
F_{\epsilon_b \kappa_b;\epsilon_a \kappa_a}^{J-1}
\left( \frac{\kappa_b-\kappa_a}{J} - 1 \right)
+ G_{\epsilon_b \kappa_b;\epsilon_a \kappa_a}^{J-1}
\left( \frac{\kappa_b-\kappa_a}{J} + 1 \right) \right]
\nonumber \\ [0mm]
&& \mbox{} - {\rm i}^{J+1} \sqrt{2J+1}
\left< J, 1, 0, \nu \left| J,\nu \right>  \right.
d^J_{\nu} \left( \kappa_b,m_b;-\kappa_a, m_a \right) \,
\sqrt{\frac{2J+1}{J(J+1)}}
\nonumber \\ [0mm]
&& \mbox{} \rule{7mm}{0mm} \times (\kappa_b+\kappa_a) \left[
F_{\epsilon_b \kappa_b;\epsilon_a \kappa_a}^J
+ G_{\epsilon_b \kappa_b;\epsilon_a \kappa_a}^J \right]
\nonumber \\ [0mm]
&& \mbox{} + {\rm i}^{J+2} \sqrt{2J+3}
\left< J+1, 1, 0, \nu \left| J,\nu \right>  \right.
d^J_{\nu} \left( \kappa_b,m_b;\kappa_a, m_a \right) \,
\sqrt{J+1}
\nonumber \\ [0mm]
&& \mbox{} \rule{7mm}{0mm} \times
\left. \left[
F_{\epsilon_b \kappa_b;\epsilon_a \kappa_a}^{J+1}
\left( \frac{\kappa_b -\kappa_a}{J+1} + 1 \right) +
G_{\epsilon_b \kappa_b;\epsilon_a \kappa_a}^{J+1}
\left( \frac{\kappa_b-\kappa_a}{J+1} - 1 \right) \right] \right\}.
\label{5.24}\eeqa

\begin{table}[h!]
\caption{\rm Clebsch--Gordan coefficients
$\left< J_1, 1, M-\mu, \mu | J , -M \right>$
\citep[see, \eg,][]{Rose1995}.
\label{tab1}}
\vskip 4mm
\begin{center}
\begin{tabular}{lcc} \hline\hline
$J_1=$ & $\mu=\pm 1$ & $\mu = 0$ \rule[-3mm]{0mm}{9mm}\\ \hline
$J-1$ \rule{3mm}{0mm} & \rule[+1mm]{0mm}{9mm}
$\displaystyle{ \left[ \frac{(J\mp M-1)(J\mp M)}{2 J(2J-1)}
\right]^{1/2}}$ & \rule{3mm}{0mm}
$\displaystyle{ \left[ \frac{(J-M)(J+M)}{J(2J-1)}
\right]^{1/2}}$ \rule{3mm}{0mm} \\ [6mm]
$J$ & \rule{3mm}{0mm}
$\displaystyle{ \mp \left[ \frac{(J\mp M)(J\pm M+1)}{2J(J+1)}
\right]^{1/2}}$ \rule{3mm}{0mm} &
$\displaystyle{ - \,\frac{M}{[\, J(J+1)]^{1/2}}}$ \\ [6mm]
$J+1$ &
$\displaystyle{ \left[ \frac{(J\pm M+1)(J\pm M+2)}{2(J+1)(2J+3)}
\right]^{1/2}}$ &
$\displaystyle{ - \, \left[ \frac{(J-M+1)(J+M+1)}{(J+1)(2J+3)}
\right]^{1/2}}$
\rule[-7mm]{0mm}{4mm} \\ \hline\hline
\end{tabular} \end{center} \end{table}

\noindent Inserting the values of the Clebsch--Gordan coefficients (see, Table
\ref{tab1}),
\beq
\begin{array}{rclrcl}
\langle J-1, 1, 0, 0 | J, 0 \rangle &=&
\displaystyle{\sqrt{\frac{J}{2J-1}},}\rule{10mm}{0mm}&
\langle J-1, 1, 0, \pm 1 | J, \pm1 \rangle &=&
\displaystyle{\sqrt{\frac{J+1}{2(2J-1)}},} \\ [4mm]
\langle J, 1, 0, 0 | J, 0 \rangle &=&
\displaystyle{0,}&
\langle J, 1, 0, \pm 1 | J, \pm1 \rangle &=&
\displaystyle{\mp\sqrt{\frac{1}{2}},} \\ [4mm]
\langle J+1, 1, 0, 0 | J, 0 \rangle &=&
\displaystyle{-\sqrt{\frac{J+1}{2J+3}},}&
\langle J+1, 1, 0, \pm 1 | J, \pm1 \rangle &=&
\displaystyle{\sqrt{\frac{J}{2(2J+3)}},}
\end{array}
\label{5.25}\eeq
after simple algebraic manipulations we obtain
\beqa
G_0
&=&\sum_{J} {\rm i}^J d^J_{0} \left( \kappa_b,m_b;\kappa_a, m_a \right)
\nonumber \\ [2mm]
&& \mbox{} \left\{ (\kappa_b - \kappa_a) \left(
F_{\epsilon_b \kappa_b;\epsilon_a \kappa_a}^{J-1} +
G_{\epsilon_b \kappa_b;\epsilon_a \kappa_a}^{J-1} \right)
- J \left (F_{\epsilon_b \kappa_b;\epsilon_a \kappa_a}^{J-1} -
G_{\epsilon_b \kappa_b;\epsilon_a \kappa_a}^{J-1} \right)
\rule{0mm}{5mm}\right.
\nonumber \\ [2mm]
&& \mbox{} \left. + (\kappa_b - \kappa_a) \left(
F_{\epsilon_b \kappa_b;\epsilon_a \kappa_a}^{J+1} +
G_{\epsilon_b \kappa_b;\epsilon_a \kappa_a}^{J+1} \right)
+(J+1) \left (F_{\epsilon_b \kappa_b;\epsilon_a \kappa_a}^{J+1} -
G_{\epsilon_b \kappa_b;\epsilon_a \kappa_a}^{J+1} \right)
\rule{0mm}{5mm}\right\}
\label{5.26}\eeqa
and
\beqa
G_{\pm 1} &=& \sum_{J=1}^\infty {\rm i}^{J}
\frac{2J+1}{\sqrt{2 J(J+1)}} \left\{  \rule{0mm}{7mm}
\frac{J(J+1)}{2J+1}
d^J_{\pm 1} \left( \kappa_b,m_b;\kappa_a, m_a \right) \right.
\nonumber \\ [2mm]
&& \mbox{} \rule{7mm}{0mm} \times \left[
\frac{\kappa_b-\kappa_a}{J} \left(
F_{\epsilon_b \kappa_b;\epsilon_a \kappa_a}^{J-1}
+ G_{\epsilon_b \kappa_b;\epsilon_a \kappa_a}^{J-1} \right)
- \left(
F_{\epsilon_b \kappa_b;\epsilon_a \kappa_a}^{J-1}
- G_{\epsilon_b \kappa_b;\epsilon_a \kappa_a}^{J-1} \right)
\right.
\nonumber \\ [2mm]
&& \mbox{} \rule{7mm}{0mm} \left.
- \frac{\kappa_b-\kappa_a}{J+1} \left(
F_{\epsilon_b \kappa_b;\epsilon_a \kappa_a}^{J+1}
+ G_{\epsilon_b \kappa_b;\epsilon_a \kappa_a}^{J+1} \right)
- \left(
F_{\epsilon_b \kappa_b;\epsilon_a \kappa_a}^{J+1}
- G_{\epsilon_b \kappa_b;\epsilon_a \kappa_a}^{J+1} \right)
\right]
\nonumber \\ [2mm]
&& \mbox{} \pm {\rm i}
d^J_{\pm 1} \left( \kappa_b,m_b;-\kappa_a, m_a \right) \,
\left. (\kappa_b+\kappa_a) \left(
F_{\epsilon_b \kappa_b;\epsilon_a \kappa_a}^J
+ G_{\epsilon_b \kappa_b;\epsilon_a \kappa_a}^J \right) \rule{0mm}{7mm}
\right\}
\label{5.27}\eeqa

Following the tradition, we introduce the radial integrals,
\beqa
^{\rm l}{\cal R}_{\epsilon_b \kappa_b;\epsilon_a \kappa_a}^{J}
&=& \frac{1}{2J+1} \left[ \rule{0mm}{5mm} (\kappa_b-\kappa_a) \,
\left( F_{\epsilon_b \kappa_b;\epsilon_a \kappa_a}^{J-1}
+ G_{\epsilon_b \kappa_b;\epsilon_a \kappa_a}^{J-1} \right)
- J \left( F_{\epsilon_b \kappa_b;\epsilon_a \kappa_a}^{J-1}
- G_{\epsilon_b \kappa_b;\epsilon_a \kappa_a}^{J-1} \right) \right.
\nonumber \\ [2mm]
&& \mbox{} + \left. \rule{0mm}{5mm}
(\kappa_b-\kappa_a) \,
\left( F_{\epsilon_b \kappa_b;\epsilon_a \kappa_a}^{J+1}
+ G_{\epsilon_b \kappa_b;\epsilon_a \kappa_a}^{J+1} \right)
+ (J+1) \left( F_{\epsilon_b \kappa_b;\epsilon_a \kappa_a}^{J+1}
- G_{\epsilon_b \kappa_b;\epsilon_a \kappa_a}^{J+1} \right) \right]
\nonumber \\
\label{5.28}\eeqa
\beqa
^{\rm e}{\cal R}_{\epsilon_b \kappa_b;\epsilon_a \kappa_a}^{J}
&=& \frac{J(J+1)}{2J+1}
\left[\rule{0mm}{5mm}
\frac{\kappa_b-\kappa_a}{J} \,
\left( F_{\epsilon_b \kappa_b;\epsilon_a \kappa_a}^{J-1}
+ G_{\epsilon_b \kappa_b;\epsilon_a \kappa_a}^{J-1} \right)
- \left( F_{\epsilon_b \kappa_b;\epsilon_a \kappa_a}^{J-1}
- G_{\epsilon_b \kappa_b;\epsilon_a \kappa_a}^{J-1} \right) \right.
\nonumber \\ [2mm]
&& \left.
- \frac{\kappa_b-\kappa_a}{J+1} \,
\left( F_{\epsilon_b \kappa_b;\epsilon_a \kappa_a}^{J+1}
+ G_{\epsilon_b \kappa_b;\epsilon_a \kappa_a}^{J+1} \right)
- \left( F_{\epsilon_b \kappa_b;\epsilon_a \kappa_a}^{J+1}
- G_{\epsilon_b \kappa_b;\epsilon_a \kappa_a}^{J+1} \right)
\rule{0mm}{5mm}\right],
\label{5.29}\eeqa
and
\beqa
^{\rm m}{\cal R}_{\epsilon_b \kappa_b;\epsilon_a \kappa_a}^{J}
&=& (\kappa_a+\kappa_b) \,
\left( F_{\epsilon_b \kappa_b;\epsilon_a \kappa_a}^J
+ G_{\epsilon_b \kappa_b;\epsilon_a \kappa_a}^J \right).
\label{5.30}\eeqa
The superscripts ``l'', ``e'' and ``m'' stand for ``longitudinal'',
``electric'' and ``magnetic'', respectively, because these radial
integrals also arise in an alternative treatment based on the multipole
expansion of the radiation field \citep[see, \eg,][]{Scofield1975,
Scofield1978}.  Expressions \req{5.33} are more suited for numerical
evaluation than equivalent forms of these integrals given in the
literature. The latter are obtained by inserting the integrals
\req{5.19} and making use of the recurrence relations of the spherical
Bessel functions \citep{AbramowitzStegun1974}. This procedure gives
\beqa
^{\rm l}{\cal R}_{\epsilon_b \kappa_b;\epsilon_a \kappa_a}^{J}
&=& \int_0^\infty \d r \left\{ (\kappa_b-\kappa_a) \,
\left[P_{\epsilon_b \kappa_b}(r) \, Q_{\epsilon_a \kappa_a}(r)
+ \rule{0mm}{4mm}
Q_{\epsilon_b \kappa_b}(r) \, P_{\epsilon_a \kappa_a}(r) \right]
\frac{1}{qr}\, j_{J} (qr)  \right.
\nonumber \\ [2mm]
&& \left. \mbox{} - \left [P_{\epsilon_b \kappa_b}(r)
\, Q_{\epsilon_a \kappa_a}(r)
- \rule{0mm}{4mm}
Q_{\epsilon_b \kappa_b}(r) \, P_{\epsilon_a \kappa_a}(r) \right]
\frac{\d }{\d (qr)} \, j_{J} (qr) \rule{0mm}{5mm}\right\}\, ,
\label{5.31}\eeqa
\beqa
^{\rm e}{\cal R}_{\epsilon_b \kappa_b;\epsilon_a \kappa_a}^{J}
&=& \int_0^\infty \d r \left\{\rule{0mm}{5mm}
(\kappa_b-\kappa_a) \,
\left[P_{\epsilon_b \kappa_b}(r) \, Q_{\epsilon_a \kappa_a}(r)
+ \rule{0mm}{4mm}
Q_{\epsilon_b \kappa_b}(r) \, P_{\epsilon_a \kappa_a}(r) \right] \right.
\nonumber \\ [2mm]
&& \mbox{} \rule{5mm}{0mm} \times \left[ \frac{\d j_{J}(qr)}{\d (qr)}
+ \frac{1}{qr} j_{J}(qr) \right]
\nonumber \\ [2mm]
&& \mbox{} - \left.
\left[P_{\epsilon_b \kappa_b}(r) \, Q_{\epsilon_a \kappa_a}(r)
- \rule{0mm}{4mm}
Q_{\epsilon_b \kappa_b}(r) \, P_{\epsilon_a \kappa_a}(r) \right]
\frac{J(J+1)}{qr}\, j_J(qr) \right\} \, , \rule{10mm}{0mm}
\label{5.32}\eeqa
and
\beqa
^{\rm m}{\cal R}_{\epsilon_b \kappa_b;\epsilon_a \kappa_a}^{J}
&=& (\kappa_b+\kappa_a) \, \int_0^\infty \d r
\left[P_{\epsilon_b \kappa_b}(r) \, Q_{\epsilon_a \kappa_a}(r)
+ \rule{0mm}{4mm}
Q_{\epsilon_b \kappa_b}(r) \, P_{\epsilon_a \kappa_a}(r) \right]
 j_{J}(qr) \, . \rule{10mm}{0mm}
\label{5.33}\eeqa

We can thus write
\beq
G_0 = \sum_{J=0}^{\infty} {\rm i}^{J} \, (2J+1) \,
d^{J}_{0}(\kappa_b, m_b;\kappa_a, m_a)
\, {^{\rm l}{\cal R}_{\epsilon_b \kappa_b;\epsilon_a \kappa_a}^{J}}
\label{5.34}\eeq
and
\beqa
G_{\pm 1} &=& \sum_{J=1}^{\infty} {\rm i}^{J} \,
\frac{2J+1}{\sqrt{2J(J+1)}} \,
\left\{
\rule{0mm}{5mm}
d^{J}_{\pm 1}(\kappa_b, m_b;\kappa_a, m_a) \,
{^{\rm e}{\cal R}_{\epsilon_b \kappa_b;\epsilon_a \kappa_a}^{J}}
\right.
\nonumber \\ [2mm]
&& \left. \pm  {\rm i} \,
d^{J}_{\pm 1}(\kappa_b, m_b;-\kappa_a, m_a) \,
{^{\rm m}{\cal R}_{\epsilon_b \kappa_b;\epsilon_a \kappa_a}^{J}}
\rule{0mm}{5mm}\right\}.
\label{5.35}\eeqa
It should be noted that the coefficients $d^{J}_{M}(\kappa_b,
m_b;\kappa_a, m_a)$ vanish unless $M=m_b-m_a$. Moreover, the product of
$d^{J}_{M}(\kappa_b, m_b;\kappa_a, m_a)$ and $d^{J}_{M}(\kappa_b,
m_b;-\kappa_a, m_a)$ equals zero, Eq.\ \req{C.10d}, which implies that
one of these two coefficients is necessarily null. That is, at least
three of the four coefficients in the expression in curly braces on the
right-hand side of Eq.\ \req{5.35} vanish.

To simplify the formulas, we introduce the quantities
\begin{subequations}
\label{5.36}
\beqa
{^{\rm l}{\cal Y}_{\epsilon_b \kappa_b;\epsilon_a \kappa_a}^{J}}
&\equiv &  \sqrt{2J+1} \,
\left<\ell_b \1o2 j_b || {\bf C}^{(J)} || \ell_a
\1o2 j_a\right>
{^{\rm l}{\cal R}_{\epsilon_b \kappa_b;\epsilon_a \kappa_a}^{J}}
\label{5.36a}\\ [2mm]
{^{\rm e}{\cal Y}_{\epsilon_b \kappa_b;\epsilon_a \kappa_a}^{J}}
&\equiv & \sqrt{\frac{2J+1}{2J(J+1)}} \,
\left<\ell_b \1o2 j_b || {\bf C}^{(J)} || \ell_a
\1o2 j_a\right>
{^{\rm e}{\cal R}_{\epsilon_b \kappa_b;\epsilon_a \kappa_a}^{J}}
\label{5.36b}\\ [2mm]
{^{\rm m}{\cal Y}_{\epsilon_b \kappa_b;\epsilon_a \kappa_a}^{J}}
&\equiv & \sqrt{\frac{2J+1}{2J(J+1)}} \,
\left<\ell_b \1o2 j_b || {\bf C}^{(J)} || \overline{\ell}_a
\1o2 j_a\right>
{^{\rm m}{\cal R}_{\epsilon_b \kappa_b;\epsilon_a \kappa_a}^{J}}.
\label{5.36c}\eeqa
\end{subequations}
Considering the symmetry properties of the Clebsch--Gordan coefficients,
we can write
\begin{subequations}
\label{5.37}
\beqa
G_{0} &=& \sum_{J=0}^{\infty}
{\rm i}^J \, (2J+1) \,
\frac{1}{\sqrt{2j_b+1}}
\langle j_a, J, m_a, 0 | j_b, m_b \rangle \; \frac{1}{\sqrt{2J+1}}
{^{\rm l}{\cal Y}_{\epsilon_b \kappa_b;\epsilon_a \kappa_a}^{J}}
\nonumber \\ [2mm]
&=& (-1)^{j_a-m_a} \sum_{J=0}^{\infty}
{\rm i}^J \,
\langle j_a, j_b, m_a, -m_b | J, 0 \rangle \;
{^{\rm l}{\cal Y}_{\epsilon_b \kappa_b;\epsilon_a \kappa_a}^{J}}.
\label{5.37a}\eeqa
and
\beqa
G_{\pm 1} &=& \sum_{J=1}^{\infty}
{\rm i}^J \, \frac{2J+1}{\sqrt{2J(J+1)}} \,
\frac{1}{\sqrt{2j_b+1}} \,
\langle j_a, J, m_a, \pm 1 | j_b, m_b \rangle
\nonumber \\ [2mm]
&& \mbox{} \times
\sqrt{\frac{2J(J+1)}{2J+1}}
\left\{
{^{\rm e}{\cal Y}_{\epsilon_b \kappa_b;\epsilon_a \kappa_a}^{J}}
\pm {\rm i} \;
{^{\rm m}{\cal Y}_{\epsilon_b \kappa_b;\epsilon_a \kappa_a}^{J}}
\rule{0mm}{5mm}\right\}
\nonumber \\ [2mm]
&=& (-1)^{j_a-m_a} \sum_{J=1}^{\infty}
{\rm i}^J \,
\langle j_a, j_b, m_a, -m_b | J, \mp 1 \rangle
\left\{
{^{\rm e}{\cal Y}_{\epsilon_b \kappa_b;\epsilon_a \kappa_a}^{J}}
\pm {\rm i} \;
{^{\rm m}{\cal Y}_{\epsilon_b \kappa_b;\epsilon_a \kappa_a}^{J}}
\rule{0mm}{5mm}\right\} \rule{15mm}{0mm}
\label{5.37b}\eeqa
\end{subequations}

Finally, inserting the expressions \req{5.37} into \req{5.15},
the matrix elements \req{5.11} can be obtained as
\beqa
M_{ba} &=& \zetab \dotprod {\bf G} = \sum_\nu (-1)^\nu \zeta_{-\nu}
G_{\nu}\, .
\label{5.38}\eeqa
Since electromagnetic potentials are expressed in the Coulomb gauge, the
photon polarization vector $\zetab$ is perpendicular to the wave vector
${\bf q}=q\hat{\bf z}$ and, consequently, $\zeta_0 =0$. When  photon
polarization states are represented in the basis of linear polarization,
$\hat\epsilonb_1= \hat{\bf x}$ and $\hat\epsilonb_2= \hat{\bf y}$), the
spherical components of these vectors are $x_{\pm} = \xib_{\pm 1}
\dotprod \hat{\bf x} = \mp 1/\sqrt{2}$ and $y_{\pm} = \xib_{\pm 1}
\dotprod \hat{\bf y} = - {\rm i}/\sqrt{2}$, respectively. Alternatively,
photon polarization states can be expressed in the basis of circular
polarization $\hat\epsilonb_1= -\xib_{+1}$ (right-handed) and
$\hat\epsilonb_1= \xib_{-1}$ (left-handed), which has the advantage of
having simpler spherical components.

\subsection{Transition probabilities for closed subshells \label{sec5.2}}

We can now calculate the averaged sums of squares of the transition
matrix elements, Eqs.\ \req{5.1} and \req{5.2}.  Because the operators
in these matrix elements are irreducible tensor operators, the
Wigner-Eckart theorem \citep[see, \eg,][]{Edmonds1960} implies that
these averaged sums can be expressed as sums of squares of the
corresponding reduced matrix elements, which are independent of the
magnetic quantum numbers, multiplied by radial integrals.

For the longitudinal interaction we have
\beq
{\cal F}_{ba} = \sum_{m_b,m_a} \left| F_{ba} \right|^2.
\label{5.39}\eeq
Using the sum rule \req{C.10i}, we obtain
\beqa
{\cal F}_{ba} &=& \sum_{m_b,m_a}
\sum_{\lambda'\mu'} (-{\rm i})^{\lambda'} (2 \lambda'+1) \,
C_{\lambda'\mu'} (\hat{\bf q})
\, d^{\lambda'}_{\mu'} (\kappa_b, m_b; \kappa_a, m_a)
\, R_{\epsilon_b \kappa_b; \epsilon_a,\kappa_a}^{\lambda'}(q)
\nonumber \\ [2mm]
&& \mbox{} \times
\sum_{\lambda\mu} {\rm i}^{\lambda} (2 \lambda+1) \,
C_{\lambda \mu}^\ast (\hat{\bf q})
\, d^\lambda_\mu (\kappa_b, m_b; \kappa_a, m_a)
\, R_{\epsilon_b \kappa_b; \epsilon_a,\kappa_a}^{\lambda}(q)
\nonumber \\ [2mm]
&=& \sum_{\lambda} (2 \lambda+1) \, \left< \ell_a \1o2
j_a|| {\bf C}^{(\lambda)} || \ell_b \1o2 j_b\right>^2 \, \left[
R_{\epsilon_b \kappa_b; \epsilon_a,\kappa_a}^{\lambda}(q) \right]^2
\sum_\mu C_{\lambda\mu}^\ast (\hat{\bf q}) C_{\lambda\mu} (\hat{\bf q}).
\nonumber\eeqa
Now, recalling that the addition theorem of spherical harmonics
implies the equality
\beq
\sum_\mu C_{\lambda\mu}^\ast (\hat{\bf q}) C_{\lambda\mu} (\hat{\bf q})
= 1,
\label{5.40}\eeq
we can write
\beqa
{\cal F}_{ba}
&=& \sum_{\lambda} (2\lambda+1) \,
\left< \ell_b \1o2 j_b||
{\bf C}^{(\lambda)} ||
\ell_a \1o2 j_a\right>^2 \,
\left[
R_{\epsilon_b \kappa_b; \epsilon_a,\kappa_a}^{\lambda}(q) \right]^2
\nonumber \\ [2mm]
&=& \sum_{\lambda}
\left[
{\cal X}_{\epsilon_b \kappa_b; \epsilon_a,\kappa_a}^{\lambda}(q)
\right]^2.
\label{5.41}\eeqa
This result shows that ${\cal F}_{ba}$ is independent of the direction
of the vector ${\bf q}$, as we have anticipated from general
symmetry arguments [see our comment after Eq.\ \req{4.49}].

{\allowdisplaybreaks
The sum of squared transition-matrix elements for the
transverse interaction, Eq.\ \req{5.2}, is
\beqa
{\cal G}_{ba} &=& \sum_{m_a,m_b} \left| M_{ba}
\right|^2 = \sum_{m_a,m_b} \left(
\sum_\nu (-1)^{\nu}\,\zeta_{-\nu} \, G_{\nu} \right)^\ast
\left( \sum_{\nu'} (-1)^{\nu'}\, \zeta_{-\nu'}
G_{\nu'} \right)
\nonumber \\ [2mm]
&=& \sum_{\nu, \nu'} (-1)^{\nu+\nu'}\,\zeta_{-\nu}^\ast \,
\zeta_{-\nu'}
\sum_{m_a,m_b}
G_{\nu}^\ast G_{\nu'} \, .
\label{5.42}\eeqa
We note that the only dependence on the quantum numbers $m_a$ and $m_b$
is through the coefficients $d^L_{M}(\kappa_1,m_1;\kappa_2,m_2)$. The
summation over these quantum numbers is performed easily by using
Eqs.\ \req{C.10}. We thus obtain,
\begin{subequations}
\label{5.43}
\beq
\sum_{m_a,m_b} G_0^\ast G_0 =
\sum_{J=0}^{\infty} (2J+1)
\left< \ell_b \1o2 j_b|| {\bf C}^{(J)} ||
\ell_a \1o2 j_a\right>^2
\left[ {^{\rm l}{\cal R}_{\epsilon_b \kappa_b;
\epsilon_a \kappa_a}^{J}} \right]^2
= \sum_{J=0}^{\infty}
\left[ {^{\rm l}{\cal Y}_{\epsilon_b \kappa_b;
\epsilon_a \kappa_a}^{J}} \right]^2,
\label{5.43a}\eeq
\beq
\sum_{m_a,m_b} G_0^\ast G_\pm = 0,
\label{5.43b}\eeq
\beqa
\sum_{m_a,m_b} G_\pm^\ast G_\pm &=&
\sum_{J=1}^{\infty} \,
\frac{2J+1}{2J(J+1)} \left\{
\left< \ell_b \1o2 j_b|| {\bf C}^{(J)}
|| \ell_a \1o2 j_a\right>^2
\left[{^{\rm e}{\cal R}_{\epsilon_b \kappa_b;\epsilon_a \kappa_a}^{J}}
\right]^2
\nonumber \right. \\ [2mm]
&& \rule{30mm}{0mm} \left. +
\left< \ell_b \1o2 j_b|| {\bf C}^{(J)} ||
\overline{\ell}_a \1o2 j_a\right>^2
\left[{^{\rm m}{\cal R}_{\epsilon_b \kappa_b;\epsilon_a \kappa_a}^{J}}
\right]^2 \right\}
\nonumber \\ [2mm]
&=&
\sum_{J=1}^{\infty} \,
\left\{
\left[{^{\rm e}{\cal Y}_{\epsilon_b \kappa_b;\epsilon_a \kappa_a}^{J}}
\right]^2
+
\left[{^{\rm m}{\cal Y}_{\epsilon_b \kappa_b;\epsilon_a \kappa_a}^{J}}
\right]^2 \right\},
\label{5.43c}\eeqa
and
\beq
\sum_{m_a,m_b} G_\pm^\ast G_\mp = 0,
\label{5.43d}\eeq
\end{subequations}
where $\overline{\ell}_a$ is the orbital angular momentum quantum number
corresponding to $-\kappa_a$. Therefore,
\beqa
\lefteqn{
{\cal G}_{ba} =
|\zeta_{0}|^2 \sum_{m_a,m_b} G_{0} G_{0}^\ast
+|\zeta_{-1}|^2 \sum_{m_a,m_b} G_{+1} G_{+1}^\ast
+|\zeta_{+1}|^2 \sum_{m_a,m_b} G_{-1} G_{-1}^\ast
}
\nonumber \\ [2mm]
&=& |\zeta_{0}|^2
\sum_{J}
\left[ {^{\rm l}{\cal Y}_{\epsilon_b \kappa_b;\epsilon_a \kappa_a}^{J}}
\right]^2
+ \left[ |\zeta_{-1}|^2 + |\zeta_{+1}|^2 \right]
\sum_{J} \, \left\{
\left[{^{\rm e}{\cal Y}_{\epsilon_b \kappa_b;\epsilon_a \kappa_a}^{J}}
\right]^2
\left[{^{\rm m}{\cal Y}_{\epsilon_b \kappa_b;\epsilon_a \kappa_a}^{J}}
\right]^2 \right\}\, . \rule{10mm}{0mm}
\label{5.44}\eeqa
Because the photon polarization vector $\zetab$ is perpendicular
to the wave vector ${\bf q}=q\hat{\bf z}$, $\zeta_0
=0$, the first term on the right-hand side of Eq.\ \req{5.44} vanishes,
and
\beqa
{\cal G}_{ba} &=&
\left[ |\zeta_{-1}|^2 + |\zeta_{+1}|^2 \right]
\sum_{J} \,
\left\{
\left[{^{\rm e}{\cal Y}_{\epsilon_b \kappa_b;\epsilon_a \kappa_a}^{J}}
\right]^2
+
\left[{^{\rm m}{\cal Y}_{\epsilon_b \kappa_b;\epsilon_a \kappa_a}^{J}}
\right]^2 \right\} \, .
\label{5.45}\eeqa
The definition \req{4.49} of the TGOS involves these average matrix
elements with $\zetab = \hat{\bf x}$, which corresponds to $\zeta_{\pm
1} = \xib_{\pm 1} \dotprod \hat {\bf x} = \mp 1/\sqrt{2}$.
}


\subsection{GOS and TGOS of closed subshells \label{sec5.3}}

We can now write explicit formulas for the GOS and the TGOS of a closed
subshell $n_a\kappa_a$ with ionization energy $E_a = -
\epsilon_{n_a\kappa_a}$. Introducing the result \req{5.41}, the
longitudinal GOS, Eq.\ \req{4.48}, is expressed as
\beqa
\frac{\d f_a (Q,W)}{\d W} &=& \frac{W}{Q} \; \frac{2(Q+\me
c^2)}{Q+2\me c^2}\, \frac{k_b}{\epsilon_b \pi}
\nonumber \\ [2mm]
&& \rule{5mm}{0mm}\times
\sum_{\kappa_b}
\sum_{\lambda} (2\lambda+1) \, \left< \ell_b \1o2
j_b|| {\bf C}^{(\lambda)} || \ell_a \1o2 j_a\right>^2 \, \left[
R_{\epsilon_b \kappa_b; n_a \kappa_a}^{\lambda}(q) \right]^2
\nonumber \\ [2mm]
&+& \frac{W}{Q} \;  \frac{2(Q+\me c^2)}{Q+2\me c^2}\,
\sum_{n_b,\kappa_b}
\delta(W -\epsilon_{n_b \kappa_b}+\epsilon_{n_a \kappa_a})
\nonumber \\ [2mm]
&& \rule{5mm}{0mm}\times
\sum_{\lambda} (2\lambda+1) \, \left< \ell_b \1o2
j_b|| {\bf C}^{(\lambda)} || \ell_a \1o2 j_a\right>^2 \, \left[
R_{n_b \kappa_b; n_a \kappa_a}^{\lambda}(q) \right]^2,
\label{5.46}\eeqa
with the radial integrals
\beq
R_{\epsilon_b \kappa_b; n_a \kappa_a}^{\lambda}(q) =
\int_0^\infty
\left[ P_{\epsilon_b\kappa_b}(r) P_{n_a\kappa_a}(r)
+ Q_{\epsilon_b\kappa_b}(r) Q_{n_a\kappa_a}(r) \right] \,
j_{\lambda}(qr) \, \d r\, .
\label{5.47}\eeq
Similarly, with the aid of \req{5.45} the transverse generalized
oscillator strength (TGOS), Eq.\ \req{4.49}, can be written as
\beqa
\frac{\d g_a(Q,W)}{\d W} &=& \frac{2(Q+\me
c^2)}{W}\, \frac{k_b}{\epsilon_b \pi}
\nonumber \\ [2mm]
&& \rule{5mm}{0mm}\times
\sum_{\kappa_b}
\sum_{J} \,
\frac{2J+1}{2J(J+1)} \left\{
\left< \ell_b \1o2 j_b|| {\bf C}^{(J)}
|| \ell_a \1o2 j_a\right>^2
\left[{^{\rm e}{\cal R}_{\epsilon_b \kappa_b;n_a \kappa_a}^{J}}
(q)\right]^2
\nonumber \right. \\ [2mm]
&& \rule{30mm}{0mm} \left. +
\left< \ell_b \1o2 j_b|| {\bf C}^{(J)} ||
\overline{\ell}_a \1o2 j_a\right>^2
\left[{^{\rm m}{\cal R}_{\epsilon_b \kappa_b;n_a \kappa_a}^{J}}
(q)\right]^2 \right\}
\nonumber \\ [2mm]
&+&  \frac{2(Q+\me
c^2)}{W} \sum_{n_b,\kappa_b}  \, \delta(W -\epsilon_{n_b \kappa_b}+\epsilon_{n_a \kappa_a})
\nonumber \\ [2mm]
&& \rule{5mm}{0mm}\times
\sum_{J} \,
\frac{2J+1}{2J(J+1)} \left\{
\left< \ell_b \1o2 j_b|| {\bf C}^{(J)}
|| \ell_a \1o2 j_a\right>^2
\left[{^{\rm e}{\cal R}_{n_b \kappa_b;n_a \kappa_a}^{J}}
(q)\right]^2
\nonumber \right. \\ [2mm]
&& \rule{30mm}{0mm} \left. +
\left< \ell_b \1o2 j_b|| {\bf C}^{(J)} ||
\overline{\ell}_a \1o2 j_a\right>^2
\left[{^{\rm m}{\cal R}_{n_b \kappa_b;n_a \kappa_a}^{J}}
(q)\right]^2 \right\} \, ,
\label{5.48}\eeqa
with the radial integrals
\begin{subequations}
\label{5.49}
\beqa
^{\rm e}{\cal R}_{\epsilon_b \kappa_b;n_a \kappa_a}^{J}(q) &=&
\frac{J(J+1)}{2J+1}
\left[\rule{0mm}{5mm}
- \frac{\kappa_b-\kappa_a}{J} \,
(F_{\epsilon_b \kappa_b;n_a \kappa_a}^{J-1}
+ G_{\epsilon_b \kappa_b;n_a \kappa_a}^{J-1})
+ (F_{\epsilon_b \kappa_b;n_a \kappa_a}^{J-1}
- G_{\epsilon_b \kappa_b;n_a \kappa_a}^{J-1}) \right.
\nonumber \\ [2mm]
&& \left.
+ \frac{\kappa_b-\kappa_a}{J+1} \,
(F_{\epsilon_b \kappa_b;n_a \kappa_a}^{J+1}
+ G_{\epsilon_b \kappa_b;n_a \kappa_a}^{J+1})
+ (F_{\epsilon_b \kappa_b;n_a \kappa_a}^{J+1}
- G_{\epsilon_b \kappa_b;n_a \kappa_a}^{J+1}) \rule{0mm}{5mm}\right]
\label{5.49a}\eeqa
and
\beqa
^{\rm m}{\cal R}_{\epsilon_b \kappa_b;n_a \kappa_a}^{J} (q) &=&
(\kappa_a+\kappa_b) \,
(F_{\epsilon_b \kappa_b;n_a \kappa_a}^J
+ G_{\epsilon_b \kappa_b;n_a \kappa_a}^J) \, ,
\label{5.49b}\eeqa
\end{subequations}
where
\beqa
F_{\epsilon_b \kappa_b;n_a \kappa_a}^J &=& \int_0^\infty
P_{\epsilon_b \kappa_b}(r) \,  Q_{n_a \kappa_a}(r) \,
j_J(qr) \, \d r,
\nonumber \\ [2mm]
G_{\epsilon_b \kappa_b;n_a \kappa_a}^J &=& \int_0^\infty
Q_{\epsilon_b \kappa_b}(r) \,  P_{n_a \kappa_a}(r) \,
j_J(qr) \, \d r.
\label{5.50}\eeqa

Incidentally, we can now obtain the optical oscillator strength (OOS),
Eq.\ \req{4.51}, as the $Q\rightarrow 0$ limit of the GOS. Using the
expansion of the spherical Bessel functions for small arguments
\citep[see][]{AbramowitzStegun1974},
\beq
j_\ell(x) = \frac{x^\ell}{(2\ell+1)!!}
\left[ 1 - \frac{x^2/2}{1! (2\ell+3)}
+ \frac{(x^2/2)^2}{2! (2\ell+3) (2\ell+5)} - \cdots \right],
\label{5.51}\eeq
we have
\beq
\lim_{q\rightarrow 0}
R_{\epsilon_b\kappa_b;n_a\kappa_a}^{\lambda}(q) =
\frac{q^\lambda}{(2\lambda+1)!!} \int_0^\infty
\left[ P_{\epsilon_b\kappa_b}(r) P_{n_a\kappa_a}(r)
+ Q_{\epsilon_b\kappa_b}(r) Q_{n_a\kappa_a}(r) \right] \,
r^\lambda \, \d r.
\label{5.52}\eeq
Because of the orthogonality of the initial and final orbitals, these
integrals vanish for $\lambda=0$. The lowest-order non-vanishing
contributions are from the dipole terms ($\lambda=1$) and give
\beqa
\frac{\d f_a (W)}{\d W} &\equiv&
\lim_{Q\rightarrow 0} \frac{\d f_a(Q,W)}{\d W} =
\frac{W \, 2\me}{3 \hbar^2}\,
\frac{k_b}{\epsilon_b\pi} \,
\sum_{\kappa_b}
\left< \ell_b \1o2 j_b|| {\bf C}^{(1)} ||
\ell_a \1o2 j_a\right>^2 \,
\left[ R_{\epsilon_b\kappa_b;n_a\kappa_a}^{\rm dip} \right]^2,
\nonumber \\ [2mm]
&& \rule{-15mm}{0mm}
+ \frac{W \, 2\me}{3 \hbar^2}\,
\sum_{n_b, \kappa_b} \, \delta(W -\epsilon_b+\epsilon_{n_a \kappa_a})
\,
\left< \ell_b \1o2 j_b|| {\bf C}^{(1)} ||
\ell_a \1o2 j_a\right>^2 \,
\left[ R_{n_b\kappa_b;n_a\kappa_a}^{\rm dip} \right]^2, \rule{10mm}{0mm}
\label{5.53}\eeqa
with
\beqa
R_{\epsilon_b\kappa_b;n_a\kappa_a}^{\rm dip} =
\int_0^\infty
\left[ P_{\epsilon_b\kappa_b}(r) P_{n_a\kappa_a}(r)
+ Q_{\epsilon_b\kappa_b}(r) Q_{n_a\kappa_a}(r) \right] \,
r \, \d r.
\label{5.54}\eeqa
We recall that in the $Q\rightarrow 0$ limit the TGOS also coincides
with the optical oscillator strength (see Section \ref{sec4.2}).

\subsection{Numerical methods \label{sec5.4}}

The theory described above has been implemented in the Fortran program
{\sc gosat}, which computes the GOS and the TGOS for closed subshells of
atoms and positive ions using central-field orbitals calculated by
solving numerically the Dirac equation for the DHFS self-consistent
potential. The GOSs for given values of $W$ and $Q$ are calculated as
the sums of series where each term is the squared product of a reduced
matrix element and a radial integral, Eqs.\ \req{5.46} and \req{5.48}.
The integrands [Eqs.\ \req{5.47} and \req{5.49}] are products of the
radial functions of the initial and the final orbitals and a spherical
Bessel function, which must be evaluated for a sufficiently dense grid
of radii extending from $r=0$ up to an outer radius where the radial
functions of the bound state practically vanish.

The calculation of the GOS for ionizing collisions is based on a
previous program, originally developed by \citet{Segui2003}, which has
been improved and extended to include the calculation of the TGOS.  The
program {\sc gosat} calculates also the GOS and the TGOS for excitations
of the active subshell electrons to discrete bound levels. As in the
original code of Segui \etal, radial Dirac functions are calculated
using the subroutine package {\sc radial}
\citep{SalvatFernandezVarea2019}, which implements a numerical algorithm
that effectively avoids the accumulation of truncation errors.  From a
comparison with analytical non-relativistic hydrogenic GOSs, Segui
\etal\ estimated that the GOSs given by their program were accurate to
five significant digits for relatively wide ranges of the energy
transfer $W$ and the recoil energy $Q$. A similar analysis indicates
that {\sc gosat} is slightly more accurate.

The {\sc radial} subroutines (and any other numerical solution
algorithm) have difficulties to solve the Dirac equation for highly
excited bound orbitals with small binding energies, because the radial
Dirac equations have to be integrated outwards up to the outer turning
point, which is very far from the nucleus. In principle, the {\sc
radial} subroutines are able to determine radial functions of bound
states with principal quantum number $n_b$ up to $\sim 35$, if the
radial grid extends up to sufficiently large $r$ and is dense enough. In
the present calculations, radial functions are evaluated with a grid of
15,000 radii, which are spaced non-uniformly (logarithmically near the
origin and uniformly at large radii) to accurately describe the fast
oscillations of the radial functions near the nucleus as well as the
slower oscillations at large radii. Even with such generous grids, it is
not possible to compute GOSs for transitions to final bound orbitals
with $n_b$ larger than about 30. The {\sc radial} subroutines also have
difficulties to calculate wave functions of very slow free electrons,
with kinetic energies less than $\sim 10^{-4}$ atomic units, and
moderate orbital angular momenta, because the radial wave equations need
to be integrated up to large radii beyond the turning point of the
centrifugal barrier ($r_{\rm TP} \sim \ell / k$) before they can be
properly normalized.  That is, neither the GOS nor the TGOS can be
directly evaluated for energy losses $W$ very close to the ionization
energy $E_{a}$.

\begin{figure}[hbt]
\begin{center}
\includegraphics*[width=11.5cm]{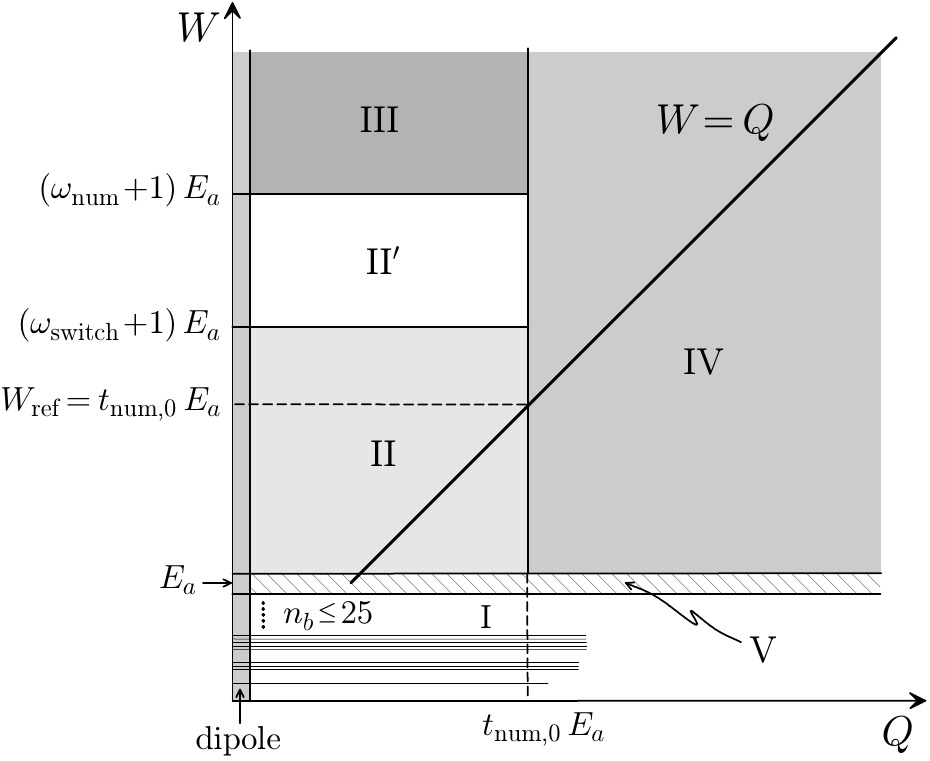}
\caption{Regions of the $(Q,W)$ plane where different strategies for
calculating the GOSs are used (not to scale). The left band, at low
recoil energies, represents the dipole region, where the GOS and the
TGOS reduce to the OOS.
\label{fig3}}
\end{center}\end{figure}

The program {\sc gosat} calculates the OOS and the GOSs for each
individual electron subshell. The numerical calculation is performed only
for limited $W$ and $Q$ ranges, and suitable extrapolations need to be
applied to generate the GOS densities outside those ranges. It is worth
noting that the accuracy of the calculated DDCSs is determined not only
by the accuracy with which the numerical GOSs are calculated, but also
by the reliability of the algorithms used to interpolate and extrapolate
the numerical GOS tables. Figure \ref{fig3} displays regions of
the $(Q,W)$ plane where different calculation schemes are adopted.
The characteristic energies in this diagram are given in units of the
ionization energy $E_a$ of the active electron subshell. For $Q < 10^{-4}
E_a$ the dipole approximation is applicable, \ie, the GOS and the TGOS
are practically equal to the OOS.

\subsubsection{Excitation to bound states \label{sec5.4.1}}

The longitudinal GOS for excitation to bound levels is [see Eqs.\
\req{5.46} and \req{5.47}]
\beqa
\frac{\d f_a^{\rm exc}(Q,W)}{\d W} &=&
\sum_{n_b,\kappa_b}
\delta(W -\epsilon_{n_b \kappa_b}+\epsilon_{n_a \kappa_a})
\, f_{ba}(Q)
\label{5.55}\eeqa
with
\beq
f_{ba}(Q) \equiv
\frac{W}{Q} \; \frac{2(Q+\me c^2)}{Q+2\me c^2}
\sum_{\lambda} (2\lambda+1) \, \left< \ell_b \1o2
j_b|| {\bf C}^{(\lambda)} || \ell_a \1o2 j_a\right>^2 \, \left[
R_{n_b \kappa_b; n_a \kappa_a}^{\lambda}(q) \right]^2
\label{5.56}\eeq
The TGOS for excitation is [Eqs.\ \req{5.48} to \req{5.50}]
\beq
\frac{\d g_a^{\rm exc}(Q,W)}{\d W} =
\sum_{n_b,\kappa_b}
\delta(W -\epsilon_{n_b \kappa_b}+\epsilon_{n_a \kappa_a})
\, g_{ba}(Q)
\label{5.57}\eeq
with
\beqa
g_{ba}(Q) &\equiv&
\frac{2(Q+\me c^2)}{W} \sum_{J} \,
\frac{2J+1}{2J(J+1)} \left\{
\left< \ell_b \1o2 j_b|| {\bf C}^{(J)}
|| \ell_a \1o2 j_a\right>^2
\left[{^{\rm e}{\cal R}_{n_b \kappa_b;n_a \kappa_a}^{J}}
\right]^2
\nonumber \right. \\ [2mm]
&& \rule{30mm}{0mm} \left. +
\left< \ell_b \1o2 j_b|| {\bf C}^{(J)} ||
\overline{\ell}_a \1o2 j_a\right>^2
\left[{^{\rm m}{\cal R}_{n_b \kappa_b;n_a \kappa_a}^{J}}
\right]^2 \right\}.
\label{5.58}\eeqa

The program {\sc gosat} calculates the GOS and TGOS for discrete
transitions from the active subshell $n_a\kappa_a$ to levels $\epsilon_{n_b
\kappa_b}$ with $n_b \le 25$ (region I in Fig.\ \ref{fig3}). Pauli's
exclusion principle forbids transitions to fully occupied levels (closed
shells). Nevertheless, {\sc gosat} calculates the GOSs for transitions
to all discrete energy levels, including those with energies
$\epsilon_{n_b \kappa_b}$ less than $\epsilon_{n_a \kappa_a}$, which are
normally occupied and for which the functions $f_{ba}(Q)$ and
$g_{ba}(Q)$ are negative. The GOS and the TGOS are computed for a
logarithmic grid of 512 recoil energies, spanning the range from $Q =
10^{-4} E_a$ up to the largest $Q$ value which can be calculated with
the adopted radial grid of 15,000 radii; beyond this recoil energy the
GOS is exceedingly small, usually smaller than $\sim \, 10^{-10}$ times
its maximum value. Using a logarithmic grid of recoil energies is
advantageous here, because it allows the subsequent integration of the
DDCS over $Q$ to be performed easily, \eg, by Simpson's method.

\begin{figure}[htbp]
\begin{center}
\includegraphics*[width=7.5cm]{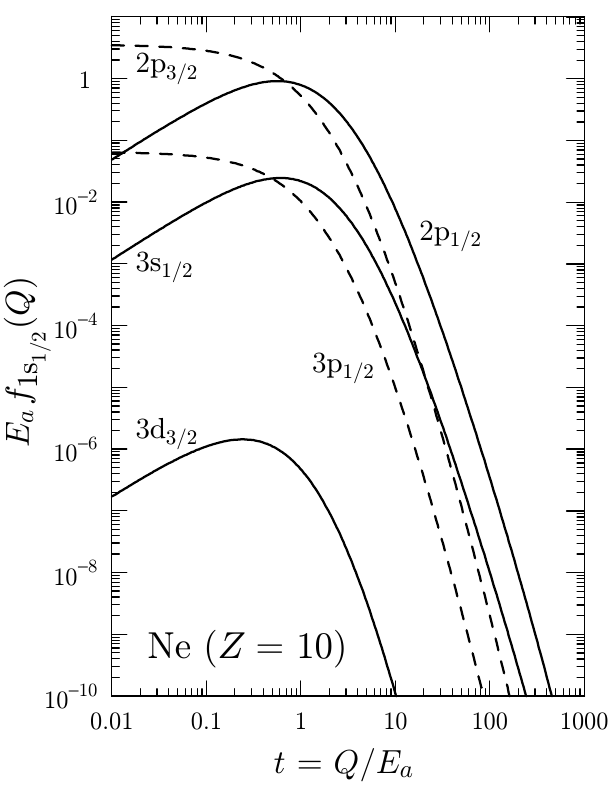} \hfill
\includegraphics*[width=7.5cm]{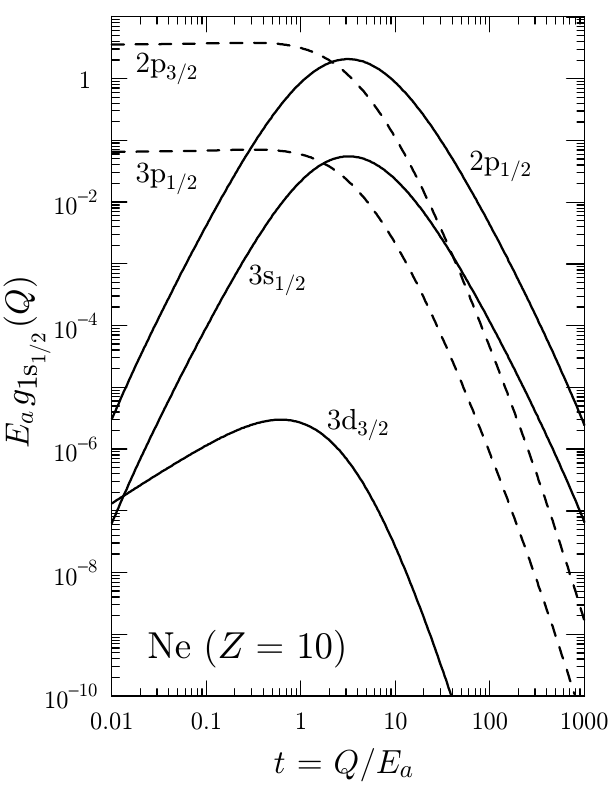}
\includegraphics*[width=7.5cm]{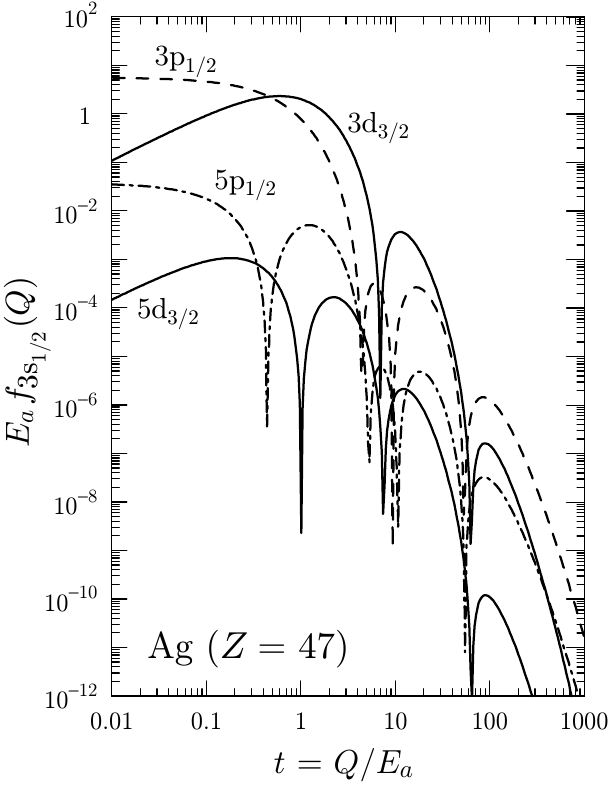} \hfill
\includegraphics*[width=7.5cm]{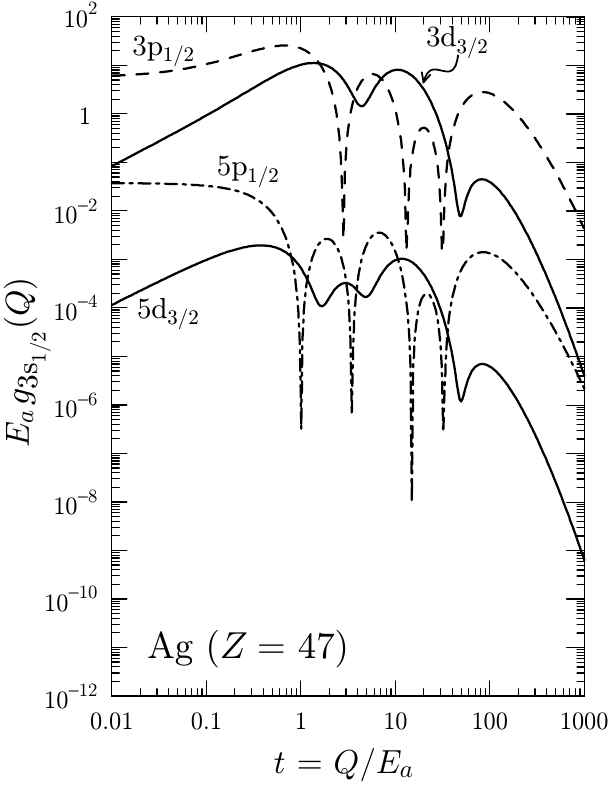}
\caption{\rm Longitudinal (left) and transverse (right) GOSs for
excitations of electrons from the K ($1s_{1/2}$) shell of neon and from
the M1 ($3s_{1/2}$) subshell of silver to bound levels. Each curve
represents the GOS for the indicated final level as a function of the
reduced recoil energy $t=Q/E_a$.
\label{fig4}}
\end{center}\end{figure}

Figure \ref{fig4} shows the GOS and TGOS for several transitions of
electrons from the K ($1s_{1/2}$) shell of neon and from the M1
($3s_{1/2}$) subshell of silver atoms to bound levels of higher energies.
Note that the GOS and the TGOS have identical values at $Q=0$. In the
case of optically-allowed transitions (\ie, transitions with
non-vanishing OOSs) these two functions are nearly constant for values
of $Q$ up to about $0.01 E_a$. Generally, for $Q$ larger than about
$E_a$ the TGOS is larger than the GOS, and decreases more slowly when $Q$
increases.

\subsubsection{Ionizing collisions \label{sec5.4.2}}

\begin{figure}[htbp]
\begin{center}
\includegraphics*[width=7.5cm]{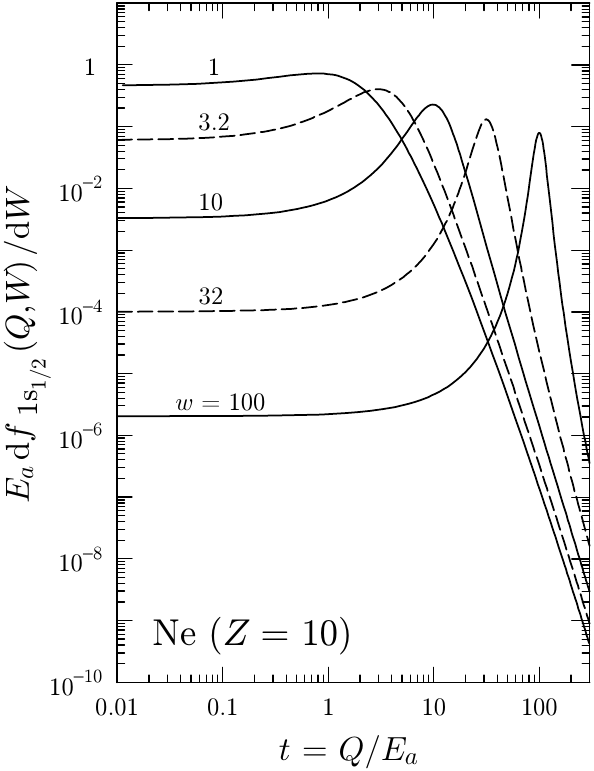} \hfill
\includegraphics*[width=7.5cm]{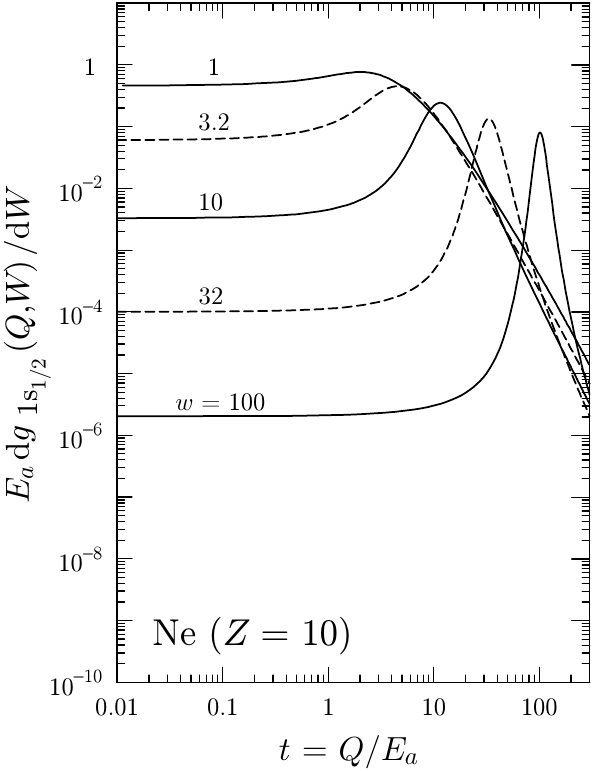} \\ [3mm]
\includegraphics*[width=7.5cm]{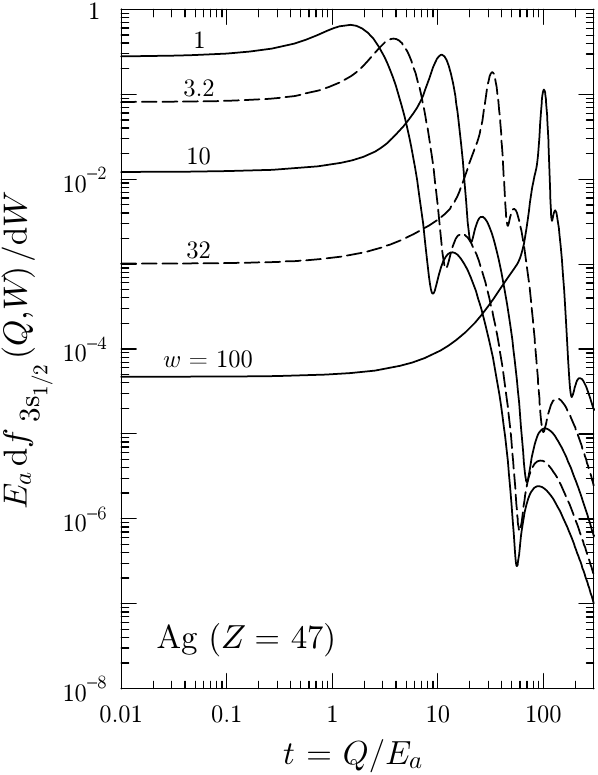} \hfill
\includegraphics*[width=7.5cm]{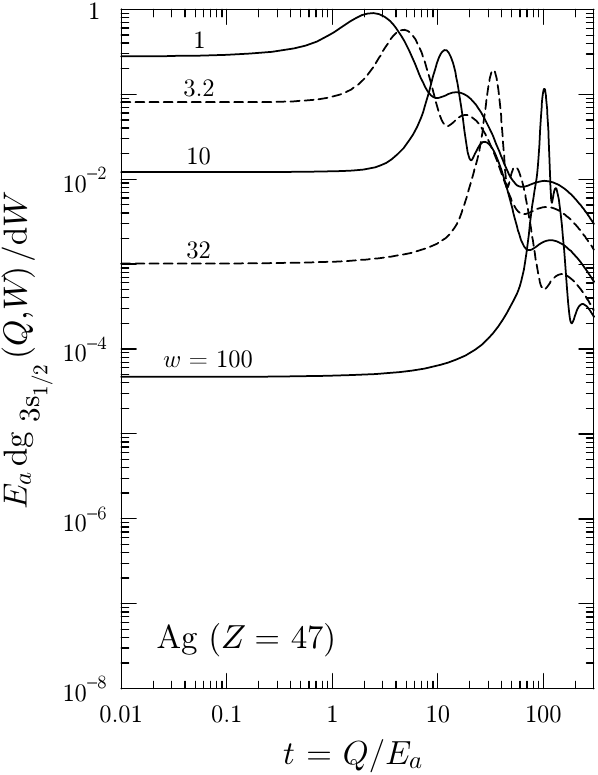}
\caption{\rm Longitudinal (left) and transverse (right) GOSs for
ionization of the K
($1s_{1/2}$) shell of neon and for the M1 ($3s_{1/2}$) subshell of silver.
Each curve represents the GOS for the indicated
value of $w=(W/E_a)-1$ as a function of $t=Q/E_a$.
\label{fig5}}
\end{center}\end{figure}

In the case of ionizing collisions, with $W>E_a$, the program {\sc
gosat} calculates the GOS and the TGOS for a grid of discrete values of
the reduced variables $t\equiv Q/E_a$ and $w\equiv(W/E_a)-1$. The GOSs
at arbitrary points $(t,w)$ are evaluated from these numerical tables by
using suitable interpolation/extrapolation schemes. The $w$ grid is
logarithmic, with 20 points per decade, and extends from $10^{-5}$ up to
a value $w_{\rm num}$, of the order of 500, where the fast oscillations
of the product of radial functions and spherical Bessel functions cannot
be followed accurately with the adopted radial grid. The $t$ grid, which
is determined independently for each value of $w$, spans the interval
from $10^{-4}$ up to a value $t_{\rm num} \sim 100$ (regions II and II'
in Fig.\ \ref{fig3}), or greater if so required by the adopted
interpolation scheme (see below). For a given $w$, the $t$-grid consists
of 150 points that are unevenly distributed, with the higher
concentration in regions where the tabulated function has stronger
curvature, to allow accurate cubic-spline log-log interpolation in $t$.
Examples of calculated GOSs for the K ($1s_{1/2}$) shell of neon and the
M1 ($3s_{1/2}$) subshell of silver are shown in Fig.\ \ref{fig5}.

The (longitudinal) GOS for ionization can be represented as a smooth
surface on the $(Q,W)$ plane, the so-called Bethe surface
\citep{Inokuti1971}. As illustrated in Fig.\ \ref{fig5}, for energy losses $w$
larger than about $2E_a$, the GOS has a prominent maximum at $w\sim t+1$
(\ie, at $W \sim Q$), the Bethe ridge \citep{Inokuti1971}, which corresponds
to collisions with relatively large momentum transfers (close
collisions). The TGOS can be represented in a similar way; its Bethe
surface is analogous to that of the GOS, one differentiating
feature being that beyond the Bethe ridge, \ie, for $Q>W$, the TGOS
decreases more slowly than the GOS. When $w$ increases, the position of
the Bethe ridge shifts to larger $t$ values. Hence, near the ridge the
GOS and the TGOS vary rapidly along the directions of both the $w$ and
the $t$ axes. The shape of the GOSs is a manifestation of the structure
of the radial functions of the initial state. The ionization GOS
may have minima at the right of the Bethe
ridge (\ie, for $Q\gtrsim W$). The number of those minima is equal to
the number of nodes of the radial functions, $n_{\rm r} =
n_a-\ell_a-1$, which is frequently referred to as the radial quantum
number.

\begin{figure}[bth]
\begin{center}
\includegraphics*[width=7.5cm]{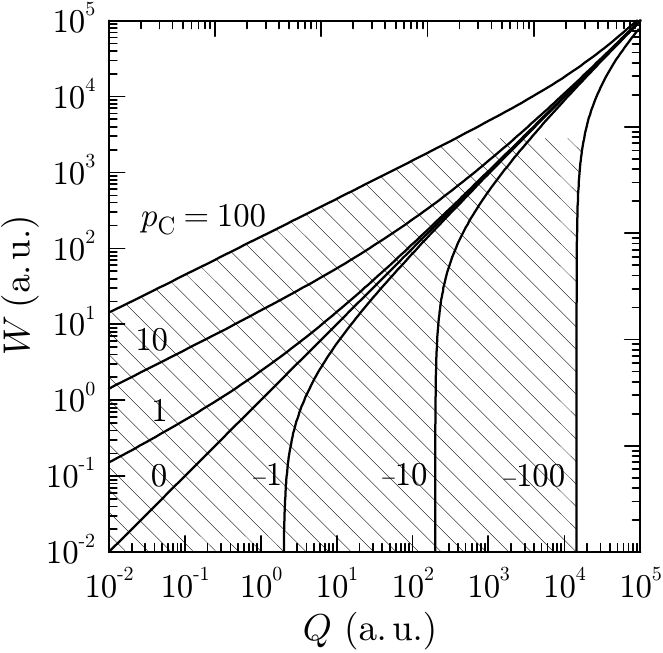} \rule{5mm}{0mm}
\includegraphics*[width=7.5cm]{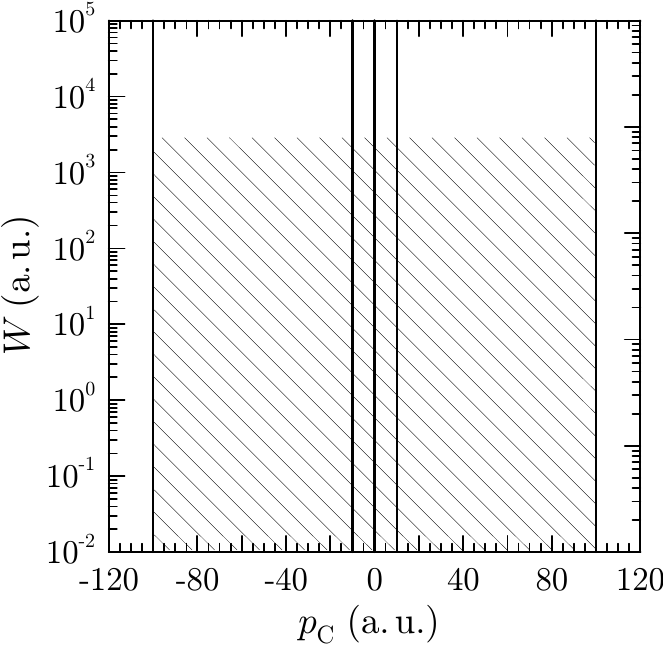}
\caption{Curves of constant $p_{\rm C}$ values in the $(Q,W)$ plane
(left) and corresponding lines in the $(p_{\rm C},W)$ plane (right).
The hatched region in the left plot, limited
by the curves $p_{\rm C}(Q,W)=-100$ and $+100$, transforms into the
hatched rectangle in the $(p_{\rm C},W)$ plane (right plot). The
Bethe ridge, $Q=W$, transforms into the line $p_{\rm C}=0$. All
quantities are in atomic units (a.u.).
\label{fig6}}
\end{center}\end{figure}

To devise an efficient interpolation scheme, it is advantageous to
introduce a transformation that renders the position of the GOS maximum,
and those of the possible minima, nearly constant with $w$, thus
reducing the fast variation of the GOS with $w$. Previous work on the
relationship between the PWBA and the impulse approximation
\citep{Segui2002} indicates that the transformation from the GOS to the
$W$-dependent Born-Compton profile meets our needs. The longitudinal (L)
and transverse (T) Born-Compton profiles, defined by Eq.\ (55) of
\citet{Segui2002}, are
\begin{subequations}
\label{5.59}
\beq
J^{\rm L}_a (W;p_{\mathrm{C}}) =
\frac{c\sqrt{Q(Q+2\me c^{2})}}{Q+\me c^{2}} \,
\frac{Q(1+Q/2\me c^{2})}{W(1+W/2\me c^{2})} \,
\frac{\d f_a (Q,W)}{\d W},
\label{5.59a}\eeq
and
\beq
J^{\rm T}_a (W;p_{\mathrm{C}}) =
\frac{c\sqrt{Q(Q+2\me c^{2})}}{Q+\me c^{2}} \,
\frac{Q(1+Q/2\me c^{2})}{W(1+W/2\me c^{2})} \,
\frac{\d g_a (Q,W)}{\d W},
\label{5.59b}\eeq
\end{subequations}
respectively. The variable $p_{\mathrm{C}}$, which in the impulse
approximation represents the minimum momentum of the target electron for
which the kinematical constraints imposed by momentum and energy
conservation are fulfilled, is defined as
\begin{subequations}
\label{5.60}
\beq
p_{\mathrm{C}} =
- \frac{1}{2c} \left( \sqrt{Q(Q+2\me c^{2})}
- W \sqrt{ 1 +
\frac{(2\me c^{2})^{2}}{Q(Q+2\me c^{2}) -W^{2} }}
\ \right) \quad \mbox{if $W\le Q$}
\label{5.60a}\eeq
and
\beqa
p_{\rm C} & = &
\frac{W-Q}{2c\sqrt{Q(Q+2\me c^{2})}} \left[ 2(Q+\me c^2)
\rule{0mm}{6mm}\right.
\nonumber \\ [2mm]
&& \mbox{} \left. +
\frac{[3\me c^2 (Q+\me c^{2})+ Q^{2}](W-Q)}{
(Q+2\me c^{2})\me c^{2}} \right] \quad \mbox{if $W>Q$}.
\label{5.60b}\eeqa
\end{subequations}
Figure \ref{fig6} displays curves in the $(Q,W)$ plane that correspond to
various $p_{\rm C}$ values. The hatched region in the left plot, limited
by the curves $p_{\rm C}(Q,W)=-100$ and $+100$, transforms into the
hatched rectangle in the $(p_{\rm C},W)$ plane (right plot). Note that
the diagonal $W=Q$ of the $(Q,W)$ plane corresponds to the vertical line
$p_{\rm C}=0$ of the $(p_{\rm C},W)$ plane. Moreover, for relatively
large $W$ values, where the Bethe ridge becomes prominent, a relatively
narrow $Q$ interval is stretched into a wide $p_{\rm C}$ interval.

\begin{figure}[ph]
\begin{center}
\includegraphics*[width=7.5cm]{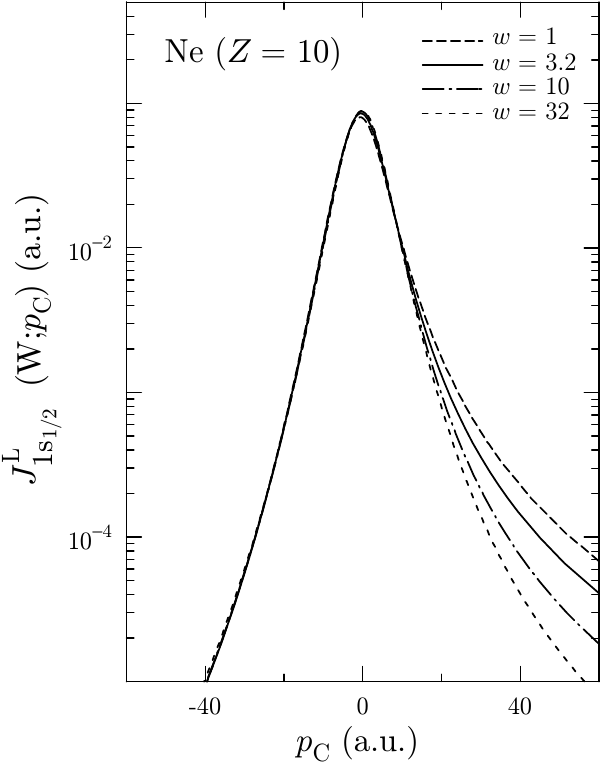} \hfill
\includegraphics*[width=7.5cm]{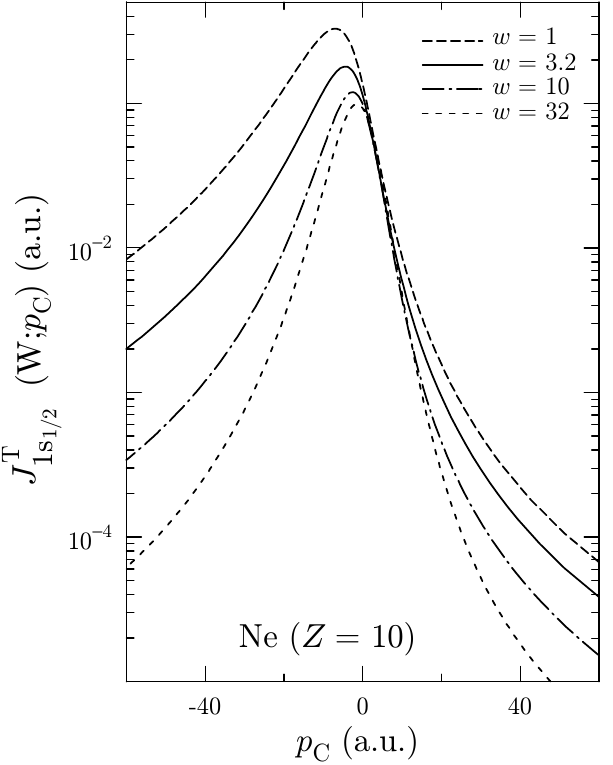}
\includegraphics*[width=7.5cm]{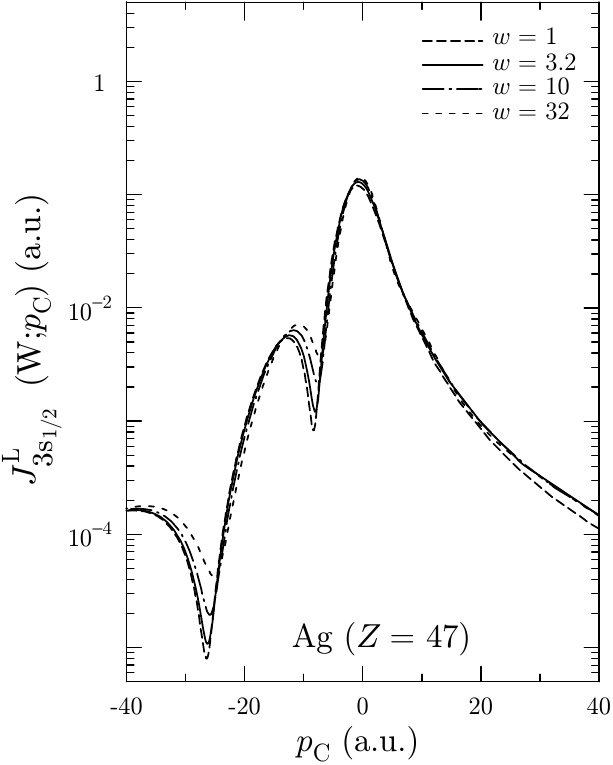} \hfill
\includegraphics*[width=7.5cm]{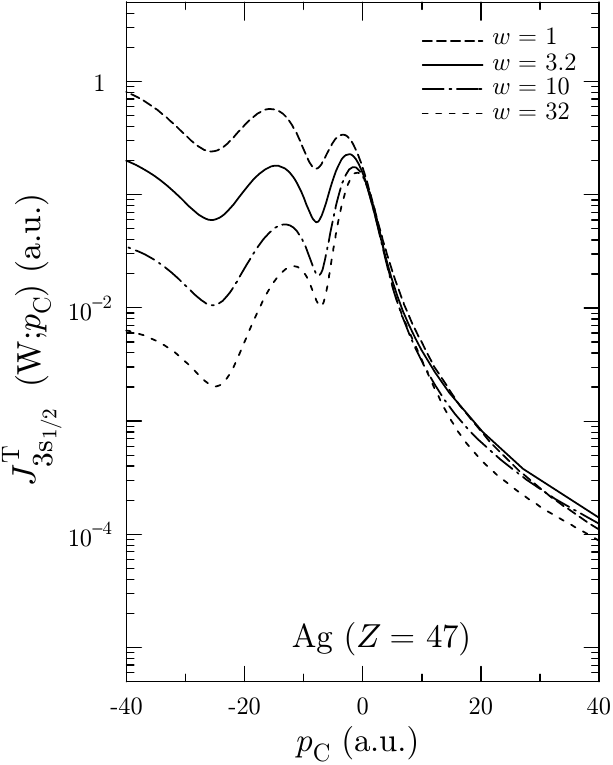}
\caption{Longitudinal (L) an transverse (T) Born-Compton profiles, Eqs.\
\req{5.59}, of the K ($1s_{1/2}$) shell of neon and of the M1
($3s_{1/2}$) subshell of silver, in atomic units. Each curve represents the
profile for the indicated value of $w=(W/U)-1$ as a function of the
variable $p_{\rm C}$, Eqs.\ \req{5.60}.
\label{fig7}}
\end{center}\end{figure}

The Born-Compton profiles obtained from the GOSs of Fig.\ \ref{fig5} are
displayed in Fig.\ \ref{fig7}. The smooth variation of these profiles
with $w$ is evident (cf.\ Figs.\ \ref{fig5} and \ref{fig7}). Hence, for
ionizations involving moderate energy and momentum transfers, say with
$w \in (10^{-5}, w_{\rm switch})$ and $t \in (10^{-4},
t_{\rm num,0})$ (region II in Fig.\ \ref{fig3}), interpolation of the
Born-Compton profiles gives errors that are much smaller than the direct
interpolation of the GOS and the TGOS. To determine the GOSs at a given
point $(Q,W)$ within region II, we use lin-log cubic spline
interpolation of the Born-Compton profiles in $p_{\rm C}$ and 4-point
log-log Lagrange interpolation in $w$, and we obtain the GOSs by
inverting the transformations \req{5.59}. To allow proper interpolation
in $w$, the numerical Born-Compton profiles need to be tabulated within
a rectangle $(p_{\rm C,min},p_{\rm C,num}) \times (0,w_{\rm num})$,
where the limits of the $p_{\rm C}$ interval do not depend on $w$ (as
shown in the right panel of Fig.\ \ref{fig6}). For $w=10^{-5}$, the GOSs
are calculated for $t$ values ranging from $t_{\rm min}=10^{-4}$ up to a
certain value $t_{\rm num,0}$. The latter determines the left end,
$p_{\rm C,min}$, of the $p_{\rm C}$ interval. For higher
energy-losses, the numerical GOSs are calculated in the interval from
$t_{\rm min}(w) = 10^{-4}$ up to a certain value $t_{\rm num}(w)$,
slightly larger than $t_{\rm num,0}$, for which $p_{\rm C}(Q,W)$
equals $p_{\rm C,min}$. Since $t_{\rm num}(w)$ increases with $w$,
the calculation for $t$ up to $t_{\rm num}(w)$ is possible only for
$w$ up to a certain value $w_{\rm switch}$, which is determined by
the program; typically $w_{\rm switch}$ is about $2\,
t_{\rm num,0}$.

For $t \in (10^{-4}, t_{\rm num,0})$ and $w \in (w_{\rm switch},
w_{\rm num}$) [region II$'$ in Fig.\ \ref{fig3}] the GOS and the
TGOS are smooth, slowly increasing functions of $Q$. In this region,
these two functions are evaluated directly from their numerical tables,
by log-log cubic spline interpolation in $t$ and 4-point lin-log
Lagrange interpolation in $w$. In regions II and II$'$, the numerical
error introduced by the interpolations in $t$ and $w$ is estimated to
cause variations in the calculated DDCS that are less than about 0.01 percent.

As mentioned above, the numerical calculation of the GOSs for values of
$t$ spanning the complete interval $(0,t_{\rm num,0})$ is possible
only up to the reduced energy loss $w_{\rm num}$, because the finite
spacing of the radial grid does not allow the radial integrals to be
computed with enough accuracy. With our radial grids, $w_{\rm num}$
is usually larger than $1.5 \, t_{\rm num,0}$. For reduced energy
losses larger than $w_{\rm num}$, where the GOSs are generally very
small, the calculation of the OOS is still reliable, because the
evaluation of dipole-matrix elements, Eq.\ \req{5.54}, does not involve
a Bessel function and it is more robust than that of the GOS and the
TGOS. Numerical results indicate that the dipole approximation is
applicable when $Q/W \lesssim 0.01$, regardless of the ionization energy
of the active electron subshell, see Fig.\ \ref{fig5}. Owing to the lack of
more accurate values, for $W$ larger than $(w_{\rm num}+1) E_a$, and for
$t \in (10^{-4}, t_{\rm num,0})$, [region III in Fig.\ \ref{fig3}]
the GOS and the TGOS are set equal to the OOS. Since the dipole approximation
is accurate only for $Q \ll W$, this introduces a discontinuity in the
GOSs at $W=(w_{\rm num}+1) E_a$, which is appreciable only for
moderate and large recoil energies.

This scheme allows the calculation of the GOSs for $t < t_{\rm num,0}$
and for any $W$, except for a narrow interval about the ionization
threshold $E_a$ with end points at $W_1 \simeq \epsilon_{26, \kappa_b} +
E_a$ and $W_2 = 1.00001 E_a$ (region V in Fig.\ \ref{fig3}). The theory
of quantum defects \citet[see, \eg,][]{JohnsonCheng1979, Seaton1983}
implies that the GOS for excitations to weakly bound levels (\ie, near
the ionization threshold) can be approximated by an ``average''
continuum distribution that at $W=E_a$ joins smoothly with the
continuous GOS for ionization. Since the contribution to the Bethe sum
of this narrow interval is small (of the order of $10^{-5}$ or so), we
will assume that for $W\in (W_1,W_2)$ the GOS and the TGOS are constant
with $W$, \ie, we set $\d f (Q,W)/\d W = \d f (Q,W_2)/\d W$, and
similarly for the TGOS.


\subsection{The Bethe sum rule \label{sec5.5}}

As in the previous sections, we consider collisions with the
$2|\kappa_a|=2j_a+1$ electrons of a closed subshell $n_a \kappa_a$. Assume for a
moment that electrons can make transitions from the initial energy level
$\epsilon_{n_a\kappa_a}$ to any other one-electron level
$\epsilon_{n_b\kappa_b}$, including those corresponding to occupied
shells. In the non-relativistic theory the
(longitudinal) GOS satisfies the Bethe sum rule
\beq
\int_0^\infty \frac{\d f_a (Q,W)}{\d W} \, \d W
= 2|\kappa_a| \qquad \forall Q,
\label{5.61}\eeq
where the integral extends over all possible transitions. This sum rule
plays a central role in the derivation of the conventional asymptotic
formula for the stopping power \citep[see, \eg,][]{Fano1963}. However,
the theoretical study of Cohen (\citeyear{Cohen2003, Cohen2003b}) shows
that relativistic corrections cause deviations of the atomic GOS from
the Bethe sum rule that are appreciable at small $Q$, of the order of
2.5 \%, for high-$Z$ elements.  The failure of the Bethe sum rule
implies that the asymptotic formula for the stopping cross section needs
to be modified (see Section \ref{sec9}).

{\allowdisplaybreaks
The relativistic generalization of the integral \req{5.61} is
the Bethe sum
\beq
S_0(a;Q) \equiv
\int_0^\infty \frac{\d f_a (Q,W)}{\d W} \, \d W \, .
\label{5.62}\eeq
From our numerical GOS tables, we calculate the Bethe sum as
\beq
S_0(a;Q) = \sum_{b\ne a} \,
f_{ba}(Q)
+ \int_{W_1}^\infty
\frac{\d f_a (Q,W)}{\d W} \, \d W \, ,
\label{5.63}\eeq
where the discrete summation extends over all transitions to final bound
levels $\epsilon_{n_b\kappa_b} \ne \epsilon_{n_a\kappa_a}$ with $n_b \le 25$, including those that are
occupied; transitions to higher bound levels (with $W>W_1$) are
described by extending the continuum GOS to energy losses less than
$E_a$ (see Section \ref{sec5.4.2}). Note that the GOS for transitions to levels
below $\epsilon_{n_a\kappa_a}$ (with $W<0$) is negative. To check the
global accuracy of our calculated Bethe sum, we have performed
calculations of that sum from ``non-relativistic'' GOSs obtained by
running our code with the speed of light $c$ replaced with a value 1,000
times larger; the resulting sum was found to agree with the number of
electrons for recoil energies up to the largest value attainable,
$Q_{\rm num}$, the relative differences being less than $\sim
10^{-4}$ in all cases.

Let us write
\beq
S_0(a;Q) = 2|\kappa_a| \left[ 1 - \Delta(a;Q)
\right],
\label{5.64}\eeq
where $\Delta(a;Q)$ is the relativistic departure from the Bethe sum
rule. The functions $S_0(a;Q)$ for the K shell of several elements and
for various subshells of gold ($Z=79$) atoms, calculated from the numerical
GOS tables for $Q$ in the interval from 0 to $Q_{\rm
num}=t_{\rm num,0}E_a$, are displayed in Figs.\ \ref{fig8} and
\ref{fig9}, respectively. In the case of the K ($1s_{1/2}$) shell, the
departure $\Delta(a;Q)$ increases with the atomic number, that is, with
the ionization energy of the active target electrons. The results
plotted in Figs.\ \ref{fig8} and \ref{fig9} show that $\Delta(a;Q)$ is
larger for the K shell and decreases, roughly in
accordance with the principal quantum number of the active subshell. The
extreme case considered in the plots is the K shell of einsteinium
($Z=99$), for which the relative departure is about 30 \%, that is, the
sum \req{5.63} lacks more than half an electron (!). Because $S_0(a;Q)$
includes transitions to all one-electron levels, its deviation from the
value $2|\kappa_a|$ is the result of purely relativistic effects; an
equivalent non-relativistic calculation gives null deviation.

\begin{figure}[htb]
\begin{center}
\includegraphics*[width=11cm]{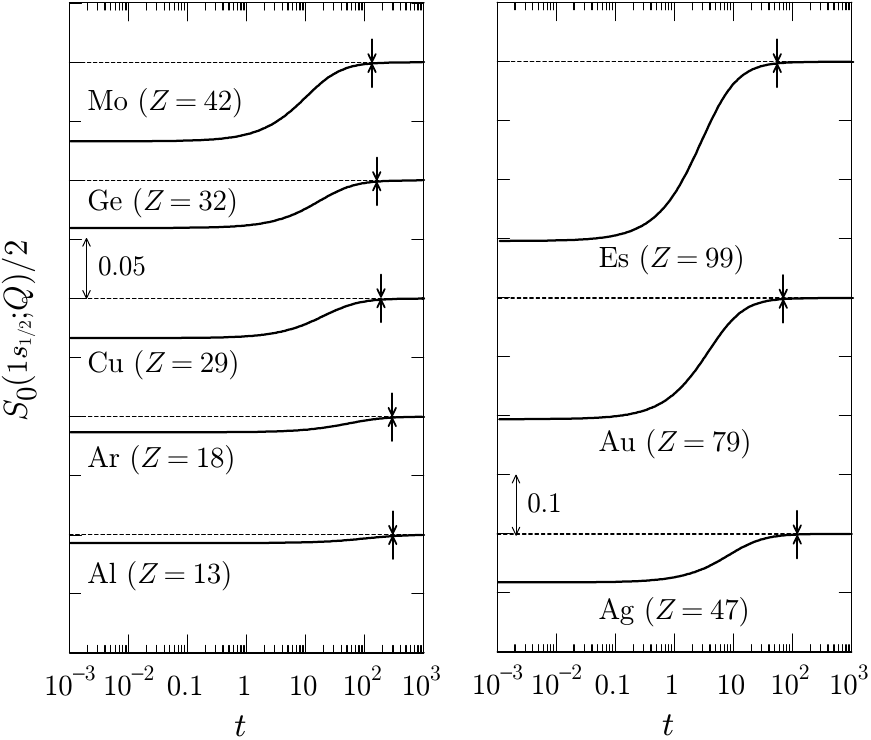}
\caption{Normalized sums $S_0(a;Q)/2$ for the K ($1s_{1/2}$) shell of
the indicated atoms, as functions of the reduced recoil energy
$t=Q/E_a$. Each curve is plotted with its own vertical axis, which is
shifted an arbitrary number of divisions to accommodate several elements
in the same plot, and all curves are drawn with the same scale. The axis
division length is given in the plot. The dashed horizontal lines
indicates the asymptotic value of each curve, which is equal to unity.
The arrow marks are at $t=t_{\rm num,0}$. For smaller $t$ values the
curves were calculated numerically; for $t>t_{\rm num,0}$, they
represent the analytical fit \req{5.69}.
\label{fig8}}
\end{center}\end{figure}

\begin{figure}[th!]
\begin{center}
\includegraphics*[width=11.0cm]{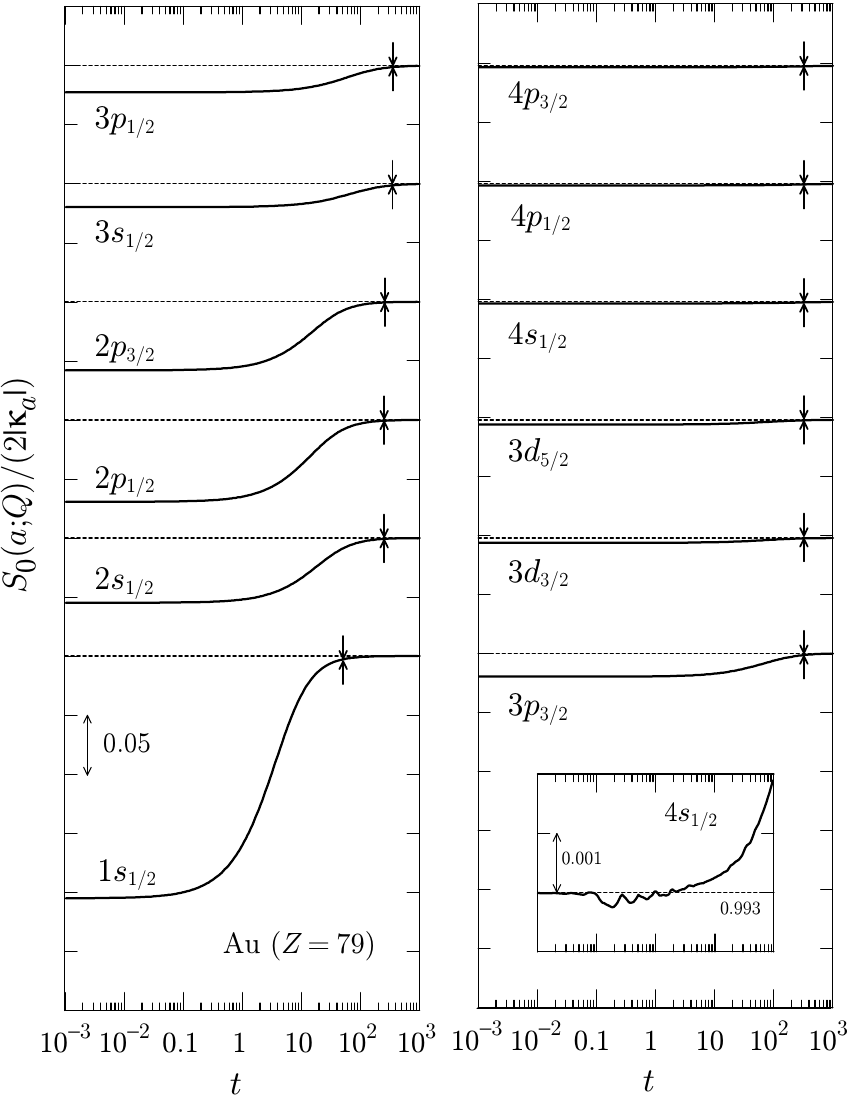}
\caption{Normalized sum $S_0(a;Q)/(2|\kappa_a|)$ for various
subshells of gold atoms ($Z=79$), as functions of the reduced recoil energy
$t=Q/E_a$. Details are the same as in Fig. \ref{fig8}. It
is clear that the deviation from the Bethe sum rule
$\Delta(a;Q) \equiv 1-S_0(n; Q)/(2|\kappa_a|)$
decreases when the principal quantum number $n_a$ increases, \ie, when
the ionization energy decreases. The inset shows the sum for the
$4s_{1/2}$ subshell in an expanded scale, to reveal numerical errors.
\label{fig9}}
\end{center}\end{figure}

Cohen (\citeyear{Cohen2003}, \citeyear{Cohen2003b}) used the Foldy-Wouthuysen transformation, and an
expansion in powers of the atomic potential-energy operator, to obtain a
perturbative expression of the relativistic deviation $\Delta(a;Q)$
valid for all $Q$. It should be noted that Cohen did not make any
specific assumption about the nature of the atomic states and, hence his
analysis should be valid for exact many-electron states. Cohen's result,
including first- and second-order corrections, and adapted to our
independent-electron model, can be expressed as [see Eqs.\ (2.11) and
(4.1) in Cohen, \citeyear{Cohen2003b}]
\beqa
\Delta (a;Q) &=&  \frac{2}{3 \me c^2 (1+Q/\me c^2)^2}
\left( 1 + \frac{3}{2(1+Q/\me c^2)^2} \right)
\left< \psi_a \left| {\cal K}^{\rm nr} \right| \psi_a \right>
\nonumber \\ [2mm]
&& \mbox{} + \frac{7}{\me c^2 (1+Q/\me c^2)^8}
\left< \psi_a \left|
{\cal K} - {\cal K}^{\rm nr} \right| \psi_a \right>
\label{5.65}\eeqa
where ${\cal K}= c\widetilde{\alpha} \dotprod {\bf p} +
(\widetilde{\beta}-1) \me c^2$ and ${\cal K}^{\rm nr} \equiv {\bf
p}^2/2\me$ are, respectively, the relativistic and nonrelativistic
kinetic energy operators. Incidentally, for $Q=0$, the first term on the
right-hand side, $5 \left< {\cal K}^{\rm nr} \right> /3\me c^2$, agrees with
the result obtained by \citet{Levinger1957} for the
relativistic dipole sum rule. The expectation
values of the operators in Eq.\ \req{5.65} can be calculated easily,
because the orbitals $\psi_a$ are solutions of the Dirac equation,
$$
\left[ c\widetilde{\alphab} \dotprod {\bf p} +
(\widetilde{\beta}-1) \me c^2 + V(r) \right] \psi_a({\bf r}) =
\epsilon_a \psi_a({\bf r}).
$$
Evidently,
\beqa
\left< \psi_a \left| {\cal K} \right| \psi_a \right>
&=& \left< \psi_a \left|
\epsilon_a - V(r) \right| \psi_a \right>
\nonumber \\ [2mm]
&=& \epsilon_a - \int_0^\infty
\left\{ [P_{n_a \kappa_a}(r)]^2 +
[Q_{n_a \kappa_a}(r)]^2 \right\} V(r) \, \d r.
\label{5.66}\eeqa
Recalling that $(\widetilde{\alphab} \dotprod {\bf p})^2 = {\bf p}^2$,
we have
\beqa
{\bf p}^2 \psi_a({\bf r}) &=& (\widetilde{\alphab} \dotprod
{\bf p})^2 \psi_a ({\bf r})= c^{-2}
\left[ \epsilon_a - (\widetilde{\beta}-1) \me c^2 - V(r)  \right]^2
\psi_a({\bf r})
\nonumber \\ [2mm]
&=& c^{-2}
\left( \begin{array}{cc}
\epsilon_a - V(r) & 0 \\ [2mm]
0 & \epsilon_a + 2\me c^2 - V(r)
\end{array} \right)^2 \psi_a({\bf r}),
\label{5.67}\eeqa
and
\beqa
&& \! \! \! \! \! \! \!
\left< \psi_a \left|
{\cal K}^{\rm nr} \right| \psi_a \right>
= \frac{1}{2\me}  \left< \psi_a \left|
{\bf p}^2 \right| \psi_a \right>
\nonumber \\ [2mm]
&=& \frac{1}{2\me c^2} \int_0^\infty
\left\{\rule{0mm}{4mm}
[\epsilon_a - V(r)]^2 P_{n_a \kappa_a}^2(r) +
[\epsilon_a + 2\me c^2 - V(r)]^2 Q_{n_a \kappa_a}^2(r) \right\}
\d r \, . \rule{10mm}{0mm}
\label{5.68}\eeqa
The dipole (Thomas--Reiche--Kuhn) sums $S_0(a;0)$ for neutral atoms
obtained from our numerical OOSs are in qualitative agreement with
estimates given in graphical form by \citet{Cohen2003}, who used
expectation
values of the kinetic energy operators obtained from non-relativistic
Hartee-Fock-Slater atomic orbitals. To our surprise, when accurate
expectation values $\left< \psi_a \left| {\cal K}^{\rm nr} \right|
\psi_a \right>$ and $\left< \psi_a \left| {\cal K} - {\cal
K}^{\rm nr} \right| \psi_a \right>$ calculated from relativistic DHFS
orbitals are inserted, the formulas \req{5.64} and \req{5.65} give
values of $S_0(a;Q)$ that differ markedly from our numerical
calculations for intermediate- and large-$Z$ elements. This discrepancy
indicates  either that Cohen's perturbative treatment is appropriate
only for light elements, for which the atomic potential $V(r)$ is
relatively weak, or that the expectation values in Eq.\ \req{5.65}
differ much from the corresponding ones calculated in the
Foldy-Wouthuysen representation, contrarily to what is suggested by
\citet{Cohen2003b}.

For $Q>Q_{\rm num,0}=t_{\rm num,0} E_a$, where the numerical
calculation of the GOS would be too demanding, the longitudinal GOS will
be approximated through a Born-Compton profile, as described in the next
Section. For the sake of internal consistency, we need a reliable
estimate of the function $S_0(a;Q)$ for $Q>Q_{\rm num}$, which will
be used to ``normalize'' the GOS estimated from the Born-Compton profile. Taking
Cohen's formula as a guide, we have found that our numerical results for
$Q$ near $Q_{\rm num}$ are described very closely by the following
analytical expression,
\beq
S_0(a;Q) = 2|\kappa_a| \left[ 1 -
\frac{a_1}{(1+a_2 t)^2} \right], \qquad t=Q/E_a.
\label{5.69}\eeq
The parameters $a_1$ and $a_2$, which are characteristic of each subshell,
are determined from a lest-squares fit of this formula to the numerical
values of $S_0(a;Q)$ in a certain interval of $t$ values relatively
close to $t_{\rm num,0}$. Because natural splines have null second
derivative at the endpoints of the interpolated table, our $Q$-interpolation scheme
is not very accurate near $Q_{\rm num,0}$, where inaccuracies of the
spline manifest by small fluctuations of the calculated $S_0(a;Q)$
values. Our computer program first determines a value $Q_{\rm
num,1}=t_{\rm num,1} E_a$ ($t_{\rm num,1} < t_{\rm num,0}$) at which the
calculated sum $S_0(a;Q_{\rm num,1})$ is judged to be sufficiently
accurate. The program then computes $S_0(a;Q)$ for a grid of ten $t$
values covering the interval $(t_{\rm num,1}-20, t_{\rm num,1})$
uniformly, and uses the resulting table to do the fitting. The
continuous curves in Figs.\ \ref{fig8} and \ref{fig9} for
$t>t_{\rm num,1}$ represent the fitted formula \req{5.69}. In the
following, we will assume that this formula provides an accurate
representation of the sum $S_0(a; Q)$ for $Q>Q_{\rm num,1}$.

It is worth observing that when the curves in Figs.\ \ref{fig8} and
\ref{fig9} are displayed in an expanded scale show small
irregularities in the interval $t<t_{\rm num,1}$, where the sum
$S_0(a;Q)$ is calculated numerically by integrating the tabulated (and
interpolated) GOS. As an example, the inset in Fig.\ \ref{fig9}
displays the details of the Bethe sum for the $4s_{1/2}$ subshell of gold.
The irregularities are the result of accumulated numerical errors, which
are seen to be quite small, on the relative order of $10^{-4}$. We may
then expect that the calculated cross sections presented below, which
are obtained by computing similar integrals, are affected by numerical
uncertainties of the same magnitude.

Let us now consider the integral
\beqa
T_0 (a; Q) &\equiv& \int_0^\infty
\frac{\d g_{a} (Q,W)}{\d W} \, \d W
\nonumber \\ [2mm]
&=& \sum_{b\ne a} \, g_{ba}(Q)
+ \int_{W_1}^\infty
\frac{\d g_a (Q,W)}{\d W} \, \d W \, ,
\label{5.70}\eeqa
which is the analogue of the Bethe sum, Eq.\ \ref{5.63}, for the TGOS.
Figures \ref{fig10} and \ref{fig11} display the sums \req{5.70} for
various subshells of iron and gold atoms, respectively, calculated by
integrating the numerical TGOS. For intermediate values of $Q$, the sum
$T_0 (a; Q)$ varies wildly. The cause for this uneasy behavior is the
presence of the factor $W$ in the denominator of the TGOS definition,
Eq.\ \req{5.48}, which magnifies the contributions from transitions with
small $W$. Interestingly, in the optical limit, $Q\rightarrow 0$, the
TGOS coincides with the GOS (see Section \ref{sec4.2}), and therefore
$T_0 (a; 0) = S_0 (a; 0)$. This is so because in that limit the squared
matrix elements in Eq.\ \req{5.49} are proportional to $W$.  It is
evident that the behavior of the transverse sum varies from subshell to
subshell and it seems difficult to be modeled in a simple way. This
apparently conflicts with the assumption, adopted in many calculations
of total cross sections and stopping cross sections \citep[see,
\eg,][]{FernandezVarea2005}, that the TGOS can be replaced with the GOS.
In fact, we have verified numerically that the difference between the
TGOS and the GOS does affect the calculated stopping cross section for
the innermost subshells of heavy elements, but the contribution of these
subshells to the atomic stopping cross section is much smaller than
those of outer subshells and the effect on the atomic cross sections is
not very important. As the violent oscillations of the transverse sum
$T_0(a; Q)$ are mostly caused by transitions to neighboring bound
levels, the TGOS sum for ionization (\ie, excluding contributions from
excitations to bound levels) varies with $Q$ in a smoother way, but not
monotonically (see Figs.\ \ref{fig10} and \ref{fig11}).

\begin{figure}[t!]
\begin{center}
\includegraphics*[width=14.50cm]{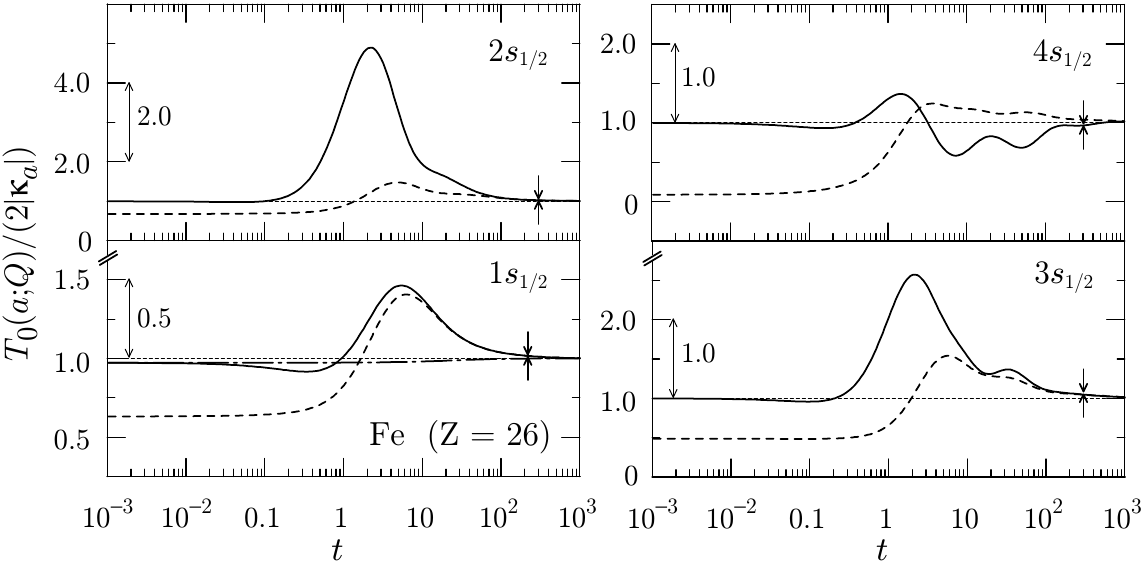}
\caption{Normalized transverse sums $T_0(a;Q)/(2|\kappa_a|)$ (continuous
curves) for various subshells of iron atoms, as functions of the reduced
recoil energy $t=Q/E_a$. The dashed curves represent the contribution of
ionizing transitions (with $W>E_a$). The arrow marks are at
$t=t_{\rm num,1}$. For smaller $t$ values the curves were calculated
numerically. For $t>t_{\rm num,1}$, the solid curve represents the
analytical fit \req{5.77}. The dot-dashed curve in the diagram of the K
($1s_{1/2}$) shell represents the longitudinal sum
$S_0(a;Q)/(2|\kappa_a|)$, Eq.\ \req{5.64}.
\label{fig10}}
\end{center}\end{figure}

\begin{figure}[h!]
\begin{center}
\includegraphics*[width=14.0cm]{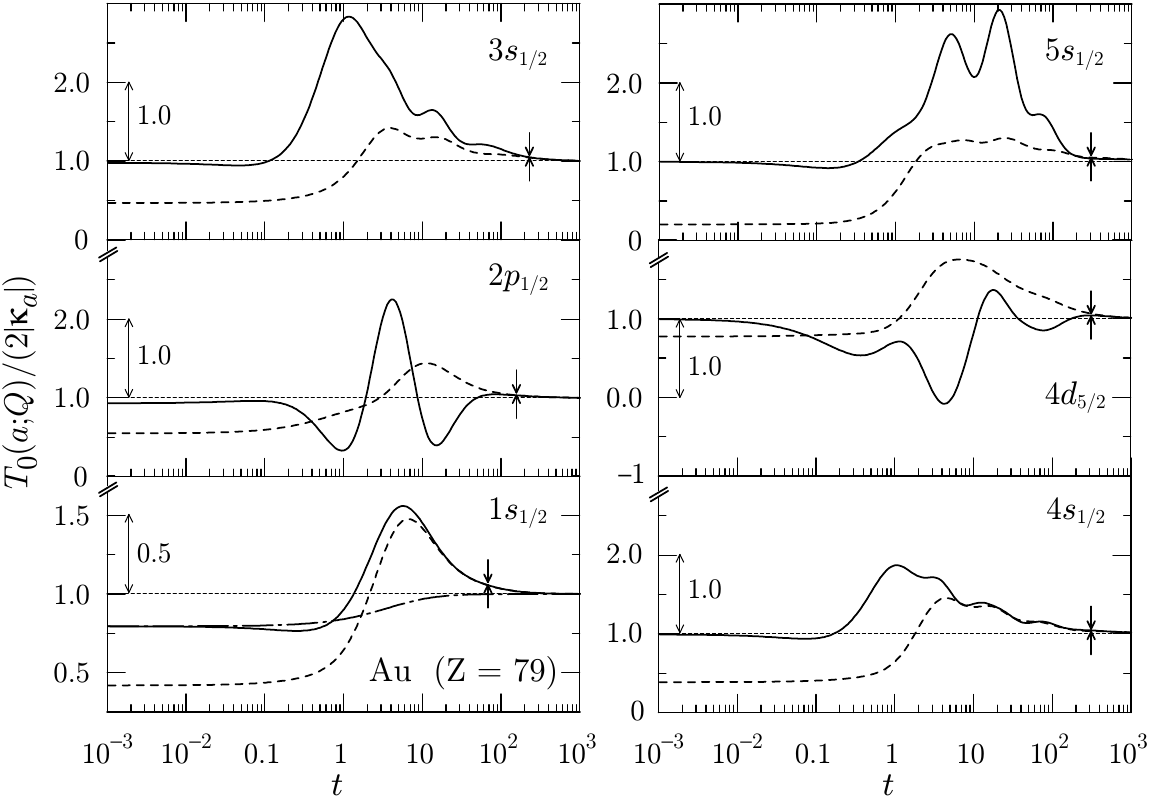}
\caption{Normalized transverse sums $T_0(a;Q)/(2|\kappa_a|)$ (continuous
curves) for various subshells of gold atoms, as functions of the reduced
recoil energy $t=Q/E_a$. Details are the same as in Fig.\ \ref{fig10}.
\label{fig11}}
\end{center}\end{figure}

{\allowdisplaybreaks
Although \citet{Cohen2003} did not explicitly consider the sum \req{5.70},
he provided intermediate results in a calculation of a more involved sum
from which we can derive an analytical expression for $T_0 (a;
Q)$ valid for large values of $Q$. To this end, we express the sum
\req{5.70} as
\beq
T_0 (a; Q) = \sum_{\epsilon_b}
\frac{2(Q+\me
c^2)}{W}
\sum_{m_a,m_b}  \left| \left< \psi_b
\left| \rule{0mm}{4mm} \widetilde {\alpha}_x \exp\left( {\rm i}
{\bf q} \dotprod {\bf r} \right) \right| \psi_a
\right> \right|^2,
\label{5.71}\eeq
where we have used the definition \req{4.49} of the TGOS and simplified
the notation by replacing the integration symbol with a formal sum over
(discrete) final levels $\epsilon_b = \epsilon_a + W$. For large enough
$Q$, the TGOS is appreciable only when $Q-W$ is small.
We can then introduce the following expansion in powers of $Q-W$,
\beq
\frac{1}{W} = \frac{1}{Q+(W-Q)} = \frac{1}{Q} \left[ 1 + \frac{Q-W}{Q}
+ \left(\frac{Q-W}{Q}\right)^2 + \cdots \right],
\label{5.72}\eeq
and write
\beq
T_0 (a; Q) = \frac{2(Q+\me c^2)}{Q}
 \left[ {\cal T}^{(0)}(Q) + \frac{1}{Q} {\cal T}^{(1)}(Q)
+ \frac{1}{Q^2} {\cal T}^{(2)}(Q) - \cdots \right] \, .
\label{5.73}\eeq
The quantities
\beq
{\cal T}^{(\nu)} (Q) = \sum_{n_b,\kappa_b} (Q-W)^\nu
\sum_{m_a,m_b}  \left| \left< \psi_b
\left| \rule{0mm}{4mm} \widetilde {\alpha}_x \exp\left( {\rm i}
{\bf q} \dotprod {\bf r} \right) \right| \psi_a
\right> \right|^2
\label{5.74}\eeq
were studied by Cohen using his Foldy-Wouthuysen
perturbative expansion. He derived the following analytical expressions
[Eqs.\ (5.12) in \citeauthor{Cohen2003}, \citeyear{Cohen2003}]:
\begin{subequations}
\label{5.75}
\beqa
{\cal T}^{(0)}(Q) &=& 2 |\kappa_a| \,
\frac{Q}{2(Q+\me c^2)}
\nonumber \\ [2mm]
&& \mbox{} + 2|\kappa_a| \, \frac{\left< \psi_a \left| {\cal
K}^{\rm nr} \right| \psi_a \right> }{6(Q+\me c^2)^5} \left[ 3Q^2
(Q+2\me c^2)^2 + 4 \me^2 c^4 (Q+\me c^2)^2 \right],
\label{5.75a}\\ [2mm]
{\cal T}^{(1)}(Q) &=& 2|\kappa_a| \, \frac{Q \left< \psi_a \left| {\cal
K}^{\rm nr} \right| \psi_a \right> }{6(Q+\me c^2)^4} \left[ 5 (Q+\me
c^2)^3 - 2 \me^2 c^4 (Q+\me c^2) - 3 \me^3 c^6 \right], \rule{15mm}{0mm}
\label{5.75b}\\ [2mm]
{\cal T}^{(2)}(Q) &=& 2|\kappa_a| \, \frac{\me c^2 Q^2 \left< \psi_a
\left| {\cal K}^{\rm nr} \right| \psi_a \right> }{3(Q+\me c^2)^3} \,
(Q+2 \me c^2)\, ,
\label{5.75c}\eeqa
\end{subequations}
where $\left< \psi_a \left| {\cal K}^{\rm nr} \right| \psi_a \right>
$ is the expectation value of the non-relativistic kinetic energy
operator for the active target electron,
Eq.\ \req{5.68}. Inserting the expressions \req{5.75} into \req{5.73},
we obtain
\beqa
T_0 (a; Q) &=&
2|\kappa_a| \left\{ 1 - \frac{\left< \psi_a \left| {\cal K}^{\rm nr}
\right| \psi_a \right> }{3 \me c^2 (1+Q/\me c^2)^4} \right.
\nonumber \\ [2mm]
&& \left. \times \left[
8 \left(\frac{Q}{\me c^2}\right)^4
+ 34 \left(\frac{Q}{\me c^2}\right)^3
+ 52  \left(\frac{Q}{\me c^2}\right)^2
+ 31 \, \frac{Q}{\me c^2}
+ 8 \right] \right\} , \rule{15mm}{0mm}
\label{5.76}\eeqa
which is expected to be valid only for large $Q$.
}

Similarly to the case of the longitudinal GOS, for recoil energies $Q$
larger than $Q_{\rm num,0}$, where the numerical calculation is not
doable, we will approximate the TGOS by means of the Born-Compton
profile. Again, we need a good estimate of the sum $T_0 (a; Q)$ to
renormalize the profile. However, from the calculated values of the
function $T_0(a;Q)$ for $Q<Q_{\rm num,1}$ we see that this function has
a complicated structure which extends beyond $Q_{\rm num,1}$, and it is
not possible to match the numerical data and the (asymptotic) expression
\req{5.76}. Owing to lack of additional information, we will simply
assume that for $Q>Q_{\rm num,1}$, the TGOS sum can be represented in
the form
\beq
T_0(a;Q) = \sum_{b\ne a} \, g_{ba}(Q) + 2|\kappa_a| \left[ 1 -
\frac{a_3}{(1+a_4 t)^2} \right] , \qquad t=Q/E_a
\label{5.77}\eeq
where the value of the parameters $a_3$ and $a_4$ are determined from a
least-squares fit of the numerical values of the function $T_0(a;Q)$,
excluding the contribution from transitions to bound levels with $n\le
25$, in the interval $(t_{\rm num,1}-20,t_{\rm num,1})$. For L and
outer subshells, transitions to lower-energy (occupied) bound levels have
transverse (and longitudinal) GOSs that are negative and extend somewhat
beyond $t_{\rm num,1}$. The difference between the solid
and dashed curves in Figs.\ \ref{fig10} and \ref{fig11} [\ie, the
discrete sum in Eq.\ \req{5.77}] is mostly due to these transitions to
inner subshells. In practical calculations of inelastic cross sections,
transitions to occupied final states are forbidden by the exclusion
principle and the TGOS for $t>t_{\rm num,0}$ practically reduces to
the ionization part.
}


\subsection{The structure of the Bethe ridge \label{sec5.6}}

As described above, the numerical calculation of the GOS and the TGOS
from Eqs.\ \req{5.46} and \req{5.48} is performed only for $t$ and $w$
values up to $t_{\rm num,0}$ and $w_{\rm num}$,
respectively (regions II, II$'$ and III in Fig.\ \ref{fig3}). For $t > t_{\rm
num,0}$ and $w > 10^{-5}$ (region IV in Fig.\ \ref{fig3}), the GOS and
the TGOS take appreciable values only in the vicinity of the Bethe ridge
($W \sim Q$). In this region, we invert the equalities \req{5.59a} and
\req{5.59b} and set
\begin{subequations}
\label{5.78}
\beq
\frac{\d f^{\rm ridge}_{a}(Q,W)}{\d W} = A_{\rm norm} (Q) \,
\frac{(Q+\me c^{2})}{c\sqrt{Q(Q+2\me c^{2})}} \,
\frac{W(W +2 \me c^{2})}{Q(Q + 2\me c^{2})} \,
J^{\rm L}_{a}(W_{\rm ref};p_{\mathrm{C}})
\label{5.78a}\eeq
and
\beq
\frac{\d g^{\rm ridge}_{a}(Q,W)}{\d W} = B_{\rm norm} (Q) \,
\frac{(Q+\me c^{2})}{c\sqrt{Q(Q+2\me c^{2})}} \,
\frac{W(W + 2\me c^{2})}{Q(Q + 2\me c^{2})} \,
J^{\rm T}_{a}(W_{\rm ref};p_{\mathrm{C}})\, ,
\label{5.78b}\eeq
\end{subequations}
where $J^{\rm L}_{a}(W_{\rm ref};p_{\mathrm{C}})$ and $J^{\rm
T}_{a}(W_{\rm ref};p_{\mathrm{C}})$ are, respectively, the longitudinal
and transverse Born-Compton profiles evaluated for a fixed energy-loss
$W_{\rm ref}$, which we set as $W_{\rm ref} = t_{\rm num,0} E_a$.
Since the Born-Compton profiles $J^{\rm L,T}(W;p_{\rm C})$ vary
slowly with $W$, by using the profiles at $W_{\rm ref}$ we are
reducing possible discontinuities of the composite GOS and TGOS in the
neighborhood of the point $W=Q=W_{\rm ref}$. The calculated
profiles  $J^{\rm
L,T}_{a}(W_{\rm ref};p_{\mathrm{C}})$ are tabulated on a grid of 256
points, unevenly distributed and with a denser spacing where the
profiles have stronger curvature. Continuous profiles are obtained from
these tables by lin-log natural cubic-spline interpolation in
$p_{\rm C}$ (\ie, the interpolated function is of the type shown in
Fig.\ \ref{fig7}). Extrapolation towards larger values of $p_{\rm
C}$, where the profiles decrease monotonically, does not introduce any
significant distortions.  However, we refrain from extrapolating to
lower $p_{\rm C}$ values, because the structure of the profiles at the
left of the maximum may not be fully covered by the calculated numerical
tables, and this may lead to wrong extrapolations (\eg, to profiles that
diverge for large negative $p_{\rm C}$ values). Hence, for $p_{\rm C}$
values less than the smallest grid point, the Born-Compton profile is
set to zero. This amounts to setting the GOSs to zero for $Q$ much
larger than $W$, which is acceptable because kinematical factors make
the DCS small in that region.

The quantities $A_{\rm norm} (Q)$ and $B_{\rm norm} (Q)$ in Eqs.\
\req{5.78} are
$Q$-dependent normalization factors, which we determine as follows. For
the longitudinal GOS we require that, for $Q> Q_{\rm num,1}$ [the
value defined in the comments after Eq.\ \req{5.69}]
\beq
\int_0^\infty \frac{\d f_{a} (Q,W)}{\d W} \, \d W = S_0(a; Q),
\label{5.79}\eeq
where $S_0(a; Q)$ is the one-subshell sum given by Eq.\
\req{5.69}. That is,
\beq
\int_{E_a}^\infty \frac{\d f^{\rm ridge}_{a}(Q,W)}{\d W} \,
\d W = S_0(a; Q) - \sum_{b\ne a} \,
f_{ba}(Q) \, .
\label{5.80}\eeq
Since $Q_{\rm num,1}$ is about $50 E_a$ or larger, for
$Q>Q_{\rm num,1}$ the summation over bound levels on the right-hand
side gets contributions only from transitions to levels $\epsilon_{n_b
\kappa_b}$ that are far below the initial level, for which $W$ is large
and negative; the corresponding GOSs, $f_{ba}(Q)$, are also negative.
Transitions to levels near or above the initial level have exceedingly
small GOS values for $Q>Q_{\rm num}$. The requirement \req{5.79} is
important to ensure consistency of the asymptotic formula for the
stopping power, which will be derived in Section \ref{sec9}, and of the
calculated subshell correction (Section \ref{sec11}).

The modeling of the transverse GOS for recoil energies larger than
$Q_{\rm num,1}$ is more uncertain, because of the lack of a precise
expression for the sum $T_0(a;Q)$. The normalization factor $B_{\rm
norm}(Q)$ for $Q > Q_{\rm num,1}$ is set by requiring that [see
Eq.\ \req{5.70}]
\beq
\int_0^\infty \frac{\d g^{\rm ridge}_{a}(Q,W)}{\d W} \, \d W
= T_0(a; Q) - \sum_{b\ne a} \, g_{ba}(Q) \, ,
\label{5.81}\eeq
where $T_0(a;Q)$ is given by Eq.\ \req{5.77}. That is,
\beq
\int_0^\infty \frac{\d g^{\rm ridge}_{a}(Q,W)}{\d W} \, \d W
= 2|\kappa_a| \left[ 1 - \frac{a_3}{(1+a_4 t)^2} \right].
\label{5.82}\eeq

\begin{figure}[hp!]
\begin{center}
\includegraphics*[width=11cm] {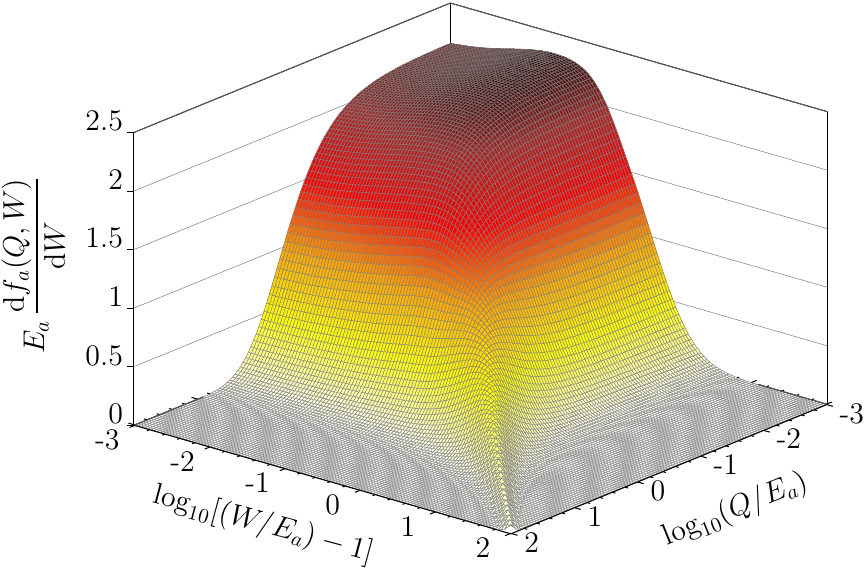} \\ [2mm]
\includegraphics*[width=11cm] {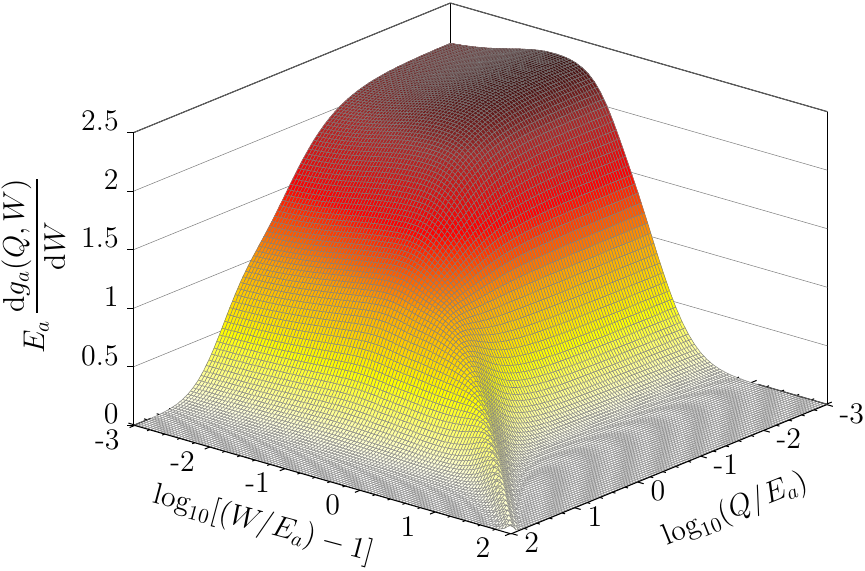} \\ [2mm]
\includegraphics*[width=6.5cm]{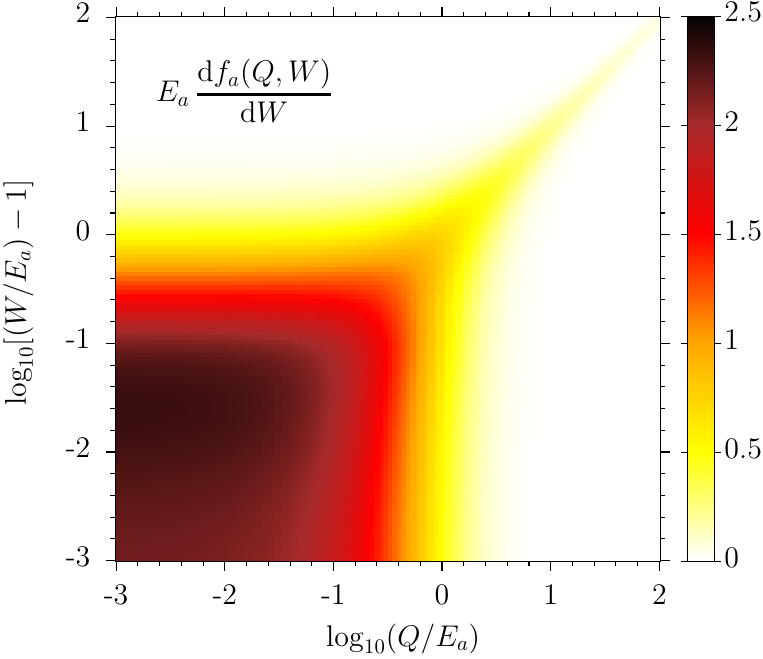} \rule{3mm}{0mm}
\includegraphics*[width=6.5cm]{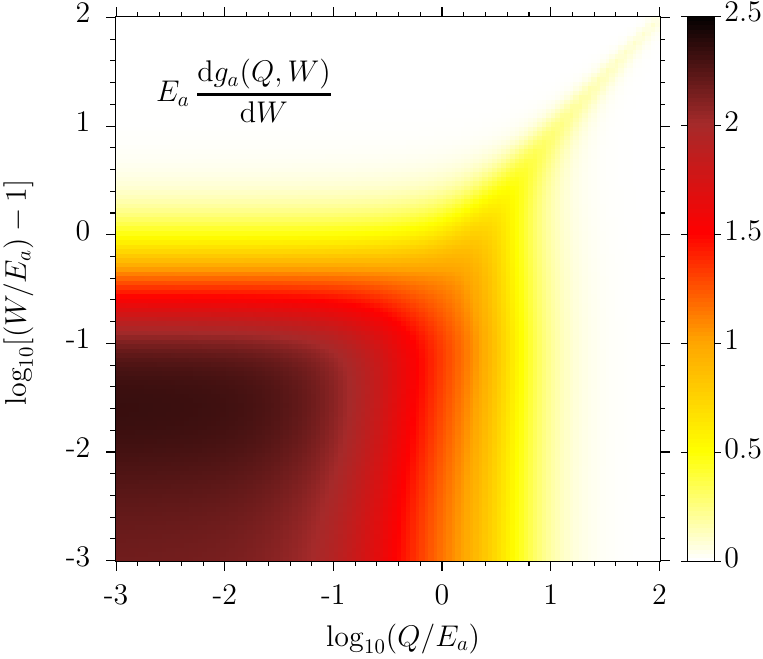}
\caption{The GOS and the TGOS for ionization of the K shell ($1s_{1/2}$)
of the neon atom ($Z=10$), represented as Bethe
3-dimensional surfaces (top and middle) and as color-level diagrams
(bottom); cf. Fig.\ \ref{fig5}.
\label{fig12}}
\end{center}\end{figure}

\begin{figure}[hp!]
\begin{center}
\includegraphics*[width=11cm] {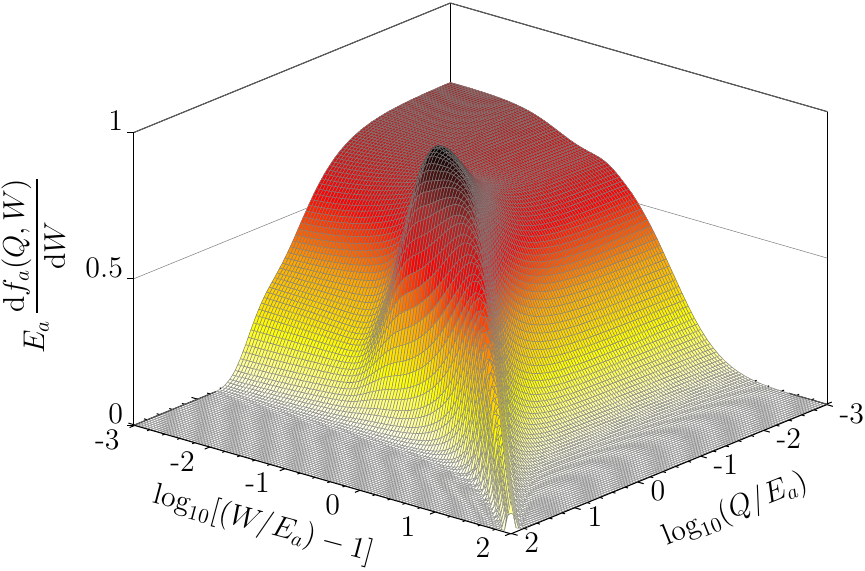} \\ [2mm]
\includegraphics*[width=11cm] {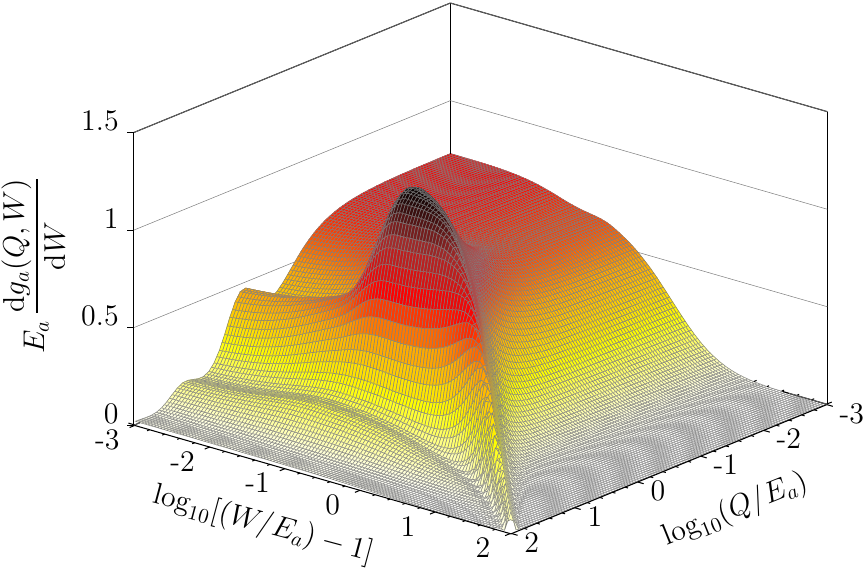} \\ [2mm]
\includegraphics*[width=6.5cm]{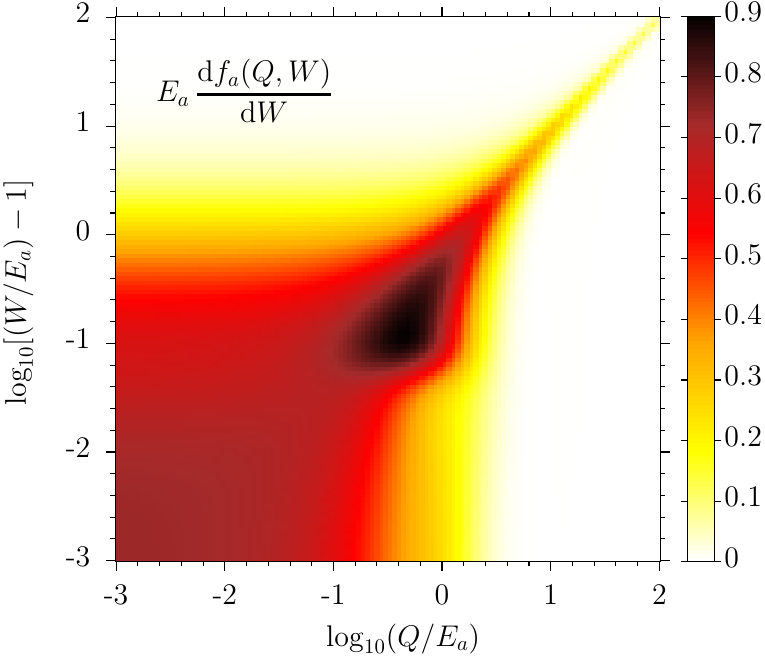} \rule{3mm}{0mm}
\includegraphics*[width=6.5cm]{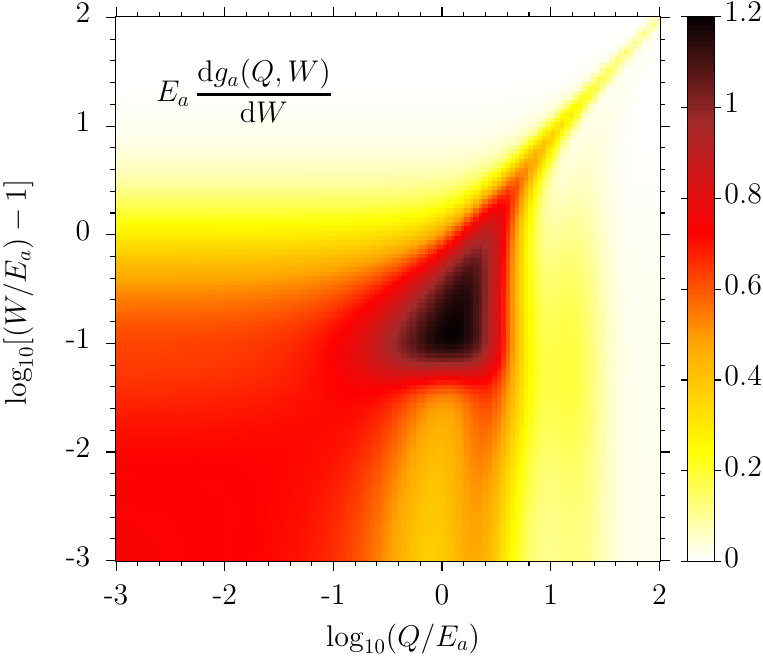}
\caption{The GOS and the TGOS for ionization of the M1 subshell ($3s_{1/2}$)
of the silver atom ($Z=47$), represented as Bethe
3-dimensional surfaces (top and middle) and as color-level diagrams
(bottom); cf. Fig.\ \ref{fig5}.
\label{fig13}}
\end{center}\end{figure}

Summarizing, to define a complete model of the GOS and the TGOS, for
arbitrary values of $Q=t E_a$ and $W=(w+1)E_a$, we combine calculated
numerical tables, which are limited to the intervals $t \in (10^{-4},
t_{\rm num,0} )$ and $w \in (10^{-5}, w_{\rm num})$, with simple
scaled models outside these ranges. Specifically, for $w >
w_{\rm num}$ and $t<t_{\rm num,1}$ (region III of Fig.\ \ref{fig3})
the GOS and the TGOS are set equal to the OOS, \ie, they are assumed to
be constant with $Q$. This approximation has a negligible effect on the
Bethe sums $S_0(a;Q)$ and $T_0(a;Q)$, because the GOS and the TGOS take
exceedingly small values in that region. For $t>t_{\rm num,1}$ (zone
IV in Fig.\ \ref{fig3}) the GOS and the TGOS are obtained from the
normalized Born-Compton profiles, Eqs.\ \req{5.78}. The calculated GOSs
and TGOSs for ionization of the K shell of neon ($Z=10$) and the M1
subshell of silver ($Z=47$) are displayed in Figs.\ \ref{fig12} and
\ref{fig13}, respectively.

The approximations introduced by our extrapolation models are tailored
to provide composite GOS models that are continuous, except at the lines
$Q=Q_{\rm num,1} = t_{\rm num,1} E_a$ and
$W=(w_{\rm num}+1)E_a$, and compliant with the Bethe sums. Indeed,
the extrapolated GOSs are more accurate than other approximations used
in stopping calculations, where the Bethe ridge is replaced with a
structureless delta function or with the GOS obtained from symmetric
Compton profiles calculated from Hartree--Fock orbitals \citep[see,
\eg,][]{Segui2002}.


\section{Cross sections for inelastic collisions with atoms \label{sec6}}
\setcounter{equation}{0}

The DDCS for inelastic collisions with an atom or ion is obtained by
adding the contributions of the different atomic-electron subshells.
That is, the atomic DDCS takes the same form as the DDCS of
individual subshells, Eq.\ \req{4.46}, with atomic GOS and TGOS given by
\begin{subequations}
\label{6.1}
\beq
\frac{\d f (Q,W)}{\d W} = \sum_{a} \frac{q_a}{2|\kappa_a|}
\frac{\d f_{a} (Q,W)}{\d W}
\label{6.1a}\eeq
and
\beq
\frac{\d g (Q,W)}{\d W} = \sum_{a} \frac{q_a}{2|\kappa_a|}
\frac{\d g_{a} (Q,W)}{\d W}\, ,
\label{6.1b}\eeq
\end{subequations}
where $\d f_{a}(Q,W)/\d W$ and $\d g_{a}(Q,W)/\d W$ are, respectively,
the GOS and the TGOS of the subshell $n_a\kappa_a$ and the sums extend over
the occupied subshells of the ground-state configuration. We recall that
the calculations of the GOS and the TGOS described above assumed that
all active subshells $n_a\kappa_a$ are closed (all orbitals are occupied)
and that the final levels $n_b\kappa_b$ are empty. In the case of an
open-subshell $n_a\kappa_a$ with $q_a$ electrons ($q_a < 2|\kappa_a|$), its
contribution to the atomic GOSs is approximated by the product of the
fractional occupancy, $q_a/2|\kappa_a|$, and the GOS of the closed
shell. Transitions of electrons to empty orbitals of partially filled
subshells $n_b\kappa_b$ are also possible. These transitions are accounted
for approximately by assuming that the contribution of each empty
orbital in the final bound level is the same as for an empty subshell. That
is, if the subshell $n_b\kappa_b$ contains $q_b < 2|\kappa_b|$ electrons,
the contributions to the GOS and the TGOS of transitions to the level
$\epsilon_{n_b\kappa_b}$, are multiplied by $1-q_b/2 |\kappa_b|$. Thus,
angular integrals and spin sums corresponding to transitions from and to
open subshells are evaluated by using the same formulas as for closed
shells.

\begin{figure}[tbhp]
\begin{center}
\includegraphics*[width=10cm]{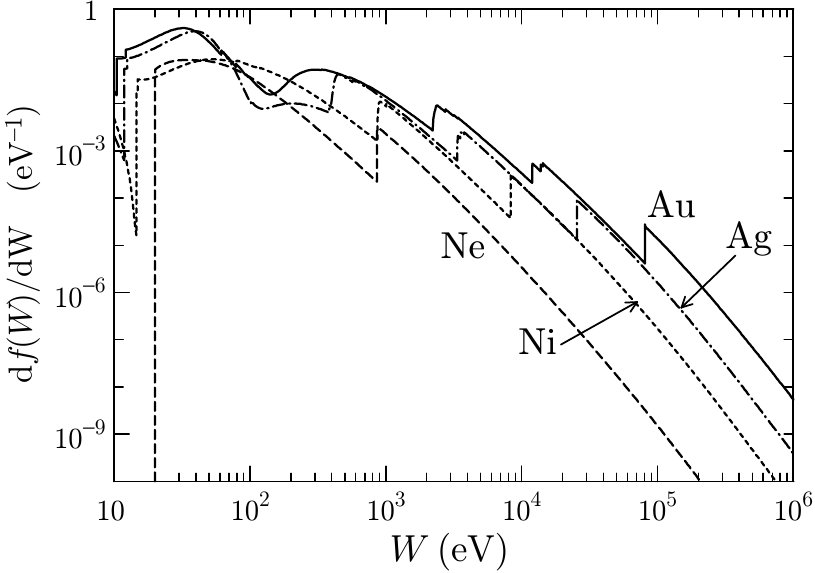}
\caption{Atomic OOSs of the indicated elements, as functions of the
energy loss $W$.
\label{fig14}}
\end{center}\end{figure}

Figure \ref{fig14} displays the OOSs (\ie, the GOS at $Q=0$) for
ionization of atoms of the elements neon, nickel, silver and gold. Note
that contributions from excitations to bound levels, which consist of a
set of discrete resonances with energy losses less than the ionization
energy $E_a$, are not included in the plot. The OOS is proportional to
the cross section for the photoelectric absorption of photons of energy
$W$, $\sigma^{\rm ph,dip}(W)$, calculated from the dipole approximation
\citep[see, \eg,][]{FanoCooper1968, SabbatucciSalvat2016},
\beq
\frac{\d f(W)}{\d W} \simeq \frac{\me c}{2\pi^2 e^2 \hbar}\,
\sigma^{\rm ph,dip}(W).
\label{6.2}\eeq
In the same way as the photoabsorption cross section, the OOS exhibits
discontinuities (``absorption edges'') at the ionization energies of the
electron subshells, $W=E_a$.

In non-relativistic theory \citep[see, \eg,][]{Inokuti1971,
Inokuti1978}, the atomic longitudinal GOS satisfies the Bethe sum rule,
\beq
\int_0^\infty \frac{\d f(Q,W)}{\d W} \, \d W = Z\, ,
\label{6.3}\eeq
which in the limit $Q=0$, reduces to the familiar dipole sum rule.
As shown in Section \ref{sec5.5}, these sum rules do not hold for the
relativistic GOS. We have
\beqa
\int_0^\infty \frac{\d f(Q,W)}{\d W} \, \d W &=& \sum_a
\frac{q_a}{2|\kappa_a|} \left[
{\sum_{b\ne a}} \, \frac{2|\kappa_b| - q_b}{2|\kappa_b|} f_{ba}(Q)
+ \int_{E_a}^\infty
\frac{\d f_{a}(Q,W)}{\d W} \, \d W
\right]\, ,
\label{6.4}\eeqa
where the summation over $b$ extends to all bound levels $\epsilon_{n_b
\kappa_b}$ different from $\epsilon_{n_a\kappa_a}$. Note that
transitions to filled subshells, with $q_b = 2 |\kappa_b|$, do not
contribute. Evidently,
\beqa
\int_0^\infty \frac{\d f(Q,W)}{\d W} \, \d W
&=& \sum_a
\frac{q_a}{2|\kappa_a|} \left[
{\sum_{b\ne a}} f_{ba}(Q)
+ \int_{E_a}^\infty
\frac{\d f_{a}(Q,W)}{\d W} \, \d W
\right]
\nonumber \\ [2mm]
&& \mbox{} - \sum_a
\frac{q_a}{2|\kappa_a|}
{\sum_{b\ne a}} \frac{q_b}{2|\kappa_b|} f_{ba}(Q)\, .
\label{6.5}\eeqa
Because $f_{ba}(Q) = -
f_{ab}(Q)$, the last summation vanishes, and,
\beqa
\int_0^\infty \frac{\d f(Q,W)}{\d W} \, \d W &=&
\sum_a \frac{q_a}{2|\kappa_a|} \, S_0(a;Q) \equiv S_0(Q)\, .
\label{6.6}\eeqa
where
\beq
S_0(a;Q) \equiv \sum_{b\ne a} \,
f_{ba}(Q)
+ \int_{E_a}^\infty
\frac{\d f_{a}(Q,W)}{\d W} \, \d W
\label{6.7}\eeq
is the one-subshell sum defined by Eq.\ \req{5.63}.
As shown in Section \ref{sec5.5}, $S_0(a;Q)$ is a monotonously
increasing function of $Q$. Hence the value $S_0(0)$ equals the total optical
oscillator strength, which is slightly smaller than $Z$. The difference
$1-S_0(0)/Z$ is negligible for low-$Z$ elements, for which the
non-relativistic theory is applicable, and increases with the atomic
number attaining values of the order of 0.025 for the heavier elements
($Z \sim 100$) [see Fig.\ \ref{fig26}].

It is instructive to consider the Bethe sum \req{6.4} from a slightly
different perspective. The Bethe sum rule (when it is expected to apply)
is useful for checking the consistency of experimental or semiempirical
GOSs. Of course, these GOSs only account for allowed transitions to
unoccupied final levels. The Bethe sum of a measured GOS would then
correspond to removing from the summation within brackets in Eq.\
\req{6.4} all transitions to closed subshells, which are forbidden by
Pauli's principle. The removed transitions from a given subshell
$n_a\kappa_a$ may be to outer subshells (with higher energy, positive GOS)
and to inner subshells (with lower energy, negative GOS). For inner (outer)
shells, the sum of the remaining terms will be smaller (larger) than the
original sum, because the removed terms are preferentially positive
(negative). Therefore, when considering real GOSs we will find that part
of the GOS is transferred from inner to outer subshells. This transfer of
GOS, which leaves the function $S_0(Q)$ unaltered, is independent of the
more fundamental deviations induced by relativistic effects (see Section
\ref{sec5.5}), which produce a net reduction of $S_0(Q)$.


\subsection{Atomic double-differential cross sections \label{sec6.1}}

\begin{figure}[thbp]
\begin{center}
\includegraphics*[width=7.5cm]{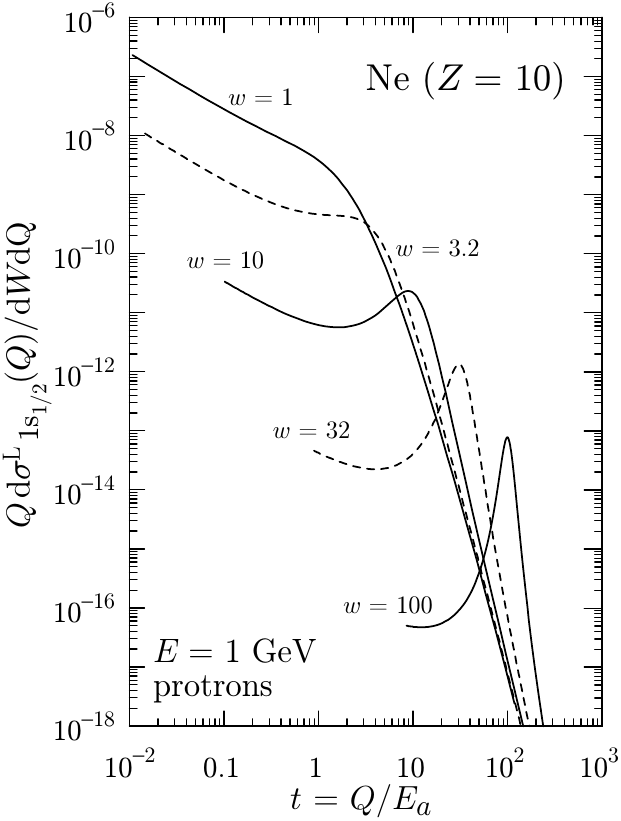} \hfill
\includegraphics*[width=7.5cm]{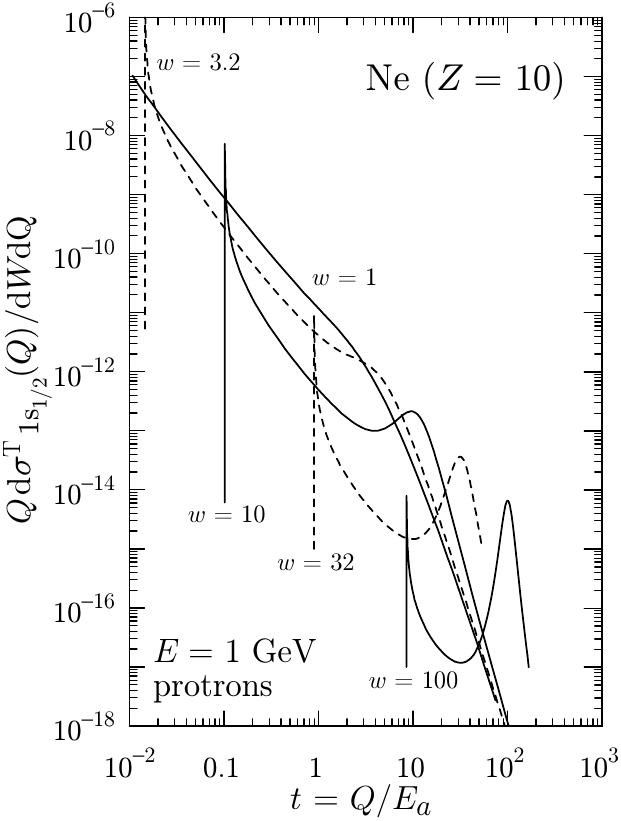} \\ [5mm]
\includegraphics*[width=7.5cm]{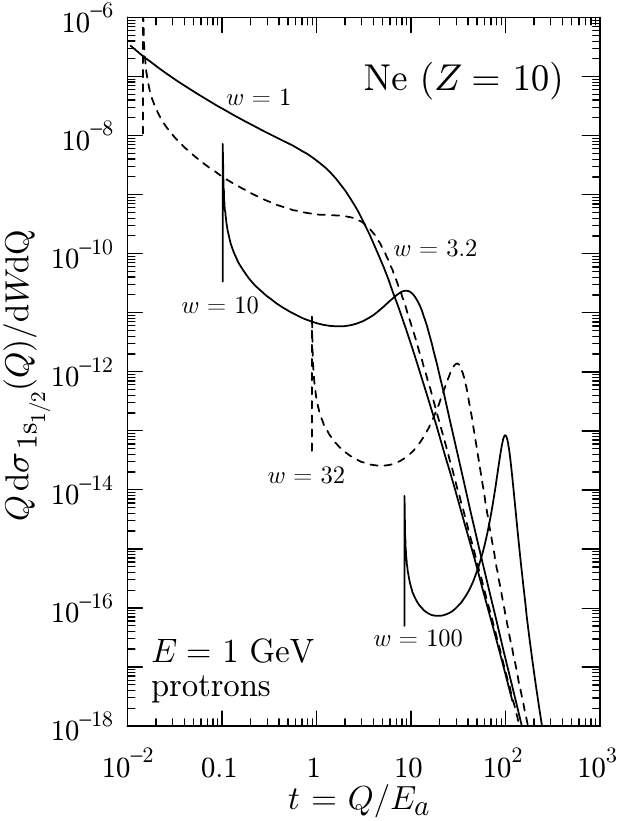}
\caption{DDCSs for longitudinal and transverse interactions, and their
sum, for ionization of the K shell ($1s_{1/2}$) of neon atoms ($Z=10$) by
impact of protons with 1 GeV kinetic energy, as functions of the reduced
recoil energy $t=Q/E_a$.
\label{fig15}}
\end{center}\end{figure}

\begin{figure}[thbp]
\begin{center}
\includegraphics*[width=7.5cm]{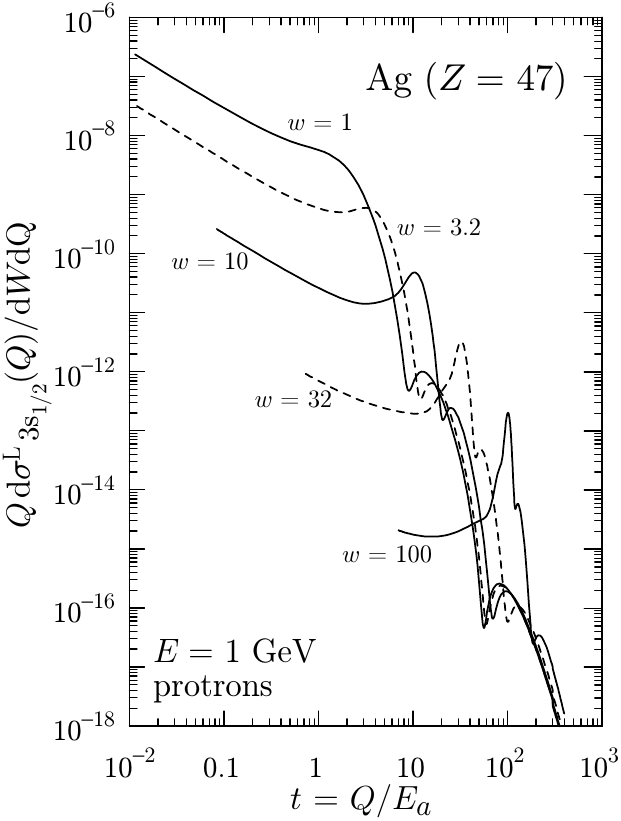} \hfill
\includegraphics*[width=7.5cm]{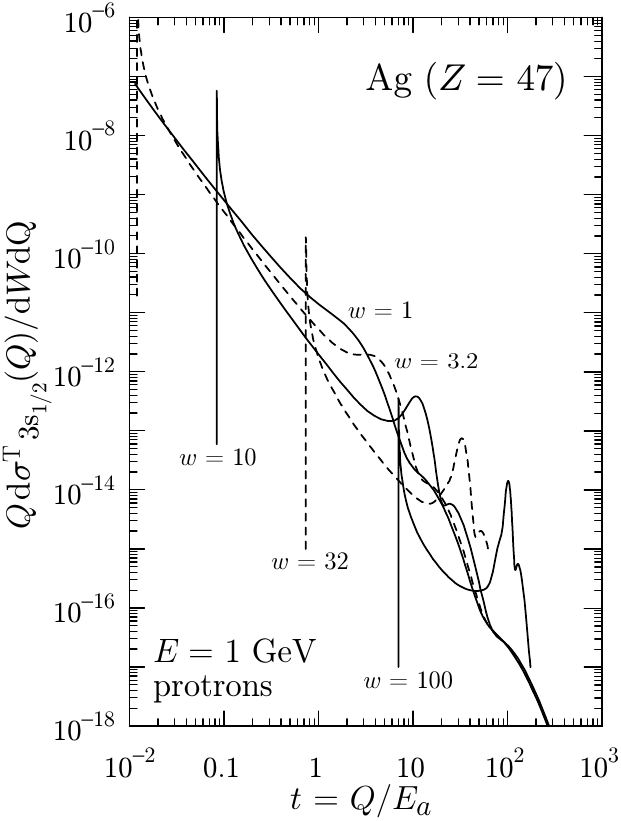} \\ [5mm]
\includegraphics*[width=7.5cm]{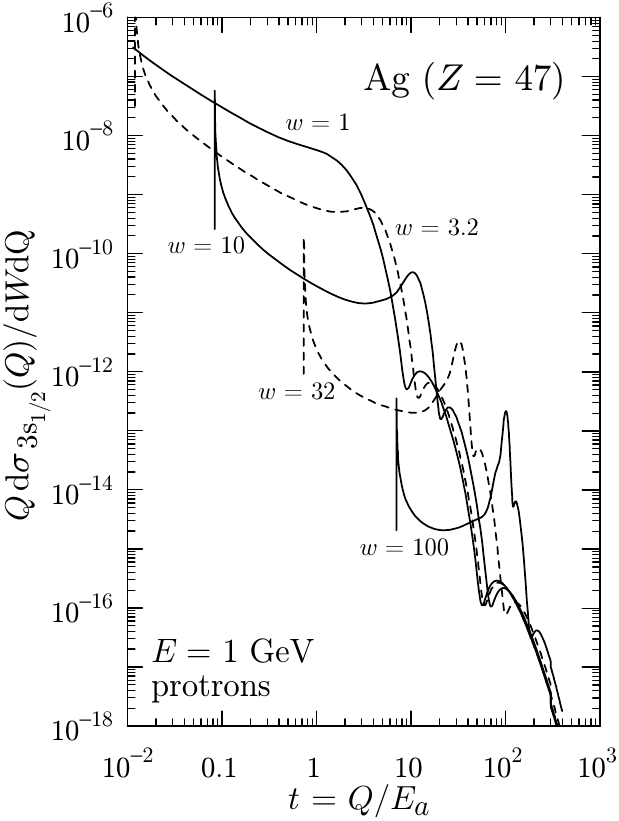}
\caption{DDCSs for longitudinal and transverse interactions, and their
sum, for ionization of the M1 subshell ($3s_{1/2}$) of silver atoms
($Z=47$) by impact of protons with 1 GeV kinetic energy, as functions of
the reduced recoil energy $t=Q/E_a$.
\label{fig16}}
\end{center}\end{figure}

{\allowdisplaybreaks
The atomic DDCS is obtained as the sum of expressions \req{4.46} for the
various electron subshells. For calculation convenience, it can be split
into contributions from longitudinal and transverse interactions,
\beq
\frac{\d^2 \sigma}{\d W \, \d Q} = \frac{\d^2 \sigma^{\rm L}}{\d
W \, \d Q} + \frac{\d^2 \sigma^{\rm T}}{\d W \, \d Q} \, ,
\label{6.8}\eeq
with
\beqa
\frac{\d^2 \sigma^{\rm L}}{\d W \, \d Q}
&=& \frac{2\pi Z_0^2 e^4}{\me v^2} \,
\frac{2\me c^2}{WQ(Q+2\me c^2)}
\nonumber \\ [2mm]
&& \mbox{} \times
\left\{ 1 - \frac{4(E+Mc^2)W - W^2 + Q(Q+2\me c^2)}{4\,
(E+M c^2)^2} \right\}
\frac{\d f (Q,W)}{\d W}
\label{6.9}\eeqa
and
\beqa
\frac{\d^2 \sigma^{\rm T}}{\d W \, \d Q}
&=& \frac{2\pi Z_0^2 e^4}{\me v^2} \,
\frac{2\me c^2 W}{[Q(Q+2\me c^2)-W^2]^2} \,
\nonumber \\ [2mm]
&& \mbox{} \times
\left( \beta^2 \sin^2\theta_{\rm r}
+ \left\{ \frac{Q(Q+2\me c^2)- W^2 }{2(E+ M c^2)^2}\,
\right\} \right) \frac{\d g (Q,W)}{\d W}\, ,
\label{6.10}\eeqa
where
\beq
\beta^2 \sin^2\theta_{\rm r} = \beta^2
- \frac{W^2}{Q(Q+2\me c^2)} \left( 1 + \frac{ Q(Q+2\me c^2)-W^2}
{2 W (E+M c^2)} \right)^2 \, .
\label{6.11}\eeq
We note that the DDCS \req{6.8} is proportional to the squared charge of
the projectile, which is a consequence of treating the interaction to
first order. More interestingly, the mass $M$ of the particle appears
only in the terms in curly brackets, which are appreciable only for
``hard'' collisions, with relative large values of the energy loss or
the recoil energy. For ``soft'' collisions, with small and moderate $W$
and $Q$, the DDCS is proportional to $Z_0^2$ and to $v^{-2}$ and
independent of the projectile mass.

Figures \ref{fig15} and \ref{fig16} display the DDCSs (calculated by
the codes {\sc gosat} and  {\sc pwacs}, see Section \ref{sec8}), together with
the contributions of longitudinal and transverse interactions, for
ionization of the K shell ($1s_{1/2}$) of neon and the M1 subshell
($3s_{1/2}$) of silver by impact of protons with $E=10^9$ eV, for the
indicated values of the reduced energy loss, $w=(W/E_a)-1$, as functions
of the reduced recoil energy $t=Q/E_a$. While the DDCS for longitudinal
interactions has a structure similar to that of the GOS, the DDCS for
transverse interactions develops a prominent narrow peak at $Q_-$, the
lower end of the allowed recoil-energy interval. Because of this peak,
care needs to be exercised in the numerical evaluation of integrals of
the DDCS over $Q$ for high-energy projectiles. For transitions with
small and moderate energy losses (say, for $w \lesssim 10$), the atomic
GOS and TGOS (as functions of $Q$) are essentially constant for $Q$ less
than $\sim 0.001 W$ (see Figs.\ \ref{fig5}, \ref{fig11} and
\ref{fig12}). In this low-$Q$ range, integrals of the DDCS can be
evaluated analytically by using the dipole approximation (see Section
\ref{sec9.1}).
}

\subsection{Angular differential cross section \label{sec6.2}}

The recoil energy $Q$ is a convenient variable to obtain compact
expressions of the DDCS. In practice, however, one may need to express
the DDCS in terms of the scattering angle $\theta$. For practical
purposes, specially for Monte Carlo simulations of charged-particle
transport, it is advantageous to measure angular deflections in terms of
the variable
\beq
\mu \equiv (1-\cos\theta)/2 = \sin^2(\theta/2)\, ,
\label{6.12}\eeq
which is related to the recoil energy through (see Appendix \ref{secA})
\beq
Q(Q+2\me c^{2}) = (cp-cp')^{2} + 4\, cp\, cp'\, \mu,
\label{6.13}\eeq
where $p$ and $p'$ are, respectively, the momenta of the projectile
before and after the interaction,
\beq
cp = \sqrt{E(E+2 M c^2)} \qquad \mbox{and} \qquad
cp' = \sqrt{(E-W)(E-W+2 M c^2)}\, .
\label{6.14}\eeq
The DDCS, differential in $W$ and $\mu$, is
\beq
\frac{\d^{2}\sigma}{\d W\,\d \mu} =
\frac{\d^{2}\sigma}{\d W\,\d Q} \frac{\d Q}{\d \mu} =
\frac{\d^{2}\sigma}{\d W\,\d Q} \,
\frac{2\, cp \, cp'}{Q+\me c^2}
\label{6.15}\eeq
with
\beq
Q = \sqrt{(cp-cp')^2 + 4 \, c p \, cp'  \mu + \me^2 c^4}-\me c^2\, .
\label{6.16}\eeq

The angular DCS is obtained upon integration of the DDCS over $W$,
\beq
\frac{\d \sigma}{\d \mu} = \int_0^{W_{\rm max}}
\frac{\d^{2}\sigma}{\d W\,\d Q} \,
\frac{2\, cp \, cp'}{Q+\me c^2}\, \d W\, .
\label{6.17}\eeq
Note that the usual angular DCS, per unit solid angle, is
\beq
\frac{\d \sigma}{\d \Omega} = \frac{1}{4\pi} \frac{\d \sigma}{\d \mu} .
\label{6.18}\eeq

For projectiles with not too high energies, we can derive a simple
expression for the angular DCS as follows. For these projectiles, the
contribution of transverse interactions is small, and the DDCS, Eq.\
\req{6.8}, reduces to the longitudinal term, Eq.\ \req{6.9},
\beq
\frac{\d^2 \sigma}{\d W \, \d Q} \simeq
\frac{2\pi Z_0^2 e^4}{\me v^2} \,
\frac{2\me c^2}{WQ(Q+2\me c^2)}
\, \frac{\d f (Q,W)}{\d W} \, ,
\label{6.19}\eeq
where we have disregarded the factor in braces, which is close to unity.
The calculation of the integral \req{6.17} is still
complicated, because $cp'$, $Q$ and the GOS depend on $W$. However, if
the energy $E$ of the projectile is substantially larger than the typical
excitation energies of the atom, we have $p'\simeq p$, and Eq.\
\req{6.13} simplifies to
\beq
Q(Q+2\me c^2) \simeq 4 (cp)^2 \mu = 4 \, E(E+2Mc^2) \mu  \, ,
\label{6.20}\eeq
an expression that is independent of $W$. As noted by
\citet{Inokuti1971},
this relation is valid within the region of the ($Q,W$) plane where the
curves of Fig.\ \ref{fig36} are straight vertical lines. Under this
circumstance,
$$
\frac{\d Q}{\d \mu} = \frac{2E(E+2Mc^2)}{Q+\me c^2} =
\frac{Q(Q+2Mc^2)}{(Q+\me c^2)2\mu}\, ,
$$
and
$$
\frac{\d^{2}\sigma}{\d W\,\d \mu} \simeq
\frac{\d^{2}\sigma}{\d W\,\d Q} \frac{\d Q}{\d \mu} =
\frac{2\pi Z_0^2 e^4}{\me v^2} \, \frac{\me c^2}{Q+\me c^2} \,
\frac{1}{Q \mu}
\, \frac{Q}{W} \, \frac{\d f (Q,W)}{\d W} \, .
$$
Assuming that $Q \ll \me c^2$ in the region where the GOS takes
appreciable values, the second factor on the right-hand side
becomes unity, and the relation \req{6.20} implies that
$$
Q = \frac{2E(E+2 Mc^2) \, \mu}{\me c^2} \, .
$$
Then we can write
$$
\frac{\d^{2}\sigma}{\d W\,\d \mu} =
\frac{\pi Z_0^2 e^4}{\beta^2} \, \frac{1}{E(E+2Mc^2)} \,
\frac{1}{\mu^2}
\, \frac{Q}{W} \, \frac{\d f (Q,W)}{\d W} \, ,
$$
and, expressing $\beta$ in the form \req{A.2}, we obtain
\beqa
\frac{\d^{2}\sigma}{\d W\,\d \mu} =
\pi Z_0^2 e^4 \left( \frac{E+M c^2}{E(E+2Mc^2)} \right)^2
\frac{1}{\mu^2}
\, \frac{Q}{W} \, \frac{\d f (Q,W)}{\d W} \, .
\label{6.21}\eeqa
The angular DCS is
$$
\frac{\d\sigma}{\d \mu} \simeq
\pi Z_0^2 e^4 \left( \frac{E+M c^2}{E(E+2Mc^2)} \right)^2
\frac{1}{\mu^2}
\, Q \int_0^{W_{\rm max}}
\frac{1}{W} \, \frac{\d f (Q,W)}{\d W} \, \d W \, .
$$
As the integrand decreases rapidly with $W$, the upper limit of the
integral can be set to infinity. This leads to the \citet{Morse1932} formula
\beqa
\frac{\d\sigma}{\d \mu} \simeq
\pi Z_0^2 e^4 \left( \frac{E+M c^2}{E(E+2Mc^2)} \right)^2
 \, \frac{1}{\mu^2} \,
S_{\rm inc}(Q)\, .
\label{6.22}\eeqa
The function
\beq
S_{\rm inc}(Q) \equiv Q \int_0^\infty \frac{1}{W} \,
\frac{\d f (Q,W)}{\d W} \, \d W
\label{6.23}\eeq
is known as the incoherent-scattering function \citep{Hubbell1975,
Hubbell1975-erratum}. This function vanishes at $Q=0$, because the integral is
finite, and increases monotonically to reach a saturation value equal to
$Z$, because for large $Q$ the GOS becomes $Z \delta(W-Q)$ [see Appendix
\ref{secD}]. The incoherent-scattering function is employed in approximate
calculations of scattering of x-rays \citep{WallerHartree1929},
electron-positron pair production by gamma rays, and bremsstrahlung
emission by electrons \citep{WheelerLamb1939, Tsai1974}. The present
derivation indicates that such calculations involve a wealth of
simplifications, which effect on the resulting cross sections is
difficult to estimate.

Tables of incoherent-scattering functions $S_{\rm inc}(Q)$ for free
atoms of the elements $Z=1$ to 100 have been published by
\citet{Hubbell1975}, for a grid of 45 values of the variable
\beq
x = \frac{q}{4\pi} \, 10^{-8} \; {\rm cm} = 20.6074 \, \frac{\hbar
q}{\me c} \, .
\label{6.24}\eeq
Expanded tabulations, for a denser grid of $x$ values, are included in the
Evaluated Photon Data Library (EPDL) of \citet{Cullen1997}. These
incoherent-scattering functions were calculated from the expression
\beq
S_{\rm inc}(Q) = \left\langle \Psi_{0}
\left| \sum_{j=1}^{Z} \sum_{k=1}^{Z}
\exp[{\rm i}{\bf q}\dotprod({\bf r}_{k}-{\bf r}_{j})] \right|
\Psi_{0} \right\rangle
- \left| \left\langle \Psi_{0}
\left| \sum_{j=1}^{Z} \exp(-{\rm i}{\bf q}\dotprod{\bf r}_{j}) \right|
\Psi_{0} \right\rangle \right|^{2},
\label{6.25}\eeq
where $\Psi_0$ is the ground-state wave function of the target atom and
the summations run over the $Z$ atomic electrons. For hydrogen ($Z=1$),
$S_{\rm inc}(Q)$ was calculated using the hydrogenic ground-state
wave function. For $Z=1$ to 6, configuration-interaction wave functions,
which account for electron-correlation effects, were employed. For all
other elements, from $Z=7$ to 100, $S_{\rm inc}(Q)$ was
calculated using non-relativistic Hartree--Fock wave functions.

\begin{figure}[h]
\begin{center}
\includegraphics*[width=7.50cm]{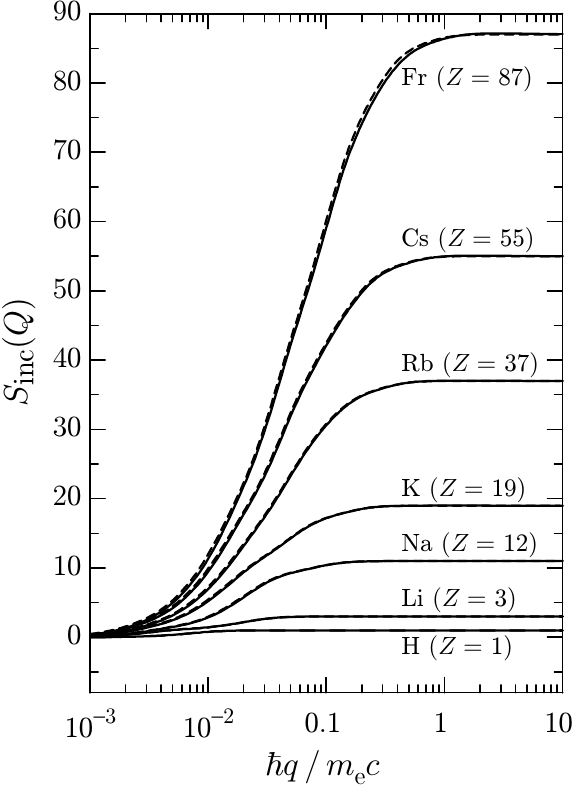}
\caption{Incoherent scattering functions $S_{\rm inc}(Q)$ of free
atoms of the alkali metals, as functions of the dimensionless variable
$\hbar q/ \me c$. The dashed curves are from the EPDL \citep{Cullen1997};
the solid curves were obtained from the numerical integration of
the GOS, Eq.\ \req{6.23}.
\label{fig17}}
\end{center}\end{figure}

Figure \ref{fig17} compares the incoherent-scattering functions of the
alkali atoms, taken from the EPDL, with those obtained by numerical
integration of our GOSs. Because of the
differences between the adopted wave functions, our results are not
expected to coincide with Hubbell \etal's data. Nonetheless, the
comparison shows that the incoherent-scattering function depends rather
weakly on the specifics of the atomic wave function and, moreover, it
serves to verify the accuracy of our integration and interpolation
algorithms. Indeed, visual inspection of the calculated $S_{\rm inc}(Q)$
functions was useful to reveal inconsistencies in the preliminary
versions of our GOS database. In the case of hydrogen, however, our wave
function is essentially the same as in Hubbell \etal's calculation, and
the corresponding incoherent-scattering functions are found to agree to
4 decimal places. For elements with $Z\ge 7$ and for intermediate recoil
energies, the non-relativistic functions of Hubbell \etal\ have values
that are systematically higher than ours. The differences seem to
increase with $Z$, and are of the same order of magnitude as the
relativistic departures from the Bethe sum rule.

\subsection{Energy-loss DCS and its integrals \label{sec6.3}}

For a given energy loss $W$, the allowed recoil energies are limited to
the interval $(Q_-,Q_+)$ with [Eq.\ \req{A.21}]
\beqa
Q_{\pm} &= &\sqrt{ (cp \pm cp')^2 + \me^2 c^4} - \me c^2
\nonumber \\ [4mm]
&=& \sqrt{
\left[ \sqrt{E(E+2mc^2)} \pm \sqrt{(E-W)(E-W+2mc^2)}\right]^2 + \me^2
c^4 } - \me c^2\, . \rule{12mm}{0mm}
\label{6.26}\eeqa
Conversely, for a given recoil energy $Q< Q_+$, the maximum allowed
energy loss is [see Eq.\ \req{A.27}]
\beq
W_{\rm m}(Q) = E + Mc^2 - \sqrt{\left[ \sqrt{E(E+2Mc^2)} -
\sqrt{Q(Q+2\me c^2)} \right]^2 + M^2 c^4}\, .
\label{6.27}\eeq
Note that the maximum energy loss $W_{\rm m}(Q)$ increases with the
energy of the projectile and at high energies tends to the energy of the
vacuum photon line [see Fig.\ \ref{fig38}],
\beq
W_0(Q)= \sqrt{Q(Q+2\me c^2)}\, .
\label{6.28}\eeq

The energy-loss DCS (\ie, the DCS as a function of only the energy loss
$W$) is obtained by integrating the DDCS over the recoil energy,
\beqa
\frac{\d \sigma}{\d W}
= \int_{Q_{-}}^{Q_{+}}\frac{\d^2 \sigma}{\d W \,\d Q} \, \d Q\, .
\label{6.29}\eeqa
The energy-loss DCS is defined only for energy losses that are less than
$W_{\rm max}=E$. For energy
losses much smaller than $E$ we have [see Eq.\ \req{A.25}],
\beq
Q_- = W^2 / (2 \me c^2 \beta^2)\, ,
\label{6.30}\eeq
which is independent of the mass of the projectile. Moreover, $Q_+$ is
much larger than $W$, so that the GOS effectively vanishes at the upper
end of the integration interval. This implies that the energy-loss DCS
for energy losses much smaller than $E$ depends on the mass $M$ of the
projectile only through the factors that appear in braces in Eqs.\
\req{6.9} and \req{6.10}. This is not true when the energy loss is
comparable to $E$, because then the values of both $Q_-$ and $Q_+$ do
depend on $M$.

For practical purposes, it is useful to introduce the following
integrals (moments) of the energy-loss DCS,
\beq
\sigma^{(i)} \equiv
\int_{0}^{W_{\mathrm{max}}} W^{i} \, \frac{\d\sigma}{\d W} \, \d W\, .
\label{6.31}\eeq
Notice that $\sigma^{(0)}$ is the total inelastic cross section.
$\sigma^{(1)}$ and $\sigma^{(2)}$ are known as the stopping cross
section and the energy straggling cross section (for inelastic
collisions), respectively. Recalling that the probability density of
the energy loss $W$ in a single collision is
\beq
p_1(W) = \frac{1}{\sigma^{(0)}} \, \frac{\d \sigma}{\d W}\, ,
\label{6.32}\eeq
we can write
\beq
\sigma^{(i)} =
\sigma^{(0)} \int_{0}^{W_{\mathrm{max}}} W^{i} \, p_1(W) \, \d W =
\sigma^{(0)} \, \langle W^{i} \rangle\, ,
\label{6.33}\eeq
where $\langle W^{i}\rangle$ is the average value of $W^{i}$ in a
single collision.

Let us consider that our fast projectile is moving in a monoatomic gas
of the element of atomic number $Z$ with mass density $\rho_{\rm M}$.
The number of atoms per unit volume is
\beq
{\cal N} = \frac{N_{\rm A} \rho_{\rm M}}{A_{\rm w}}\, ,
\label{6.34}\eeq
where $N_{\rm A} = 6.023 \times 10^{23}$ mol$^{-1}$ is Avogadro's
number,
and $A_{\rm w}$ is the molar mass (g/mol) of the element.
The mean free path $\lambda$ for inelastic collisions is given by
\beq
\lambda = 1/(\mathcal{N} \sigma^{(0)})\, .
\label{6.35}\eeq
Its inverse, $\lambda^{-1}= \mathcal{N} \sigma^{(0)}$, is the
probability of interaction per unit path length of the projectile in the
gas.
The (collision) stopping power $S$ and the energy straggling
parameter $\Omega^{2}$ are defined by
\beq
S = \mathcal{N}\sigma^{(1)} = \frac{\langle W\rangle}{\lambda}
\label{6.36}\eeq
and
\beq
\Omega^{2} = \mathcal{N} \sigma^{(2)} =
\frac{\langle W^{2}\rangle}{\lambda}\, ,
\label{6.37}\eeq
respectively. Evidently, the stopping power gives the average energy
loss per unit path length. The product $\Omega^{2}\,\d s$ is the
variance of the energy distribution of an originally monoenergetic beam
after a short path length $\d s$ \citep[see,
\eg,][]{Salvat2019}.


\subsection{Interactions with large recoil energies \label{sec6.4}}

\begin{figure}[h]
\begin{center}
\includegraphics*[width=8.5cm]{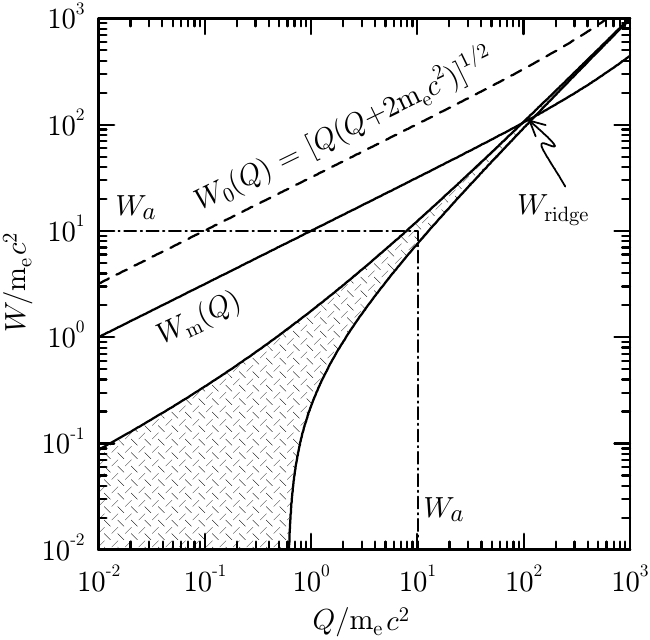}
\caption{
Schematic representation of the Bethe surface. The hatched area
represents the region where the GOS takes non-vanishing values. When
$Q$, or $W$, is larger than $W_a$ the GOS reduces to the Bethe ridge,
which peaks near the line $Q=W$. The dashed curve is the vacuum photon
line, $W_0(Q)$, Eq.\ \req{6.28}. The curve $W_{\rm m}(Q)$ represents the
expression \req{6.27} for a proton with $E=50$ MeV. The intersection of
this curve with the line $W=Q$ gives the largest possible energy
transfer $W_{\rm ridge}$ in collisions with stationary free electrons.
\label{fig18}}
\end{center}\end{figure}

Let $W_a$ be an energy loss much larger than the ionization energy $E_a$
of the active electron, such that the integral of the OOS on the
interval $(W_a, \infty)$ is negligible in comparison with the number
$2|\kappa_a|$ of electrons in the subshell (see Fig.\ \ref{fig18}).
Then, for recoil energies larger than $W_a$, the GOS and the TGOS of the
subshell can be approximated by the delta function, $2 |\kappa_a| \,
\delta(W-Q)$. That is, collisions with $Q>W_a$ can be described as
binary collisions with free electrons, and the corresponding energy-loss
DCS (per electron) is given by the formulas \req{D.27} and \req{D.28},
\beqa
\frac{\d \sigma_{\rm free}}{\d W}
&\simeq& \frac{2\pi Z_0^2 e^4}{\me
v^2} \, \frac{1}{W^2} \, F_{\rm rel}(W)
\label{6.38}\eeqa
with
\beqa
F_{\rm rel}(W) &=& 1 -
\frac{\left[(2E-W+2 M c^2)\me c^2 +M^2c^4+\me^2 c^4\right]W}
{2\me c^2(E+M c^2)^2} \, .
\label{6.39}\eeqa

The maximum allowed energy loss is determined by the intersect of the curve
$W_{\rm m} (Q)$, Eq.\ \req{6.27}, with the Bethe ridge ($W=Q$), which
occurs at a point where $W$ has the value
\beq
W_{\rm ridge} = \frac{2\me c^2 \beta^2}{1-\beta^2} \, R \qquad
\mbox{with} \qquad
R \equiv
\left[ 1+\left(\frac{\me}{M}\right)^2+ \frac{2}{\sqrt{1 - \beta^2}}
\, \frac{\me}{M} \right]^{-1}\, .
\label{6.40}\eeq
Notice that, when $M=\me$, $W_{\rm ridge}=E$.
For projectiles heavier than the electron ($M \gg \me$) with energies
less than, or comparable to, their rest energies $Mc^2$, $R \sim 1$ and
\beq
W_{\rm ridge}\simeq \frac{2\me c^2 \beta^2}{1-\beta^2}\, .
\label{6.41}\eeq
Of course, for energies of the order of $Mc^2$ or larger, the complete
expression \req{6.40} must be used.

The kinematical factor \req{6.39} can be written as
\beq
F_{\rm rel}(W) = 1 - \beta^2 \frac{W}{W_{\rm ridge}} +
\frac{1-\beta^2}{2 M^2 c^4} \, W^2 \, .
\label{6.42}\eeq
Then, the DDCS for large-$Q$ excitations, $Q>W_a$, of the active
electron subshell takes the form
\beqa
\frac{\d^2 \sigma_{a,Q>W_a}}{\d Q\, \d W}
&\simeq& \left( \frac{2\pi Z_0^2 e^4}{\me
v^2} \, \frac{1}{W^2} \right) \left[
1 - \beta^2 \frac{W}{W_{\rm ridge}} +
\frac{1-\beta^2}{2 M^2 c^4} \, W^2
\right] 2 |\kappa_a| \,
\delta(W-Q)\, ,
\label{6.43}\eeqa
and the contribution of those excitations to the energy-loss DCS is
\beqa
\frac{\d \sigma_{a,Q>W_a}^{\rm free}}{\d W}
&=& 2|\kappa_a| \, \left( \frac{2\pi Z_0^2 e^4}{\me v^2} \,
\frac{1}{W^2} \right) \left[
1 - \beta^2 \, \frac{W}{W_{\rm ridge}} + \frac{1-\beta^2}{2M^2 c^4}\, W^2
\right] \Theta(W_{\rm ridge} - W_a)\, , \rule{10mm}{0mm}
\label{6.44}\eeqa
where $\Theta(x)$ is the unit step function ($=0$ for $x < 0$ and $=1$
for $x > 0$). The factor in parenthesis, is the non-relativistic Thomson
energy-loss DCS, but with the relativistic speed $v = \beta c$. The
factor in square brackets accounts for the remaining relativistic
corrections.

The corresponding contributions to the integrated cross
sections $\sigma^{(i)}$ are
\begin{subequations}
\label{6.45}
\beq
\sigma_{a,Q>W_a}^{(0)} =  2|\kappa_a| \,
 \frac{2\pi Z_0^2 e^4}{\me v^2}
\left[ - \frac{1}{W} - \beta^2 \,\frac{\ln W}{W_{\rm ridge}}
+ \frac{1-\beta^2}{2M^2 c^4} \, W
\right]_{W_a}^{W_{\rm ridge}}\, ,
\label{6.45a}\eeq
\beq
\sigma_{a,Q>W_a}^{(1)} =  2|\kappa_a| \,
\frac{2\pi Z_0^2 e^4}{\me v^2}
\left[ \ln W - \beta^2 \, \frac{W}{W_{\rm ridge}}
+ \frac{1-\beta^2}{2M^2 c^4} \, \frac{W^2}{2}
\right]_{W_a}^{W_{\rm ridge}} \,
\label{6.45b}\eeq
and
\beq
\sigma_{a,Q>W_a}^{(2)} =  2|\kappa_a| \,
\frac{2\pi Z_0^2 e^4}{\me v^2}
\left[ W - \beta^2 \, \frac{W^2}{2W_{\rm ridge}}
+ \frac{1-\beta^2}{2M^2 c^4} \, \frac{W^3}{3}
\right]_{W_a}^{W_{\rm ridge}} \, .
\label{6.45c}\eeq
\end{subequations}
Note that these formulas apply only when $W_{\rm ridge}>W_a$; otherwise
$\sigma^{(i)}_{a,Q>W_a}=0$.


\section{Collisions of electrons and positrons \label{sec7}}
\setcounter{equation}{0}

Collisions of electrons ($Z_0=-1$, $M=\me$) with atoms differ from those
of heavy particles in that the projectile is indistinguishable from the
atomic electrons and, consequently, interactions are affected by
exchange effects (such as re-arrangement collisions and interference
between direct and exchange $T$-matrix elements). Exchange effects occur
also in inelastic collisions of positrons ($Z_0=+1$, $M=\me$). The
reason is that the (active) electron-positron pair can undergo
annihilation followed by recreation of a new pair, a process that
coexists with ordinary scattering.  Exchange effects then arise from the
indistinguishability of the target electron from the electrons in
virtual states of negative energy (the Dirac sea). In the energy range
where the PWBA is expected to be reliable, the total cross section is
known to be fairly insensitive to these effects \citep[see,
\eg,][]{BoteSalvat2008}. However, electron exchange introduces
appreciable modifications in the stopping power and the energy
straggling of both electrons and positrons. It is difficult to make
allowance of exchange within the PWBA, mainly because the projectile
plane waves are not orthogonal to the bound and free orbitals of the
active electron.  Exchange corrections can be calculated consistently
only in the case of collisions with free electrons at rest (because both
the projectile and the target are then described by plane waves). The
corresponding exchange-corrected PWBA DCSs for projectile electrons and
positrons are given by simple analytical formulas, which were derived by
\citet{Moller1932} and \citet{Bhabha1936}, respectively. The common
approach adopted to account for the effect of electron exchange on the
stopping power of electrons and positrons is based on the fact that, for
energy transfers $W$ much larger than the ionization energy $E_a$, the
target electron can be regarded as free. Thus, the well-known Bethe
formula for the stopping power \citep{RohrlichCarlson1954, ICRU37} is
obtained by assuming that the energy-loss DCS for large-$W$ collisions
can be approximated by the M\o ller or Bhabha formulas.


\subsection{Cross sections for electrons \label{sec7.1}}

In the case of collisions of non-relativistic electrons, exchange
effects can be described approximately by using the \citet{Ochkur1964}
approximation \citep[see also][]{Ochkur1964, Ochkur1965, Rudge1968},
which consists of multiplying the GOS by the factor
\beq
F_{\rm Ochkur} (Q,W) =
1+\left(\frac{Q}{\langle {\cal K} \rangle+ E-W}\right)^{2}
-\frac{Q}{\langle {\cal K} \rangle +E-W}\, ,
\label{7.1}\eeq
where $\langle {\cal K} \rangle$ is the kinetic energy of the target
electron. We notice that the Ochkur factor \req{7.1} reduces to unity
when $Q=0$, \ie, exchange effects are negligible for small-$Q$
collisions. This peculiarity does make sense because, in a
semi-classical picture, interactions with small momentum transfers
correspond to distant collisions (with large impact parameters) in which
the two particles retain their identities. In ionizing collisions, we
have two (indistinguishable) free electrons in the final state, and it
is natural to consider the fastest as the ``primary''. Consequently, the
largest allowed energy loss in ionization is
\beq
W_{\rm max} = (E-\epsilon_{n_a\kappa_a})/2 = (E + E_a)/2\, ,
\label{7.2}\eeq
where $E_a$ is, as usual, the ionization energy of the target electron.
In the case of excitation to bound levels, the ``primary'' electron is
the one that remains free after the collisions and, therefore, the
maximum energy loss is $W_{\rm max}=E$.

Within the PWBA exchange effects can be accounted for rigorously only in
the case of collisions with free electrons. The energy-loss DCS for
collisions with a free electron at rest is given by the
\citet{Moller1932} formula,
\beqa
\frac{\d \sigma_{\rm M\o ller}}{\d W} & = &
\frac{2 \pi e^4}{\me v^2} \frac{1}{W^{2}}
\, F_{\rm M\o ller} (W) \, ,
\label{7.3}\eeqa
where
\beq
F_{\rm M\o ller} (W) =
1  + \left( \frac{W}{E-W} \right)^{2}
- \frac{(1-b_0)W}{E-W} + \frac{b_0 W^{2}}{E^{2}}
\label{7.4}\eeq
with
\beq
b_0 = \left( \frac{E}{E+\me c^{2}} \right)^{2}
= \left( \frac{\gamma-1}{\gamma} \right)^{2} = \left( 1 -
\sqrt{1-\beta^2} \right)^2\, .
\label{7.5}\eeq
Here, $\gamma$ is the total energy of the projectile electron in units
of its rest energy, Eq.\ \req{A.3},
$$
\gamma \equiv \frac{1}{\sqrt{1-\beta^2}} = \frac{E + \me c^2}{\me c^2}\, .
$$

Let us note that the DDCS for collisions with free stationary electrons
($\langle {\cal K} \rangle=0$) can be expressed in the form
\beqa
\frac{\d^2 \sigma_{\rm M\o ller}}{\d Q \, \d W} & = &
\frac{2 \pi e^4}{\me v^2} \frac{1}{W^{2}}
\, F_{\rm OM} (Q,W) \, \delta(W-Q)\, ,
\label{7.6}\eeqa
with
\beq
F_{\rm OM} (Q,W) =
1  + \left( \frac{Q}{\langle {\cal K} \rangle+E-W} \right)^{2}
- \frac{(1-b_0)Q}{\langle {\cal K} \rangle+E-W} +
\frac{b_0Q^{2}}{(\langle {\cal K} \rangle+E)^{2}}\, .
\label{7.7}\eeq
The last expression differs from the conventional M\o ller
factor \req{7.4} in that the numerators contain $Q$ instead of $W$. We have
introduced this seemingly arbitrary replacement to get an expression
analogous to $F_{\rm Ochkur}(Q,W)$, Eq.\ \req{7.1}, and such that
$F_{\rm OM} (Q,W)$ reduces to unity for $Q=0$. Of course, this
formal modification does not alter the DDCS for collisions with free
electrons at rest. In the non-relativistic limit ($b_0\rightarrow
0$), the factor $F_{\rm OM}(Q,W)$ reduces to the Ochkur factor.

In the case of binary collisions of an electron with a free electron
at rest, considering the two electrons as distinguishable, the PWBA
yields the DDCS [see Eqs.\ \req{6.38} and \req{6.39}]
\beqa
\frac{\d^2 \sigma_{\rm free}}{\d Q \d W}
&=& \left( \frac{2\pi e^4}{\me
v^2} \, \frac{1}{W^2} \right) \, F_{\rm rel}(W)
\, \delta(Q-W)\,
\label{7.8}\eeqa
with
\beq
F_{\rm rel}(W) = 1 - \frac{(2E-W+4 \me c^2)W}{2(E+\me c^2)^2}\, .
\label{7.9}\eeq

To account for exchange effects in electron-atom collisions, we will
multiply the DDCS by correcting factors. In the cases of excitations to
bound states and of low-$Q$ ionizing transitions, with $Q \le E_a$, we will
adopt Ochkur's approximation, Eq.\ \req{7.1}. Specifically, we will
multiply the DDCS for longitudinal interactions by the factor \req{7.1},
but leave the DDCS for transverse interactions unchanged, because these
are mostly distant interactions for which exchange effects are expected to be
negligible. That is,
\begin{subequations}
\label{7.10}
\beq
\frac{\d^2 \sigma_a ({\rm e}^-)}{\d W \, \d Q} =
F_{\rm Ochkur} (Q,W) \, \frac{\d^2 \sigma^{\rm L}_a}{\d
W \, \d Q} + \frac{\d^2 \sigma^{\rm T}_a}{\d W \, \d Q}
\qquad
\mbox{for $W \le E_a$, or $W>E_a$ and $Q<E_a$,}
\label{7.10a}\eeq
where the DDCSs on the right-hand side are the longitudinal and
transverse parts given by Eqs.\ \req{6.9} and \req{6.10}.
For large-$Q$ ionizations, with $Q>E_a$ and $W>E_a$,
exchange effects will be accounted by multiplying the DDCS by the factor
$F_{\rm OM}(Q,W)/F_{\rm rel}(W)$, \ie,
\vspace*{-4mm}
\beq
\frac{\d^2 \sigma_a ({\rm e}^-)}{\d W \, \d Q} =
\frac{F_{\rm OM}(Q,W)}{F_{\rm rel} (W)} \left( \frac{\d^2
\sigma^{\rm L}_a}{\d W \, \d Q} + \frac{\d^2 \sigma^{\rm T}_a}{\d W \,
\d Q}\right) \qquad \mbox{for $W>E_a$ and $Q>E_a$\, .}
\label{7.10b}\eeq
\end{subequations}
Thus, for recoil energies larger $W_a$, where the GOS reduces to the
delta function, $2|\kappa_a| \delta(W-Q)$, the DDCS \req{7.10} reduces
to the M\o ller DDCS \req{7.3}, multiplied  by the number $2|\kappa_a|$
of electrons in the subshell.

\allowdisplaybreaks{
The energy-loss DCS for large-$Q$ collisions with $Q>W_a$ is given by
the M\o ller formula,
\beqa
\frac{\d \sigma_{a,Q>W_a}^{\rm free}}{\d W} & = &
\frac{2 \pi e^4}{\me v^2} \, 2|\kappa_a| \, \frac{1}{W^{2}}
\left[ 1  + \left( \frac{W}{E-W} \right)^{2}
- \frac{(1-b_0)W}{E-W} + \frac{b_0 W^{2}}{E^{2}} \right]\, .
\label{7.11}\eeqa
The contributions of these collisions to the total, stopping and
energy-straggling cross sections are given by the following analytical
expressions,
\begin{subequations}
\label{7.12}
\beqa
\sigma_{a,Q>W_a}^{(0)}  ({\rm e}^-) & = &
\frac{2 \pi e^4}{\me v^2} \,  2|\kappa_a| \,
\left[ -\frac{1}{W} + \frac{1}{E-W}
+ \frac{1-b_0}{E}
\ln \left( \frac{E-W}{W}\right) + \frac{b_0 W}{(E)^{2}}
\right]_{W_a}^{W_{\rm max}} , \rule{13mm}{0mm}
\label{7.12a}\eeqa
\beqa
\sigma_{a,Q>W_a}^{(1)}  ({\rm e}^-)& = & \frac{2 \pi e^4}{\me
v^2} \,  2|\kappa_a| \,
\left[ \ln W + \frac{E}{E-W}
+ (2-b_0)\ln(E-W) + \frac{b_0 W^{2}}{2(E)^{2}}
\right]_{W_a}^{W_{\rm max}},  \nonumber \\
\label{7.12b}\eeqa
and
\beqa
\sigma_{a,Q>W_a}^{(2)} ({\rm e}^-) & = & \frac{2 \pi e^4}{\me v^2} \,
 2|\kappa_a| \,
\left[ (3-b_0)W + \frac{E^2}{E-W}
+ (3-b_0)E \ln(E-W) + \frac{b_0 W^3}{3E^2}
\right]_{W_a}^{W_{\rm max}}.  \nonumber \\ 
\label{7.12c}\eeqa
\end{subequations}
Of course, these formulas apply only when $W_{\rm max}>W_a$; otherwise
$\sigma^{(i)}_{Q>W_a}=0$.
}


\subsection{Cross sections for positrons \label{sec7.2}}

The scattering of positrons by electrons at rest has been investigated
by \citet{Bhabha1936}, who found that the energy-loss DCS for inelastic
collisions of a positron having kinetic energy $E$ with a free
stationary electron is given by
\beq
\frac{\d \sigma_{{\rm Bhabha}}}{\d W} =
\frac{2 \pi e^4}{\me v^2} \, \frac{1}{W^{2}}
\, F_{\rm Bhabha} (W) \, ,
\label{7.13}\eeq
where
\beq
F_{\rm Bhabha} (W) = 1 - b_{1}\frac{W}{E} +
b_{2}\left(\frac{W}{E}\right)^{2} - b_{3}\left(\frac{W}{E}\right)^{3} +
b_{4}\left(\frac{W}{E}\right)^{4}\, ,
\label{7.14}\eeq
with
\beqa
b_{1} & = & \left( \frac{\gamma-1}{\gamma} \right)^{2}
\frac{2(\gamma+1)^{2}-1}{\gamma^{2}-1}, \nonumber \\[2mm]
b_{2} & = & \left( \frac{\gamma-1}{\gamma} \right)^{2}
\frac{3(\gamma+1)^{2}+1}{(\gamma+1)^{2}}, \nonumber \\[2mm]
b_{3} & = & \left( \frac{\gamma-1}{\gamma} \right)^{2}
\frac{2\gamma(\gamma-1)}{(\gamma+1)^{2}}, \nonumber \\[2mm]
b_{4} & = & \left( \frac{\gamma-1}{\gamma} \right)^{2}
\frac{(\gamma-1)^{2}}{(\gamma+1)^{2}}\, .
\label{7.15}\eeqa
The maximum allowed energy loss is $W_{\rm max} = E$, because the
positron and the electron are distinguishable. The Bhabha formula
accounts for the effects of electron exchange and of the competing
process of annihilation followed by recreation. Note that in the
non-relativistic limit ($\gamma \rightarrow 1$) $b_i=0$.

We see that, in collisions of positrons with free electrons, exchange
effects are appreciable only for relativistic projectiles (such that
$\gamma$ is appreciably larger than unity) and for collisions with relatively large energy transfers. That
is, most of the effect occurs for large-$W$ collisions, for which the
Bhabha DCS is expected to provide a fairly good approximation. In the
case of collisions with electrons bound in atoms and ions, it is
difficult to modify the PWBA to include exchange effects. Following the
usual practice in stopping theory, we will give a partial account of
these effects by simply multiplying the DDCS for high-$W$ excitations,
for which $Q \simeq W$, by a factor that yields the Bhabha DCS \req{7.13}
at large $Q$, leaving the DDCS for low- and intermediate $Q$ unchanged.
Specifically, we will set
\beq
\frac{\d^2 \sigma_a ({\rm e}^+)}{\d W \, \d Q} = \left\{
\begin{array}{ll}
\displaystyle{
\frac{\d^2 \sigma_a}{\d W \, \d Q}  }
& \mbox{if $Q<E_a$\, ,} \\ [6mm]
\displaystyle{\frac{F_{\rm Bhabha}(W)}{F_{\rm rel} (W)} \,
\frac{\d^2 \sigma_a}{\d W \, \d Q} \; \; \; } & \mbox{if
$Q> E_a$\, ,}
\end{array}\right.
\label{7.16}\eeq
where the DDCSs on the right-hand side is given by Eqs.\ \req{4.46} and
\req{4.47}, and
$F_{\rm rel}(W)$ is defined by Eq.\ \req{7.9}.

The DDCS for high-$Q$ collisions, with $Q>W_a$, of positrons with
the electrons of a subshell $n_a\kappa_a$ is given by the Bhabha formula
\req{7.13} multiplied by the number $2|\kappa_a|$ of electrons in the
shell,
\beq
\frac{\d \sigma_{a,Q>W_a}^{\rm free}}{\d W} =
\frac{2 \pi e^4}{\me v^2} \, 2|\kappa_a| \, \frac{1}{W^{2}}
\left[ 1 - b_{1}\frac{W}{E} +
b_{2}\left(\frac{W}{E}\right)^{2} - b_{3}\left(\frac{W}{E}\right)^{3} +
b_{4}\left(\frac{W}{E}\right)^{4} \right] \, .
\label{7.17}\eeq
The contributions of these collisions to the total, stopping and
energy-straggling cross sections can then be calculated analytically and
are
\begin{subequations}
\label{7.18}
\beq
\sigma^{(0)}_{a,Q>W_a} ({\rm e}^+) =
\frac{2 \pi e^4}{\me v^2} \,  2|\kappa_a| \,
\left[ -\frac{1}{W} - b_{1}\frac{\ln W}{E} + b_{2}\frac{W}{E^{2}} -
b_{3}\frac{W^{2}}{2E^{3}} + b_{4}\frac{W^{3}}{3E^{4}}
\right]_{W_a}^{E},
\label{7.18a}\eeq
\beq
\sigma^{(1)}_{a,Q>W_a} ({\rm e}^+) =
\frac{2 \pi e^4}{\me v^2} \,  2|\kappa_a| \,
\left[ \ln W - b_{1}\frac{W}{E} + b_{2}\frac{W^{2}}{2E^{2}} -
b_{3}\frac{W^{3}}{3E^{3}} + b_{4}\frac{W^{4}}{4E^{4}}
\right]_{W_a}^{E},
\label{7.18b}\eeq
and
\beq
\sigma^{(2)}_{a,Q>W_a} ({\rm e}^+) =
\frac{2 \pi e^4}{\me v^2} \,  2|\kappa_a| \,
\left[ W - b_{1}\frac{W^2}{2E} + b_{2}\frac{W^{3}}{3E^{2}} -
b_{3}\frac{W^{4}}{4E^{3}} + b_{4}\frac{W^{5}}{5E^{4}}
\right]_{W_a}^{E},
\label{7.18c}\eeq
\end{subequations}
if $E>W_a$; otherwise they all vanish.


\section{The programs {\sc gosat} and {\sc pwacs} \label{sec8}}
\setcounter{equation}{0}

The theory and numerical methods described in Section \ref{sec5} have been
tailored to allow accurate calculation of the GOS and the TGOS of
neutral atoms and ions for relatively wide ranges of the energy loss $W$
and the recoil energy $Q$. The basic information needed for the
calculation of the GOSs is the potential $V(r)$ felt by the atomic
electrons. As already indicated, in the present study we use the
DHFS self-consistent potential for the
ground-state configuration, which is calculated with the program {\sc
dhfs} of \citet{SalvatFernandezVarea2019}. The calculations
reported here pertain to collisions with neutral atoms. The program {\sc
dhfs} can also calculate the DHFS potential of positive ions, from
which we can compute the GOSs for collisions with positive ions by
following the same procedure (and using the same programs) as for
neutral atoms.

The calculation of Dirac radial functions of one-electron states in the
central potential $V(r)$ is performed by means of the subroutine package
{\sc radial} \citep{Salvat1995, SalvatFernandezVarea2019}. These
subroutines implement a robust solution method which, for the DHFS
potential, allows the calculation of bound states with principal quantum
numbers $n$ up to about 35, and of free states with energies larger than
about $10^{-4}$ atomic units.  This method consists in representing the
potential as a natural cubic spline $\widetilde{V}(r)$ that interpolates
a table $V(r_i)$ ($r_1=0 < r_2 < \cdots < r_u$). That is, between each
interval $[r_i,r_{i+1}]$ the function $\widetilde{V}(r)$ is a cubic
polynomial that matches the values of $V(r)$ at the grid points and,
moreover, $\widetilde{V}(r)$ and its first and second derivatives are
continuous in the whole interval $[r_0,r_u]$. The Dirac radial wave
functions in each interval $[r_i,r_{i+1}]$ are then given by exact
series expansions, whose coefficients satisfy recurrence relations
determined by the coefficients of the spline. These series are summed up
to the prescribed tolerance, thus effectively avoiding round-off errors.
The radial functions can be calculated for an arbitrary radial grid,
different from the grid where the potential is tabulated, with up to
15,000 grid points.

The code {\sc gosat} computes the GOS and the TGOS of closed electron
subshells using the formulas given in Section \ref{sec5.2}. The reduced
matrix elements of Racah tensors are evaluated from their analytical
expression
$$
\left<\ell_1 \1o2 j_1|| {\bf C}^{(L)} ||
\ell_2 \1o2 j_2\right>
= \upsilon(L,\ell_1,\ell_2) \, \sqrt{2j_2+1} \;
\langle L j_2 0 \1o2 | j_1 \1o2 \rangle\, .
\eqno{\rm \req{C.5}} $$
The calculation of the radial
integrals \req{5.47} and \req{5.49} is delicate, because the integrands
are rapidly oscillating functions. To ensure accuracy of these
integrals, the radial functions are evaluated at the radii of a dense
grid, such that the Bessel function and the radial function of each
state involved has at least
20 points in a wavelength. Thus, the rapid oscillations of the
integrands are accurately reproduced when using 6-point Lagrange
interpolation \citep{AbramowitzStegun1974} and, hence, the radial
integrals can be evaluated accurately by using 6-point Lagrange
quadrature. For a given energy loss $W$, the final-state radial
functions are calculated and stored in memory, and for each recoil
energy $Q$ a table of values of the Bessel functions at the points of
the radial grid is evaluated quite rapidly using their recurrence
relations \citep{AbramowitzStegun1974}. The integration
interval extends from $r=0$ up to a point $r_N$ where the radial
functions of the initial bound state take negligibly small values. When
the energy loss $W$ increases, the energy $\epsilon_b = W-E_a$ of the
final state of the active electron increases and, therefore, its
wavelength decreases. As a consequence, because the number of points in
the radial grid cannot exceed 15,000, the calculations for a given value
of $Q$ cannot be performed beyond a certain energy loss
$W_{\rm num}=(w_{\rm num}+1) E_a$. Similarly, the wavelength of
the Bessel function decreases when the recoil energy $Q$ increases, and
hence calculations cannot be performed for recoil energies larger than
some limiting value $Q_{\rm num}(W)$, which depends on the energy
loss $W$. Of course, the values $W_{\rm num}$ and $Q_{\rm num}$
can be increased by enlarging the dimension of the radial grid, at the
expense of slowing down the calculation. The summation of the partial-wave
series \req{5.46} and \req{5.48} is performed by adding terms with
increasing values of $|\kappa_b|$ and $\lambda$; the summation is
discontinued when the relative contribution of the last six added terms
is less than $10^{-9}$. For ionizing transitions, the code allows the
calculation of final state radial functions and integrals with
$|\kappa_b|$ up to 300, a value that can be increased by modifying the
array dimension in the source program. As the convergence of the
partial-wave series becomes slower when $Q$ increases, with this fixed
number of terms the series do not converge for recoil energies beyond a
certain upper value, which is determined by the program.

As indicated in Section \ref{sec5.4.2}, in the case of ionization, the
code {\sc gosat} generates a table of the GOS and the TGOS for a
logarithmic grid of reduced energy losses, $w=(W/E_a)-1$, and for a
non-uniform grid of reduced recoil energies $t=Q/E_a$ which, for each
reduced energy loss $w$, extends from $t=10^{-4}$ up to a maximum value
$t_{\rm num}(w)$ determined by the requirements on the spacing and
the dimension of the radial grid. The program also generates a table of
the OOS for the same grid of reduced energy losses, but extending to
higher $W$, and tables of the longitudinal and transverse Born-Compton profiles
for $W_{\rm ref} = t_{\rm num,0} E_a$ and for a dense grid of
values of the variable $p_{\rm C}$, to allow accurate cubic-spline
interpolation in $p_{\rm C}$. These profiles are used to calculate
the GOS and TGOS for $Q>t_{\rm num,0} E_a$ (see Section
\ref{sec5.6}). The program {\sc gosat} generates also the GOS and TGOS
for discrete transitions from the subshell $n_a\kappa_a$ to bound levels
$\epsilon_{n_b\kappa_b}$ with $\epsilon_b \ne \epsilon_a$ and $n_b \le
25$, including also those to levels that are occupied, which are needed
for computing the one-shell Bethe sum $S_0(a;Q)$, Eq.\ \req{5.62}.

A complete database of longitudinal and transverse GOSs for all
subshells of neutral atoms ($Z=1$ to 99) in their ground-state
configurations has been generated by using the DHFS self-consistent
potentials calculated previously with the DHFS program of
\citet{SalvatFernandezVarea2019}. The calculations were performed by
running {\sc gosat} in four simultaneous processes or user threads on an
Intel Core i7-8550 computer at 1.99 GHz for about 280 hours, which
amounted to a total processing time of about 1200 hours. The resulting
GOS database allows the calculation of the atomic DDCS given by Eqs.\
\req{6.8} to \req{6.11} for inelastic collisions of charged particles
heavier then the electron, and of the DDCS for collisions of electrons
and positrons according to the approximations described in section
\ref{sec7}.

We have written a second Fortran program, named {\sc pwacs}, that
calculates the energy-loss DCS and the integrated cross sections
$\sigma^{(i)}$ (for heavy charged particles, electrons and positrons)
from the GOSs tables generated by the {\sc gosat} program. The
{\sc pwacs} program computes the cross sections for the individual electron
subshells $n_a\kappa_a$, and adds the results of the various subshells to
obtain the atomic cross sections. Excitations to discrete levels
correspond to sharp resonances with well defined excitation energies,
$W_{ba} = \epsilon_{n_b\kappa_b} - \epsilon_{n_a\kappa_a}$. For each
shell, the total cross sections for excitation from the (closed) subshell
to empty or partially filled bound levels $\epsilon_{n_b\kappa_b}$ with
$n_b \le 25$ is calculated as
\beq
\sigma_{ba}^{\rm exc}
= \frac{2|\kappa_b|-q_b}{2|\kappa_b|}
\int_{Q_-}^{Q_+} \frac{\d \sigma^{\rm exc}_{ba}}{\d Q} \, \d Q \, ,
\label{8.1}\eeq
where $q_b$ is the number of electrons in the final level ($=0$ for an
empty level), and the DCS $\d \sigma^{\rm exc}_{ba}/\d Q$ is obtained
by a trivial integration over $W$ of the general expression \req{4.46}
with the GOSs corresponding to the discrete transition, which are given
by Eqs.\ \req{5.55} to \req{5.58}. The limits $Q_-$ and $Q_+$ of the
kinematically allowed interval of recoil energies are given by Eq.\
\req{6.26}. Note that the energy-loss DCS for excitations to bound
levels can be expressed as
\beq
\frac{\d \sigma_{ba}^{\rm exc}}{\d W} =
\sigma_{ba}^{\rm exc} \, \delta(W-\epsilon_{n_b\kappa_b} +
\epsilon_{n_a\kappa_a}).
\label{8.2}\eeq
The energy-loss DCS for excitations of a closed subshell $n_a\kappa_a$
involving energy losses greater than the effective threshold energy
$W_1=\epsilon_{26,\kappa_b} + E_a$ (see Section \ref{sec5.4.2}) is
obtained from the continuous GOSs, which are calculated from the
numerical tables generated by {\sc gosat} by using the
interpolation/extrapolation methods described in Sections \ref{sec5.4.2}
and \ref{sec5.6}.  The continuous energy-loss DCS is obtained from the
DDCS given by Eq.\ \req{4.46} as
\beq
\frac{\d \sigma_{a}^{\rm cont}}{\d W}
= \int_{Q_-}^{Q_+} \frac{\d^2 \sigma_a}{\d W \d Q} \, \d Q \, .
\label{8.3}\eeq

The integrals over the recoil energy are not trivial to calculate,
because of the fast variation of the transverse DDCS with $Q$ near the
lower limit of the integral, and because of the narrowness of the Bethe
ridge for large $W$. We evaluate these integrals by splitting the
integration interval in at least two parts. Firstly, we consider the
contribution of the $Q$-interval from from $Q_-$ to
$Q_{\rm num}=t_{\rm num,0} E_a$, where the GOSs are calculated by
interpolation of their numerical tables. Secondly, we evaluate the
contribution of the interval from $Q_{\rm num}$ up to $Q_+$, with the
GOSs represented in terms of the Born-Compton profiles (see Section
\ref{sec5.5}). These two contributions are calculated by means of an adaptive
20-point Gauss quadrature method, which implements a bisection scheme to
estimate and reduce the numerical error, and which yields the value of
the integral with a prescribed accuracy. This algorithm gets stuck at
the discontinuities of the integrand and its first derivatives (which
cannot be approximated by a polynomial). By splitting the integration
interval at $Q_{\rm num}$, where our numerical GOS may have a
discontinuity, we ensure that only smooth functions will be passed to
the integration subroutine. Finally, for $W>W_a$ (with $W_a \sim 10^4
E_a$, Section
\ref{sec6.4} and Fig\ \ref{fig18}) the target electrons
are considered as free and stationary and the GOS is represented as $2
|\kappa_a|\, \delta(W-Q)$, so that the energy-loss DCS is given by the
simple analytical expressions \req{6.44}, \req{7.11} and \req{7.17}. It
is found that the calculated energy-loss DCS for energy losses $W
\lesssim W_a$ differs slightly from the energy-loss DCS for collisions
with free electrons, because of the finite width of the Bethe ridge. To
account partially for this effect at $W>W_a$, and to avoid a (usually
small) discontinuity in the energy-loss DCS at $W_a$, we set
\beq
\frac{\d \sigma_{a,W>W_a}^{\rm cont}}{\d W}
= \frac{W}{W+\delta}\,  \frac{\d \sigma_{a,W>W_a}^{\rm free}}{\d W}
\, ,
\label{8.4}\eeq
where the last factor on the right-hand side is the energy-loss DCS for
collisions with free stationary electrons [Eqs.\ \req{6.44}, \req{7.11}
and \req{7.17}] and the parameter $\delta$ is determined by requiring
that this formula reproduces the value of the energy-loss DCS at $W \sim
0.75 W_a$ calculated by numerical integration as described above. This
procedure effectively removes the discontinuity of the energy-loss DCS
at $W=W_a$, and yields the correct free-electron result in the limit of
very large $W$. In the cases of projectile electrons and positrons, the
$Q$ interval is further split at $Q=E_a$, where the DDCSs \req{7.10} and
\req{7.16} have an extra discontinuity.

\begin{figure}[tbh]
\begin{center}
\includegraphics*[width=10cm]{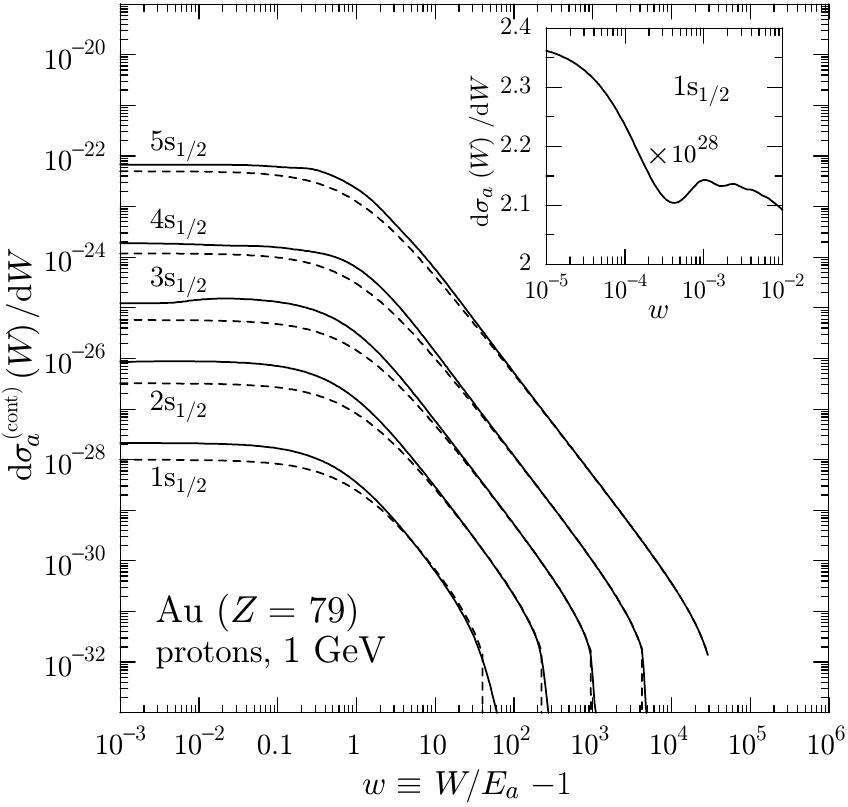}
\caption{Energy-loss DCS for ionization of various electron subshells of
gold atoms by impact of 1 GeV protons, as functions of the reduced
energy loss $w=(W/E_a)-1$. The solid curves represent DCSs obtained
by numerical integration of the DDCS; the dashed curves are the
energy-loss DCSs for collisions with stationary free electrons, Eq.\
\req{6.44}. The inset shows the fairly complicated structure of the
energy-loss DCS of the K ($1s_{1/2}$) shell near the ionization
threshold.
\label{fig19}}
\end{center}\end{figure}

Figure \ref{fig19} shows the energy-loss DCSs for ionization of different
electron subshells of gold atoms by impact of protons with 1 GeV kinetic
energy. The solid curves represent the numerical DCSs, calculated from
Eqs.\ \req{8.3} and \req{8.4}, and the dashed curves represent the
energy-loss DCSs for collisions with $2|\kappa_a|$ free stationary
electrons, Eq.\ \req{6.44}. For energy transfers near the ionization
threshold, the variation of the numerical energy-loss DCS with $W$ is
not monotonic, and its fluctuations reflect the structure of the GOS.
To ensure accurate interpolation, and integration, we consider the
energy-loss DCS as a function of the reduced energy-loss, $w=(W/E_a)-1$,
and use a logarithmic grid of $w$ values to properly reproduce the fast
variations the DCS near threshold. It is interesting to note that the
free-electron DCS is quite accurate for energy losses larger than about
$20 E_a$. Moreover, for smaller $W$, the free-electron DCS is always
less than the numerical DCS, because the former does not account for
distant interactions. This last peculiarity can be used to devise
simplified (but still realistic) interaction models for, \eg, Monte
Carlo simulation of the transport of charged particles in matter.

The energy-loss DCS of the entire atom is
\beq
\frac{\d \sigma}{\d W} = \sum_a \frac{q_a}{2|\kappa_a|}
\left[ \frac{\d \sigma_{a}^{\rm cont}}{\d W} +
\sum_{b\ne a} \sigma_{ba}^{\rm exc} \, \delta(W-\epsilon_{n_b\kappa_b} +
\epsilon_{n_a\kappa_a})  \right] \, ,
\label{8.5}\eeq
where the summation runs over all occupied subshells and $q_a$ is the
number of electrons in subshell $n_a\kappa_a$. The integrated atomic cross
sections $\sigma^{(i)}$, Eq.\ \req{6.31}, are given by
\beq
\sigma^{(i)} = \sum_a \frac{q_a}{2|\kappa_a|}
\left[ [\sigma_a^{\rm cont}]^{(i)} +
\sum_{b\ne a} \sigma_{ba}^{\rm exc} \, (E_a+\epsilon_{n_b\kappa_b})^i
\right].
\label{8.6}\eeq
with
\beq
[\sigma_a^{\rm cont}]^{(i)} =
\int_{W_1}^{W_{\rm max}} W^i \, \frac{\d \sigma_{a}^{\rm cont}}{\d
W} \, \d W.
\label{8.7}\eeq
To calculate the last integral, the energy-loss DCS $\d \sigma_{a}^{\rm
cont}/\d W$ is tabulated for a logarithmic grid of reduced energy losses
$w=W/E_a-1$, which includes the grid points $w_j$ where the numerical
GOSs are tabulated. The calculation of the energy-loss DCS at the grid
points $w_j$ is faster and more accurate, because it does not require
interpolation in $w$. Moreover, errors in the latter interpolation would
become apparent when plotting the energy-loss DCS as a function of $W$;
numerical data do not show traces of such errors. The integrals of the
energy-loss DCS, Eq.\ \req{8.7}, are evaluated using Simpson's rule
\citep{AbramowitzStegun1974} after changing the integration variable to
$x=\ln W$.

As indicated above, the first term in the right-hand side of Eq.\
\req{8.5} accounts for ionization ($W>E_a$) as well as for excitations
to the quasi-continuum of highly excited levels $\epsilon_{n_b\kappa_b}$
with $n_b > 25$ (region V in Fig.\ \ref{fig3}), which starts at the
energy loss $W_1=\epsilon_{26,\kappa_b} + E_a$. To obtain the integrated
cross sections for ionization (\ie, for energy transfers $W$ greater
than the ionization threshold $E_a$) the lower limit $W_1$ of the
integrals should be replaced with $E_a$. That is,
\beq
[\sigma^{\rm ion}]^{(i)} = \sum_a \frac{q_a}{2|\kappa_a|} \,
[\sigma_a^{\rm ion}]^{(i)}
\label{8.8}\eeq
with
\beq
[\sigma_a^{\rm ion}]^{(i)} =
\int_{E_a}^{W_{\rm max}} W^i \, \frac{\d \sigma_{a}^{\rm cont}}{\d
W} \, \d W.
\label{8.9}\eeq

The code {\sc pwacs} calculates tables of the integrated cross sections
$\sigma^{(i)}$ with $i=0$, 1 and 2 (total cross section, stopping cross
section and energy-straggling cross section, respectively) for a given
projectile (proton, muon, electron, positron, ...) as functions of its
kinetic energy $E$. {\sc pwacs} also delivers tables of the energy-loss
DCS of individual electron subshells, and its integrals for specific
energies of the projectile.  To run {\sc pwacs} for a given atom or ion,
we need to have precalculated the corresponding GOSs, OOS and
Born-Compton profiles.  These functions are generated by running the
program {\sc gosat} which, in turn, needs the DHFS potential generated
by the code {\sc dhfs}. The distribution package includes text (ascii)
files with the DHFS potentials of neutral atoms with $Z=1$ (hydrogen) to
$Z=99$ (einsteinium). We have used {\sc gosat} with these potentials to
calculate the GOS, TGOS, OOS and Born-Compton profiles for all the
subshells of the ground-state configurations of these elements. With these
data at hand, the code {\sc pwacs} allows the calculation of integrated
cross sections and other interaction data with modest computer times, of
the order of 10 seconds, per each active subshell and each value of the
kinetic energy of the projectile.

\begin{figure}[tp]
\begin{center}
\includegraphics*[width=7.5cm]{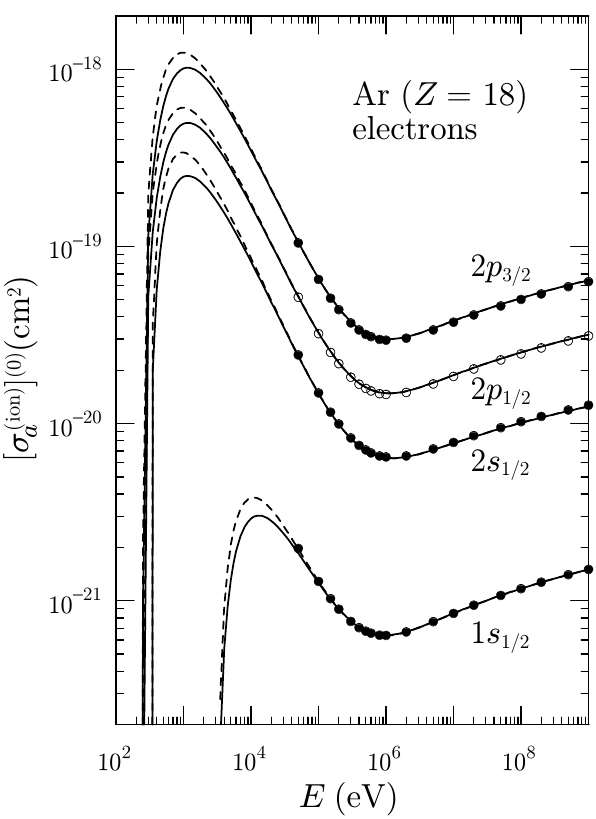} \rule{5mm}{0mm}
\includegraphics*[width=7.5cm]{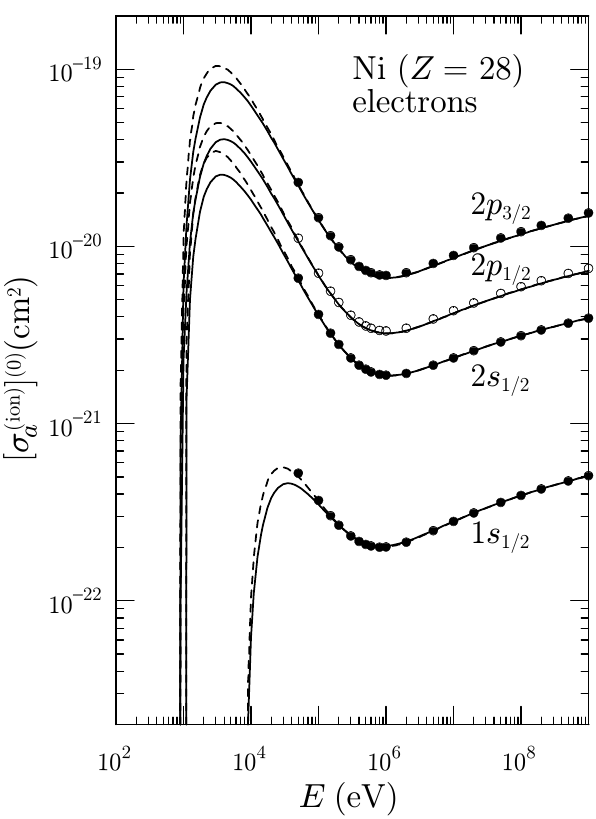} \\
\includegraphics*[width=7.5cm]{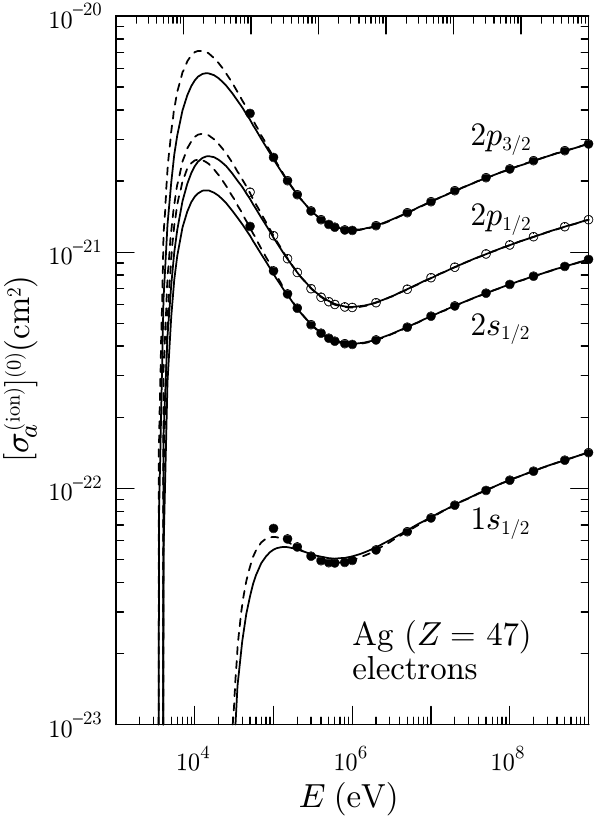} \rule{5mm}{0mm}
\includegraphics*[width=7.5cm]{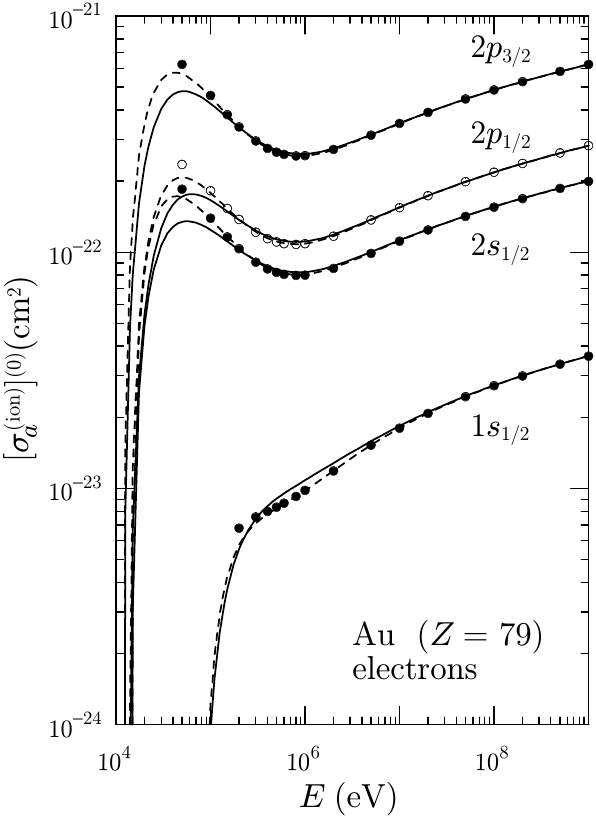}
\caption{Total cross sections for ionization of the K shell and L subshells of
argon, nickel, silver and gold by impact of electrons as functions of
the kinetic energy of the projectiles. The solid curves were generated
by the code {\sc pwacs} from the electron DDCS described in Section
\ref{sec7.1}.
The dashed curves were calculated considering the projectile electrons
as distinguishable from the target ones; circles represent the results
from PWBA calculations by \citet{Scofield1978}, obtained under the same
assumptions.
\label{fig20}}
\end{center}\end{figure}

We give here a few examples of calculation results. Figure
\ref{fig20} displays total cross sections for ionization of K and L
subshells of gold atoms by impact of electrons, calculated considering the
projectile electrons as distinguishable from the ones in the target.
Our subshell cross sections for electron impact ionization should be close
to those calculated by \citet{Scofield1978}, who used the PWBA formulated in
the Lorentz gauge. Indeed, our code yields ionization cross sections in
good agreement with those of Scofield, except for projectile electrons
with kinetic energies less than about $20 E_a$, probably because of
differences in the interpolation and integration methods adopted in the
two calculations. The solid curves in Fig\
\ref{fig20} represent the cross sections calculated with the exchange
corrections described in Section \ref{sec7.1}. The effect of exchange is a
reduction of the total cross section, which is appreciable  only for
kinetic energies less that about $30 E_a$.

\begin{figure}[tp]
\begin{center}
\includegraphics*[width=7.5cm]{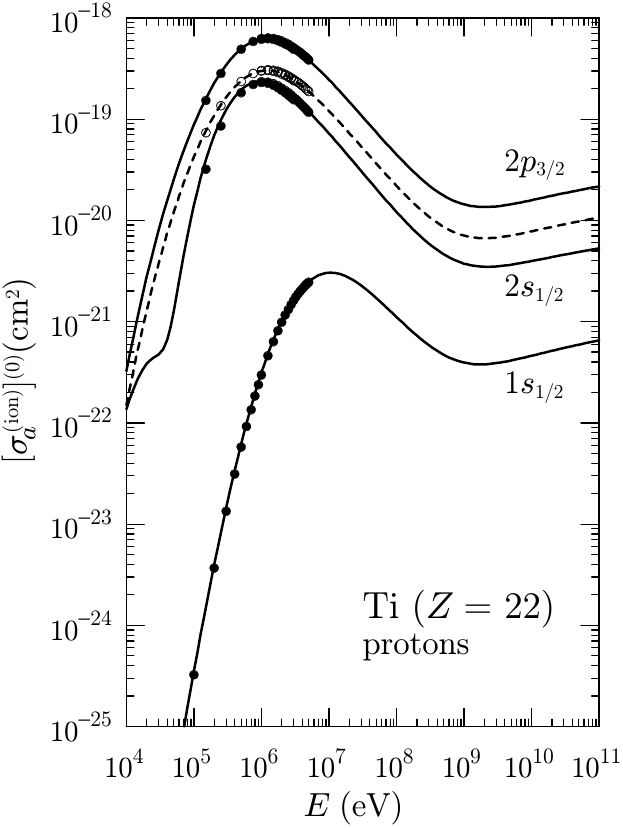} \rule{5mm}{0mm}
\includegraphics*[width=7.5cm]{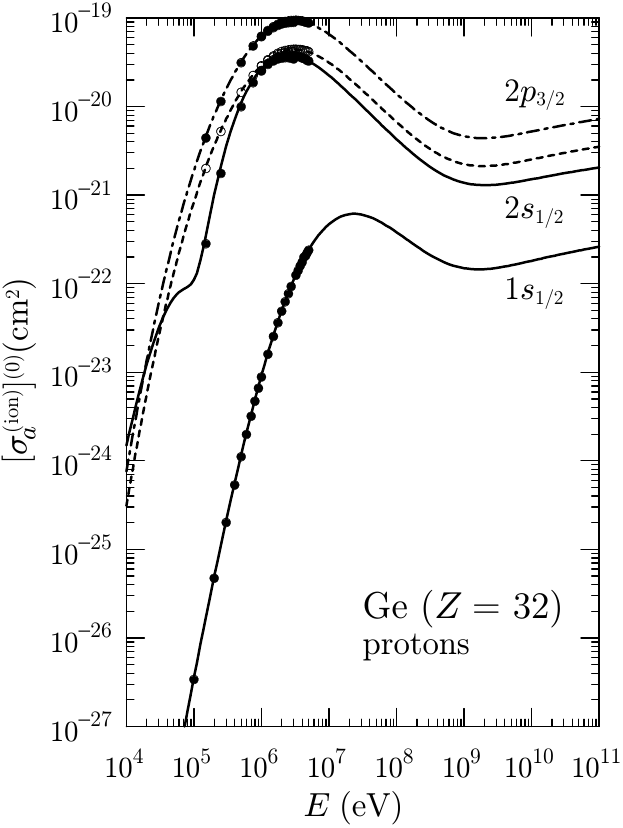} \\
\includegraphics*[width=7.5cm]{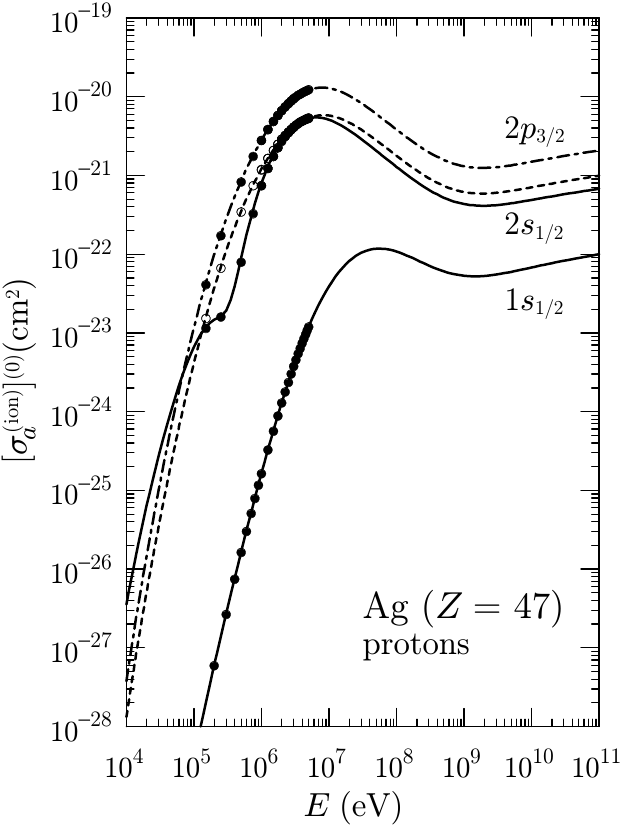} \rule{5mm}{0mm}
\includegraphics*[width=7.5cm]{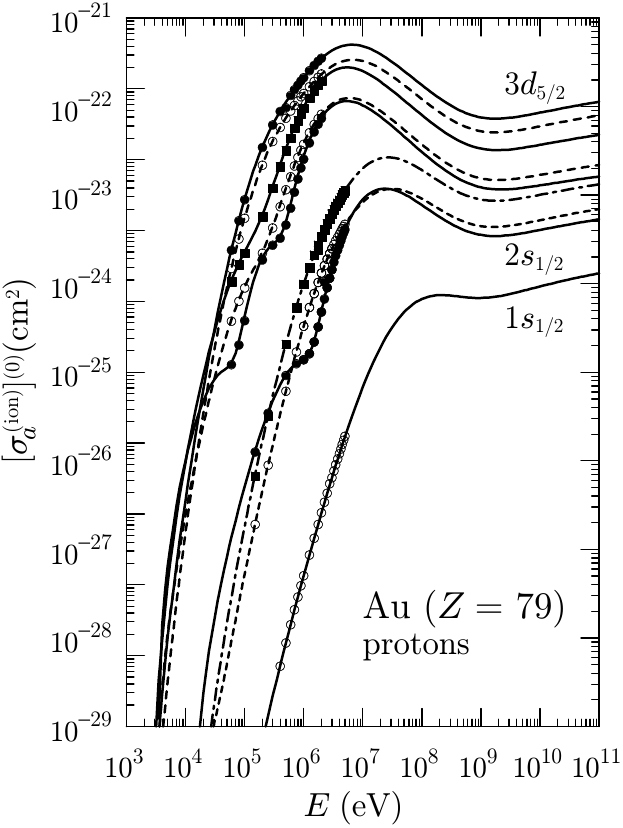}
\caption{Total cross sections for ionization of the indicated (K, L and
M) subshells of titanium, germanium, silver and gold by impact of protons
as functions of the kinetic energy of the projectile. The curves were
generated by the code {\sc pwacs}. Circles are results from PWBA
calculations by \citet{ChenCrasemann1985, ChenCrasemann1989} using the
DHFS potential and neglecting the contribution from transverse
interactions.
\label{fig21}}
\end{center}\end{figure}

\citet{ChenCrasemann1985, ChenCrasemann1989} calculated total cross
sections for the ionization of the K shell and L subshells of 28 elements by
impact of protons with energies from about 0.1 to 5 MeV, and for the M
subshells of 15 elements and protons with energies from 0.06 to 2 MeV,
using the relativistic PWBA with the DHFS potential. They neglected the
contribution of transverse interactions, which is negligible in this
energy range (see below), and introduced corrections to account for
low-energy effects (change in binding energy due to polarization and
Coulomb deflection, see Section \ref{sec11}). Their uncorrected PWBA
cross sections for the elements titanium, germanium, silver and gold are
compared with the results from the {\sc pwacs} code in Fig.\
\ref{fig21}. Note that the two calculations were performed with the
same potential and, consequently, the small discrepancies found likely
arise from differences in the computational algorithms.

\begin{figure}[h!]
\begin{center}
\includegraphics*[width=7.5cm]{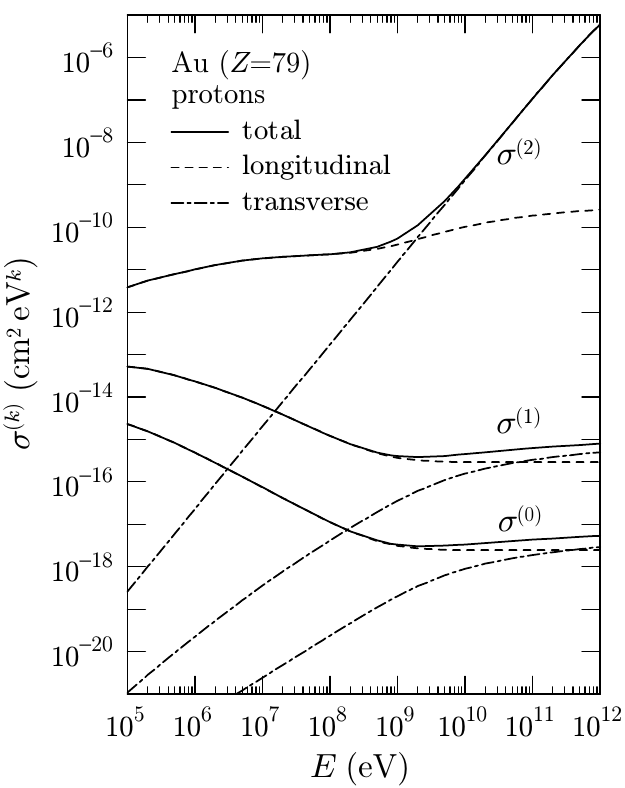} \rule{5mm}{0mm}
\includegraphics*[width=7.5cm]{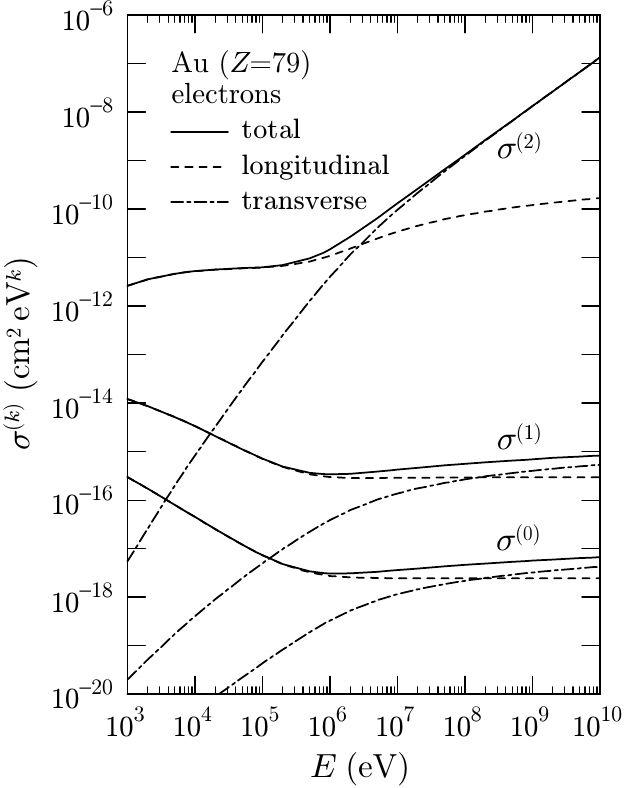}
\caption{Integrated cross sections $\sigma^{(i)}$ for inelastic
collisions of protons and electrons with gold atoms as functions of the
kinetic energy of the projectile. The dashed and dot-dashed curves
represent the contributions from longitudinal and transverse
interactions, respectively.
\label{fig22}}
\vspace*{-6mm}
\end{center}\end{figure}

\begin{figure}[hp!]
\begin{center}
\includegraphics*[width=7.5cm]{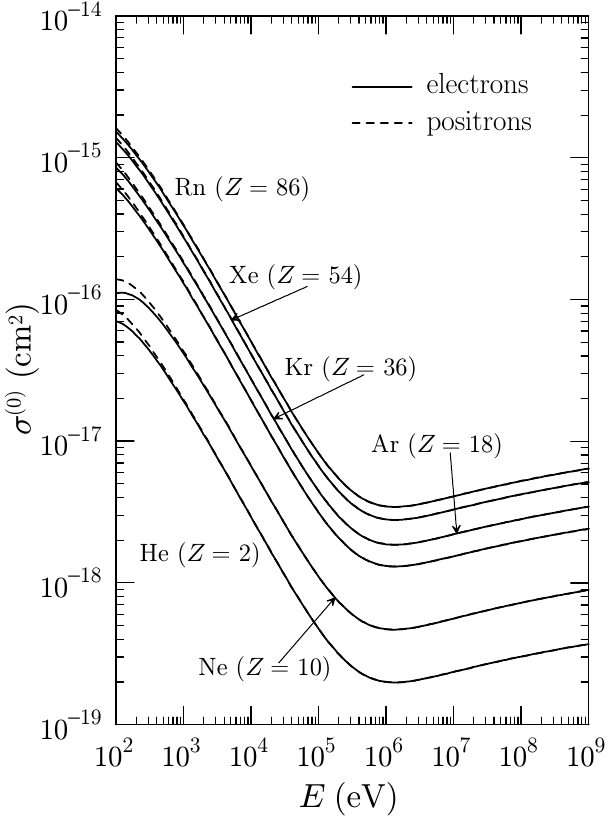} \\ [2mm]
\includegraphics*[width=7.5cm]{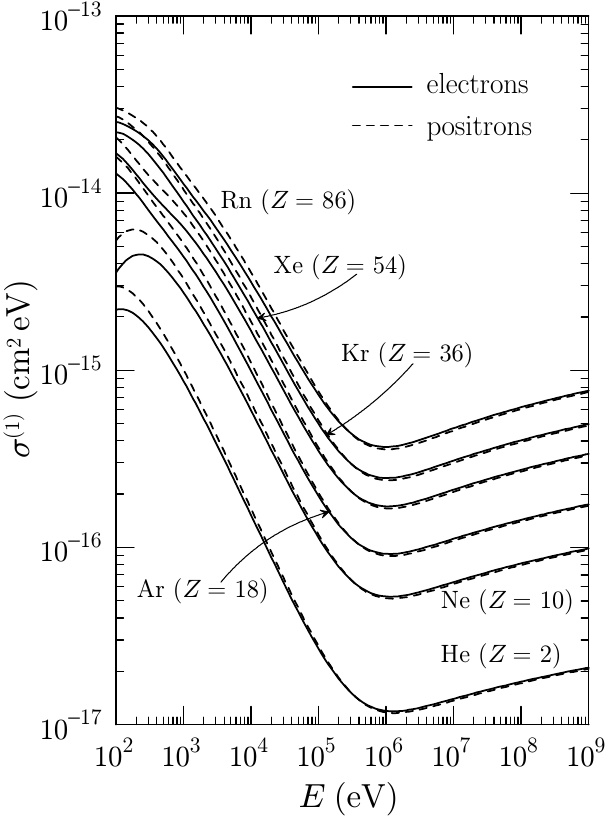} \rule{5mm}{0mm}
\includegraphics*[width=7.5cm]{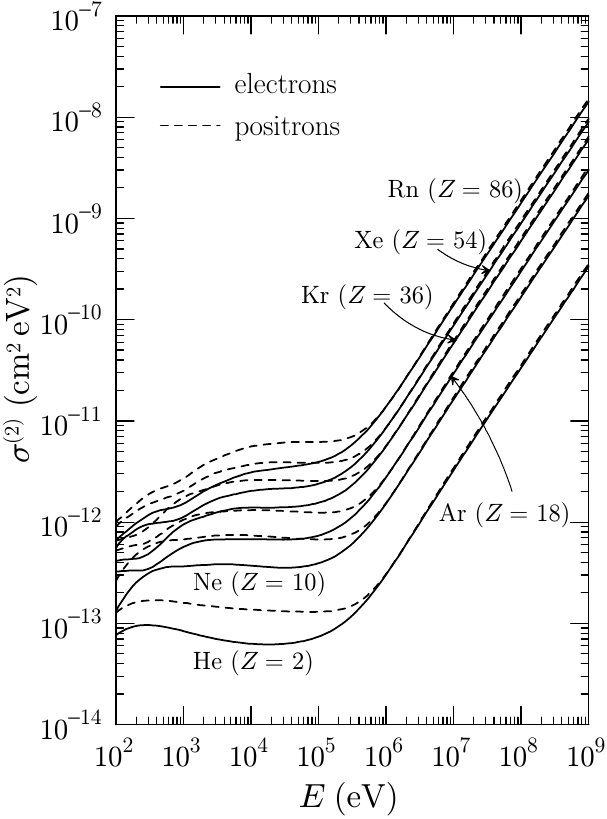}
\caption{Integrated cross sections for inelastic
collisions of electrons (solid) and positrons (dashed) with noble-gas
atoms, as functions of the kinetic energy of the projectile.
\label{fig23}}
\end{center}\end{figure}

The very detailed information generated by the {\sc pwacs} code can be
used to illustrate basic aspects of the interaction. Thus, Fig.\
\ref{fig22} displays integrated cross sections for inelastic collisions
of protons and electrons with gold atoms as functions of the kinetic
energy of the projectile. The figure also shows the contributions from
longitudinal and transverse interactions. It is interesting to observe
that the transverse contributions increase monotonically with energy;
they become appreciable for energies of about 500 keV for electrons and
about 1 GeV for protons. In general, for projectiles of mass $M$, the
effect of transverse interactions is negligible when their kinetic
energy is less than about $Mc^2$. The energy dependence
of transverse interactions (exchange of virtual photons) is similar to
that of the related process of emission of real (bremsstrahlung) photons
\citep[see, \eg,][]{SalvatFernandezVarea2009}.

As a final example of results computed with the code {\sc pwacs},
Fig.\ \ref{fig23} shows integrated cross sections for inelastic
collisions of electrons and positrons with noble-gas atoms, as functions
of the kinetic energy $E$ of the projectile. The total cross sections,
$\sigma^{(0)}$, for the two particles are practically equal for energies
larger than about 1 keV. For energies up to about 0.5 MeV, the stopping
and energy-straggling cross sections for positrons are larger than those
for electrons. The differences are largely due to the convention, in
ionizing electron collisions, of considering the ``projectile'' as the
fastest of the two free electrons after the collision. Consider, for
instance, a hard collision of the projectile electron with a
distinguishable bound electron, with ionization energy $E_a$, which
involves an energy transfer $W$ larger than $(E+E_a)/2$.  After the
collision we have two electrons with kinetic energies $E-W$ and $W-E_a$.
A collision with the same energy transfer between indistinguishable
electrons would also yield two free electrons with those energies, but
then we would call the primary the one with the larger energy, $W-E_a$,
and consider that the energy loss is $E-(W-E_a)$, which is less than
$W$. Thus, by picking the fastest of the two electrons, we effectively
reduce the energy loss in hard interactions, and this affects the
integrated cross sections $\sigma^{(1)}$ and $\sigma^{(2)}$.


\section{Bethe asymptotic formulas \label{sec9}}
\setcounter{equation}{0}

\citet{Bethe1930, Bethe1933} \citep[see also][]{Fano1963} derived simple
and accurate analytical formulas for the stopping cross section and the
total cross section for high-energy projectiles, which are among the
most useful formulas in radiation physics. Their usefulness stems from
the fact that each formula contains only two parameters,
which are characteristic of each element or material. These parameters
can be inferred from experimental measurements of the stopping power and
the total cross section. Thus, the Bethe formulas with empirically
determined parameters provide reliable values of these cross sections
for different kinds of charged particles, and in a wide energy range,
for molecules and condensed materials for which first-principles
calculations are not available or possible. Although the Bethe formulas
are asymptotic (\ie, valid only for projectiles with very high
energies), they remain fairly accurate down to moderately low energies
(see Section \ref{sec9.3}). A great deal of work, both theoretical and
experimental, has been devoted to determining suitable corrections to
improve the accuracy of these formulas, and to extend its range of
validity (see, \eg, the reviews by \citeauthor{Inokuti1971},
\citeyear{Inokuti1971}, and \citeauthor{Ahlen1980},
\citeyear{Ahlen1980}).

In principle, the asymptotic formulas give the lowest-order terms of the
expansions of $\sigma^{(i)}(E)$ in powers of $E^{-1}$. It is important
to note that at intermediate and low energies these formulas may be less
accurate than alternative semi-empirical formulas; the advantage of the
asymptotic formulas is that they can be systematically improved (\ie,
extended to lower energies) by adding terms of higher orders in $E^{-1}$
\citep[see, \eg,][]{KimInokuti1971, FernandezVarea1993}. In this
section, we derive the asymptotic formulas for $\sigma^{(i)}$ ($i=0$, 1
and 2) using a method similar to the one adopted in
\citeauthor{Fano1963}'s (\citeyear{Fano1963}) review. The conventional
derivation of the stopping power formula makes explicit use of the Bethe
sum rule [\ie, the Bethe sum $S_0(a; Q)$ is assumed to be equal to $Z$
the number of electrons in the atom for all $Q$], which is correct only
for sufficiently large recoil energies (see Section \ref{sec5.5}). In
our derivation we allow deviations from that sum rule for small and
intermediate recoil energies.

For simplicity, we will derive first asymptotic formulas for excitations
of a single closed electron (sub)shell $n_a\kappa_a$ by charged particles
(with charge $Z_0e$ and mass $M$) moving with kinetic energy $E$ much
larger than the subshell ionization energy $E_a$. We start from the DDCS
given by Eq.\ \req{4.46},
\beqa
\frac{\d^2 \sigma_{a}}{\d W \, \d Q} &=&
\frac{2\pi Z_0^2 e^4}{\me v^2} \left[ \frac{2\me c^2}{WQ(Q+2\me c^2)}
\;  A_{\rm L} \, \frac{\d f_a (Q,W)}{\d W} \right.
\nonumber \\ [2mm]
&& \mbox{} \rule{11mm}{0mm}
+ \left. \frac{2\me c^2 W}{[Q(Q+2\me c^2)-W^2]^2} \;
A_{\rm T} \, \frac{\d g_a (Q,W)}{\d W} \right]
\label{9.1}\eeqa
with
\beq
A_{\rm L} =  1 - \frac{4(E+Mc^2)W - W^2 + Q(Q+2\me c^2)}{4\,
(E+M c^2)^2}
\label{9.2}\eeq
and
\beqa
A_{\rm T} &=&
\beta^2
- \frac{W^2}{Q(Q+2\me c^2)} \left( 1 + \frac{ Q(Q+2\me c^2)-W^2}
{2 W (E+M c^2)} \right)^2
\nonumber \\ [2mm]
&& \mbox{} + \frac{Q(Q+2\me c^2)- W^2 }{2(E+ M c^2)^2}\, .
\label{9.3}\eeqa
For projectiles with sufficiently high energy, the relevant energy
transfers are such that $W \ll E$ and the minimum allowed recoil energy
$Q_-$ is given by Eq.\ \req{A.25},
\beq
Q_- (Q_- + 2 \me c^2) = W^2 / \beta^2,
\label{9.4}\eeq
where we have disregarded terms containing factors $W/E$. Within
the same approximation, and for small and moderate recoil energies, the
maximum allowed energy loss is [Eq.\ \req{A.28}]
\beq
W_{\rm m}(Q) = \beta \sqrt{Q(Q+2\me c^2)}.
\label{9.5}\eeq
When the energy of the projectile increases, both
$Q_-(W)$ and $W_{\rm m}(Q)$ tend towards the vacuum photon line, Eq.\
\req{A.33},
$$
W_0(Q) = \sqrt{Q(Q+2\me c^2)}.
$$
The recoil energy at this line is
\beqa
Q_0 (W) &=& \me c^2 \left[ \sqrt{1 + \left(\frac{W}{\me c^2}
\right)^2} -1 \right]
\nonumber \\ [2mm]
&\simeq& \frac{W^2}{2 \me c^2} \left[ 1 - \left( \frac{W}{2\me
c^2} \right)^2 + 2 \left( \frac{W}{2\me
c^2} \right)^4 - \cdots \right].
\label{9.6}\eeqa
That is, recoil energies less than $Q_0(W)$ and energy losses larger
than $W_0(Q)$ are not attainable in inelastic collisions. For
the most probable excitations, $W$ is much less than $2\me c^2$ and
$Q_0(W)$ is closely approximated by the first term of the expansion
\req{9.6}, \ie, $Q_0(W) \simeq W^2/2\me c^2$.

Following \citet{Fano1963}, we evaluate the integrals of the DDCS
\req{9.1} approximately by considering various ranges of $Q$.  For
recoil energies smaller than about $0.01 \, W$, the GOS and the TGOS
practically coincide with the optical oscillator strength (OOS), \ie,
the dipole approximation is applicable. Hence, we may introduce a cutoff
recoil energy $Q_1 \simeq 0.001 \, E_a$, and consider that for $Q <Q_1$
both the GOS and the TGOS are approximately equal to the OOS, $\d
f_a(W)/\d W$. For recoil energies between $Q_1$ and a certain value
$Q_2$, which we assume much larger than the binding energy of the atomic
electrons, the maximum allowed energy transfer $W_{\rm m}(Q)$ is
sufficiently large to include the practical totality of the GOS, \ie, to
exhaust the Bethe sum. In our calculations
we set $Q_2 = {\rm max}(10^4 E_a,10^5 E_{\rm h})$, where $E_{\rm h}$ is
the Hartree energy. We have verified that for recoil energies higher
than $Q_2$ the GOSs reduce to the Bethe ridge and the Bethe sum rule
\req{5.61} is satisfied. Then, for $Q>Q_2$ the GOSs
reduce to the Bethe ridge and the DDCS can be approximated by that of
binary collisions with free electrons at rest, which amounts to
replacing the GOS and the TGOS with $2 |\kappa_a| \, \delta(W-Q)$. The
maximum allowed energy-loss in binary collisions is [Eq.\ \req{6.40}]
\beq
W_{\rm ridge} = \frac{2\me c^2 \beta^2}{1-\beta^2} \, R \qquad
\mbox{with} \qquad
R \equiv
\left[ 1+\left(\frac{\me}{M}\right)^2+ \frac{2}{\sqrt{1 - \beta^2}}
\, \frac{\me}{M} \right]^{-1}\, .
\label{9.7}\eeq
Although an impact parameter is not defined in the quantum formulation,
the classical picture suggests that large (small) momentum transfers
roughly correspond to small (large) impact parameters; indeed,
interactions with small and large recoil energies are frequently
referred to as distant and close interactions, respectively.

\vspace*{3mm}

\noindent $\bullet$ {\bf Interactions with small and intermediate $Q$}
\\ [2mm]
The contribution of (distant) interactions with $Q<Q_2$ to the
energy-loss DCS can be estimated by assuming that $Q_2 \ll 2 \me c^2$.
Notice that this assumption is questionable in the case of the innermost
subshells of heavy elements, whose K shells have binding energies $E_a$
of the order of 100 keV ($\sim 0.2 \me c^2$).

The DDCS for longitudinal distant interactions can be approximated as
\beq
\frac{\d^2 \sigma_{a}^{\rm L,d}}{\d W \, \d Q}
\simeq \frac{2\pi Z_0^2 e^4}{\me v^2} \,
\frac{1}{WQ} \, \frac{\d f_a(Q,W)}{\d W} \, \Theta(Q_2-Q),
\label{9.8}\eeq
where we have neglected terms proportional to
$(E+Mc^2)^{-1}$ (which do not contribute in the asymptotic limit) and
we have inserted the unit step function $\Theta(x)$ ($=1$ is $x>0$
and $=0$ otherwise), to indicate that this DDCS vanishes if $Q > Q_2$.
The corresponding energy-loss DCS is
\beq
\frac{\d \sigma_{a}^{\rm L,d}}{\d W} =
\int_{Q_-}^{Q_2} \d Q \; \frac{\d^2 \sigma_{a}^{\rm L,d}}{\d W \, \d Q}
= \frac{2\pi Z_0^2 e^4}{\me v^2} \,
\frac{1}{W}
\int_{Q_-}^{Q_2} \frac{\d Q}{Q} \, \frac{\d f_a(Q,W)}{\d W} \, \Theta(Q_2-Q).
\label{9.9}\eeq
It is convenient to remove the energy dependence of the lower limit of
this integral by writing
\beq
\frac{\d \sigma_{a}^{\rm L,d}}{\d W} =
\frac{2\pi Z_0^2 e^4}{\me v^2} \,
\frac{1}{W} \left(
\int_{Q_0}^{Q_2} \frac{\d Q}{Q} \, \frac{\d f_a(Q,W)}{\d W}
- \int_{Q_0}^{Q_-} \frac{\d Q}{Q} \, \frac{\d f_a(Q,W)}{\d W} \right)
\Theta(Q_2-Q).
\label{9.10}\eeq
where $Q_0$ is the recoil energy of the vacuum photon line, Eq.\
\req{9.6}.
If the kinetic energy of the projectile is high enough, the second
integral in \req{9.10} involves only small recoil energies, for which the
dipole approximation is applicable, and
\beq
\int_{Q_0}^{Q_-} \frac{\d Q}{Q} \, \frac{\d f_a (Q,W)}{\d W}
\simeq \frac{\d f_a (W)}{\d W} \int_{Q_0}^{Q_-} \frac{\d Q}{Q}
= \frac{\d f_a (W)}{\d W} \ln\left(\frac{Q_-}{Q_0}\right) \simeq
- \, \frac{\d f_a (W)}{\d W} \ln\beta^2.
\label{9.11}\eeq
where we have used the approximations $Q_- \simeq W^2 / (2\me c^2
\beta^2)$ and $Q_0 \simeq W^2 / (2 \me c^2)$. Note that
the latter is valid only for energy losses such that $W \ll 2 \me c^2$
and, therefore, the following formulas have limited accuracy for the
inner subshells of heavy elements, whose GOSs are not negligible for
$W \gtrsim 2 \me c^2$. Thus we can write
\beq
\frac{\d \sigma_{a}^{\rm L,d}}{\d W} =
\frac{2\pi Z_0^2 e^4}{\me v^2} \,
\frac{1}{W} \left(
\int_{Q_0}^{Q_2} \frac{\d Q}{Q} \, \frac{\d f_a(Q,W)}{\d W}
+ \frac{\d f_a(W)}{\d W} \, \ln \beta^2\right)
\Theta(Q_2-Q).
\label{9.12}\eeq

The DDCS of transverse distant interactions is obtained from Eq.\
\req{9.1} after removing terms proportional to $(E+Mc^2)^{-1}$,
\beqa
\frac{\d^2 \sigma_a^{\rm T,d}}{\d W \, \d Q}
&=& \frac{2\pi Z_0^2 e^4}{\me v^2} \,
\frac{2\me c^2 W}{[Q(Q+2\me c^2)-W^2]^2}
\nonumber \\ [2mm]
&& \mbox{} \times \left(
\beta^2 - \frac{W^2}{Q(Q+2\me c^2)} \right)
\frac{\d g_a (Q,W)}{\d W}\, \Theta(Q_2-Q).
\label{9.13}\eeqa
The contribution of these excitations to the energy-loss DCS is
\beqa
\frac{\d \sigma_a^{\rm T,d}}{\d W} &=&  \int_{Q_-}^{Q_2} \d Q \;
\frac{\d^2 \sigma_a^{\rm T,d}}{\d W \, \d Q}
= \frac{2\pi Z_0^2 e^4}{\me v^2}
\int_{Q_-}^{Q_2} \d Q \;
\frac{2\me c^2 W}{[Q(Q+2\me c^2)-W^2]^2}
\nonumber \\ [2mm]
&& \mbox{} \times \left(
\beta^2 - \frac{W^2}{Q(Q+2\me c^2)} \right)
\frac{\d g_a (Q,W)}{\d W} \, \Theta(Q_2-Q).
\label{9.14}\eeqa
Following \citet{Fano1963}, because the DDCS \req{9.13} decreases
rapidly with $Q$, we will replace the TGOS with the OOS (optical
approximation). In addition, to allow the analytical evaluation of the
integral, we multiply the integrand by a factor $(Q+\me c^2)/\me c^2$,
which approaches unity for small $Q$. This gives
\beqa
\frac{\d \sigma_a^{\rm T,d}}{\d W} &=&
\frac{2\pi Z_0^2 e^4}{\me v^2} \, \frac{\d f_a (W)}{\d W}
\int_{Q_-}^{Q_2} \d Q \;
\frac{2(Q+\me c^2) W}{[Q(Q+2\me c^2)-W^2]^2}
\nonumber \\ [2mm]
&& \mbox{} \times \left(
\beta^2 - \frac{W^2}{Q(Q+2\me c^2)} \right)
 \, \Theta(Q_2-Q).
\label{9.15}\eeqa
To evaluate the integral we introduce the angle $\vartheta_{\rm r}$
defined by
\beq
\cos^2 \vartheta_{\rm r}(Q) \equiv \frac{W^2/\beta^2}{Q(Q +2\me c^2)}.
\label{9.16}\eeq
For fast projectiles, with $\beta \simeq 1$, $\vartheta_{\rm r}$
approaches the recoil angle $\theta_{\rm r}$ [see Fig.\ \ref{fig35}
and Eq.\ \req{A.16}], which is the angle between the vectors ${\bf p}$
and ${\bf q}$. Since
\beq
\frac{\d(\cos^2\vartheta_{\rm r})}{\d Q} = - \frac{2(Q+\me
c^2)W^2/\beta^2} {[Q(Q+2\me c^2)]^2}\, ,
\label{9.17}\eeq
we have
\beq
\frac{\d \sigma_{a}^{\rm T,d}}{\d W}
= \frac{2\pi Z_0^2 e^4}{\me v^2} \, \frac{\d f_a(W)}{\d W} \,
\frac{1}{W} \int_{Q_-}^{Q_2} \d Q \;
\left[- \, \left\{ \frac{\beta^4 (1-\cos^2\vartheta_{\rm r})}
{(1-\beta^2 \cos^2\vartheta_{\rm r})^2} \right\}
\frac{\d(\cos^2\vartheta_{\rm r})}{\d Q}
\right] .
\label{9.18}\eeq
The function in curly brackets equals unity at $\cos^2\vartheta_{\rm r}
=0$, which corresponds to large $Q$ values, and vanishes at
$\cos^2\vartheta_{\rm r}(Q_-)=1$; this function has a single maximum at
$\cos^2\vartheta_{\rm r}=2 - \beta^{-2}$, the width of which decreases
when the speed of the particle increases. At high energies, the
sharpness of this maximum makes the numerical calculation of the
integral of the transverse DDCS over $Q$ difficult. With the optical
approximation the dependence of the GOSs on $Q$ is removed and the
integral over $Q$ can be calculated. This gives
\beqa
\frac{\d \sigma_{a}^{\rm T,d}}{\d W}
&=& \frac{2\pi Z_0^2 e^4}{\me v^2} \, \frac{\d f_a(W)}{\d W}
\, \frac{1}{W}
\nonumber \\ [2mm]
&& \mbox{} \times \left[ - \frac{\beta^2-1}{1-\beta^2 \cos^2
\vartheta_{\rm r}}
+ \ln \left( 1-\beta^2 \cos^2\vartheta_{\rm r} \right)
\right]_{1}^{\cos \vartheta_{\rm r}((Q_2)} .
\label{9.19}\eeqa
When the energy of the projectile is sufficiently high, the most probable
distant interactions involve energy transfers that are much
less than $W_{\rm m}(Q_2)$, for which $Q_- \ll Q_2$ and
$\cos^2\vartheta_{\rm r}(Q_2) \simeq 0$. To get simpler formulas,
we shall set $\cos^2\vartheta_{\rm r}(Q_2) \simeq 0$ in Eq.\
\req{9.19}; this amounts to extending the integral over $Q$ values
larger than $Q_2$ or, equivalently, to removing the $\Theta(Q_2-Q)$
function in Eq.\ \req{9.15}. Thus, expression \req{9.19} simplifies
to
\beq
\frac{\d \sigma_{a}^{\rm T,d}}{\d W}
= \frac{2\pi Z_0^2 e^4}{\me v^2}
\left[ - \beta^2 - \ln \left( 1-\beta^2 \right)
\right] \frac{1}{W} \, \frac{\d f_a(W)}{\d W} \, .
\label{9.20}\eeq

The energy-loss DCS for distant interactions can now be expressed as
\beqa
\frac{\d \sigma_a^{\rm d}}{\d W} &=&
\frac{\d \sigma_a^{{\rm L,d}}}{\d W}
+ \frac{\d \sigma_a^{{\rm T,d}}}{\d W}
\nonumber \\ [2mm]
&\simeq&  \frac{2\pi Z_0^2 e^4}{\me v^2}
\left\{ \frac{1}{W} \left(
\int_{Q_0}^{Q_2} \frac{\d Q}{Q} \, \frac{\d f_a(Q,W)}{\d W}
+ \frac{\d f_a(W)}{\d W} \, \ln \beta^2\right) \Theta(Q_2-Q)
\right.
\nonumber \\ [2mm]
&& \mbox{} \left. + \left[ - \beta^2 - \ln \left( 1-\beta^2 \right) \right]
\frac{1}{W} \, \frac{\d f_a(W)}{\d W} \right\} .
\label{9.21}\eeqa
The integrated cross sections for distant interactions are given by
\beqa
\left[ \sigma_{a}^{\rm d} \right]^{(i)}
&=& \int_0^{W_{\rm m}(Q_2)} W^i \frac{\d \sigma_{a}^{\rm d}}{\d W}
\, \d W
\nonumber \\ [2mm]
&=&  \frac{2\pi Z_0^2 e^4}{\me v^2}
\left[ \ln \left( \frac{\beta^2}{1-\beta^2} \right) - \beta^2
\right]  \int_0^{W_{\rm m}(Q_2)}
W^{i-1} \, \frac{\d f_a(W)}{\d W} \, \d W
\nonumber \\ [2mm]
&& \mbox{} + \frac{2\pi Z_0^2 e^4}{\me v^2}
\int_0^{W_{\rm m}(Q_2)} \d W \; W^{i-1}
\int_{Q_0}^{Q_2} \frac{\d Q}{Q} \, \frac{\d f_a(Q,W)}{\d W} \, .
\label{9.22}\eeqa
Strictly, this result is valid only for particles different from the
electron and the positron. For projectile electrons we should account
for exchange effects, and in the case of positrons also for
annihilation-recreation [see Section \ref{sec7}]. Notice that for low
recoil energies these corrections are expected to be small. Because the
integrals analogous to those in Eq.\ \req{9.22} are difficult to
calculate when including these corrections, we shall assume that the
expression \req{9.22} is valid also for electrons and positrons, and
consider the neglected contributions as part of the shell corrections
(see Section \ref{sec11}).

To evaluate the integral
\beq
{\cal X}_i \equiv
\int_0^{W_{\rm m}(Q_2)} \d W \; W^{i-1}
\int_{Q_0}^{Q_2} \frac{\d Q}{Q} \, \frac{\d f_a(Q,W)}{\d W}
\label{9.23}\eeq
we first separate the low-$Q$ interval where the dipole approximation is
valid ($Q<Q_1$),
\beqa
{\cal X}_i &\simeq&
\int_0^{W_{\rm m}(Q_2)} \d W \; W^{i-1}
\int_{Q_0}^{Q_1} \frac{\d Q}{Q} \, \frac{\d f_a(W)}{\d W}
+ \int_0^{W_{\rm m}(Q_2)} \d W \; W^{i-1}
\int_{Q_1}^{Q_2} \frac{\d Q}{Q} \, \frac{\d f_a(Q,W)}{\d W}
\nonumber \\ [2mm]
&=&
\int_0^{W_{\rm m}(Q_2)} \d W \; W^{i-1}
\ln \left( \frac{Q_1}{Q_0} \right) \frac{\d f_a(W)}{\d W}
+ \int_0^{W_{\rm m}(Q_2)} \d W \; W^{i-1}
\int_{Q_1}^{Q_2} \frac{\d Q}{Q} \, \frac{\d f_a(Q,W)}{\d W}\, .
\nonumber \eeqa
Recalling that $W_{\rm m}(Q_2)$ is assumed to be sufficiently large to
exhaust the Bethe sum, it can be replaced with $\infty$. We can then
exchange the order of the integrals in the second term and write
\beqa
{\cal X}_i &=&
\int_0^\infty \d W \; W^{i-1}
\ln \left( \frac{Q_1}{Q_0} \right) \frac{\d f_a(W)}{\d W}
+ \int_{Q_1}^{Q_2} \frac{\d Q}{Q} \, S_{i-1} (a;Q),
\nonumber \eeqa
where
\beq
S_{i}(a;Q) \equiv
\int_0^\infty W^{i} \, \frac{\d f_a(Q,W)}{\d W}\, \d W.
\label{9.24}\eeq
Introducing the approximation $Q_0=W^2/(2\me c^2)$ we have
\beq
{\cal X}_i = S_{i-1}(a) \ln \left( 2\me c^2 \, Q_1 \right) - 2
S_{i-1}(a) \, \ln[I_{i-1}(a)]
+ \int_{Q_1}^{Q_2} \frac{\d Q}{Q} \, S_{i-1} (a;Q),
\label{9.25}\eeq
with $S_{i}(a)=S_{i}(a;0)$, and
\beq
S_{i}(a) \, \ln[I_{i}(a)] = \int_0^\infty W^{i} \, \ln W \,
\frac{\d f_a(W)}{\d W} \, \d W.
\label{9.26}\eeq
Now we can write the cross section for distant interactions as
\beqa
\left[ \sigma_{a}^{\rm d} \right]^{(i)}
&=& \frac{2\pi Z_0^2 e^4}{\me v^2} \left\{
S_{i-1}(a)
\left[ \ln \left( \frac{\beta^2}{1-\beta^2} \right) - \beta^2 \right]
+ 2 \, S_{i-1}(a) \ln \left( \frac{2\me c^2}{I_{i-1}(a)} \right)
\right.
\nonumber \\ [2mm]
&& \mbox{} \left. +
S_{i-1}(a) \ln \left( \frac{Q_1}{2\me c^2} \right)
+ \int_{Q_1}^{Q_2} \frac{\d Q}{Q} \, S_{i-1} (a;Q)
\right\}.
\label{9.27}\eeqa

\vspace*{3mm}

\noindent $\bullet$ {\bf Close interactions} \\ [2mm]
Interactions with $Q>Q_2$
will be described as binary collisions with stationary free electrons. The
corresponding energy-loss DCS is given by
\beq
\frac{\d \sigma_{a}^{\rm c}}{\d W}
= \frac{2\pi Z_0^2 e^4}{\me v^2} \, 2|\kappa_a| \, \frac{1}{W^2} \,
F_{\rm rel}(W) \, \Theta(W-Q_2) \, \Theta(W_{\rm ridge}-W) \, ,
\label{9.28}\eeq
where $W_{\rm ridge}$ is the maximum allowed energy-loss [Eq.\
\req{9.7}]. The factor $F_{\rm rel}(W)$ is
\begin{subequations}
\label{9.29}
\beq
F_{\rm rel}(W) =
1 - \beta^2 \, \frac{W}{W_{\rm ridge}} + \frac{1-\beta^2}{2M^2 c^4}\, W^2
\label{9.29a}\eeq
for projectiles other than electrons and positrons [Eq.\ \req{6.42}],
\beq
F_{\rm rel}(W) = F_{\rm M\o ller}(W) =
 1  + \left( \frac{W}{E-W} \right)^{2}
- \frac{(1-b_0)W}{E-W} + \frac{b_0 W^{2}}{E^{2}}
\label{9.29b}\eeq
for electrons [Eq.\ \req{7.4}], and
\beq
F_{\rm rel}(W) = F_{\rm Bhabha}(W) =
1 - b_{1}\frac{W}{E} +
b_{2}\left(\frac{W}{E}\right)^{2} - b_{3}\left(\frac{W}{E}\right)^{3} +
b_{4}\left(\frac{W}{E}\right)^{4}
\label{9.29c}\eeq
for positrons [Eq.\ \req{7.14}].
\end{subequations}
It is worth recalling that these DCSs include contributions of both
longitudinal and transverse interactions.

The contributions of close interactions to the integrated cross sections
are
\beqa
\left[ \sigma_{a}^{\rm c} \right]^{(i)}
&=& \int_{Q_2}^{W_{\rm ridge}} W^i \frac{\d \sigma_{a}^{\rm d}}{\d W}
\, \d W
\nonumber \\ [2mm]
&=& \frac{2\pi Z_0^2 e^4}{\me v^2} \, 2|\kappa_a|
\int_{Q_2}^{W_{\rm ridge}} W^{i-2} \, F_{\rm rel}(W) \, \d W.
\label{9.30}\eeqa
These quantities are given by the analytical expressions derived
in Sections \ref{sec6.4} and \ref{sec7}. In the high-energy limit when
$W_{\rm ridge} \gg Q_2$, these expressions reduce to the following
forms: \\
\noindent For particles other than electrons and positrons,
\begin{subequations}
\label{9.31}
\beqa
\left[ \sigma_{a}^{\rm c} \right]^{(0)} &\simeq&
\frac{2\pi Z_0^2 e^4}{\me v^2} \, 2|\kappa_a| \left[ \frac{1}{Q_2}
\right] ,
\label{9.31a} \\ [2mm]
\left[ \sigma_{a}^{\rm c} \right]^{(1)} &\simeq&
\frac{2\pi Z_0^2 e^4}{\me v^2} \, 2|\kappa_a|
\left[ \ln \left( \frac{2\me c^2 \beta^2 R}{(1-\beta^2)\, Q_2}
\right) - \beta^2
+ \left( \frac{\me}{M} \beta^2 \gamma R\right)^2 \right],
\label{9.31b} \\ [2mm]
\left[ \sigma_{a}^{\rm c} \right]^{(2)} &\simeq&
\frac{2\pi Z_0^2 e^4}{\me v^2} \, 2|\kappa_a|
\left[ - Q_2
+ (2-\beta^2) \, \frac{\me c^2 \, \beta^2 }{1-\beta^2} R
+ \frac{4}{3} \, \frac{\me^2}{M^2} \, \beta^6 \gamma^4 \, \me c^2 R^3
\right]. \rule{10mm}{0mm}
\label{9.31c}\eeqa
\end{subequations}
\noindent For electrons,
\begin{subequations}
\label{9.32}
\beqa
\left[ \sigma_{a}^{\rm c}({\rm e}^-)\right]^{(0)}
&\simeq& \frac{2 \pi e^4}{\me v^2} \,  2|\kappa_a| \left[
\frac{1}{Q_2} \right] ,
\label{9.32a} \\ [2mm]
\left[ \sigma_{a}^{\rm c}({\rm e}^-)\right]^{(1)}
&\simeq& \frac{2 \pi e^4}{\me v^2} \,  2|\kappa_a| \,
\left[ \ln \left( \frac{E}{2 Q_2} \right) + 1
- (2-b_0) \ln 2 + \frac{b_0}{8} \right],
\label{9.32b} \\ [2mm]
\left[ \sigma_{a}^{\rm c}({\rm e}^-)\right]^{(2)}
&\simeq& \frac{2 \pi e^4}{\me v^2} \,  2|\kappa_a|
\left[ - Q_2+  E \left( \frac{5}{2} - (3-b_0) \ln 2 - \frac{11 b_0}{24}
\right) \right]. \rule{10mm}{0mm}
\label{9.32c}\eeqa
\end{subequations}
\noindent For positrons.
\begin{subequations}
\label{9.33}
\beqa
\left[ \sigma_{a}^{\rm c}({\rm e}^+) \right]^{(0)}
&\simeq& \frac{2 \pi e^4}{\me v^2} \,  2|\kappa_a|
\left[ \frac{1}{Q_2} \right] ,
\label{9.33a}\\ [2mm]
\left[ \sigma_{a}^{\rm c}({\rm e}^+)\right]^{(1)}
&\simeq& \frac{2 \pi e^4}{\me v^2} \,  2|\kappa_a| \,
\left[ \ln \left(\frac{E}{Q_2} \right)
- b_{1}+ \frac{b_2}{2} -
\frac{b_3}{3} + \frac{b_4}{4} \right],
\label{9.33b}\\ [2mm]
\left[ \sigma_{a}^{\rm c}({\rm e}^+)\right]^{(2)}
&\simeq& \frac{2 \pi e^4}{\me v^2} \,  2|\kappa_a| \,
\left[ - Q_2 + E \left( 1 - \frac{b_{1}}{2} + \frac{b_{2}}{3} -
\frac{b_{3}}{4} + \frac{b_{4}}{5} \right) \right]. \rule{10mm}{0mm}
\label{9.33c}\eeqa
\end{subequations}

In the following we shall limit our considerations to the case of
particles other than electrons and positrons; the results for the latter
particles are obtained by simply replacing the expressions within square
brackets in the Eqs.\ \req{9.31} with the analogous expressions in Eqs.\
\req{9.32} and \req{9.33}.


\subsection{Integrated subshell cross sections \label{sec9.1}}

We can now derive asymptotic formulas of the integrated cross sections
$\sigma_a^{(i)}$ for excitations of a subshell $n_a\kappa_a$ with $q_a$
($\le 2|\kappa_a|$) electrons. Hereafter, the subshell GOS and TGOS are
assumed to describe only ``real'' transitions that are allowed by
Pauli's exclusion principle, that is, one-electron transitions to empty
final states. Formulas for atoms or ions will be obtained latter on by
adding the contributions from the subshells in the ground-state
configuration.


\subsubsection{Total cross section \label{sec9.1.1}}

The total cross section for distant interactions of fast projectiles is
given by Eq.\ \req{9.27},
\beqa
\left[ \sigma_{a}^{\rm d} \right]^{(0)}
&=& \frac{2\pi Z_0^2 e^4}{\me v^2} \left\{
S_{-1}(a)
\left[ \ln \left( \frac{\beta^2}{1-\beta^2} \right) - \beta^2 \right]
+ 2 \, S_{-1}(a) \ln \left( \frac{2\me c^2}{I_{-1}(a)} \right)
\right.
\nonumber \\ [2mm]
&& \mbox{} \left. +
S_{-1}(a) \ln \left( \frac{Q_1}{2\me c^2} \right)
+ \int_{Q_1}^{Q_2} \frac{\d Q}{Q} \, S_{-1} (a;Q)
\right\}.
\label{9.34}\eeqa
The total cross section is obtained by adding to this result the
contribution from close collisions, given by Eq.\ \req{9.31a}, with the
occupancy $q_a$ of the subshell,
\beq
\left[ \sigma_{a}^{\rm c} \right]^{(0)} =
\frac{2\pi Z_0^2 e^4}{\me v^2} \, q_a \, \frac{1}{Q_2}.
\label{9.35}\eeq
We have
\beqa
\sigma_{a}^{(0)} &=&
\frac{2\pi Z_0^2 e^4}{\me v^2}  \left\{  S_{-1}(a)
\left[ \ln \left( \frac{\beta^2}{1-\beta^2} \right) - \beta^2 \right]
\right.
\nonumber \\ [2mm]
&& \left. + 2 \, S_{-1}(a) \ln \left( \frac{2\me c^2}{I_{-1}(a)}
\right) + D_{-1}(a) \right\},
\label{9.36}\eeqa
with
\beq
D_{-1}(a)= S_{-1}(a) \ln \left( \frac{Q_1}{2\me c^2} \right)
+ \int_{Q_1}^{Q_2} \frac{\d Q}{Q} \, S_{-1} (a;Q) +
q_a \, \frac{1}{Q_2}.
\label{9.37}\eeq


\subsubsection{Stopping cross section \label{sec9.1.2}}

The stopping cross section for distant interactions [see Eq.\
\req{9.27}] with the electrons in the active subshell can be written as
\beqa
\left[ \sigma_{a}^{\rm d} \right]^{(1)}
&=& \frac{2\pi Z_0^2 e^4}{\me v^2} \left\{
S_{0}(a)
\left[ \ln \left( \frac{\beta^2}{1-\beta^2} \right) - \beta^2 \right]
+ 2 \, S_{0}(a) \ln \left( \frac{2\me c^2}{I_{0}(a)} \right)
\right.
\nonumber \\ [2mm]
&& \mbox{} \left. +
S_{0}(a) \ln \left( \frac{Q_1}{2\me c^2} \right)
+ \int_{Q_1}^{Q_2} \frac{\d Q}{Q} \, S_{0} (a;Q)
\right\}.
\label{9.38}\eeqa
The stopping cross section of the subshell is obtained by
adding this result and the contribution of close interactions
[Eq. \req{9.31b}] with the number $q_a$ of electrons in the subshell,
\beq
\left[ \sigma_{a}^{\rm c} \right]^{(1)} =
\frac{2\pi Z_0^2 e^4}{\me v^2} \, q_a
\left[ \ln \left( \frac{2\me c^2 \beta^2 R}{(1-\beta^2)\, Q_2}
\right) - \beta^2
+ \left( \frac{\me}{M} \beta^2 \gamma R\right)^2 \right],
\label{9.39}\eeq
to give
\beqa
\sigma_{a}^{(1)}
&=& \frac{2\pi Z_0^2 e^4}{\me v^2} \left\{ [S_0(a)+q_a]
\left[\rule{0mm}{4mm} \ln \left( \frac{\beta^2}{1-\beta^2} \right)
- \beta^2 \right] \right.
\nonumber \\ [2mm]
&& \mbox{} \left. +
2 \, S_0(a) \ln \left(\frac{2 \me c^2}{I_0(a)} \right)
+ D_0(a) + q_a \left[ \ln R
+ \left( \frac{\me}{M} \beta^2 \gamma R\right)^2
\right] \right\}, \rule{10mm}{0mm}
\label{9.40}\eeqa
with
\beq
D_0(a) =
S_0(a) \ln \left( \frac{Q_1}{2\me c^2} \right)
+ \int_{Q_1}^{Q_2} \frac{\d Q}{Q} \, S_0(a;Q) - q_a
\ln  \left( \frac{Q_2}{2\me c^2} \right)
\label{9.41}\eeq


\subsubsection{Energy-straggling cross section \label{sec9.1.3}}

The energy-straggling cross section for distant interactions with the
active subshell is [see Eq.\ \req{9.27}]
\beqa
\left[ \sigma_{a}^{\rm d} \right]^{(2)}
&=& \frac{2\pi Z_0^2 e^4}{\me v^2} \left\{
S_{1}(a)
\left[ \ln \left( \frac{\beta^2}{1-\beta^2} \right) - \beta^2 \right]
+ 2 \, S_{1}(a) \ln \left( \frac{2\me c^2}{I_{1}(a)} \right)
\right.
\nonumber \\ [2mm]
&& \mbox{} \left. +
S_{1}(a) \ln \left( \frac{Q_1}{2\me c^2} \right)
+ \int_{Q_1}^{Q_2} \frac{\d Q}{Q} \, S_{1} (a;Q)
\right\}.
\label{9.42}\eeqa
The energy-straggling cross section of the subshell is obtained by
adding to this result the contribution from close
collisions [given by Eq.\ \req{9.31c}] with the number $q_a$ of
electrons in the subshell,
\beq
\left[ \sigma_{a}^{\rm c} \right]^{(2)} =
\frac{2\pi Z_0^2 e^4}{\me v^2} \, q_a
\left[ - Q_2
+ (2-\beta^2) \, \frac{\me c^2 \, \beta^2 }{1-\beta^2} R
+ \frac{4}{3} \, \frac{\me^2}{M^2} \, \beta^6 \gamma^4 \, \me c^2 R^3
\right].
\label{9.43}\eeq
We thus get
\beqa
\sigma_{a}^{(2)} &=&
\frac{2\pi Z_0^2 e^4}{\me v^2}
\left\{ S_{1}(a) \left[
\ln \left( \frac{\beta^2}{1-\beta^2} \right) - \beta^2 \right]
+ 2 S_{1}(a) \ln \left( \frac{2 \me c^2}{I_1(a)} \right) + D_{1}(a)
\right.
\nonumber \\ [2mm]
&& \mbox{} \left.
+ q_a \left[
(2-\beta^2) \, \frac{\me c^2 \, \beta^2 }{1-\beta^2} R
+ \frac{4}{3} \, \frac{\me^2}{M^2} \, \beta^6 \gamma^4 \, \me c^2 R^3
\right] \right\},
\label{9.44}\eeqa
with
\beq
D_{1}(a) =  S_{1}(a) \ln \left( \frac{Q_1}{2\me c^2} \right)
+ \int_{Q_1}^{Q_2} \frac{\d Q}{Q} \, S_{1}(a;Q) -  q_a \, Q_2.
\label{9.45}\eeq


\subsubsection{Summary of asymptotic subshell formulas \label{sec9.1.4}}

We compile here the asymptotic subshell formulas resulting from the above
derivation, with the definitions of the involved coefficients and
function for the various types of charged projectiles.

\vspace*{3mm}
\hrule \hrule
\begin{subequations} \label{9.46}
\beqa
\sigma_{a}^{(0)} &=&
\frac{2\pi Z_0^2 e^4}{\me v^2}  \left\{  S_{-1}(a)
\left[ \ln \left( \frac{\beta^2}{1-\beta^2} \right) - \beta^2 \right]
\right.
\nonumber \\ [2mm]
&& \left. + 2 \, S_{-1}(a) \ln \left( \frac{2\me c^2}{I_{-1}(a)}
\right) + D_{-1}(a) \right\}
\label{9.46a}\eeqa
\beqa
\sigma_{a}^{(1)}
&=& \frac{2\pi Z_0^2 e^4}{\me v^2} \left\{ [S_0(a)+q_a]
\left[\rule{0mm}{4mm} \ln \left( \frac{\beta^2}{1-\beta^2} \right)
- \beta^2 \right] \right.
\nonumber \\ [2mm]
&& \mbox{} \left. +
2 \, S_0(a) \ln \left(\frac{2 \me c^2}{I_0(a)} \right)
+ D_0(a) + q_a \, f(\gamma)
\right\}
\label{9.46b}\eeqa
\beqa
\sigma_{a}^{(2)} &=&
\frac{2\pi Z_0^2 e^4}{\me v^2}
\left\{ S_{1}(a) \left[
\ln \left( \frac{\beta^2}{1-\beta^2} \right) - \beta^2 \right]
\right.
\nonumber \\ [2mm]
&& \mbox{} \left.
+ 2 \, S_{1}(a) \ln \left( \frac{2 \me c^2}{I_1(a)} \right)
+ D_{1}(a) + q_a \, \me c^2 \, g(\gamma) \right\}
\label{9.46c}\eeqa
\end{subequations}

\hrule \hrule
\vspace*{2mm}

\begin{subequations}
\label{9.47}
\beq
S_{i} (a;Q) =
\int_0^\infty W^i \, \frac{\d f_a (Q,W)}{\d W} \, \d W
\label{9.47a}\eeq
\beq
S_{i} (a) = S_{i} (a;0) =
\int_0^\infty W^i \, \frac{\d f_a (W)}{\d W} \, \d W
\label{9.47b}\eeq
\beq
\ln [I_{i}(a)] = \frac{1}{S_{i}(a)}
\int_0^\infty W^i \, \ln W \, \frac{\d f_a (W)}{\d W} \, \d W
\label{9.47c}\eeq
\end{subequations}

\vspace*{2mm}
\hrule \hrule
\vspace*{2mm}

\begin{subequations} \label{9.48}
\beq
D_{-1}(a)= S_{-1}(a) \ln \left( \frac{Q_1}{2\me c^2} \right)
+ \int_{Q_1}^{Q_2} \frac{\d Q}{Q} \, S_{-1} (a;Q) +
q_a \, \frac{1}{Q_2}
\label{9.48a}\eeq
\beq
D_0(a) =
S_0(a) \ln \left( \frac{Q_1}{2\me c^2} \right)
+ \int_{Q_1}^{Q_2} \frac{\d Q}{Q} \, S_0(a;Q) - q_a
\ln  \left( \frac{Q_2}{2\me c^2} \right)
\label{9.48b}\eeq
\beq
D_{1}(a) =  S_{1}(a) \ln \left( \frac{Q_1}{2\me c^2} \right)
+ \int_{Q_1}^{Q_2} \frac{\d Q}{Q} \, S_{1}(a;Q) -  q_a \, Q_2
\label{9.48c}\eeq

\noindent with $Q_1 \simeq 10^{-3} E_a$ and $Q_2 = 10^3 E_a$.
\end{subequations}

\vspace*{3mm}
\hrule \hrule
\vspace*{2mm}

\begin{subequations}
\label{9.49}
\noindent Electrons:
\beq
f(\gamma) = \frac{2\gamma^2-1}{\gamma^2}
+ \frac{1}{8} \left( \frac{\gamma-1}{\gamma} \right)^2
- \left[ 4 - \left( \frac{\gamma-1}{\gamma} \right)^2 \right] \ln 2
- \ln(\gamma+1)
\label{9.49a}\eeq
\beq
g(\gamma) = (\gamma-1) \left[ \frac{5}{2}
+ \frac{1 - 2\gamma - 2 \gamma^2}{\gamma^2} \, \ln 2
- \frac{11}{24} \left( \frac{\gamma-1}{\gamma} \right)^2 \right]
\label{9.49b}\eeq
\noindent Positrons:
\beq
f(\gamma) = \frac{\gamma^2-1}{12\gamma^2} \left[ 1
- \frac{14}{\gamma+1}
- \frac{10}{(\gamma+1)^2}
- \frac{4}{(\gamma+1)^3} \right]
- \ln 2 - \ln(\gamma+1)
\label{9.49c}\eeq
\beq
g(\gamma) =  (\gamma-1)
\left[ 1
-\frac{\gamma^2-1}{30\gamma^2}
\left(9 +\frac{21}{\gamma+1}
+\frac{23}{(\gamma+1)^2}
+\frac{8}{(\gamma+1)^3}
\right)  \right]
\label{9.49d}\eeq
\noindent Heavier particles:
\beq
f(\gamma) = \ln(R) + \left( \frac{\me}{M}\,
\frac{\gamma^2-1}{\gamma} \, R \right)^2
\label{9.49e}\eeq
\beq
g(\gamma) = \frac{\gamma^4-1}{\gamma^2} \, R + \frac{4}{3} \,
\frac{\me^2}{M^2}\, \frac{(\gamma^2-1)^3}{\gamma^2} \, R^3
\label{9.49f}\eeq
\end{subequations}
with $R$ defined by Eq.\ \req{9.7}. \\ [1mm]
\hrule \hrule
\vspace*{3mm}

\begin{figure}[h]
\begin{center}
\includegraphics*[width=7.5cm]{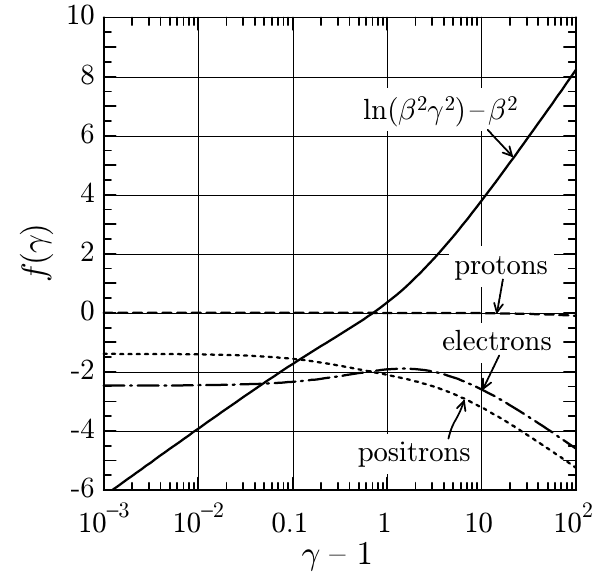} \hfill
\includegraphics*[width=7.5cm]{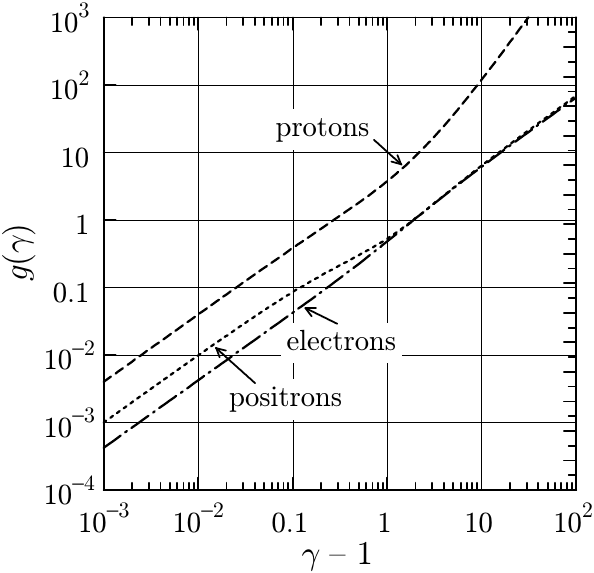}
\caption{Energy-dependent terms in the asymptotic formulas of the
stopping cross section (left) and the energy-straggling cross section
(right). The quantity $\gamma-1$ is the kinetic energy of the projectile
in units of its rest energy.
\label{fig24}}
\end{center}\end{figure}

Since the functions $f(\gamma)$ and $g(\gamma)$ depend on the mass of
the projectile, particles of equal charge and different masses moving
with the same speeds have slightly different stopping and
energy-straggling cross sections (see Fig.\ \ref{fig24}).

The above formulas reduce to their conventional forms when $S_0(a;Q)=
q_a$ for all $Q$, that  is, when the Bethe sum rule is satisfied. In
such case $D_0 =0$, but the other $D_n$ coefficients are generally not
null.

It is important to notice that our derivation of the asymptotic formulas
is based on the assumption that the GOS effectively vanishes for energy
transfers $W$ higher than the photon line. This assumption is only valid
for electron subshells with small and moderate binding energies. It {\it
does not hold} for the innermost subshells of heavy elements, for which the
predictions of the asymptotic formulas are expected to deviate from the
values obtained by numerical integration of the DDCSs. These deviations
must be considered as part of the shell correction (see Section
\ref{sec10}).


\subsection{Asymptotic formulas for atoms \label{sec9.2}}

The integrated cross sections for atoms (or ions) are obtained by adding
the contributions of the individual electron subshells, which are given by
Eqs.\ \req{9.46}.


\subsubsection{Total cross section \label{sec9.2.1}}

The asymptotic formula for the total cross section is [see Eq.\
\req{9.46a}]
\beqa
\sigma^{(0)} = \sum_a \sigma_{a}^{(0)} &=&
\frac{2\pi Z_0^2 e^4}{\me v^2}  \left\{  S_{-1}
\left[ \ln \left( \frac{\beta^2}{1-\beta^2} \right) - \beta^2 \right]
\right.
\nonumber \\ [2mm]
&& \left. + 2 \, S_{-1} \ln \left( \frac{2\me c^2}{I_{-1}}
\right) + D_{-1} \right\}
\label{9.50}\eeqa
with
\beq
S_{-1} = \sum_a S_{-1}(a) =
\int_0^\infty \frac{1}{W} \, \frac{\d f(W)}{\d W} \, \d W,
\label{9.51}\eeq
\beq
\ln[I_{-1}] = \frac{1}{S_{-1}} \sum_a S_{-1}(a)\, \ln [I_{-1}(a)]
= \frac{1}{S_{-1}} \int_0^\infty \frac{1}{W} \, \ln W \,
\frac{\d f (W)}{\d W} \, \d W,
\label{9.52}\eeq
and
\beq
D_{-1} = \sum_a D_{-1}(a).
\label{9.53}\eeq
Notice that the limits of the integral in the definition of $D_{-1}$,
Eqs.\ \req{9.48a}, depend on the binding energy $E_a$ of the electron subshell.

\begin{figure}[htb]
\begin{center}
\includegraphics*[width=7.5cm]{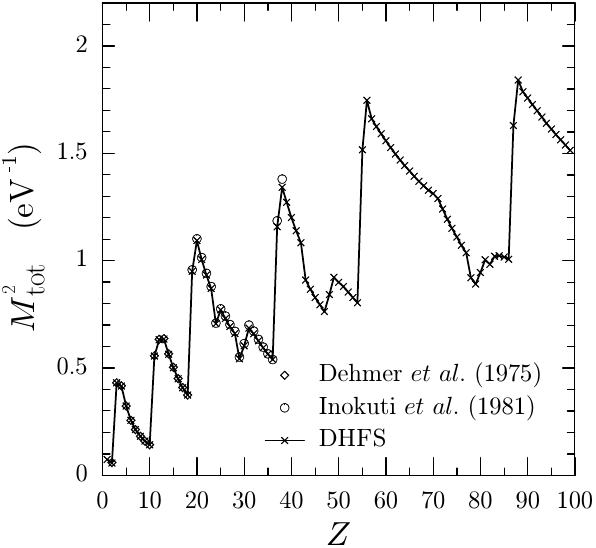} \hfill
\includegraphics*[width=7.5cm]{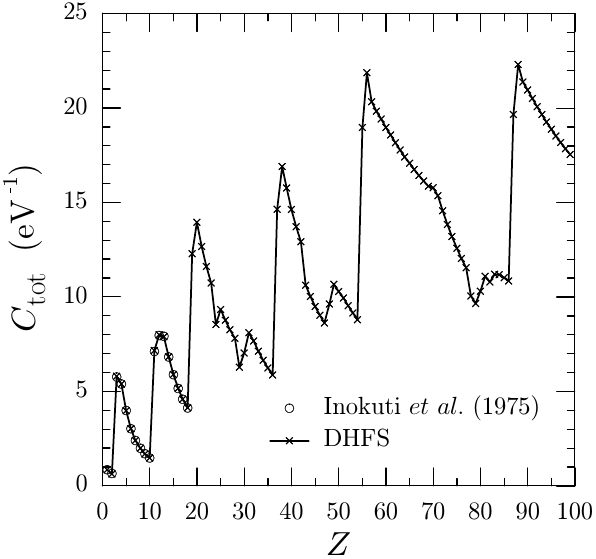}
\caption{Parameters of the asymptotic formula \req{9.54} of the total
cross section for inelastic collisions with free neutral atoms, calculated
from the present numerical GOS (crosses). The circles are values
obtained by \citet{Dehmer1975} and \citet{Inokuti1981} and
\citet{Inokuti1975} from similar atomic models.
\label{fig25}}
\end{center}\end{figure}

The expression \req{9.50} can be recast in a form similar to the
``conventional'' formula employed normally in the related literature,
\beq
\sigma^{(0)} = \frac{2\pi Z_0^2 e^4}{\me v^2} \left\{
M_{\rm tot}^2
\left[ \ln \left( \frac{\beta^2}{1-\beta^2} \right)- \beta^2  \right]
+ C_{\rm tot} \right\} \, ,
\label{9.54}\eeq
where
\beq
M^2_{\rm tot} = S_{-1}
\label{9.55}\eeq
is the total dipole-matrix element squared \citep{Inokuti1971}, and
\beq
C_{\rm tot} =
2 \, S_{-1} \ln \left( \frac{2\me c^2}{I_{-1}}
\right) + D_{-1}\, .
\label{9.56}\eeq
We see that the total cross section $\sigma^{(0)}$ depends only on the
speed and the charge of the projectile, but not on its mass.
Protons, antiprotons, electrons and positrons moving with the same
speed have the same total cross sections. This feature is in
contradistinction to the stopping and energy-straggling cross sections,
which are different for different particles with the same speed (see
below). The parameters $M^2_{\rm tot}$ and $C_{\rm tot}$ are
energy-independent constants, characteristic of the target atom or ion.
The values of these parameters for free atoms, calculated from our
numerical GOSs, are displayed in Fig.\ \ref{fig25}, which also shows the
values obtained from non-relativistic calculations by \citet{Dehmer1975}
and \citet{Inokuti1981} using the Hartree--Slater potential for the
elements up to strontium ($Z \le 38$). There is good agreement between
the two calculations, because relativistic effects are small for these
elements. It should be borne in mind that, because of the simplicity of
our central-field approximation, the asymptotic formula with these
constants may yield total cross sections that differ substantially from
their actual values. Although total cross sections are very sensitive to
aggregation effects, the formula \req{9.54}, with appropriate values of
the parameters $M^2_{\rm tot}$ and $C_{tot}$, is valid also for
inelastic collisions of charged projectiles with molecules or solids
\citep[see, \eg,][]{FernandezVarea1993}.

The formula \req{9.54} is analogous to the one derived by
\citet{Fano1954b}, who considered only ionizing collisions and used the
non-relativistic GOS for longitudinal interactions and the dipole
approximation for transverse interactions. \citet{Bethe1930} obtained
the non-relativistic analogue of this formula for atomic hydrogen. As
noted by Fano, a plot of the total cross section as a function of the
quantity $\ln(\beta^2 \gamma^2)-\beta^2$ is a straight line with slope
$M_{\rm tot}^2$ and ordinate intercept $C_{\rm tot}$. This ``Fano plot''
has been used to assess the validity of the PWBA, and as a consistency
check of experimental data \citep[see, \eg,][]{Inokuti1971, Powell2016}.


\subsubsection{Stopping cross section \label{sec9.2.2}}

\begin{figure}[htb]
\begin{center}
\includegraphics*[width=7.5cm]{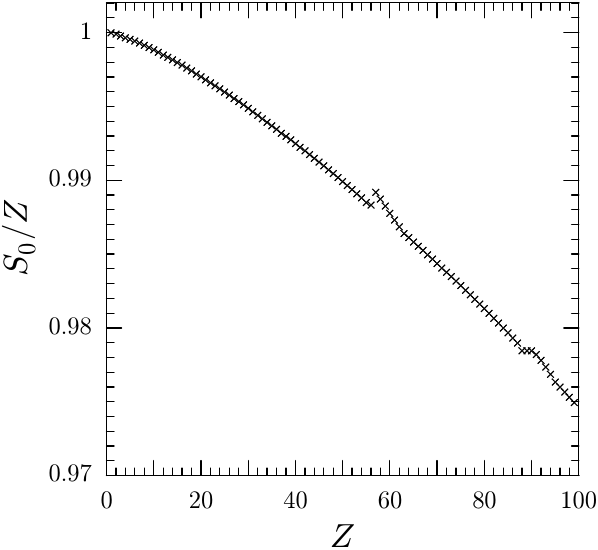} \hfill
\includegraphics*[width=7.5cm]{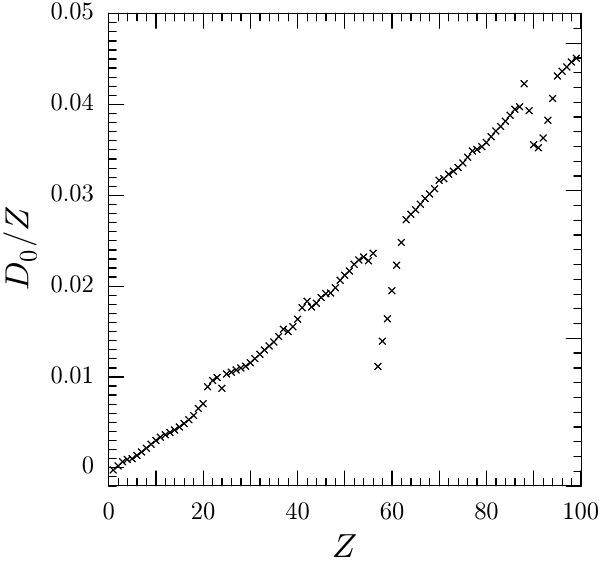}
\caption{Values of the dipole sum $S_0$ and the parameter $D_0$
calculated from the present database of numerical GOSs of neutral atoms.
For the sake of clarity, the ratios $S_0/Z$ and $D_0/Z$ are plotted
{\it versus} $Z$.
\label{fig26}}
\end{center}\end{figure}

The stopping cross section of the atom is obtained as [see Eq.\
\req{9.46b}]
\beqa
\sigma^{(1)} = \sum_a \sigma_{a}^{(1)}
&=& \frac{2\pi Z_0^2 e^4}{\me v^2} \left\{ [S_0+Z]
\left[\rule{0mm}{4mm} \ln \left( \frac{\beta^2}{1-\beta^2} \right)
- \beta^2 \right] \right.
\nonumber \\ [2mm]
&& \mbox{} \left. +
2 \, S_0 \ln \left(\frac{2 \me c^2}{I_0} \right)
+ D_0 + Z \, f(\gamma)
\right\}
\label{9.57}\eeqa
with
\beq
S_{0} = \sum_a S_{0}(a) =
\int_0^\infty \frac{\d f(W)}{\d W} \, \d W,
\label{9.58}\eeq
\beq
\ln I_{0} = \frac{1}{S_{0}} \sum_a S_{0}(a)\, \ln [I_{0}(a)]
=  \frac{1}{S_{0}} \int_0^\infty \ln W \,
\frac{\d f (W)}{\d W} \, \d W,
\label{9.59}\eeq
and
\beq
D_{0} = \sum_a D_{0}(a).
\label{9.60}\eeq
Figure \ref{fig26} shows the dipole sum $S_0$ and the parameter $D_0$
calculated from the present numerical GOSs of neutral atoms.

It is interesting to compare the result \req{9.57} with the familiar
\citet{Bethe1933, Ahlen1980, ICRU49} stopping power formula, which can be written
in the form
\beq
\sigma^{(1)}_{\rm Bethe}
= \frac{2\pi Z_0^2 e^4}{\me v^2} \, 2Z \left[
\ln \left( \frac{2 \me v^2}{I} \right)
+ \ln \left( \frac{1}{1-\beta^2} \right) - \beta^2
+ \frac{1}{2} \, f(\gamma) \right]
\label{9.61}\eeq
with the {\it mean excitation energy} $I$ defined by
\beq
\ln I = \frac{1}{Z}  \int_0^\infty \ln W \,
\frac{\d f (W)}{\d W} \, \d W.
\label{9.62}\eeq
The derivation of the Bethe formula \citep[see, \eg,][]{Fano1963} makes
explicit use of the Bethe sum rule ($S_0(Q) =Z$), which is assumed to
hold for any $Q$. Consequently, the formula is strictly valid only for
light elements, for which relativistic deviations from the sum rule are
small and do not modify appreciably the calculated stopping cross
sections. In spite of this simplification, the Bethe formula is
considered to be valid for all atomic numbers \citep[see,
\eg,][]{ICRU37,ICRU49} and it is generally used (with various correction
terms) to provide stopping powers for transport and dosimetry
calculations.

The only non-trivial parameter in the Bethe formula is the mean
excitation energy, which characterizes the stopping power of the
material. In principle, the $I$ value can be calculated from knowledge
of the OOS. The mean excitation energies $I_0$ computed from our
numerical OOSs are close to the results of non-relativistic calculations
by \citet{Dehmer1975} and \citet{Inokuti1981}. These authors used the
Hartree--Slater potential and considered elements up to strontium ($Z \le
38$). Since relativistic effects are weak for these low and moderate
atomic numbers, the two calculations yield nearly equivalent results.
The mean excitation energy depends strongly on the details of the
excitation spectrum and, therefore, we should not expect our
independent-electron approximation to give accurate $I$ values even for
free atoms. In addition, the OOS for low-energy excitations depends on
the state of aggregation of the material because the wave functions of
weakly bound and slow free electrons are modified by the presence of
neighboring atoms. Since first-principle calculations of the OOS for
condensed materials are not available, the practical alternative is to
determine the $I$ value from semi-empirical OOS models that combine
measured optical data for low-$W$ excitations with calculated OOSs of
inner electron subshells \citep[see, \eg][]{FernandezVarea2005}.

\begin{figure}[htb!]
\begin{center}
\includegraphics*[width=8.5cm]{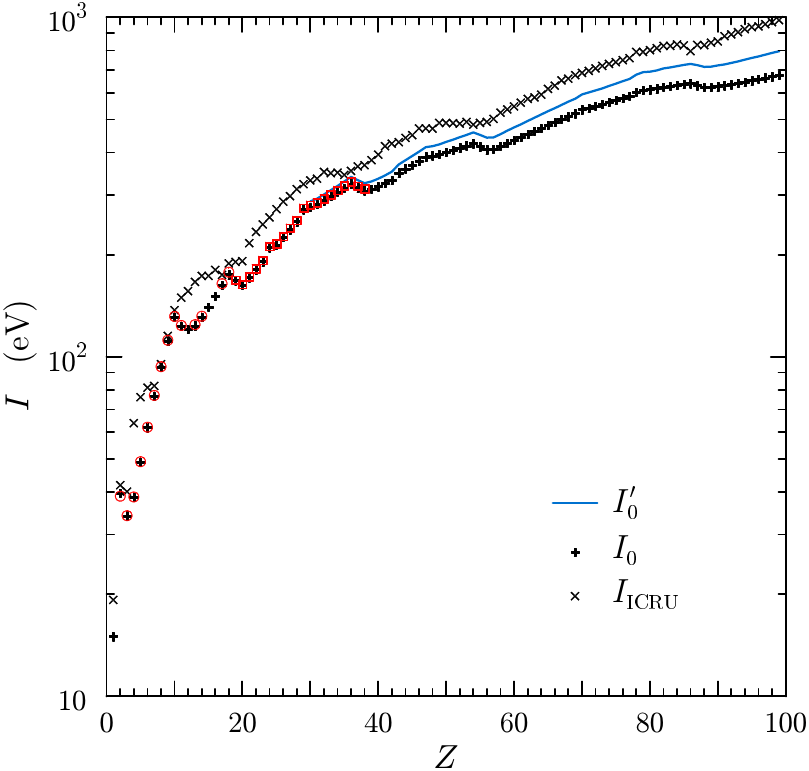}
\caption{Mean excitation energy of elemental substances {\it versus} the
atomic number $Z$. The mean excitation energies recommended in the
\citet{ICRU37} are represented by the times symbols.  The red open
circles and squares are values calculated by \citet{Dehmer1975} and
\citet{Inokuti1981} with a non-relativistic Hartree--Slater potential.
Crosses indicate our calculated $I_0$ values, and the solid curve
represents the modified mean excitation energies obtained from Eq.\
\req{9.64}.
\label{fig27}}
\end{center}\end{figure}

Because measured optical data are frequently affected by considerable
uncertainties, detailed knowledge of the OOS in the entire range of
energy transfers is available only for a few selected materials. As a
result, mean excitation energies of materials have generally been
derived from measured stopping powers, on the basis of the
Bethe formula with suitable corrections \citep[see, \eg,][and references
therein]{ICRU37}. Figure \ref{fig27} displays the mean excitation
energies of elemental materials recommended in the \citet{ICRU37},
which were inferred from a combination of stopping power measurements
and calculations for specific materials. Also shown are the values of
the parameter $I_0$ for neutral DHFS atoms calculated from the GOSs in
our database. The $I_0$ values and the ICRU empirical $I$
values are seen to vary similarly with $Z$, although the difference $I-I_0$
increases gradually with $Z$. The DHFS results are expected to be more
accurate for the noble gas atoms, with closed-subshell configurations, than
for other elements which naturally are in condensed phases or in
molecular forms. The results in Fig.\ \ref{fig27} show that our
calculated $I_0$ values are close to the experimental values for He
($Z=2$), Ne ($Z=10$), and Ar ($Z=18$), but are clearly too small for Kr
($Z=36$), Xe ($Z=54$) and Rn ($Z=86$). The difference $I-I_0$ for
high-$Z$ elements is partially due to the relativistic deviation from
the Bethe sum rule.

To verify the last assertion, we express the asymptotic formula
\req{9.57} into a form as similar as possible to that of the Bethe
formula most commonly used in the literature, Eq.\ \req{9.61}. We write
\beqa
\sigma^{(1)} &=& \frac{2\pi Z_0^2 e^4}{\me v^2} \, 2 Z \left\{
\ln \left(\frac{2 \me v^2}{I'_0} \right)
+ \ln \left( \frac{1}{1-\beta^2} \right)
- \beta^2 + \frac{1}{2} \, f(\gamma) \right.
\nonumber \\ [2mm]
&& \mbox{} \left. +
\frac{S_0-Z}{2Z} \left[
\ln \left( \frac{\beta^2}{1-\beta^2} \right)
- \beta^2 \right]
\right\},
\label{9.63}\eeqa
where we have grouped the energy-independent terms by introducing the
``modified'' mean excitation energy $I'_0$ defined by
\beq
\ln \left( \frac{2 \me c^2}{I'_0} \right) = \frac{S_0}{Z}
\, \ln \left( \frac{2 \me c^2}{I_0} \right) + \frac{D_0}{2Z}.
\label{9.64}\eeq
Evidently, Eq.\ \req{9.63} reduces to the Bethe formula when $S_0=Z$
(the optical oscillator strength satisfies the Thomas--Reiche--Kuhn sum
rule) {\it and} $D_0 = 0$ (the Bethe sum rule is valid for all $Q$).
Under these circumstances, we also have $I'_0 = I_0 = I$. The quantity
in the second line of Eq.\ \req{9.63} is generally small and can be
regarded as part of the shell correction (see Section \ref{sec10}).

Figure \ref{fig27} includes the values of $I'_0$ obtained from our DHFS
calculations, which are seen to be closer to the empirical mean
excitation energies. These results indicate that for the noble gases
heavier than Ar nearly half the difference $I-I_0$ is due to the
departure from the Bethe sum rule. We conclude that  the usual
definition of the mean excitation energy, Eq.\ \req{9.62}, should be
abandoned and replaced with Eq.\ \req{9.64}, which does account for
that departure. It is worth noticing that the term $D_0/2Z$ is much
smaller than $\ln(2\me c^2/I_0)$ and, for most practical purposes, it
may be neglected.


\subsubsection{Energy-straggling cross section \label{sec9.2.3}}

The energy-straggling cross section for collisions of high-energy
projectiles with atoms is given by [see Eq.\ \req{9.46c}]
\beqa
\sigma^{(2)} = \sum_a \sigma_{a}^{(2)} &=&
\frac{2\pi Z_0^2 e^4}{\me v^2}
\left\{ S_{1} \left[
\ln \left( \frac{\beta^2}{1-\beta^2} \right) - \beta^2 \right]
\right.
\nonumber \\ [2mm]
&& \mbox{} \left.
+ 2 \, S_{1} \ln \left( \frac{2 \me c^2}{I_1} \right)
+ D_{1} + Z \, \me c^2 \, g(\gamma) \right\},
\label{9.65}\eeqa
with
\beq
S_{1} = \sum_a S_{1}(a) =
\int_0^\infty W \, \frac{\d f(W)}{\d W} \, \d W,
\label{9.66}\eeq
\beq
\ln I_{1} = \frac{1}{S_{1}} \sum_a S_{1}(a)\, \ln [I_{1}(a)]
=  \frac{1}{S_{1}} \int_0^\infty W \, \ln W \,
\frac{\d f (W)}{\d W} \, \d W,
\label{9.67}\eeq
and
\beq
D_{1} = \sum_a D_{1}(a).
\label{9.68}\eeq
Grouping the energy-independent terms, we can write
\beqa
\sigma^{(2)} =
\frac{2\pi Z_0^2 e^4}{\me v^2}
\left\{ S_{1} \left[
\ln \left( \frac{\beta^2}{1-\beta^2} \right) - \beta^2 \right]
+ A_{\rm tot} + Z \, \me c^2 \, g(\gamma) \right\},
\label{9.69}\eeqa
where
\beq
A_{\rm tot} = 2 \, S_{1} \ln \left( \frac{2 \me c^2}{I_{1}} \right)
+ D_{1}.
\label{9.70}\eeq

\begin{figure}[htb]
\begin{center}
\includegraphics*[width=7.5cm]{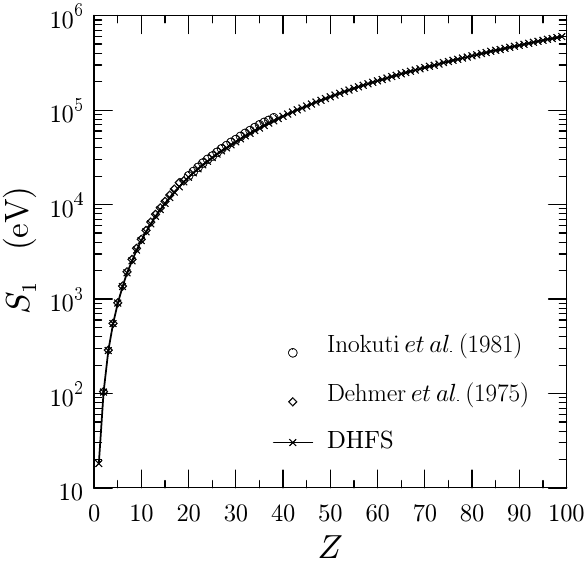}
\includegraphics*[width=7.5cm]{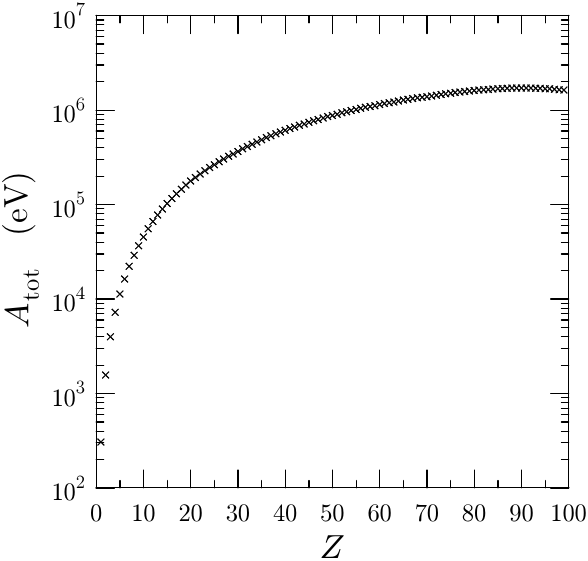}
\caption{Parameters $S_{1}$ and $A_{\rm tot}$ of the asymptotic
formula \req{9.69} of the energy-straggling cross section for inelastic
collisions with free neutral atoms, calculated from the present
numerical GOS (crosses).  The symbols are values obtained by
\citet{Dehmer1975} and \citet{Inokuti1981} from non-relativistic
calculations using the Hartree--Slater potential.
\label{fig28}}
\end{center}\end{figure}

The left panel of Fig.\ \ref{fig28} shows the values of the parameter
$S_{1}$ for free neutral atoms calculated from our numerical OOSs,
together with values obtained by \citet{Dehmer1975} and
\citet{Inokuti1981} from non-relativistic calculations with the
Hartree--Slater potential for the elements with $Z=1$ to 38. The
difference between these two data sets increases gradually with the
atomic number, probably due to the increasing importance of relativistic
effects. The right plot in Fig.\ \ref{fig28} displays the calculated
values of $A_{\rm tot}$. Because these two parameters depend strongly on
the details of the excitation spectrum, our central field approximation
is not expected to give accurate values, except possibly for noble gas
atoms. Nevertheless, the asymptotic formula \req{9.69} is valid for
molecules and solids, although the parameters are sensitive to
aggregation effects and should be determined empirically for each
material.


\subsection{Numerical results for noble gases\label{sec9.3}}

\begin{figure}[h!]
\begin{center}
\includegraphics*[width=11.5cm]{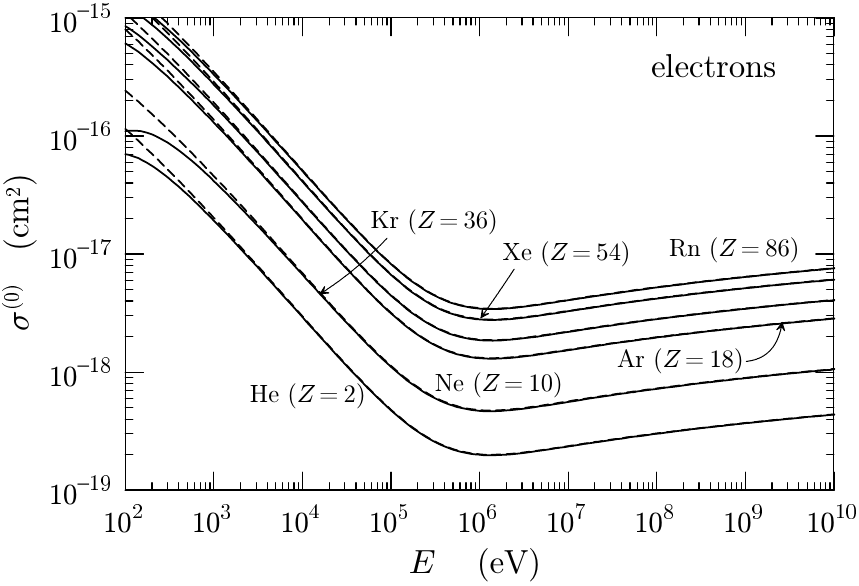}
\caption{Total cross sections, $\sigma^{(0)}$, for inelastic collisions
of electrons with noble-gas atoms, as functions of the
kinetic energy of the projectile. The solid curves represent numerical
results calculated with the program {\sc pwacs}. The dashed curves are
the predictions of the asymptotic formula \req{9.54}.
\label{fig29e}}
\end{center}\end{figure}

\begin{figure}[h!]
\begin{center}
\includegraphics*[width=11.5cm]{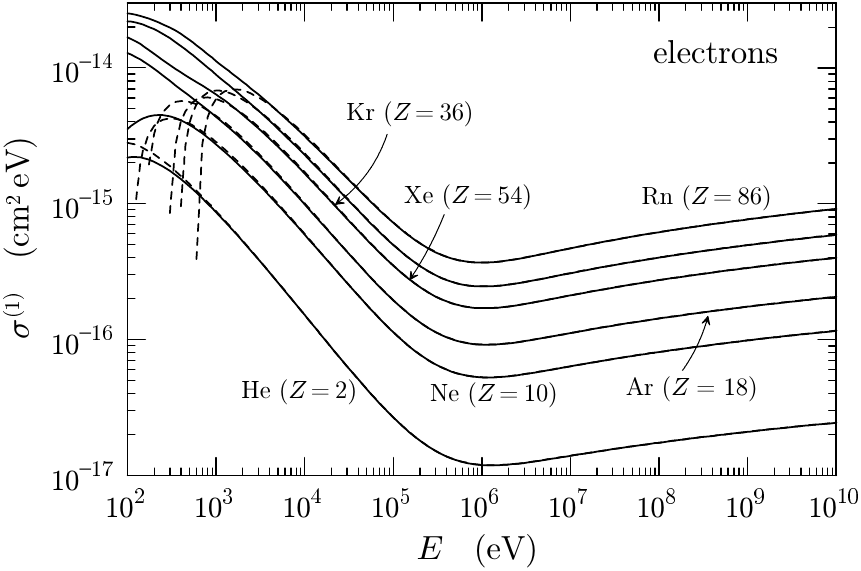}
\caption{Stopping cross sections, $\sigma^{(1)}$, for inelastic
collisions of electrons with noble-gas atoms, as functions
of the kinetic energy of the projectile. The solid curves represent
numerical results calculated with the program {\sc pwacs}. The dashed
curves are the predictions of the asymptotic formula \req{9.63}.
\label{fig30e}}
\end{center}\end{figure}

\begin{figure}[h!]
\begin{center}
\includegraphics*[width=11.5cm]{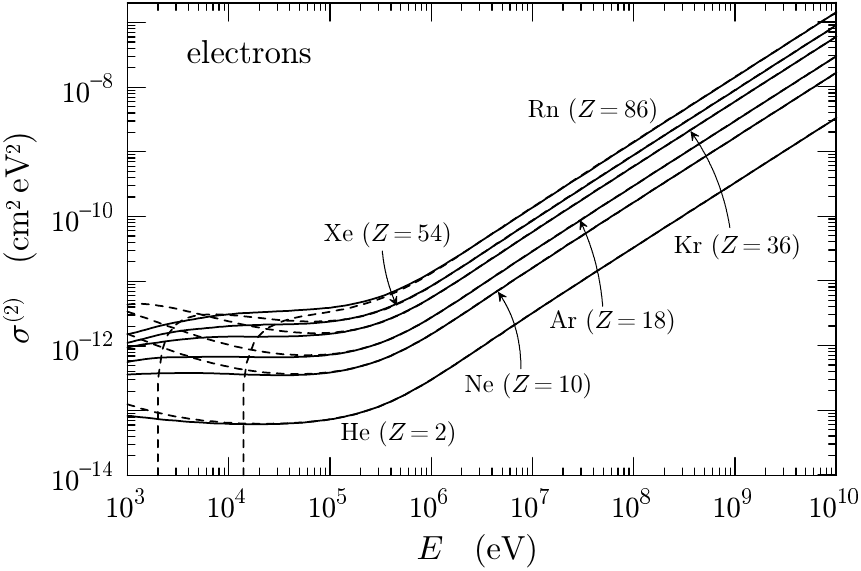}
\caption{Energy-straggling cross sections, $\sigma^{(2)}$, for inelastic
collisions of electrons with noble-gas atoms, as functions
of the kinetic energy of the projectile. The solid curves represent
numerical results calculated with the program {\sc pwacs}. The dashed
curves show the predictions of the asymptotic formula \req{9.69}.
\label{fig31e}}
\end{center}\end{figure}

The accuracy of the asymptotic formulas derived above will be assessed
by direct comparison with integrated cross sections $\sigma^{(i)}$
obtained from the numerical integration of the DDCS times $W^i$. For the
sake of concreteness, we will consider collisions with noble-gas atoms,
for which the central-field approximation is expected to be more
accurate than for atoms with open-subshell configurations. Figures
\ref{fig29e}, \ref{fig30e} and \ref{fig31e} display
 display total cross sections,
stopping cross sections and energy-straggling cross sections for
inelastic collisions of electrons as functions of the kinetic energy of
the projectile. The asymptotic formulas for $\sigma^{(i)}$ are seen to
approximate the numerical values of the total cross section, the
stopping cross section, and the energy-straggling cross sections for
kinetic energies of the projectile larger than about 1 keV, 10 keV, and
100 keV, respectively.

\begin{figure}[h!]
\begin{center}
\includegraphics*[width=11.5cm]{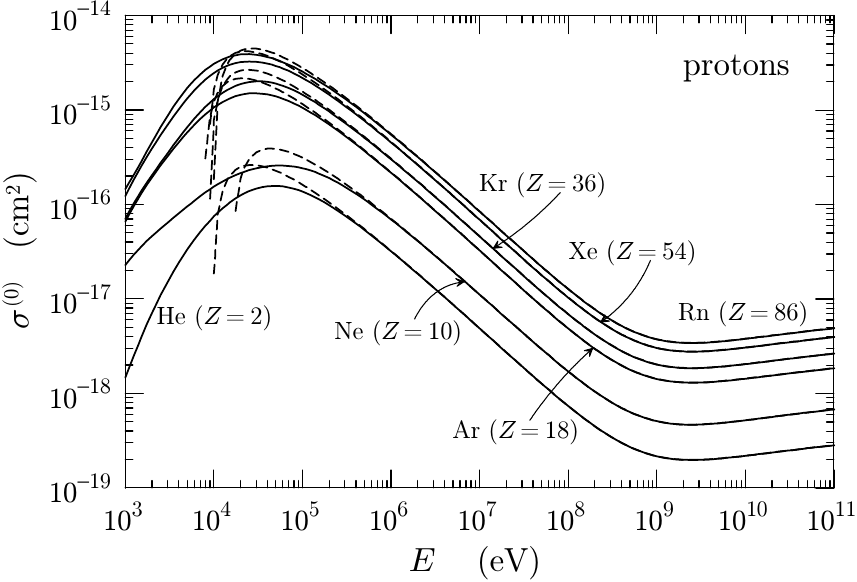}
\caption{Total cross sections, $\sigma^{(0)}$, for inelastic collisions
of protons with noble-gas atoms, as functions of the
kinetic energy of the projectile. The solid curves represent numerical
results calculated with the program {\sc pwacs}. The dashed curves
show the predictions of the asymptotic formula \req{9.54}.
\label{fig29}}
\end{center}\end{figure}

\begin{figure}[h!]
\begin{center}
\includegraphics*[width=11.5cm]{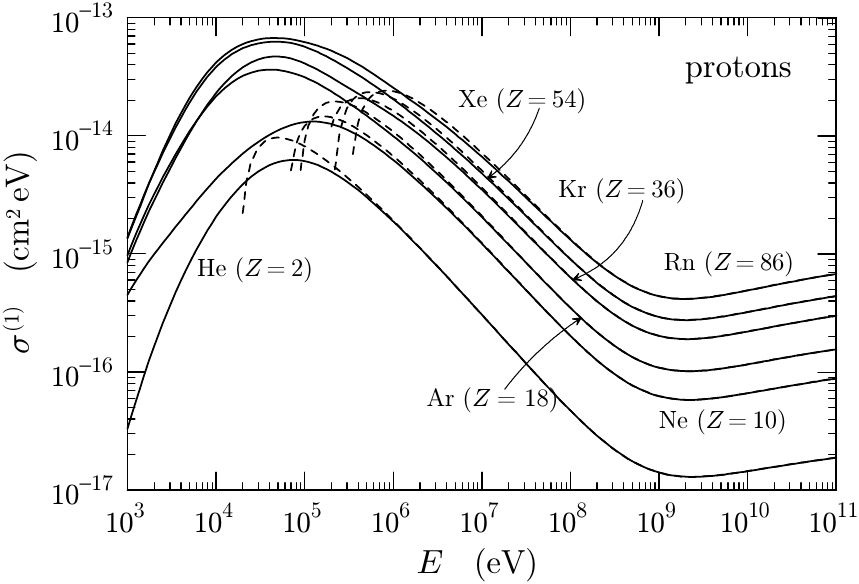}
\caption{Stopping cross sections, $\sigma^{(1)}$, for inelastic
collisions of protons with noble-gas atoms, as functions
of the kinetic energy of the projectile. The solid curves represent
numerical results calculated with the program {\sc pwacs}. The dashed
curves show the predictions of the asymptotic formula \req{9.63}.
\label{fig30}}
\end{center}\end{figure}

\begin{figure}[h!]
\begin{center}
\includegraphics*[width=11.5cm]{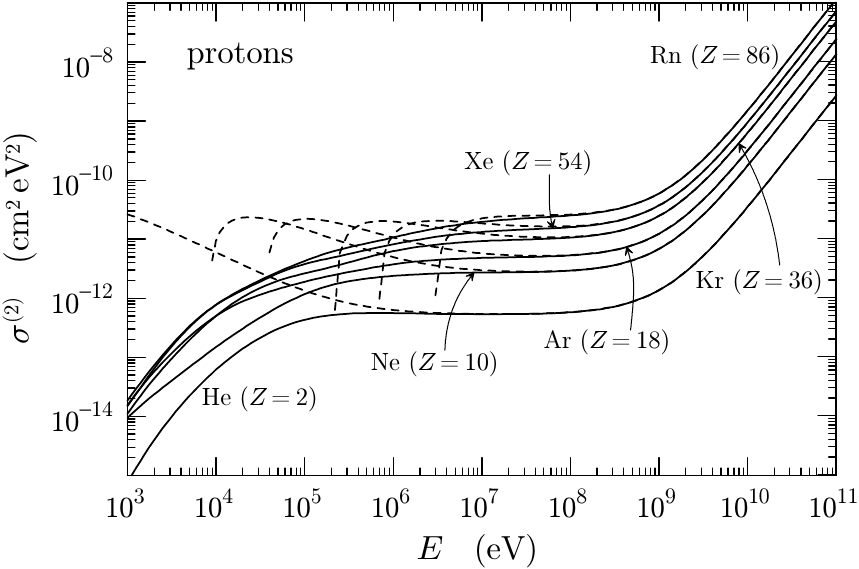}
\caption{Energy-straggling cross sections, $\sigma^{(2)}$, for inelastic
collisions of protons with noble-gas atoms, as functions
of the kinetic energy of the projectile. The solid curves represent
numerical results calculated with the program {\sc pwacs}. The dashed
curves show the predictions of the asymptotic formula \req{9.69}.
\label{fig31}}
\end{center}\end{figure}

Figures \ref{fig29}, \ref{fig30} and \ref{fig31} display total cross sections,
stopping cross sections and energy-straggling cross sections for
inelastic collisions of protons as functions of the kinetic energy of
the projectile. We see that the differences between the numerical cross
sections and the asymptotic formulas decrease smoothly when the energy
of the projectile increases beyond a certain value, as expected for
asymptotic formulas. For the total cross section, the differences become
imperceptible, on the scale of the plots, for energies higher than about
1 MeV.  Similarly, the curves of the asymptotic formulas and the
numerical values of the stopping cross sections are seen to merge at
energies of the order of 10 MeV.  The asymptotic formula for the
energy-straggling cross section agrees with the calculated values for
energies larger than about 100 MeV.

It is worth observing that the larger the order $i$ of the integrated
cross section $\sigma^{(i)}$ the higher the energy where the
corresponding formula starts yielding reasonably accurate values,
because of the increasing importance of large-$W$ interactions, part of
which correspond to excitations of electrons in inner subshells. Therefore,
the relative shell corrections (see Section \ref{sec10}) are larger for
the stopping cross section than for the total cross section, and even
larger for the energy-straggling cross section.

The parameters of the asymptotic formulas were obtained by direct
integration of the GOS by using numerical methods that are independent
of those employed in the calculation of the integrated cross sections.
The good agreement, at sufficiently high energies, between the numerical
integrated cross sections and the asymptotic formulas indicates that the
numerical algorithms used to interpolate and integrate the DDCSs remain
accurate for energies up to $10^{11}$ eV.  Nonetheless, for high-energy
projectiles, the cross sections calculated numerically by integrating
the DDCS are found to differ by up to 1~\% from the predictions of the
asymptotic formulas. These differences arise from numerical errors
accumulated during the lengthy calculations and, in the case of inner
subshells with large ionization energies, from the approximations
invoked in the derivation of the formulas. For subshells with small and
moderate ionization energies, say up to about 5 keV, and for high-energy
projectiles, the difference between the subshell integrated cross
sections obtained numerically and from the asymptotic formulas is
typically less than about 0.5 \%.


\section{Shell corrections \label{sec10}}
\setcounter{equation}{0}

The asymptotic formulas derived in the last Section result from some
drastic approximations, whose accuracy deteriorates as the energy of the
projectile decreases. The difference between the ``exact'' cross section
$\sigma^{(i)}$ (\ie, the one obtained by integrating the DDCS) and the
corresponding asymptotic formula is known as the {\it shell correction}.
This name is motivated by the fact that the largest errors in the
asymptotic formulas come from the contributions of the innermost
subshells (\ie, those with the largest ionization energies). Since the
wave functions of inner-shell electrons are quite insensitive to the
effect of aggregation, shell corrections calculated for collisions with
free atoms are expected to be approximately valid also for collisions in
condensed media.

Previous theoretical calculations of shell corrections to the asymptotic
formula for the stopping power were essentially non-relativistic.
Explicit formulas for the shell corrections in terms of the GOSs were
obtained by subtracting from the integrals that define $\sigma^{(1)}$
those effectively used to calculate its asymptotic formula \citep[see,
\eg,][]{Fano1963}. Thus, \citet{Walske1952, Walske1956} used hydrogenic
wave functions to obtain shell corrections for K shell and L subshells;
\citet{KhandelwalMerzbacher1966} and \citet{Bichsel1983} performed
similar calculations for M subshells. \citet{Bonderup1967} obtained an
atomic subshell correction from stopping powers calculated using the
local-plasma approximation of \citet{LindhardScharff1953}. More recently,
\citet{Bichsel2002} determined the corrections for the inner subshells of
aluminum and silicon by direct integration of the non-relativistic GOSs
of \citet{Manson1972}.

At this point, it is relevant to mention that the derivation of the
relativistic asymptotic formulas is less ``clean'' than that of the
non-relativistic ones. As the non-relativistic DDCS has a simpler form
(transverse interactions do not occur in the non-relativistic theory)
and the corresponding GOS satisfies the Bethe sum rule, the integrals
can be performed neatly. In the relativistic formulation, the DDCS is
more involved, it contains a transverse contribution, and the GOSs do
not satisfy the Bethe sum rule. To cope with these complications, we had
to rely on more severe assumptions, which make the direct calculation of
shell corrections from defining integrals difficult.

With our
computational tools, it is much easier to obtain the shell corrections
${\cal C}^{(i)}$ to the asymptotic formulas of the cross sections
$\sigma^{(i)}$ (with $i=0$, 1 and 2) as the differences between the
numerical integrated cross sections $\sigma_{\rm num}^{(i)}$
and the results from the asymptotic formulas.
For this purpose, we define the shell corrections ${\cal
C}^{(i)}$ by the following equations \citep{Salvat2022a}
\begin{subequations}
\label{10.1}
\beq
\sigma^{(0)} = \frac{2\pi Z_0^2 e^4}{\me v^2} \left\{
M_{\rm tot}^2
\left[ \ln \left( \frac{\beta^2}{1-\beta^2} \right)- \beta^2  \right]
+ C_{\rm tot} - {\cal C}^{(0)} \right\} ,
\label{10.1a}\eeq
\beqa
\sigma^{(1)} &=& \frac{2\pi Z_0^2 e^4}{\me v^2} \left\{
2Z \left[
\ln \left(\frac{2 \me c^2}{I'_0} \right)
+ \ln \left( \frac{\beta^2}{1-\beta^2} \right)
- \beta^2 + \frac{1}{2} \, f(\gamma) \right] \right.
\nonumber \\ [2mm]
&& \mbox{} \left.
+ \left(S_0-Z \right)
\left[ \ln \left( \frac{\beta^2}{1-\beta^2} \right) - \beta^2 \right]
- {\cal C}^{(1)} \right\}, \rule{5mm}{0mm}
\label{10.1b}\eeqa
and
\beqa
\sigma^{(2)} =
\frac{2\pi Z_0^2 e^4}{\me v^2}
\left\{ S_{+1} \left[
\ln \left( \frac{\beta^2}{1-\beta^2} \right) - \beta^2 \right]
+ A_{\rm tot} + Z \, \me c^2 \, g(\gamma) - {\cal C}^{(2)}\right\},
\label{10.1c}\eeqa
\end{subequations}
where the cross sections on the left-hand sides represent ``exact''
values obtained by direct numerical integration of the DDCS. The
expressions on the right-hand sides of Eqs. \req{10.1} with ${\cal
C}^{(i)}=0$ are the asymptotic formulas for the atomic integrated cross
sections $\sigma^{(i)}$. Evidently, each shell correction
is proportional to the ``error'' of the associated asymptotic formula,
\beq
{\cal C}^{(i)} = \left(\frac{2\pi Z_0^2 e^4}{\me v^2}\right)^{-1}
\left[ \sigma^{(i)}_{\rm asympt} - \sigma_{\rm num}^{(i)} \right].
\label{10.2}\eeq
Upon insertion of the calculated numerical cross section, this equation
determines the correction ${\cal C}^{(i)}$.

A peculiarity of the present calculation scheme is that it provides the
partial contributions of individual shells or subshells to the shell
correction of the entire atom. Hydrogenic models also allow obtaining
individual subshell contributions, but require setting a certain convention
on how to deal with the Pauli exclusion principle, which prevents
transitions of the active electron to orbitals that are occupied. One
could either disregard the exclusion principle and include transitions
to all hydrogenic energy levels or consider only transitions to empty
hydrogenic levels. Neither of the two options is realistic because
the actual orbitals and energies are only roughly approximated by the
hydrogenic model. Consequently, in the following we shall deal only with
the shell corrections to the atomic cross sections calculated with the
DHFS potential. In addition we shall limit our considerations to the
case of projectile protons, as representative of heavy particles. The
shell corrections for electrons and positrons are generally smaller than for
protons because the lowest allowed recoil energy $Q_-(W)$ is much closer
to the photon line than for heavier projectiles (see Fig.\ \ref{fig38}
in Appendix A). Indeed, comparison of the results shown in Figs.\
\ref{fig30e} and \ref{fig30} reveals that, for intermediate energies
where the asymptotic formulas approach the numerical results, the shell
correction to the stopping cross section for electrons is manifestly
smaller than for protons.

The results displayed in Fig. \ref{fig29} show that the shell correction
${\cal C}^{(0)}$ for protons with energies higher than about 1 MeV is
small. At $E= 10$ MeV, the asymptotic formula \req{10.1a} approximates
the ``exact'' total cross to an accuracy better than about 0.1 percent.
The shell correction ${\cal C}^{(2)}$ to the asymptotic formula of the
energy-straggling cross section may be neglected for proton energies
higher than about $5 Z$ MeV (see Fig.\ \ref{fig31}). At lower energies,
the asymptotic formula departs rapidly from the exact cross section and
it is preferable to use a tabulation of the numerical cross section
$\sigma^{(2)}$ instead of the asymptotic formula.

The correction ${\cal C}^{(1)}$ to the stopping power formula is
important because of the relevance of the stopping power in practical
calculations of dosimetry and charged-particle transport. The asymptotic
formula for protons (with ${\cal C}^{(1)}=0$) is found to differ from
the numerical stopping cross section by less than about 1 \% for
energies higher than about 100 MeV. At intermediate energies, say
between 1 MeV and 100 MeV, the asymptotic formula yields values larger
than the numerical ones. In the case of uranium ($Z=92$), a maximum
difference of about 7.5 \% is found at $E \sim$ 5 MeV; for gold ($Z=79$)
the largest difference is 8.3 \% at $E \sim$ 4.5 MeV. Because these
differences are only one order of magnitude larger than the estimated
numerical uncertainty of the calculated stopping cross sections, the
resulting shell correction is very sensitive to accumulated numerical
errors.

To conform with the literature, let us write the formula \req{10.1b},
as \citep[cf.\/][]{Fano1963, Ahlen1980},
\beqa
\sigma^{(1)} &=& \frac{2\pi Z_0^2 e^4}{\me v^2} \, 2 Z \left\{
\ln \left(\frac{2 \me c^2}{I'_0} \right)
+ \ln \left( \frac{\beta^2}{1-\beta^2} \right)
- \beta^2 + \frac{1}{2} \, f(\gamma) \right.
\nonumber \\ [2mm]
&& \mbox{} \left.
- \frac{Z-S_0}{2Z}
\left[ \ln \left( \frac{\beta^2}{1-\beta^2} \right) - \beta^2 \right]
- \frac{C}{Z} \right\},
\label{10.3}\eeqa
where $C/Z \equiv {\cal C}^{(1)}/2$ is the quantity usually referred to
as {\it the shell correction}. The first term in the second line of this
equation accounts for the departure from the Thomas--Reiche--Kuhn sum
rule; it does not occur in the non-relativistic theory.

\begin{figure}[ht!]
\begin{center}
\includegraphics*[width=7.5cm]{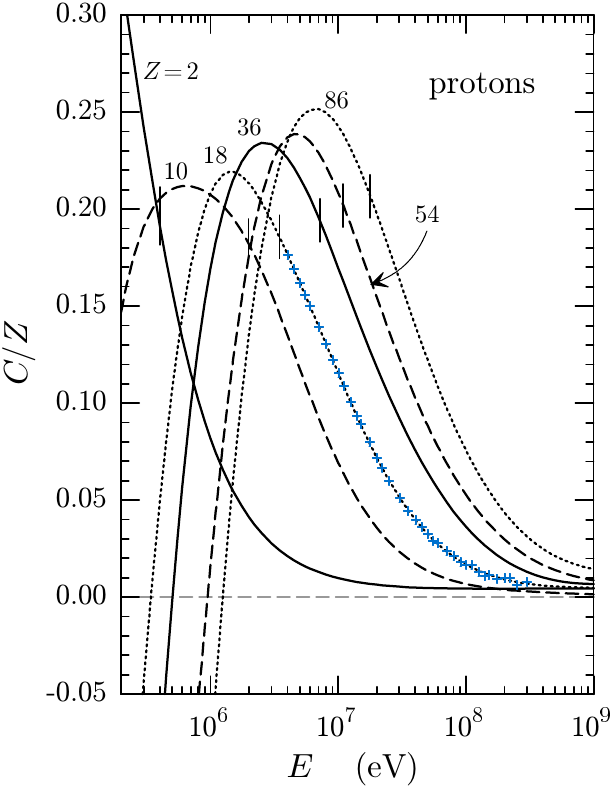}
\caption{Shell corrections $C/Z$ to the asymptotic formula of
the stopping cross section for inelastic collisions of protons
with noble-gas atoms, as functions of the kinetic energy of the projectile.
The vertical segments indicate the energies $E_{\rm cut}=0.2 Z$ MeV
above which the correction is described by the fitted formula \req{10.4}.
Crosses represent values obtained from Eq.\ \req{10.2}, with
visible fluctuations arising from numerical uncertainties of the
calculated $\sigma_{\rm num}^{(1)}$ at energies higher than $\sim 10^8$
eV.
\label{fig32}}
\end{center}\end{figure}

The program {\sc pwacs} has been run to compute tables of the integrated
cross sections $\sigma^{(0)}$, $\sigma^{(1)}$, and $\sigma^{(2)}$ for
collisions of protons with atoms of all the elements from $Z=1$ to 99.
The calculations were made for a grid of kinetic energies that extended
from 1 keV to 100 GeV, with nearly logarithmic spacing and 20 points per
decade. The resulting tables of integrated cross sections were used for
determining the shell correction $C/Z$ to the stopping cross section.
For low energies, up to $E_{\rm cut} = 0.2 \, Z$ MeV the correction was
obtained by means of Eq.\ \req{10.2}, that is, from the difference
between the stopping cross section calculated numerically and the result
of the asymptotic formula with ${\cal C}^{(1)}=0$.  Figure
\ref{fig32} displays the shell correction to the stopping cross section
for collisions of protons with noble-gas atoms as functions of the
kinetic energy of the projectile.

As mentioned above, the difference on the right-hand side of Eq.\
\req{10.2} magnifies the numerical errors accumulated throughout the
calculation of the integrated cross section $\sigma^{(1)}_{\rm num}$.
This is illustrated in Fig.\ \ref{fig32} for the case of argon and
kinetic energies higher than $E_{\rm cut}$, where the blue crosses
represent the calculated numerical values. Although the magnitude of the
errors in the calculated stopping cross sections is estimated to be less
than about 0.5~\%, these errors blur the continuous curves of the shell
correction as a function of the proton energy. To obtain a well-defined
shell correction for energies higher than $E_{\rm cut}$, we approximate
it in the form \citep{Salvat2022a}
\beq
\frac{C}{Z} = \sum_{n=1}^6 c_n \left( \frac{m_{\rm p} c^2}{E}
\right)^{n/4}
\label{10.4}\eeq
where $m_{\rm p}$ is the proton mass. By similarity with the
non-relativistic theory, here we assume that the shell correction tends
to zero at high energies.  The parameters $c_n$ ($n=1$ to 6) have been
determined through a least-squares fit to the calculated stopping cross
section $\sigma^{(1)}_{\rm num}$ for energies in the interval from
$E_{\rm cut}$ up to 1 GeV. In that energy interval, the analytical
expression \req{10.4} with the fitted parameters approaches the
calculated stopping cross sections with an accuracy generally better
than 0.05 \% for all the elements. The fitted formula effectively
averages the numerical errors and gives estimates of the shell
correction that are probably better than the values obtained numerically
from Eq.\ \req{10.2}. Assuming that $C/Z$ remains constant for energies
higher than 1 GeV, the relative difference between the Bethe formula and
the calculated $\sigma^{(1)}_{\rm num}$ values remains less than 0.3~\%
for energies up to 10 GeV. The fitting procedure provides evidence that
numerical errors accumulated throughout the calculation of integrated
cross sections are less than about 0.5~\% for energies up to 10 GeV.

\begin{figure}[h!]
\begin{center}
\includegraphics*[width=11.0cm]{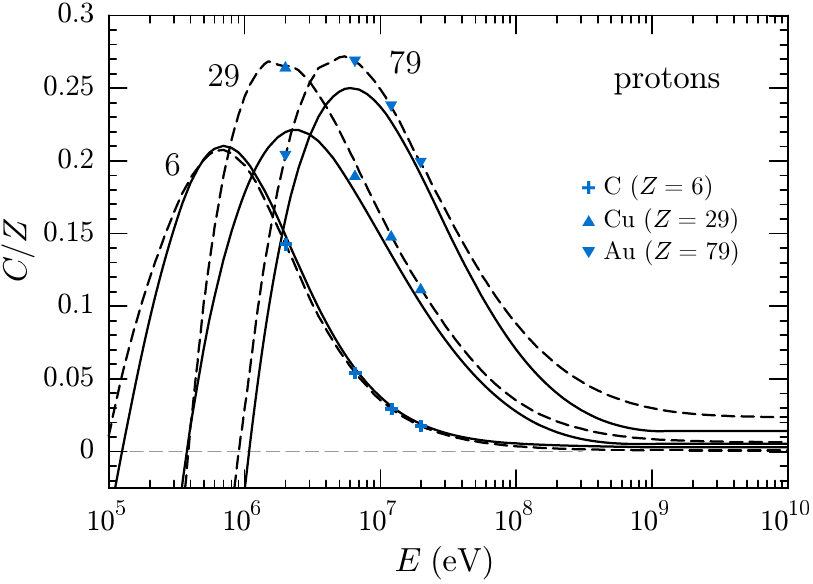} \\ [5mm]
\includegraphics*[width=11.0cm]{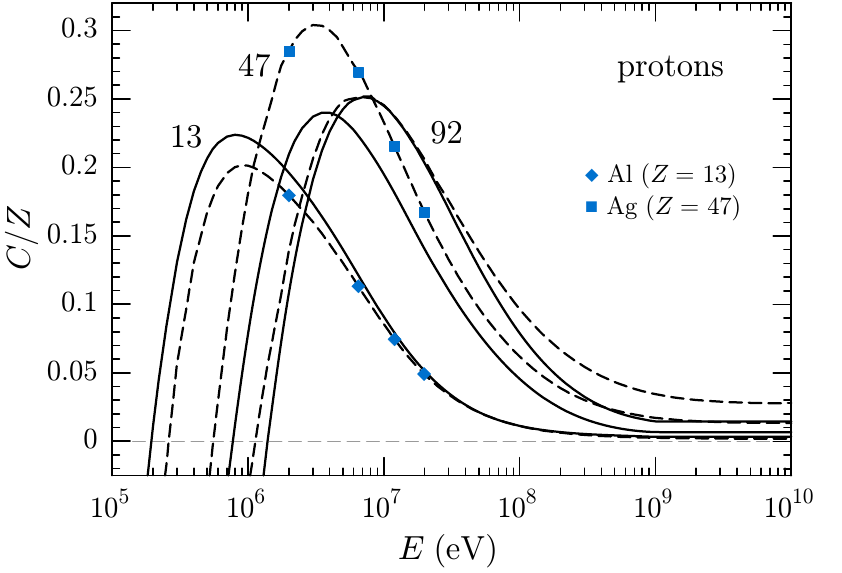}
\caption{
Shell corrections $C/Z$ to the asymptotic formula of the
stopping cross section for inelastic collisions of protons with atoms of
the elements with the indicated atomic numbers, as functions of the
kinetic energy of the projectile. Solid curves are the present
results, symbols are Bichsel's semi-empirical shell corrections given in
the \citet{ICRU37}, and the dashed curves were generated with
the program {\sc best} of \citet{BergerBichsel1994}.
\label{fig33}}
\end{center}\end{figure}

Figure \ref{fig33} displays the shell correction to the
stopping cross sections for collisions of protons with atoms of
the elements C, Al, Cu, Ag, Au, and U, as
functions of the kinetic energy of the projectile. Also shown are the
values of Bichsel's (Model 1) semi-empirical shell correction given in
the \citet{ICRU37} and calculated with the program {\sc best}
\citep{BergerBichsel1994}. It is worth noticing that Bichsel
\citep{ICRU49} derived the shell
correction {\it and} the mean excitation energy $I$ from a
multi-parametric fit of the formula [cf.\ Eq.\ \req{9.61}]
\beq
\sigma^{(1)}_{\rm Bethe}
= \frac{2\pi Z_0^2 e^4}{\me v^2} \, 2Z \left[
\ln \left( \frac{2 \me v^2}{I} \right)
+ \ln \left( \frac{1}{1-\beta^2} \right) - \beta^2
+ \frac{1}{2} \, f(\gamma) - \frac{C}{Z} \right]
\label{10.5}\eeq
to available experimental stopping-power data for the elements C, Al,
Cu, Ag, and Au. At intermediate energies near the maximum of the $C/Z$
curves, our calculated corrections agree reasonably with Bichsel's
estimates for C and U, are sensibly larger for Al, and smaller for Cu,
Ag, and Au. The differences between our shell corrections and Bichsel's
estimates are much larger than the numerical inaccuracies of our
calculated data. At least partially, these differences are caused by the
neglect of the relativistic departure from the Bethe sum rule, which is
implicit in Eq.\ \req{10.5}.

Note that here we have considered shell corrections that arise {\it
only} from inaccuracies in approximating the integrated cross sections
obtained from the PWBA. Empirical shell corrections resulting from
comparisons of the asymptotic formulas with experimental data, or with
results from more elaborate calculations (\eg, using the distorted-wave
Born approximation), would account also for the simplifications implied
by the PWBA. Effects beyond the PWBA for atoms are conventionally
introduced by adding extra terms in the asymptotic formulas, the most
relevant of those are the Barkas and Bloch corrections \citep[see,
\eg,][]{Ahlen1980, LindhardSorensen1996} and the density-effect
correction \citep{Fano1963, InokutiSmith1982}.


\section{Ionizing collisions beyond the PWBA \label{sec11}}
\setcounter{equation}{0}

The main limitation of the Born approximation is due to the neglect of
the distortion of the projectile wave functions caused by the field of
the target atom. For electrons and positrons, this distortion can be
largely accounted for by using the distorted-wave Born approximation
(DWBA), in which the projectile states are represented as distorted
plane waves. Employing the expansion \req{B.22} of the distorted waves
in terms of spherical waves, the energy-loss DCS can be calculated as a
truncated series of products of vector coupling coefficients and
Slater's radial integrals. The DWBA calculation scheme is described,
\eg, by \citet{Segui2003}. Because of the slow convergence of the
partial-wave series, this kind of calculation is only possible for
projectile electrons and positrons with kinetic energies up to about $20
E_a$. \citet{BoteSalvat2008} used an optimized computation strategy,
which combines the DWBA and the PWBA, to generate a database of
ionization cross sections for the K shell and the L and M subshells of
all the elements from hydrogen to einsteinium ($Z=1$ to 99) and for
energies of the projectile from 50 eV up to 1 GeV.  The results were
found to agree well with available experimental data \citep{Llovet2014}.

Unfortunately, the calculation of cross sections for inelastic
collisions from the DWBA is not feasible for charged particles heavier
than the electron, because the smallness of the de Broglie wavelength of
the projectile renders the calculation of free spherical waves extremely
difficult. \citet{Chen1983, ChenCrasemann1985, ChenCrasemann1989} went
beyond the PWBA by using the perturbed-stationary-state approximation of
\citet{BrandtLapicki1979}, which accounts for (1) alterations in the
binding of the active electron due to the presence of the projectile
near the nucleus of the target atom, and (2) the deflection of the
projectile path caused by the Coulomb field of the nucleus. In our PWBA
calculations of ionization cross sections for heavy projectiles, these
effects are introduced by means of semi-classical correcting factors,
which are described in the following two Subsections.


\subsection{Binding effects \label{sec11.1}}

In collisions where the projectile penetrates deep into the target atom,
the presence of the projectile modifies the binding energy of the active
electron and, in the case of positively charged projectiles, leads to a
reduction of the DCS. For the K shell and L subshells, \citet{BrandtLapicki1979}
performed a first-order perturbation analysis, assuming that the
projectile follows a straight trajectory and using hydrogenic wave
functions. They obtained the following average ionization-energy shift of
the active target electron,
\beq
\Delta E_a = \frac{2Z_0 E_a}{Z_a \Theta_a}
\left[g_a(\xi) - h_a(\xi) \right] \, ,
\label{11.1}\eeq
where $Z_a=Z-\delta_a$ is the effective nuclear charge felt by the
electrons in the unperturbed orbitals, with $\delta_{\rm K}=0.3$ and
$\delta_{{\rm L}i}=4.15$. The quantity $\Theta_a$ is the reduced ionization
energy,
\beq
\Theta_a = 2 n_a^2 E_a / (Z_a^2 E_{\rm h}),
\label{11.2}\eeq
where $E_{\rm h} = \me e^4/\hbar^2 = 27.211$ eV is the Hartree energy. The last factor in
Eq.\ \req{11.1} is a function of the dimensionless parameter
\beq
\xi \equiv \frac{Z_a E_{\rm h}}{n_a  E_a} \sqrt{\frac{\me}{M} \,
\frac{2E}{E_{\rm h}}}\, .
\label{11.3}\eeq
The $g_a(\xi)$ function is given by
\begin{subequations}
\label{11.4}
\beq
g_{\rm K}(\xi) = (1 + 9\xi + 31 \xi^2 +98 \xi^3 +12 \xi^4 + 25 \xi^5
+ 4.2 \xi^6 +0.515\xi^7)/(1+\xi)^9\, ,
\label{11.4a}\eeq
\beq
g_{\rm L1}(\xi)= (1 + 9\xi + 31 \xi^2 +49 \xi^3 +162 \xi^4 + 63 \xi^5
+ 18 \xi^6 +1.97\xi^7)/(1+\xi)^9\, ,
\label{11.4b}\eeq
\beq
g_{\rm L2,3}(\xi) = (1 + 10\xi + 45 \xi^2 +102 \xi^3 +331 \xi^4 + 6.7 \xi^5
+ 58 \xi^6 +7.8\xi^7+0.888\xi^8)/(1+\xi)^{10}\, ,
\label{11.4c}\eeq
\end{subequations}
and
\beq
h_a(\xi)=\frac{2n_a}{\Theta_a \xi^3} I \left(\frac{C_a n_a}{\xi} \right) \, ,
\label{11.5}\eeq
where $C_{\rm K} = C_{\rm L1} = 1.5$ and $C_{\rm L2,3}= 1.25$. The
function $I(x)$ is \citep{Basbas1978, ChenCrasemann1985}
\beq
I(x) =  \left\{
\begin{array}{ll}
(3\pi/4) [\ln(x^{-2}) - 1] & \mbox{for $0 < x \le 0.035$,} \\ [2mm]
\exp(-2x) (0.031 + 0.210 x^{1/2} + 0.005 x & \\ [2mm]
\rule{15mm}{0mm} -0.069 x^{3/2} + 0.324
x^2)^{-1} & \mbox{for $0.035 < x \le 3.1$,} \\ [2mm]
2 \exp(-2x) x^{-1.6} & \mbox{for $3.1 < x \le 11$,} \\ [2mm]
0 & \mbox{for $x>11$.}
\end{array} \right.
\label{11.6}\eeq

For M and outer subshells, \citet{Chen1983} considered that the effective
ionization energy is the one of the ``united'' atom (\ie, of the atom with
atomic number $Z_0+Z$); note that this assumption tends to overestimate
the effect. We shall adopt a similar approach, which avoids the need of
considering ionization energies of other atomic species. Expressing the
ionization energies of the unperturbed states as (screened hydrogenic
levels)
\begin{subequations}
\label{11.7}
\beq
E_a =  \frac{(Z-\delta_a)^2}{2n_a^2} \, E_{\rm h}\, ,
\label{11.7a}\eeq
and noting that the screening constant $\delta_a$ is nearly the same for
neighbouring elements, we can approximate the effective ionization energy
in the form
\beq
E_a' =  \frac{(Z_0+Z-\delta_a)^2}{2n_a^2} \, E_{\rm h}\, .
\label{11.7b}\eeq
\end{subequations}
This gives the following ionization-energy shift
\beq
\Delta E_a = E_a'-E_a = \frac{Z_0^2+2 Z_0 (Z-\delta_a)}{2n_a^2} \, E_{\rm
h}\, .
\label{11.8}\eeq

To take account of binding effects, we simply shift the ionization
threshold in $\Delta E_a$, \ie, the minimum energy loss needed to cause
ionization is set equal to $E_a +\Delta E_a$. We thus have
[cf.\ Eq.\ \req{8.9}]
\beq
[\sigma_a^{\rm ion}]^{(n)} =
\int_{E_a+\Delta E_a}^{W_{\rm max}}
W^n \, \frac{\d \sigma_{a}^{\rm cont}}{\d
W} \, \d W.
\label{11.9}\eeq


\subsection{Coulomb deflection \label{sec11.2}}

For projectiles with small speeds, the PWBA, and the
equivalent straight-trajectory semi-classical approximation \citep[see,
\eg,][, and references therein]{Amundsen1977}, can overestimate the ionization
cross sections by orders of magnitude because they neglect the effect of
the Coulomb field of the nucleus on the trajectory of the projectile.
In the semi-classical treatment, the energy-loss DCS for a projectile
following a classical hyperbolic orbit in the Coulomb potential of the
bare target nucleus can be obtained by multiplying the energy-loss DCS,
calculated by assuming that the projectile follows a straight
trajectory, by a correction factor. This Coulomb-deflection factor can
be approximated as \citep{BrandtLapicki1979}
\beq
F^{\rm Coul}_a (E;W) = \left( 1 - \frac{1}{3} x^{1/3} + \frac{5}{3}
x^{2/3} \right) \exp(-2\pi x) \qquad \mbox{with} \qquad
x \equiv d_0\, q_{0}\, ,
\label{11.10}\eeq
where
\beq
\hbar q_{0} = \frac{W}{v} = W \sqrt{\frac{M}{2E}}
\label{11.11}\eeq
is the minimum momentum transfer from the projectile to the active
electron, calculated from the approximate expression \req{A.37}, and
\beq
d_0 = \frac{Z_0 Z e^2}{M_{\rm red} v^2}
= \frac{M Z_0 Z e^2}{ 2 M_{\rm red} E}
\label{11.12}\eeq
is the half-distance of closest approach in a head-on collision of
the projectile with the bare (unscreened) nucleus.
$M_{\rm red} = M M_{\rm n}/(M+M_{\rm n})$ is the reduced mass of the
projectile and the target nucleus. The mass $M_{\rm n}$ of the latter is
practically equal to the atomic mass. Combining these expressions, we
have
\beq
x=d_0 q_0 =  Z_0 Z \sqrt{\frac{M^3}{8 \me M^2_{\rm red}}\,
\frac{E_{\rm h}}{E^3}} \; W.
\label{11.13}\eeq

Following \citet{BrandtLapicki1979}, to account for the Coulomb
deflection we multiply the energy-loss DCS obtained from the PWBA by the
semi-classical correction factor $F^{\rm Coul}_a (E;W)$. That is, we set
\beq
\frac{\d \sigma_a^{\rm Coul}}{\d W} = F^{\rm Coul}_a (E;W) \,
\frac{\d \sigma_a^{\rm cont}}{\d W}\, .
\label{11.14}\eeq
Thus, the ionization cross section, including the binding and
Coulomb-deflection corrections, is given by
\beq
[\sigma_a^{\rm ion}]^{(k)} =
\int_{E_a+\Delta E_a}^{W_{\rm max}}
W^k \,  F^{\rm Coul}_a (E;W) \,
\frac{\d \sigma_{a}^{\rm cont}}{\d
W} \, \d W.
\label{11.15}\eeq
The integrated cross sections calculated in this way, could be used to
determine improved shell corrections. It is worth mentioning that these
shell corrections would include part of the Barkas and Bloch
corrections \citep{Ahlen1980, LindhardSorensen1996}.


\subsection{Sample numerical results \label{sec11.3}}

\begin{figure}[tp]
\begin{center}
\includegraphics*[width=7.5cm]{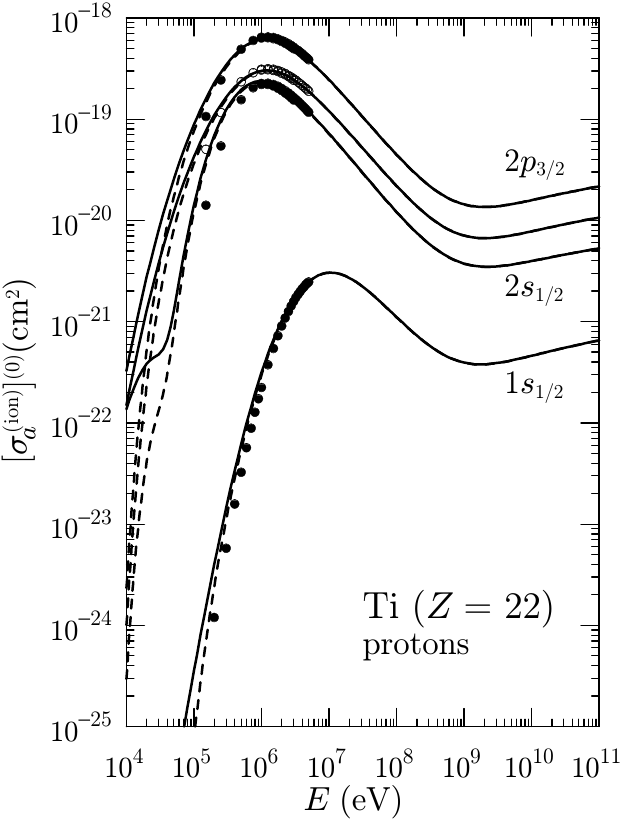} \rule{9mm}{0mm}
\includegraphics*[width=7.5cm]{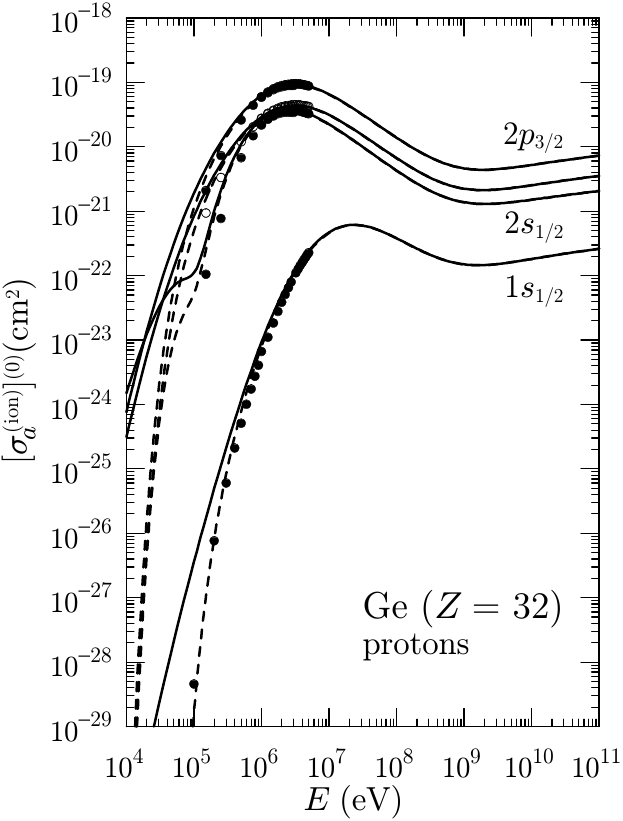} \\
\includegraphics*[width=7.5cm]{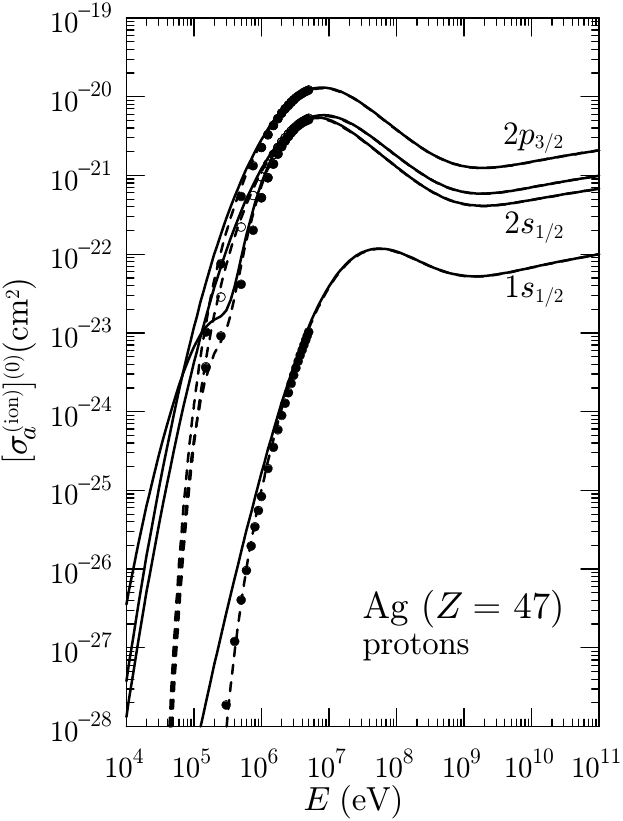} \rule{9mm}{0mm}
\includegraphics*[width=7.5cm]{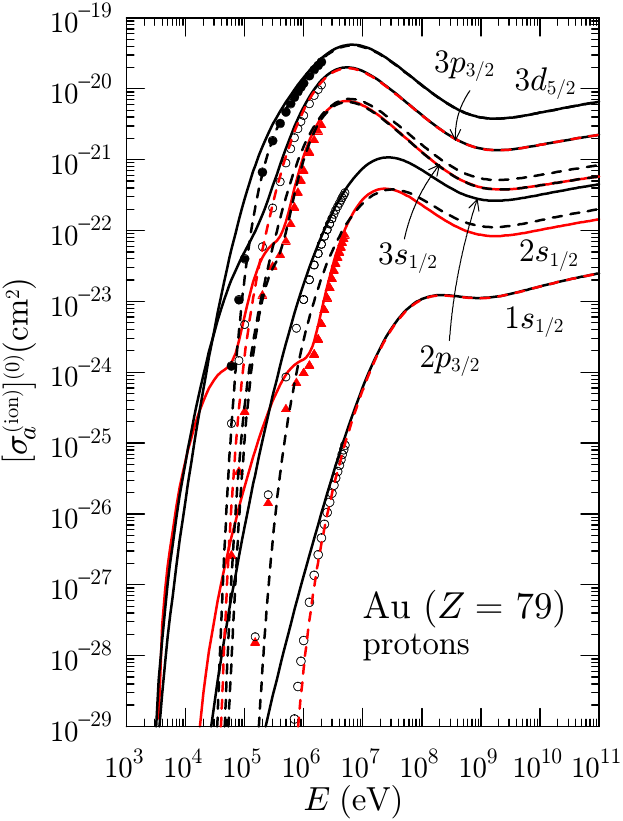}
\caption{Total cross sections for ionization of the indicated (K, L and
M) subshells of titanium, germanium, silver and gold by impact of protons
as functions of the kinetic energy of the projectile. Solid curves were
calculated by using the unmodified PWBA, the dashed curves represent
results from the PWBA with binding and Coulomb-deflection corrections,
Eq.\ \req{11.15}. Symbols are results from equivalent PWBA
calculations by Chen and Crasemann (\citeyear{ChenCrasemann1985},
\citeyear{ChenCrasemann1989}), which also include
these low-energy corrections.
\label{fig34}}
\end{center}\end{figure}

The program {\sc pwacs} allows including binding and Coulomb-deflection
effects for charged particles heavier than the electron. To give a feel
of the magnitude of the binding and Coulomb-deflection corrections,
Fig.\ \ref{fig34} displays the calculated ionization cross sections of
various electron subshells of titanium, germanium, silver and gold atoms
by impact of protons as functions of the kinetic energy of the
projectile.  The uncorrected PWBA cross sections (solid curves) are the
same as in Fig.\ \ref{fig21}. As expected, the effect of the binding and
Coulomb-deflection corrections is appreciable only for relatively low
energies, up to about 1,000~$E_a$. For comparison purposes, Fig.\
\ref{fig34} includes also values calculated by
\citeauthor{ChenCrasemann1985} (\citeyear{ChenCrasemann1985},
\citeyear{ChenCrasemann1989}) for the DHFS potential without and with
binding and Coulomb-deflection corrections (expressed in a slightly
different form).  Note the systematic shift of the ionization threshold,
which results from the binding correction.

\begin{figure}[h!]
\begin{center}
\includegraphics*[width=8.0cm]{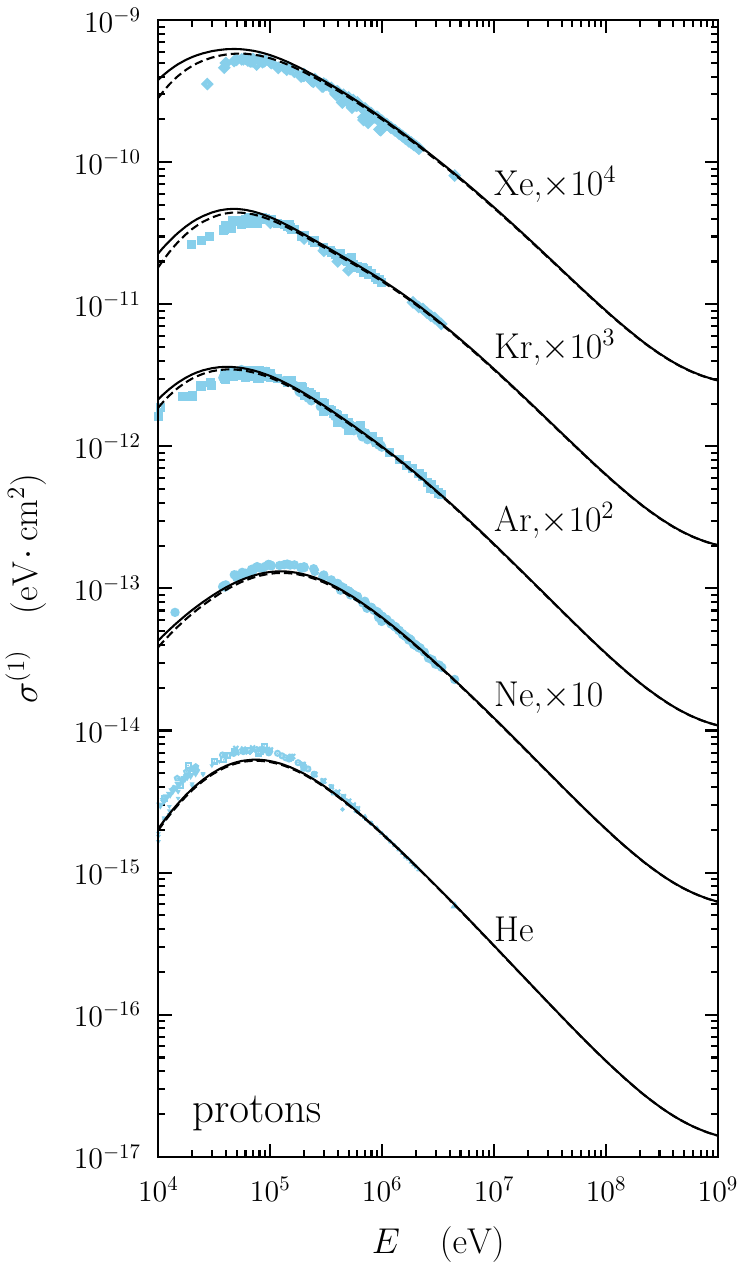}
\includegraphics*[width=8.0cm]{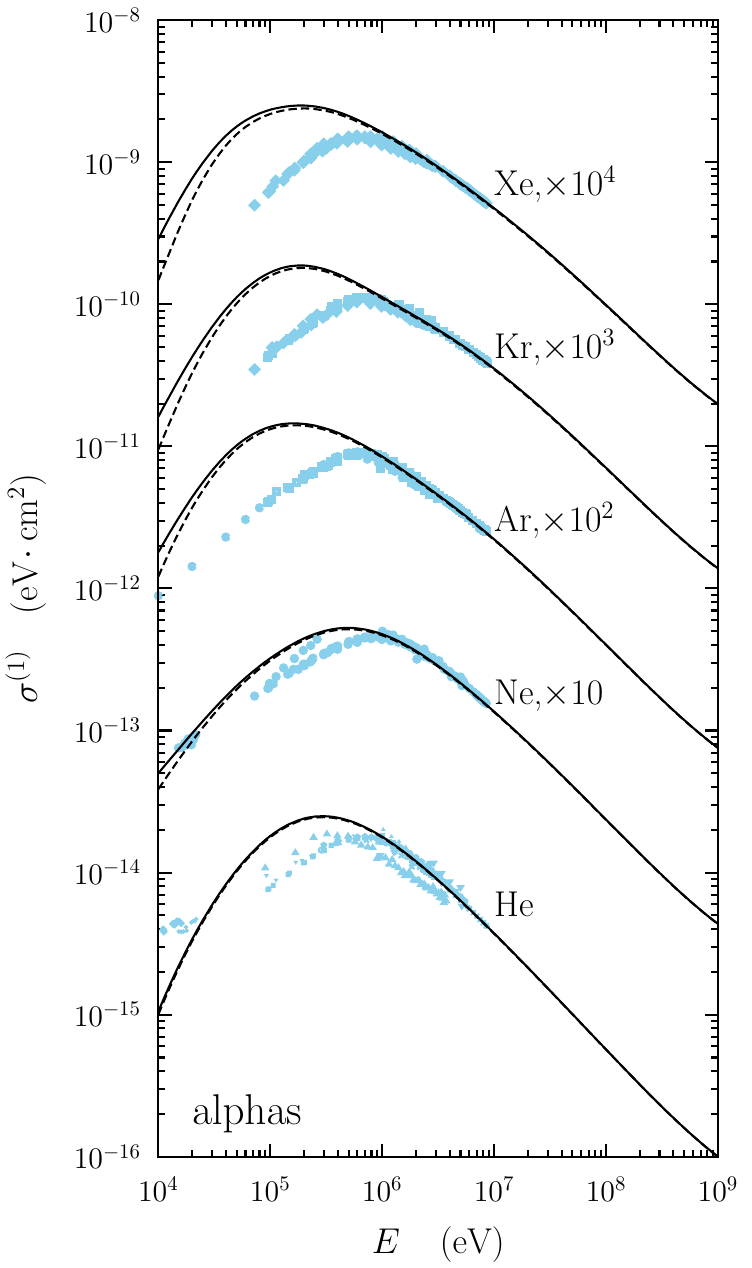}
\caption{Stopping cross sections for inelastic collisions of protons and
alpha particles with noble-gas atoms calculated by integrating the PWBA
energy-loss DCS (solid curves). The dashed curves are results
from calculations with binding and Coulomb-deflection corrections.
Symbols represent experimental data from the IAEA stopping-power
database.
\label{fig34bis}}
\end{center}\end{figure}

It is worth recalling that the relativistic PWBA is expected to be valid
only for charged projectiles with sufficiently large energies. Figure
\ref{fig34bis} compares stopping cross sections calculated from the PWBA
for collisions of protons and alpha particles with atoms of the noble
gases with experimental data from the exhaustive IAEA online database on
``Electronic Stopping Power of Matter for Ions''\footnote{This database
	is available from the IAEA web site,
\url{https://www-nds.iaea.org/stopping/index.html}. The data used here
were downloaded in March 2022.} \citep{Montanari2017}. Since the DHFS
model is expected to be accurate for atoms with closed-subshell
configurations, this comparison indicates that the theory is reliable
for protons and alphas with kinetic energies higher than about 0.75 MeV
and 5 MeV, respectively.  Notice that the effect of binding and
Coulomb-deflection corrections on the stopping cross section is small at
the energies where the theory is expected to be applicable.

\section{Conclusion \label{sec12}}
\setcounter{equation}{0}

We have presented a detailed formulation of the relativistic PWBA for
inelastic collisions of electrons and heavier charged particles with
atoms and ions, assuming that atomic wave functions can be described by
using an independent-electron model. We have derived a closed expression
for the DDCS in terms of the familiar (longitudinal) GOS and we have
introduced the TGOS, which accounts for the exchange of virtual photons.
We have avoided the approximation of small momentum transfers adopted in
Fano's (\citeyear{Fano1963}) review and by other authors. This
approximation causes
appreciable modifications in the integrated cross sections for electrons
with kinetic energies that are not much higher than the ionization
threshold.

We have developed robust numerical methods for computing tables of the
longitudinal and transverse GOSs, as well as interpolation and
extrapolation procedures to obtain the DDCS from the tabulated GOSs.
These methods have been implemented in a set of computer codes, which
have been run, using the DHFS self-consistent atomic potential, to
generate a complete database of GOSs and TGOSs for the occupied subshells
of neutral atoms from hydrogen to einsteinium ($Z=1$ -- 99) in their
ground-state configurations. We have verified that the calculated GOSs
show systematic departures from the Bethe sum rule caused by
relativistic effects, in accordance with the work of \citet{Cohen2003b}
and others.

From our GOS database, we have calculated total cross sections, stopping
cross sections and energy-straggling cross sections for inelastic
collisions of electrons, positrons and protons with free atoms by
straight numerical integration of the DDCS. Asymptotic formulas for
these integrated cross sections, taking into account the relativistic
departure from the Bethe sum rule, have been derived. Shell corrections
to the asymptotic formulas can be obtained directly from the differences
between the these formulas and the numerical integrated cross sections.

Although the assumed independent-electron model provides only a rough
approximation regarding the details of the excitation spectrum of atoms,
it is expected to give fairly accurate estimates of the departure from
the Bethe sum rule and of the shell corrections, which get the larger
contributions from excitations of inner subshells.

Finally, binding and Coulomb-deflection effects in ionizing collisions
of heavy projectiles have been accounted for by using approximate
corrections based on the work of \citet{BrandtLapicki1979} and
others.

The present report is accompanied with a database of GOSs, longitudinal
and transverse, for free atoms calculated from the DHFS potential, and
with the {\sc gosat} and {\sc pwacs} Fortran programs. These perform all
the required calculations, from the generation of tables of GOSs and
associated quantities, to the evaluation of the energy-loss DCS and its
integrals for free atoms, using the GOS database. All the results
presented in this report have been generated by using the {\sc gosat}
and {\sc pwacs} programs and our GOS database.

\newpage


\appendix

\vspace*{1mm}

\section{Kinematics of inelastic collisions \label{secA}}
\setcounter{equation}{0}

We consider here the kinematics of inelastic collisions of a charged
particle of mass $M$ and velocity ${\bf v}$ with a target atom. For
simplicity, we will work in the laboratory frame, where the target atom
is at rest. Let ${\bf p}=\hbar{\bf k}$ and $E$ be the momentum and the
kinetic energy of the projectile just before an inelastic collision, the
corresponding quantities after the collision are denoted, respectively,
by ${\bf p'}=\hbar{\bf k}'$ and $E'=E-W$; where $W$ is the energy lost
by the projectile. We recall that the kinetic energy and the momentum of
a free particle that moves with velocity ${\bf v}$ are, respectively
\beq
E = (\gamma-1) M c^2 \quad \mbox{and} \quad {\bf p} = \beta \gamma Mc
\hat{\bf v}\, ,
\label{A.1}\eeq
where
\beq
\beta = \frac{v}{c} = \sqrt{\frac{\gamma^2-1}{\gamma^2}} =
\sqrt{\frac{E(E+2Mc^2)}{(E+Mc^2)^2}}
\label{A.2}\eeq
is the speed in units of $c$ and
\beq
\gamma = \sqrt{\frac{1}{1-\beta^2}} =
\frac{E+Mc^2}{Mc^2}
\label{A.3}\eeq
is the total energy in units of the rest energy of the particle. Note
that $E$ and $p$ are related by
\beq
(cp)^2 = (E+Mc^2)^2 - M^2 c^4 = E(E+2Mc^2)\, .
\label{A.4}\eeq
For particles other than electrons, the maximum energy loss in a
collision is $W_{\rm max}=E$. In the case of electron collisions, the
indistinguishability of the projectile and the target electrons,
combined with Pauli's exclusion principle, reduces the maximum effective
energy loss to $W_{\rm max} \simeq E/2$ (see Section \ref{sec7.1}).

\begin{figure}[tb]
\begin{center}
\includegraphics*[width=9.0cm]{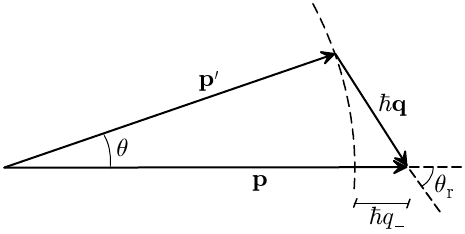}
\caption {\rm Momentum transfer and scattering angles in inelastic
collisions.
\label{fig35}}
\end{center}\end{figure}

The momentum transfer in the collision is defined as $\hbar{\bf q} \equiv
{\bf p}-{\bf p'}$. Squaring this equality we have
\beq
(\hbar q)^2 = p^2 + p'^2 - 2 p p' \cos\theta,
\label{A.5}\eeq
where $\theta=\arccos(\hat{\bf p} \dotprod \hat{\bf p}')$ is the polar
scattering angle (see Fig.\ \ref{fig35}). Recalling that the final
momentum is
\beq
p' = \frac{1}{c} \sqrt{(E-W) (E-W+2M c^2)},
\label{A.6}\eeq
we can write
\beqa
(c \hbar q)^2 & = & E(E+2 M c^2) + (E-W) (E-W+ 2 M c^2)
\nonumber \\ [2mm]
&& \mbox{} - 2 \sqrt{E(E+2 M c^2) \, (E-W) (E-W+ 2 M c^2)} \,
\cos\theta.
\label{A.7}\eeqa
Evidently, given the initial energy and the energy loss, the momentum
transfer determines the scattering angle,
\beq
\cos\theta = \frac{ E(E+2M c^2) + (E-W)(E-W+2M c^2) - (c \hbar q)^2}
{2\sqrt{ E(E+2M c^2) \; (E-W)(E-W+2M c^2)}}\, .
\label{A.8}\eeq
Let $\hbar q_-$ and $\hbar q_+$ denote the minimum and maximum momentum
transfers, which correspond to $\theta =0$ and $\theta=\pi$,
respectively. From \req{A.5} we have
\beq
\hbar q_-=p-p' \qquad \mbox{and} \qquad \hbar q_+ = p+p',
\label{A.9}\eeq
that is,
\beq
c \hbar q_\pm =  \sqrt{E(E+2Mc^2)} \pm \sqrt{(E-W)(E-W+2Mc^2)}
\label{A.10}\eeq
Notice that Eq.\ \req{A.5} can be stated as
\beqa
(\hbar q)^2 &=& (p-p')^2 + 2 p p' (1-\cos\theta)
\nonumber \\ [2mm]
&=& (\hbar q_-)^2 + 4 p p' \sin^2(\theta/2).
\label{A.11}\eeqa
For small scattering angles, such that $\sin(\theta/2) \ll 1$, we can
write
\beq
(\hbar q) ^2 \simeq (\hbar q_-)^2 + pp' \theta^2.
\label{A.12}\eeq
When $W \ll E$, the momentum $p'$ of the projectile after the collision
can be evaluated from the Taylor expansion [see Eq.\ \req{A.6}]
\beqa
p'& \simeq & p -
\frac{\d p}{\d E} \, W
+ \frac{1}{2} \, \frac{\d^{2}p}{\d E^{2}} \, W^{2}
- \frac{1}{6} \, \frac{\d^{3}p}{\d E^{3}} \, W^{3} + \cdots
\nonumber \\[2mm]
& = & p - \frac{W}{\beta c} \left[ 1 + \frac{1}{2\gamma(\gamma+1)} \,
\frac{W}{E} + \frac{1}{2(\gamma+1)^{2}} \left( \frac{W}{E} \right)^{2} +
\cdots \right],
\label{A.13}\eeqa
which leads to the following approximation for the minimum momentum
transfer
\beqa
\hbar q_- & \simeq &
\frac{W}{\beta c} \left[ 1 + \frac{1}{2\gamma(\gamma+1)} \,
\frac{W}{E} + \frac{1}{2(\gamma+1)^{2}} \left( \frac{W}{E} \right)^{2}
\right].
\label{A.14}\eeqa
Combining this result with Eq.\ \req{A.10}, and recalling that $p=\beta
\gamma M c$ [see Eq.\ \req{A.1}] and $\gamma M c^2 = E+M c^2$, we obtain
the approximation
\beqa
\sin^2(\theta/2) & = & \frac{(c \hbar q)^2 - (c\hbar q_-)^2}
{4\, cp\, cp'} \simeq
\frac{(c \hbar q)^2-\beta^{-2} W^2}{4\, [E (E+2 M c^2)- \gamma M c^2 W ]}
\nonumber \\ [2mm]
&=& \frac{(c \hbar q)^2-\beta^{-2} W^2}{4\, [E (E+2 M c^2)- (E + Mc^2) W ]} ,
\label{A.15}\eeqa
which is valid when $W \ll E$.

The angle $\theta_{\rm r}$ between ${\bf p}$ and ${\bf q}$ (see Fig\
\ref{fig35}) is called the recoil angle, because in the case of binary
collisions with free electrons at rest it corresponds to the recoil
direction of the target electron. Taking the square of the identity
${\bf p} - \hbar {\bf q} = {\bf p}'$, we obtain
\beqa
\cos\theta_{\rm r} &=& \frac{(cp)^2 - (cp')^2 + (c \hbar q)^2}
{2(cp)(c \hbar q)}
\nonumber \\ [2mm]
&=&
\frac{ E(E+2M c^2) - (E-W)(E-W+2M c^2) + (c \hbar q)^2}
{2\sqrt{ E(E+2M c^2)} \; (c \hbar q)}
\nonumber \\ [2mm]
&=&
\frac{W}{\beta (c \hbar q)} \left( 1 + \frac{ (c \hbar q)^2-W^2}
{2 W (E+M c^2)} \right)\, .
\label{A.16}\eeqa
For collisions with $W \ll E$ and $c\hbar q < W$ the following approximation
holds
\beq
\cos \theta_{\rm r} \simeq \frac{W}{\beta (c \hbar q)}\, .
\label{A.17}\eeq

The calculation of inelastic collisions from the PWBA \citep{Bethe1932,
Fano1954b} reveals that the momentum transfer ${\bf q}$ is the most natural
variable for describing the interactions. Indeed, when the DCS is
expressed in terms of the scattering angle $\theta$ instead of $\hbar q$, the
formal expressions become more complicated and fundamental physical
aspects are obscured. Following \citeauthor{Fano1954b}
(\citeyear{Fano1954b}, \citeyear{Fano1963}), we will usually
express the DCSs in terms of the recoil energy $Q$ defined as the
kinetic energy of an electron having momentum equal to the momentum
transfer, that is,
\beq
Q(Q+2 \me c^{2}) = (c \hbar q)^{2},
\label{A.18}\eeq
where $\me$ is the electron mass. Equivalently, we can write
\beq
Q = \sqrt{(c \hbar q)^2 + \me^2 c^4 }- \me c^2.
\label{A.19}\eeq
When the collision is with a free electron at rest, the energy loss is
completely transformed into kinetic energy of the recoiling electron
and, consequently, $Q=W$. For collisions with bound electrons, the
relation $Q\simeq W$ still holds for hard ionizing collisions, that is,
when the energy transfer $W$ is much larger than the ionization energy
of the target electron so that binding effects are negligible.

\begin{figure}[htbp]
\begin{center}
\includegraphics*[width=7.7cm]{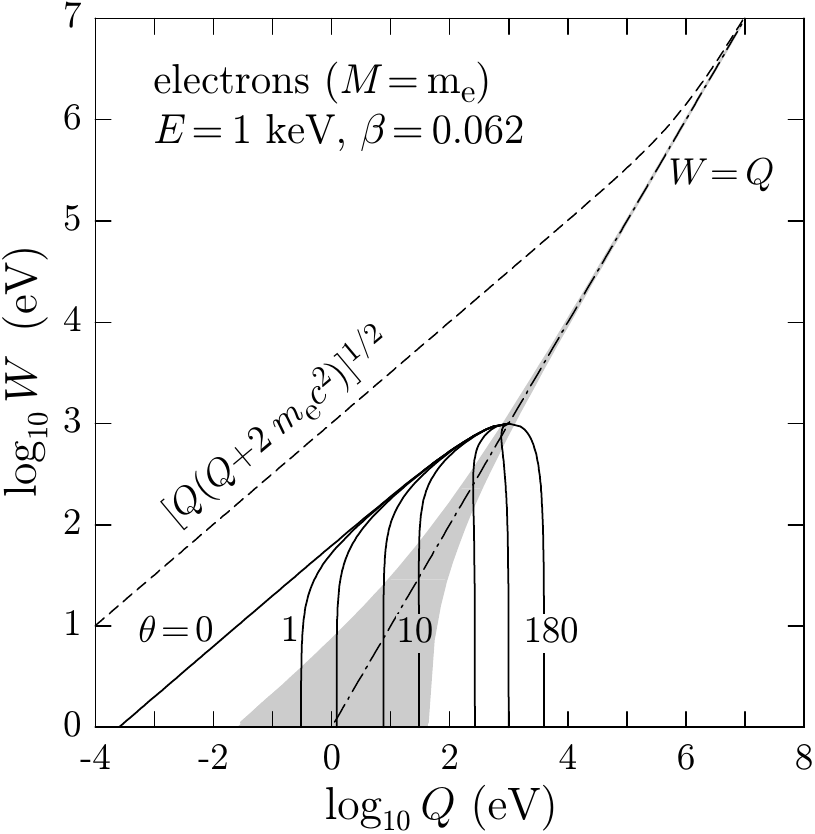} \hfill
\includegraphics*[width=7.7cm]{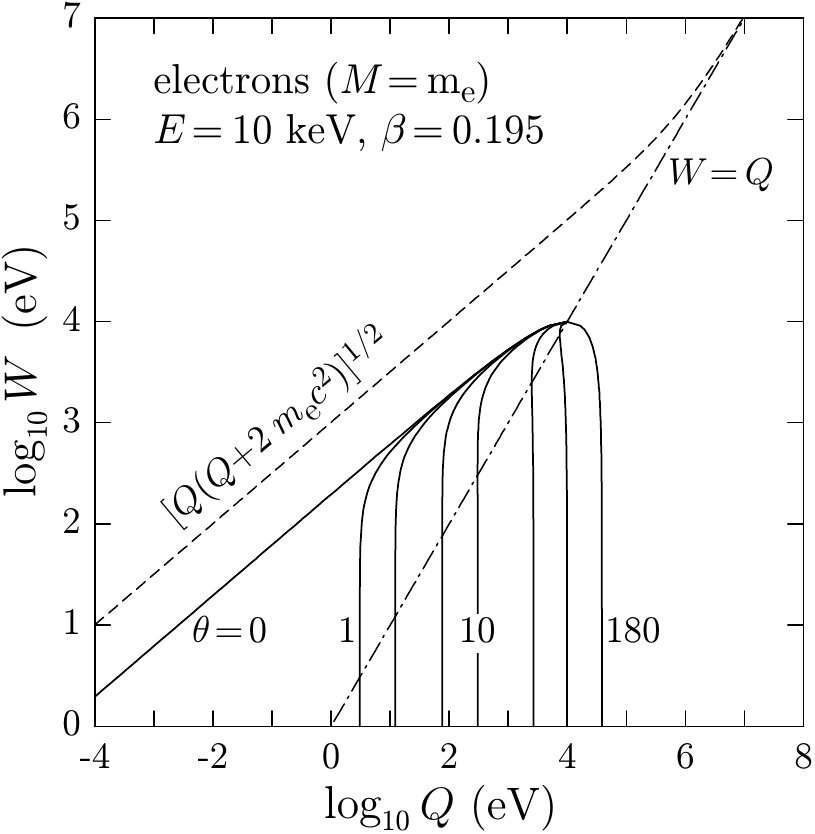} \\
\includegraphics*[width=7.7cm]{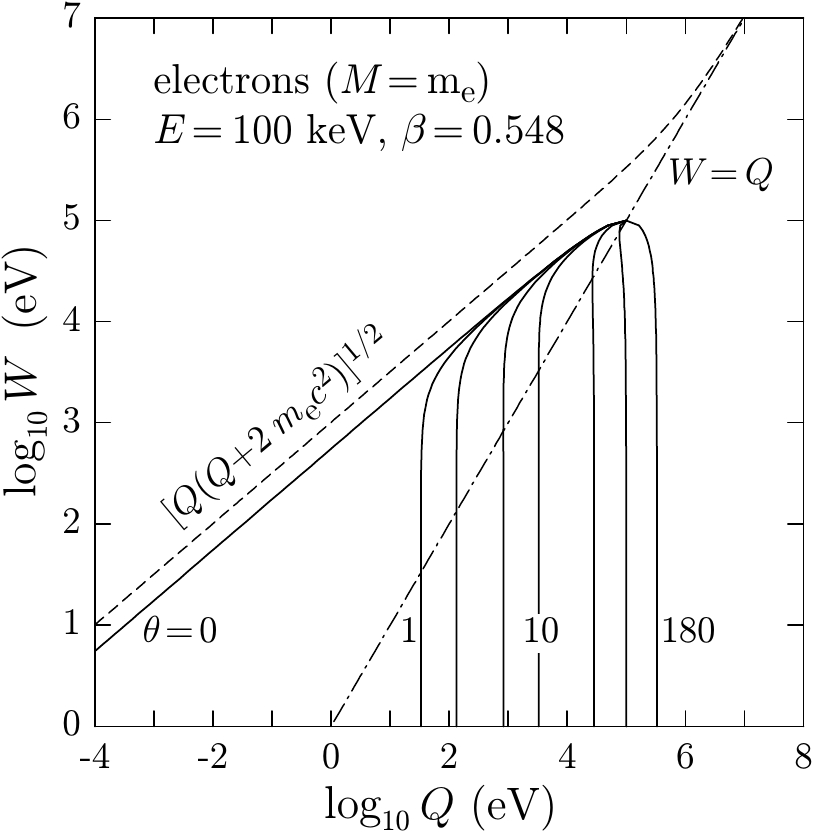} \hfill
\includegraphics*[width=7.7cm]{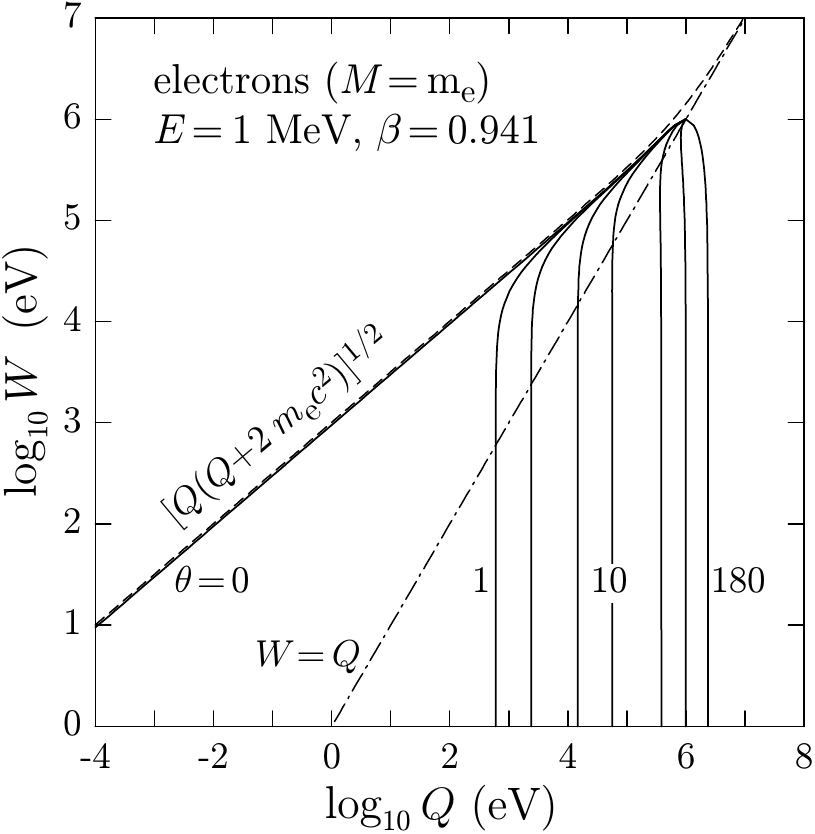}
\caption {\rm Kinematics of inelastic collisions for electrons
of the indicated energies. The curves represent the value of
the recoil energy $Q$ (abscissa) that corresponds to the energy loss $W$
(ordinate) for a given scattering angle. The displayed curves correspond
to scattering angles of 0, 1, 2, 5, 10, 30, 60, and 180 degrees. The
shaded area in the top-left plot represents the possible excitations of
a free-electron gas with a Fermi energy of 10 eV.
\label{fig36}}
\end{center}\end{figure}

Inserting the expression \req{A.7} into Eq.\ \req{A.19} we obtain
\beqa
Q & = & \left[ E(E+2 M c^2) + (E-W) (E-W+ 2 M c^2) \rule{0mm}{5mm} \right.
\nonumber \\ [2mm]
&& \mbox{} \left. - 2 \sqrt{E(E+2 M c^2) \, (E-W) (E-W+ 2 M c^2)} \,
\cos\theta + \me^2 c^4 \rule{0mm}{5mm}\right]^{1/2} - \me c^2.
\rule{10mm}{0mm}
\label{A.20}\eeqa
For a given energy loss, the kinematically allowed recoil energies lie
in the interval $Q_- < Q < Q_+$, with endpoints given by Eq.\ \req{A.19}
with $\hbar q=\hbar q_\pm$,
\beqa
Q_{\pm} &= &\sqrt{ (c\hbar q_\pm)^2 + \me^2 c^4} - \me c^2
\nonumber \\ [4mm]
&=& \sqrt{
\left[ \sqrt{E(E+2Mc^2)} \pm \sqrt{(E-W)(E-W+2Mc^2)}\right]^2 + \me^2
c^4 } - \me c^2\, . \rule{10mm}{0mm}
\label{A.21}\eeqa
Now, using the identity \req{A.11}, Eq.\ \req{A.18} can be restated as
\beq
Q(Q+2 \me c^{2}) = Q_-(Q_-+2 \me c^{2}) + 4 \, cp \, cp' \sin^2
(\theta/2).
\label{A.22}\eeq
It is worth noticing that, for $W < E$, $Q_{+}$ is larger than $W$ (see
Fig.\ \ref{fig36}). When
$W\ll E$, expression \req{A.21} is not suited for evaluating $Q_-$
since it involves the subtraction of two similar quantities. In this
case it is more convenient to use the approximate relation \req{A.14}
and calculate $Q_-$ as
\beq
Q_{-} =
\me c^{2} \left[ \sqrt{ \left(\frac{c\hbar q_-}{\me c^{2}} \right)^{2} +
1} - 1 \right]
\simeq
\me c^{2} \left( x - \frac{x^{2}}{2} + \frac{x^{3}}{2} \right)\, ,
\label{A.23}\eeq
with
\beq
x \equiv \frac{1}{2} \left( \frac{c\hbar q_-}{\me c^{2}} \right)^{2} \simeq
\frac{W^{2}}{2\beta^{2}\,\me^{2}c^{4}}
\left[ 1 + \frac{1}{2\gamma(\gamma+1)} \,
\frac{W}{E} + \frac{1}{2(\gamma+1)^{2}} \left( \frac{W}{E} \right)^{2}
\right]^{2}\, .
\label{A.24}\eeq
Alternatively, for $W \ll E$, the result \req{A.14} implies that
\beq
Q_{-}(Q_{-}+2\me c^{2}) = (c\hbar q_-)^{2} \simeq W^{2}/\beta^{2}\, .
\label{A.25}\eeq
and the relation \req{A.17} can be recast as
\beq
\cos \theta_{\rm r} \simeq \sqrt{\frac{W^2}{\beta^2 (c \hbar q)^2}}
\simeq
\sqrt{\frac{Q_{-}(Q_{-}+2\me c^{2})}{Q(Q+2\me c^{2})}}\, .
\label{A.26}\eeq

\begin{figure}[htbp]
\begin{center}
\includegraphics*[width=7.7cm]{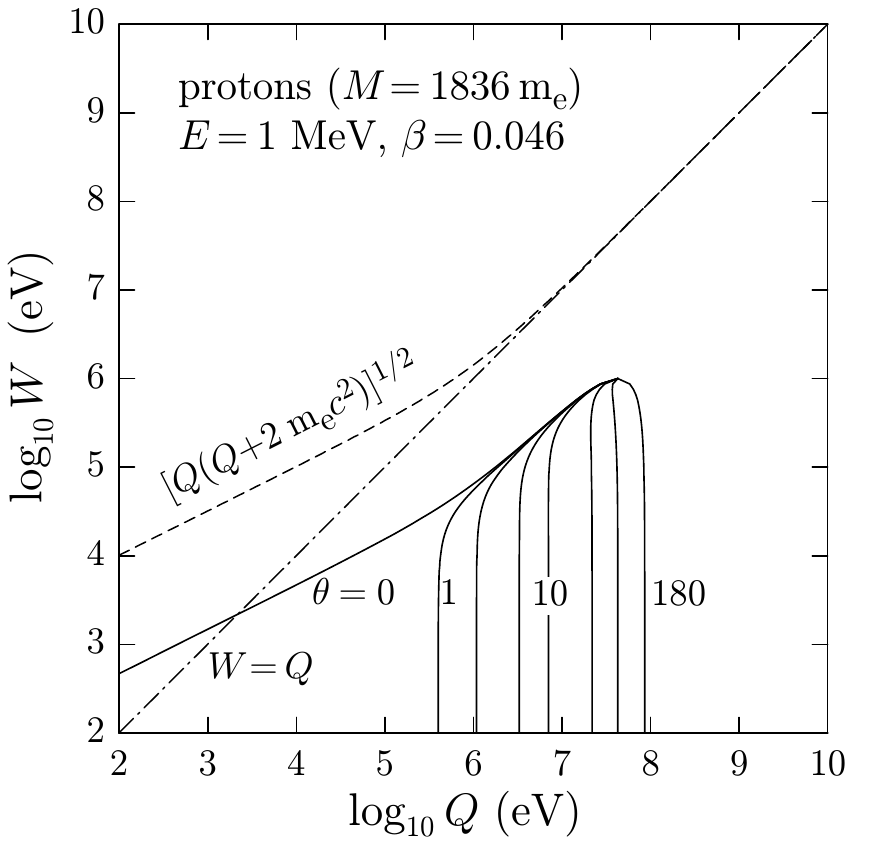} \hfill
\includegraphics*[width=7.7cm]{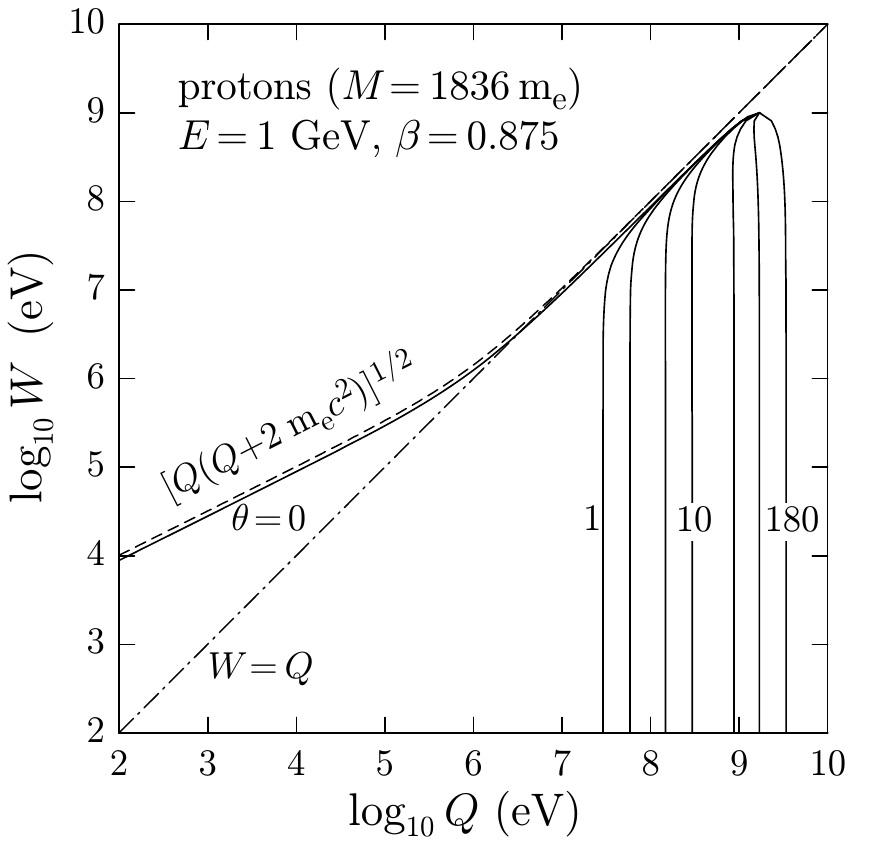}
\caption {\rm Kinematics of inelastic collisions for protons of the
indicated energies. The curves represent the value of
the recoil energy $Q$ (abscissa) that corresponds to the energy loss $W$
(ordinate) for a given scattering angle. Details are the same as in
Fig.\ \ref{fig36}.
\label{fig37}}
\end{center}\end{figure}

Equation \req{A.20} relates the energy loss, the scattering angle and the
recoil energy. Figures \ref{fig36} and \ref{fig37} display the
recoil energy as a function of the energy loss at fixed scattering
angles, $Q(W; \theta)$, for electrons and protons of the indicated
energies. It is worth noting that for energy transfers
that are much less than the energy of the projectile, the curves are
nearly vertical straight lines. That is, when the energy loss is small,
the recoil energy $Q$ is a function of only the scattering angle ($Q$ is
independent of $W$). This behavior changes when the energy loss
increases, because each curve with $\theta \lesssim 60$ deg approaches
smoothly the $\theta=0$ curve. All curves converge to a single point
when the energy loss reaches its maximum allowed value
$W_{\rm max}=E$.

\begin{figure}[thb]
\begin{center}
\includegraphics*[width=7.7cm]{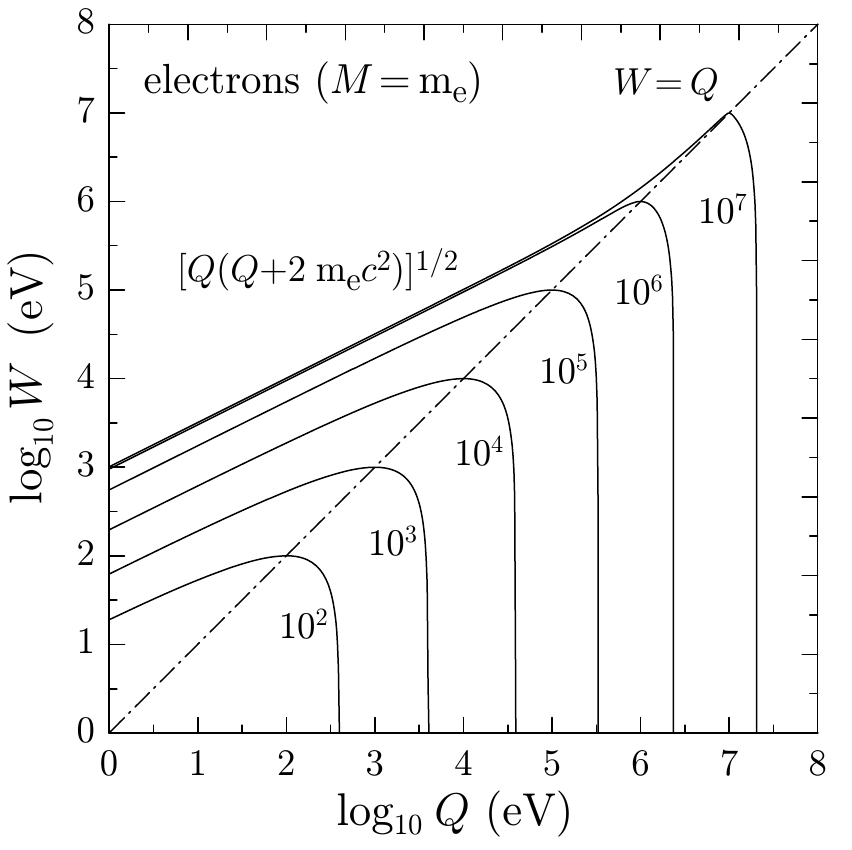}\hfill
\includegraphics*[width=7.7cm]{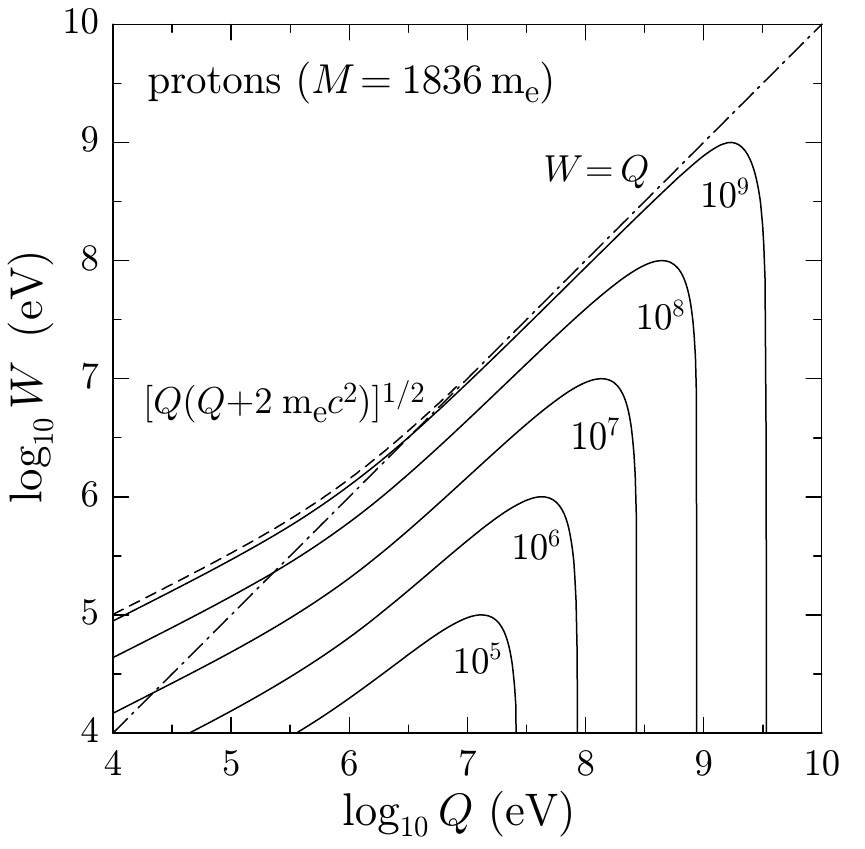}
\caption {\rm Domains of kinematically allowed transitions in the
$(Q,W)$ plane for electrons/positrons (left) and protons (right). The
curves represent the maximum allowed energy loss $W_{\rm m}(Q)$, given
by Eq.\ \req{A.16}, for projectiles with the indicated kinetic
energies (in eV). When $E$ increases, $W_{\rm m}(Q)$ approaches the
vacuum photon line, $W_0(Q) =[Q(Q+2\me c^2)]^{1/2}$, which is an absolute
upper bound for the allowed energy losses.
\label{fig38}}
\end{center}\end{figure}

From \req{A.21}, it is clear that the curves $Q=Q_-(W)$ and
$Q=Q_+(W)$ intersect at $W = E$. Thus, they define a single continuous
function $W=W_{\rm m}(Q)$ in the interval $0 < Q < Q_+(0)$. By solving
the equations $Q=Q_{\pm}(W_{\rm m})$ we obtain
\beq
W_{\rm m}(Q) = E + Mc^2 - \sqrt{\left[ \sqrt{E(E+2Mc^2)} -
\sqrt{Q(Q+2\me c^2)} \right]^2 + M^2 c^4}\, ,
\label{A.27}\eeq
which, when $W \ll E$, reduces to
\beq
W_{\rm m}(Q) \simeq \beta \sqrt{Q(Q+2\me c^2)}\, .
\label{A.28}\eeq
Now it follows that, for given values of $E$ and $Q$ [$<Q_+(0)$], the
only kinematically allowed values of the energy loss are those in the
interval $0 < W < W_{\rm m}(Q)$ (see Fig.\ \ref{fig38}). The plots in
Figs.\ \ref{fig36} to \ref{fig38} reveal a noteworthy difference
between the allowed domains for electrons and protons. For electrons and
positrons ($M=\me$) the maxima of the curves $W_{\rm m}(Q)$ coincide with their
intersections with the diagonal of the $(Q,W)$ plane, \ie, at $Q=E$. In
the case of protons, and heavier ions, the maxima of the curves occur at
recoil energies larger than $E$. It is also worth observing that the
intersection of the curve $W_{\rm m}(Q)$ with the diagonal $W=Q$, the
Bethe ridge, occurs at a point where the energy loss is
\beq
W_{\rm ridge} = 2 \beta^2 \gamma^2 \me c^2 R
\label{A.29}\eeq
with
\beq
R = \left[ 1 + \left( \frac{\me}{M} \right)^2 +
2\gamma \, \frac{\me}{M} \right]^{-1}.
\label{A.30}\eeq
For electrons and positrons, $W_{\rm ridge}=E$. For heavy
projectiles ($M \gg \me$) with energies much smaller than their rest
energy $Mc^2$, $R \simeq 1$ and
\beq
W_{\rm ridge} \simeq 2 \beta^2 \gamma^2 \me c^2
= 2 \, \frac{E (E+2 Mc^2)}{M^2 c^4} \, \me c^2 .
\label{A.31}\eeq

For a given energy loss $W$, the minimum value of the momentum transfer,
\beq
\hbar q_- \equiv c^{-1} \sqrt{Q_-(Q_-+2\me c^2)}\, ,
\label{A.32}\eeq
occurs when $\theta=0$. $\hbar q_-$ is always larger than $W/c$.  When the
energy of the projectile increases, $\beta \rightarrow 1$ and $\hbar q_-$
decreases approaching (but never reaching) the value $W/ c$. It is worth
recalling that a photon of energy $W$ in vacuum has a linear momentum $\hbar q
= W/c$ and, hence, interactions consisting of emission of bare photons
would be located on the line
\beq
W_0(Q) = c\hbar q = \sqrt{Q(Q+2\me c^2)}
\label{A.33}\eeq
of the ($Q$,$W$) plane, the so-called vacuum photon line. This line lies
outside the kinematically allowed region, \ie, the ``recoil'' energy of
the photon is less than $Q_-$ (see Figs.\ \ref{fig36}, \req{fig37} and
\ref{fig38}).  Therefore, when the target is a single atom, the
emission of photons by the projectile is not possible\footnote{In a
condensed medium, ultra-relativistic projectiles can emit real photons
(Cerenkov radiation) under certain, very restricting circumstances}.
When the energy $E$ of the projectile
increases, $Q_-$ decreases and tends to the photon line when $\beta$
tends to unity. Hence, emission of photons by ultra-relativistic
projectiles in low-density media is barely prevented by energy and
momentum conservation. Generally speaking, as the interaction involves
the exchange of a virtual photon, the DCS increases as the photon
becomes more real, that is, as we approach the photon line. For a thin
gas, this causes a gradual increase of the cross section with the
projectile energy when $\beta \rightarrow 1$.

In the non-relativistic regime ($c \rightarrow \infty$), the recoil energy is
\beq
Q \equiv (\hbar q)^{2}/2\me \, ,
\label{A.34}\eeq
and the limits of the interval ($Q_-^{\rm nr},Q_+^{\rm nr})$ of
allowed recoil energies are
\beq
Q_{\pm}^{\rm nr} = \frac{M}{\me} \left[ \sqrt{E} \pm \sqrt{E-W} \right]^{2}
\, .
\label{A.35}\eeq
The maximum energy loss for a given value of $Q$ $[< Q_+(0)]$ is
\beq
W_{\rm m}^{\rm nr} (Q) = \sqrt{\frac{\me}{M} \, Q} \left( 2 \sqrt{E}
- \sqrt{\frac{\me}{M} \, Q} \right)
= v \hbar q - \frac{(\hbar q)^2}{2 M} \, ,
\label{A.36}\eeq
where to obtain the last form we have set $E=\1o2 M v^2$. In the case of soft
collisions with $W \ll E$, we have [cf.\ Eqs.\ \req{A.14} and
\req{A.28}]
\beq
\hbar q_-^{\rm nr} \simeq \frac{W}{v}
\qquad \mbox{and} \qquad
W_{\rm m}^{\rm nr} (\hbar q) \simeq v\, \hbar q.
\label{A.37}\eeq
It is interesting to notice that $W_{\rm m}^{\rm nr}(Q)$ increases
without limit when the energy $E$ of the projectile increases. This is
in sharp contrast with the relativistic case; Eq.\ \req{A.28} shows that
$[Q(Q+2\me c^2)]^{1/2}$ is an upper bound for $W_{\rm m}(Q)$.

As mentioned above, the recoil energy $Q$ (or the squared momentum
transfer) is the most convenient variable for describing the DCSs
obtained from the PWBA \citep{Fano1963}. Nevertheless, some authors keep
using the alternative variable
\beq
Q' \equiv \frac{(\hbar q)^2}{2\me} - \frac{W^2}{2\me c^2}
= \frac{Q(Q+2\me c^2) - W^2}{2\me c^2}
\label{A.38}\eeq
introduced by \citet{Bethe1933}, because it slightly simplifies the
expression of the DDCS. However, $Q'$ is not independent of $W$ and,
consequently, the GOSs cannot be regarded as natural functions of $Q'$
and $W$. Moreover, the usage of $Q'$ instead of $Q$ complicates the
majority of kinematical formulas.


\section{Dirac wave functions \label{secB}}
\setcounter{equation}{0}

In this Appendix we briefly review some specifics of the Dirac equation,
and set the notation used in the theory sections.  The Dirac Hamiltonian
for an electron in an external electromagnetic field described by the
4-potential $({\bf A}, {\rm i} \varphi)$ is \citep[see,
\eg,][]{Rose1961}
\beq
{\cal H}_{\rm D} = c \widetilde{\alphab}
\dotprod \left( {\bf p} + \frac{e}{c} \, {\bf A} \right) +
\widetilde{\beta} \me c^2 - e \, \varphi\, ,
\label{B.1}\eeq
where ${\bf p}=-{\rm i}\hbar \nabla$ is the momentum operator, and
$\widetilde{\alphab}$ and $\widetilde \beta$ are the Dirac matrices. The
standard representation for these matrices is
\beq
\widetilde{\alphab} = \left( \begin{array}{cc}
0 & \sigmab \\
\sigmab & 0 \\ \end{array} \right), \qquad
\widetilde{\beta} = \left( \begin{array}{cc}
I_2 & 0  \\
0 & - I_2\\ \end{array} \right)\, ,
\label{B.2}\eeq
where $\sigmab$ stands for the familiar $2\times 2$ Pauli spin matrices
and $I_2$ is the $2\vecprod 2$ unit matrix. The Dirac Hamiltonian
\req{B.1} satisfies the commutation relation
\beq
\left[ {\cal H}_{\rm D} , I_4 f({\bf r},t) \right] = c \widetilde{\alphab}
\dotprod {\bf p} \, f({\bf r},t) = - {\rm i} c \hbar \,
\widetilde{\alphab} \dotprod \nablab f({\bf r},t)
\label{B.3}\eeq
for any scalar function $f({\bf r},t)$.  This
equality is useful to simplify the calculation of certain matrix
elements.

When the 4-potential does not depend on time, the time-independent Dirac wave equation takes the form
\beq
\left[ c \widetilde{\alphab}
\dotprod \left( {\bf p} + \frac{e}{c} \, {\bf A} \right) +
\widetilde{\beta} \me c^2 - e \, \varphi \right] \psi({\bf
 r}) = {\cal W} \psi({\bf r})\, ,
\label{B.4}\eeq
where the ${\cal W}$ is the energy eigenvalue, inclusive of the electron
rest energy, ${\cal W}=E+\me c^2$.

\subsection{Plane waves \label{secB.1}}

In the case of a free electron (${\bf A}=0$ and $\varphi=0$), the
Hamiltonian \req{B.1} commutes with the momentum operator and, hence,
there exists a complete set of eigenfunctions $\phi_{{\bf
k}\mu\tau}({\bf r})$ common to ${\cal H}_{\rm D}$ and ${\bf p}$. These
are the plane waves
\beq
\phi_{{\bf k} \mu \tau}({\bf r}) =
\frac{{\rm e}^{{\rm i} {\bf k}\cdot {\bf r} }}{(2\pi)^{3/2}}
U_{{\bf k} \mu \tau}\, ,
\label{B.5}\eeq
where the index $\tau$ ($=\pm 1$) denotes the sign of the energy,
$\mu=\pm 1/2$ and $U_{{\bf k} \mu \tau}$ are the following
double spinors,
\begin{subequations}
\label{B.6}
\beqa
U_{{\bf k}, \mu, +1}
&=& \left[ 1 + \frac{(c\hbar k)^2}{(|{\cal W}|
+ \me c^2)^2} \right]^{-1/2}
\left( \begin{array}{c}  I_2 \\ [2mm]
\displaystyle{+ \frac{c\hbar \, \sigmab \dotprod {\bf
k}}{|{\cal W}| +\me c^2}}
\end{array} \right) \chi_{\mu}\, ,
\label{B.6a} \\ [2mm]
U_{{\bf k}, \mu, -1}
&=& \left[ 1 + \frac{(c\hbar k)^2}{(|{\cal W}|
+ \me c^2)^2} \right]^{-1/2}
\left( \begin{array}{c}
\displaystyle{- \frac{c\hbar \, \sigmab \dotprod {\bf
k}}{|{\cal W}| +\me c^2}}
\\ [2mm] I_2
\end{array} \right) \, \chi_{\mu}\, ,
\label{B.6b}\eeqa
\end{subequations}and
\beq
\chi_{+1/2}  = \left( \begin{array}{c} 1 \\ 0 \end{array} \right)\, ,
\qquad
\chi_{-1/2} = \left( \begin{array}{c} 0 \\ 1 \end{array} \right)\, ,
\label{B.7}\eeq
are the Pauli unit spinors.
It can be easily verified that, for a given ${\bf k}$,
\beq
U_{{\bf k} \mu' \tau'}^\dagger
U_{{\bf k} \mu \tau} =
\delta_{\mu',\mu} \, \delta_{\tau',\tau} \quad \mbox{and}
\quad
\sum_{\mu,\tau}
U_{{\bf k} \mu \tau} U_{{\bf k} \mu \tau}^\dagger = I_4\, .
\label{B.8}\eeq

In the formulation of the PWBA we make explicit use of the completeness
of the plane-wave four-spinors through the usage of energy-projection
operators. For our purposes, it is convenient to introduce the operator
\beq
\Pi_{{\bf k},+1} \equiv \sum_{\mu=\pm 1/2} U_{{\bf k}, \mu, +1}
U_{{\bf k}, \mu, +1}^\dagger
=\frac{1}{2|{\cal W}|} \left( |{\cal W}| + c\hbar\,
\widetilde{\alphab} \dotprod {\bf k} + \widetilde{\beta} \me c^2 \right)
\, .
\label{B.9}\eeq
It can be easily verified that $\Pi_{{\bf k},+1}^2
= \Pi_{{\bf k},+1}$. Therefore, $\Pi_{{\bf k},+1} $ is a projection
operator: it projects spinor states on the subspace of positive energy.


\subsection{Spherical waves and distorted plane waves \label{secB.2}}

Let us now consider an electron in a central field $V(r)$ [${\bf A}=0$,
$V(r) = -e \, \varphi(r)$]. The angular
momentum operator for a Dirac particle is ${\bf J} = {\bf L} + {\bf S}$,
where ${\bf L} = -{\rm i} {\bf r} \vecprod \nabla$ is the orbital
angular momentum and ${\bf S}$ is the spin angular momentum (all angular
momenta are in units of $\hbar$). Since ${\cal H}_{\rm D}$ commutes with
${\bf J}^2$, $J_z$ and with the parity operator (${\cal P} =
\widetilde{\beta} \times\,$space inversion), there exists a complete
basis of eigenfunctions common to these four operators. These
eigenfunctions are the spherical waves, and have the form \citep{Rose1961,
Grant1965}
\beq
\psi_{\epsilon\kappa m}({\bf r}) = \frac{1}{r}
\left( \begin{array}{c}
P_{\epsilon\kappa}(r) \, \Omega_{\kappa, m} (\hat{\bf r}) \\ [1mm]
{\rm i} Q_{\epsilon\kappa}(r)
\, \Omega_{-\kappa, m} (\hat{\bf r}) \end{array} \right).
\label{B.10}\eeq
where $\Omega_{\kappa,m}(\hat{\bf r})$ are spherical spinors, and
$P_{\epsilon\kappa}(r)$
and $Q_{\epsilon\kappa}(r)$ are the
large- and small-component radial functions, which
satisfy the coupled differential equations
\beq
\begin{array}{cccc}
{\displaystyle \frac{\d P_{\epsilon\kappa}}{\d r}} & =
& {\displaystyle -\; \frac{\kappa}{r} \,P_{\epsilon\kappa}}
& {\displaystyle +\; \frac{\epsilon-V+2\me c^2}{c\hbar}} \,
Q_{\epsilon\kappa}\, , \\ [4mm]
{\displaystyle \frac{\d Q_{\epsilon\kappa}}{\d r}} & =
& \displaystyle{ -\; \frac{\epsilon-V}{c\hbar} \, P_{\epsilon\kappa}}
& {\displaystyle + \; \frac{\kappa}{r}}\, Q_{\epsilon\kappa}\, .
\end{array}
\label{B.11}\eeq
where $\epsilon={\cal W}-\me c^2$ is the electron energy, exclusive of
its rest energy.
The spherical spinors are eigenfunctions of the total angular momentum
of Pauli's theory, and are given by
\beq
\Omega_{\kappa,m}(\hat{\bf r}) \equiv
\Omega_{j,m}^{\ell}(\hat{\bf r})
= \displaystyle{\sum_{\mu=\pm 1/2}
\langle \ell, \1o2, m-\mu, \mu | j, m \rangle \;
Y_{\ell,m-\mu} (\hat{\bf r}) \, \chi_\mu} \, .
\label{B.12}\eeq
These are simultaneous eigenfunctions of $L^2$, $S_{\rm P}^2$, $J^2$ and $J_z$
with eigenvalues $\ell(\ell+1)$, $3/4$, $j(j+1)$ and $m$, respectively.
Here ${\bf S}_{\rm P} = \1o2 \sigmab$ denotes the two-dimensional Pauli spin
operator. The quantities $\langle j_1 j_2 m_1 m_2 |j,m\rangle$ are
Clebsch--Gordan coefficients, and $\chi_\mu$ are the Pauli spinors, \ie,
the eigenstates of $S_{\rm P}^2$ and $S_{{\rm P}3}$ with eigenvalues
$3/4$ and $\mu=\pm \1o2$, respectively. More explicitly,
\beqa
\Omega_{\ell\pm1/2,m}^{\ell}(\hat{\bf r}) &=&
\left( \begin{array}{c}
\langle \ell, \1o2,
m-\1o2, +\1o2 | j, m \rangle \, Y_{\ell,m-1/2}
(\hat{\bf r}) \\ [2mm]
\langle \ell, \1o2, m+\1o2,
-\1o2 | j, m \rangle \, Y_{\ell,m+1/2}
(\hat{\bf r})
\end{array} \right)
\nonumber \\ [2mm]
&=&
\frac{1}{\sqrt{2\ell+1}}
\left( \begin{array}{c}
\pm \sqrt{\ell \pm m + \1o2} \, Y_{\ell,m-1/2}
(\hat{\bf r}) \\ [2mm]
\sqrt{\ell \mp m + \1o2} \, Y_{\ell,m+1/2}
(\hat{\bf r})
\end{array} \right)\, .
\label{B.13}\eeqa
To simplify notation, it is customary to introduce the
relativistic angular momentum quantum number
\beq
\kappa = (\ell-j) (2j+1)\, ,
\label{B.14}\eeq
which specifies both the total angular momentum [$j$] and the parity
[$(-1)^\ell$] of the Dirac spherical wave,
\beq
j = |\kappa | - \1o2,\qquad \ell =
\displaystyle{j + \frac{\kappa}{2|\kappa|}}
= \left\{
\begin{array}{ll}
\kappa & \mbox{if $\kappa >0$} \\ [2mm]
-\kappa -1 \rule{3mm}{0mm} & \mbox{if $\kappa <0$}
\end{array} \right\} \, .
\label{B.15}\eeq
It is also convenient to consider the quantum number
\beq
\overline\ell \equiv \left\{
\begin{array}{ll}
-\kappa & \mbox{if $\kappa < 0$} \\ [2mm]
\kappa -1 \rule{3mm}{0mm} & \mbox{if $\kappa >0$}
\end{array} \right\} = \ell - \frac{\kappa}{|\kappa|}\, ,
\label{B.16}\eeq
which is the value of $\ell$ corresponding to $-\kappa$.
Note that the spherical wave $\psi_{E\kappa m}({\bf r})$ is not an
eigenfunction of $L^2$; the index $\ell$ used in spectroscopic notation
is the eigenvalue of the upper-component spinor and serves to indicate
the parity of the eigenstate.

The DHFS potentials occurring in the present calculations are combinations of
a short-range field and a Coulomb field,
\beq
V(r) = V_{\rm sr} + \frac{Z_{\infty}e^2}{r}\, ,
\label{B.17}\eeq
where the short-range component $V_{\rm sr}(r)$ vanishes for $r$
larger than $r_{\rm Latter}$, the onset of the Latter tail [see Eq.\
\req{2.19}], and $Z_{\infty}=-1$ for neutral atoms. Radial functions for
these potentials can be calculated numerically to high accuracy by using
the subroutine package {\sc radial} \citep{SalvatFernandezVarea2019}.
The numerical algorithm implemented in these subroutines combines a
cubic-spline interpolation of the function $rV(r)$ with local
power-series expansions of the radial functions in such a way that
truncation errors are effectively reduced. In the case of bound orbitals
($\epsilon < 0$), each discrete energy level is characterized by the
principal quantum number $n$ and the relativistic quantum number
$\kappa$. Bound orbitals calculated by {\sc radial} are normalized to
unity and, therefore, the calculated orbitals satisfy the orthonormality
relation
\beq
\int \psi^\dagger_{n'\kappa' m'} ({\bf r})
\psi_{n\kappa m} ({\bf r}) \, \d {\bf r} = \delta_{n'n}\,
\delta_{\kappa'\kappa} \, \delta_{m'm}\, .
\label{B.18}\eeq
The radial functions of free spherical waves (with $\epsilon > 0$) are
normalized in such a way that the large-component radial function
asymptotically oscillates with unit amplitude,
\beq
P_{\epsilon\kappa}(r)
\begin{array}[t]{c} \sim \\ [-2mm] \scriptstyle{ r \rightarrow \infty}
\end{array}
\sin \left(k r - \ell \frac{\pi}{2} - \eta
\ln 2k r + \delta_{\epsilon\kappa} \right)\, ,
\label{B.19}\eeq
where
\beq
k = (c\hbar)^{-1} \sqrt{\epsilon(\epsilon+2\me c^2)}
\label{B.20}\eeq
is the wave number and $\eta= Z_{\infty} e^2 \me/(\hbar^2 k)$ is the
Sommerfeld parameter. The phase shift $\delta_{\epsilon\kappa}$ is
determined numerically by integrating the radial equations from $r=0$
outward to a point further than the range $r_c$ of the  $V_{\rm sr}$
potential, and matching it at that point to a combination of the regular and
irregular Dirac--Coulomb functions. The package {\sc radial} implements
efficient algorithms for the calculation of Dirac--Coulomb wave
functions. In the limit $Z_\infty=0$, the radial Dirac--Coulomb functions
reduce to Bessel functions and, therefore, the generic algorithm is also
valid for finite-range fields. Free spherical waves normalized in the
form \req{B.19} satisfy the orthogonality relation
\beq
\int \psi^\dagger_{\epsilon'\kappa' m'} ({\bf r})
\psi_{\epsilon\kappa m} ({\bf r}) \, \d {\bf r} =
\frac{\epsilon}{k}\, \pi \, \delta(\epsilon'-\epsilon)\,
\delta_{\kappa'\kappa} \, \delta_{m'm}\, .
\label{B.21}\eeq

In collision theory, states of free particles in the initial and final
channels are described as distorted plane waves (DPWs), \ie, by solutions
of the Dirac equation for the potential $V(r)$ that asymptotically
behave as a plane wave plus an outgoing ($+$) or incoming ($-$)
spherical wave. A DPW is characterized by the wave vector ${\bf k}$ and
spin $\mu$; it can be expanded in the basis of spherical waves as
\citep[see, \eg,][]{Rose1961}
\beq
\psi_{{\bf k}\mu}^{(\pm)} ({\bf r}) = \frac{1}{k}
\, \sqrt{\frac{\epsilon+2 \me c^2}{\pi(\epsilon+\me c^2)}}
\sum_{\kappa,m} {\rm i}^\ell \, \exp \left( \pm {\rm i} \delta_{\kappa}
\right) \, \left\{ \left[ \Omega_{\kappa m} (\hat{\bf k})
\right]^\dagger \chi_{\mu} \right\} \psi_{\epsilon\kappa m} ({\bf r})
\, ,
\label{B.22}\eeq
where
\beq
\epsilon = \sqrt{(c\hbar k)^2 + (\me c^2)^2}-\me c^2
\label{B.23}\eeq
is the kinetic energy of the particle. The expansion \req{B.21} is known
as the partial-wave series. It can be easily verified that,
with the adopted normalization for free spherical waves, the DPWs satisfy
the orthogonality relation
\beq
\int
\left[ \psi_{{\bf k}'\mu'}^{(\pm)} ({\bf r}) \right]^\dagger
\psi_{{\bf k}\mu}^{(\pm)} ({\bf r}) \, \d {\bf r}
= \delta({\bf k}'-{\bf k}) \, \delta_{\mu'\mu}\, .
\label{B.24}\eeq
In the limit where the strength of the potential tends to zero ($V=0$), the phase shifts vanish and the radial functions of free states
($E>0$) reduce to regular spherical Bessel functions
\beq
\begin{array}{lll}
P^{(0)}_{\epsilon\kappa}(r) = kr\, j_{\kappa}(kr), & \quad
Q^{(0)}_{\epsilon\kappa}(r) = {\displaystyle \sqrt{\frac{\epsilon}{\epsilon+2 \me c^2}}} \;
kr\, j_{\kappa-1}(kr)
& \quad {\rm if} \; \kappa > 0\, , \\ [5mm]
P^{(0)}_{\epsilon\kappa}(r) = kr\, j_{-\kappa-1}(kr), & \quad
Q^{(0)}_{\epsilon\kappa} (r)
= {\displaystyle - \; \sqrt{\frac{\epsilon}{\epsilon+ 2 \me c^2}}} \;
kr\, j_{-\kappa}(kr) & \quad {\rm if} \; \kappa < 0\, .
\end{array}
\label{B.25}\eeq
In a more compact form, valid for any $\kappa$,
\beq
P^{(0)}_{\epsilon\kappa}(r) = kr\, j_{\ell}(kr), \quad
Q^{(0)}_{\epsilon\kappa}(r) = \frac{\kappa}{|\kappa|} {\displaystyle \sqrt{\frac{\epsilon}{\epsilon+2 \me c^2}}} \;
kr\, j_{\overline{\ell}}(kr)\, ,
\label{B.26}\eeq
where $\overline{\ell}$ is the value of $\ell$ corresponding to $-\kappa$,
Eq.\ \req{B.16}. Note that, in the $V=0$ limit, the  DPW reduces to the
positive-energy plane wave, $\psi_{{\bf k}\mu}^{(\pm)} ({\bf r})
\rightarrow \phi_{{\bf k},\mu,+1} ({\bf r})$.


\section{Matrix elements of Racah tensors \label{secC}}
\setcounter{equation}{0}

In calculations of atomic structure and radiation theory Dirac spherical
waves are employed because they allow the analytical calculation of
angular integrals in matrix elements. The evaluation of matrix elements
is simplified when operators are expressed in terms of the Racah
functions, defined by
\beq
C_{L M}(\hat{\bf r}) \equiv \sqrt{\frac{4\pi}{2L+1}} \;
Y_{L M}(\hat{\bf r})\, ,
\label{C.1}\eeq
because the set of $2L+1$ functions $C_{LM}(\hat{\bf r})$
constitute an irreducible tensor of rank $L$, ${\bf C}^{(L)}$. For
example, the Coulomb interaction between two electrons is proportional to
\beq
\frac{1}{|{\bf r}_1 - {\bf r}_2|} = \sum_{L=0}^\infty
\frac{r_<^L}{r_>^{L +1}}
\sum_{M=-L}^L C_{LM}^\ast(\hat{\bf r}_1)
\, C_{LM}(\hat{\bf r}_2)\, ,
\label{C.2}\eeq
where $r_<$ is the smaller of $r_1$ and $r_2$, and $r_>$ is the larger.
Another operator that occurs frequently in radiation studies is the
plane wave $\exp({\rm i} {\bf k} \dotprod {\bf r})$, which can be
represented by means of the Rayleigh expansion,
\beq
\exp ({\rm i} {\bf k} \cdot {\bf r})
= \sum_{L=0}^\infty \sum_{M=-L}^L
{\rm i}^{L} (2 L +1) \, j_{L}(kr)
\, C_{LM}^{\ast}(\hat{\bf k})\, C_{LM}(\hat{\bf r})\, ,
\label{C.3}\eeq
where $j_\lambda (qr)$ are spherical Bessel functions. When operators
are expressed in this way, calculations involve the matrix elements of
Racah tensors ${\bf C}^{(L)}$ for eigenstates $\Omega_{jm}^{\ell} =
\left| \ell \1o2 j m \right>$ of the total angular momentum ${\bf J}=
{\bf L}+ {\bf S}_{\rm P}$ of a spin $\1o2$ particle, [Eq.\ \req{B.12}]. By
virtue of the Wigner-Eckart theorem \citep[see, \eg,][]{Edmonds1960}, these
matrix elements can be written as
\beqa
\langle \Omega_{j_1 m_1}^{\ell_1} | C_{LM} | \Omega_{j_2m_2}^{\ell_2}
\rangle &\equiv&
\int
\left[\Omega_{j_1m_1}^{\ell_1}(\hat{{\bf r}}) \right]^\dagger\,
C_{LM}(\hat{{\bf r}}) \,
\Omega_{j_2m_2}^{\ell_2} (\hat{{\bf r}}) \, \d{\hat{\bf r}}
\nonumber \\ [2mm]
&=& \frac{1}{\sqrt{2j_1+1}} \,
\langle j_2 L m_2 M | j_1 m_1 \rangle \;
\left<\ell_1 \1o2 j_1 || {\bf C}^{(L)} || \ell_2
\1o2 j_2\right>\, .
\label{C.4}\eeqa
Using elaborate angular-momentum methods, it can be shown that
the reduced matrix element
$\left<\ell_1 \1o2 j_1|| {\bf C}^{(L)} || \ell_2
\1o2 j_2\right>$ is
given by the expression (see, \eg, Grant, 1961)
\beq
\left<\ell_1 \1o2 j_1|| {\bf C}^{(L)} ||
\ell_2 \1o2 j_2\right>
= \upsilon(L,\ell_1,\ell_2) \, \sqrt{2j_2+1} \;
\langle L j_2 0 \1o2 | j_1 \1o2 \rangle\, ,
\label{C.5}\eeq
where the factor
\beq
\upsilon(L,\ell_1,\ell_2) \equiv \left\{ \begin{array}{l}
1 \mbox{ if } L + \ell_1 + \ell_2 \mbox{ is even} \\
0 \mbox{ otherwise,}
\end{array} \right\} \equiv \1o2 \left[ 1 + (-1)^{L+\ell_1+\ell_2}
\right]\, ,
\label{C.6}\eeq
accounts for the parity selection rule.

The Clebsch--Gordan coefficient in expression \req{C.5} can be
calculated
analytically as follows \citep[see, \eg,][]{CondonOdabasi1980}. We
first note that the coefficients $\left< j_1 j_2 0 0 | j_3 0 \right>$
vanish when $j_1 + j_2 + j_3$ is odd. When this
sum is even, $j_1+j_2+j_3 = 2J$, we have \citep{Rose1995}
\beq
\left< j_1 j_2 0 0 | j_3 0 \right> = \displaystyle{(-1)^{J-j_3}
\sqrt{\frac{2j_3+1}{2J+1}}
\left[ \frac{\tau(J)}{\tau(J-j_1) \tau(J-j_2) \tau(J-j_3)}
\right]^{1/2}}\, ,
\label{C.7}\eeq
with
$$
\tau(n) = \frac{2^n (n!)^2}{(2n)!} = \frac{n!}{1\cdot 3 \cdot 5
\cdots (2n-1)}\, .
$$
Now, the coefficients $\left< L j_2 0 \1o2 | j_1 \1o2
\right>$ with integer $L$ and odd half-integer values\footnote{A
rational number $j$ is said to be an odd
half-integer if $j=(2n+1)/2$, where $n$ is an integer.} $j_2$ and $j_1$
can be evaluated using the formula \citep{Louck1958}
\beqa
\left< L j_2 0 \1o2 | j_1 \1o2 \right>
&=& \sqrt{\frac{(K-2L)(K+1)}{2j_1 (2j_2+1)}} \,
\left< L, j_2-\1o2, 0, 0 | j_1-\1o2, 0 \right>
\nonumber \\ [2mm]
&& \mbox{} +
\sqrt{\frac{(K-2j_2)(K-2j_1+1)}{2j_1 (2j_2+1)}} \,
\left< L, j_2+\1o2, 0, 0 | j_1-\1o2, 0 \right>\, ,
\label{C.8}\eeqa
where $K \equiv  L + j_2 + j_1$. Note that one of the two
Clebsch--Gordan coefficients in this expression is null; the other is
given by Eq.\ \req{C.7}.

\begin{table}[phtb]
\caption{\rm Coefficients $d^L_{m_1-m_2} (\kappa_1 m_1; \kappa_2 m_2)$,
Eq.\ \req{C.9}. With the aid of the symmetry relations \req{C.10a} to
\req{C.10c}, all the coefficients with $|\kappa_1|,|\kappa_2| \le 2$ can
be obtained from the tabulated values.
\label{table.dcoefs} }
\vskip 4mm
\begin{center}
\begin{tabular}{crrcccccc} \hline\hline
 & & & & \multicolumn{4}{c}{$L$}\rule{0mm}{5mm} \\ \cline{5-8}
$\kappa_1,\kappa_2$ & $m_1$ & $m_2$ & & 0 & 1 & 2 &
3\rule{0mm}{5mm}\rule[-2mm]{0mm}{4mm}  \\ \hline
$+1,+1$ & $+1/2$ & $-1/2$ & & 0 & 0 & 0 & 0\rule[+3mm]{0mm}{3mm} \\
      & $+1/2$ & $+1/2$ & & 1 & 0 & 0 & 0 \\ \hline
$+1,-1$ & $+1/2$ & $-1/2$ & \rule{12mm}{0mm} & 0 & $+\sqrt{2/9}$ & 0 & 0 \\
      & $+1/2$ & $+1/2$ & & 0 & $-\sqrt{1/9}$ & 0 & 0 \\  \hline
$+2,+1$ & $+1/2$ & $-1/2$ & & 0 & $+\sqrt{1/9}$ & 0 &  0 \\
      & $+1/2$ & $+1/2$ & & 0 & $+\sqrt{2/9}$ & 0 &  0 \\
      & $+3/2$ & $-1/2$ & & 0 & 0 & 0 &  0 \\
      & $+3/2$ & $+1/2$ & & 0 & $+\sqrt{3/9}$ & 0 &  0 \\  \hline
$+2,-1$ & $+1/2$ & $-1/2$ & & 0 & 0 & $+\sqrt{3/25}$ & 0 \\
      & $+1/2$ & $+1/2$ & & 0 & 0 & $-\sqrt{2/25}$ & 0 \\
      & $+3/2$ & $-1/2$ & & 0 & 0 & $+\sqrt{4/25}$ & 0 \\
      & $+3/2$ & $+1/2$ & & 0 & 0 & $-\sqrt{1/25}$ & 0 \\  \hline
$+2,+2$ & $+1/2$ & $-3/2$ & & 0 & 0 & $-\sqrt{2/25}$ & 0 \\
      & $+1/2$ & $-1/2$ & & 0 & 0 & 0 & 0 \\
      & $+1/2$ & $+1/2$ & & 1 & 0 & $+\sqrt{1/25}$ & 0 \\
      & $+1/2$ & $+3/2$ & & 0 & 0 & $-\sqrt{2/25}$ & 0 \\
      & $+3/2$ & $-3/2$ & & 0 & 0 & 0 & 0 \\
      & $+3/2$ & $-1/2$ & & 0 & 0 & $-\sqrt{2/25}$ & 0 \\
      & $+3/2$ & $+1/2$ & & 0 & 0 & $+\sqrt{2/25}$ & 0 \\
      & $+3/2$ & $+3/2$ & & 1 & 0 & $-\sqrt{1/25}$ & 0 \\  \hline
$+2,-2$ & $+1/2$ & $-3/2$ & & 0 & 0 & 0 & $-\sqrt{90/1225}$ \\
      & $+1/2$ & $-1/2$ & & 0 & $+\sqrt{8/225}$ & 0 & $+\sqrt{108/1225}$ \\
      & $+1/2$ & $+1/2$ & & 0 & $-\sqrt{1/225}$ & 0 & $-\sqrt{81/1225}$ \\
      & $+1/2$ & $+3/2$ & & 0 & $-\sqrt{6/225}$ & 0 & $+\sqrt{36/1225}$ \\
      & $+3/2$ & $-3/2$ & & 0 & 0 & 0 & $-\sqrt{180/1225}$ \\
      & $+3/2$ & $-1/2$ & & 0 & 0 & 0 & $+\sqrt{90/1225}$ \\
      & $+3/2$ & $+1/2$ & & 0 & $+\sqrt{6/225}$ & 0 & $-\sqrt{36/1225}$ \\
      & $+3/2$ & $+3/2$ & & 0 & $-\sqrt{9/225}$ & 0 &
      $+\sqrt{9/1225}$\rule[-3mm]{0mm}{3mm} \\
\hline\hline
\end{tabular}
\end{center} \end{table}

It is useful to introduce the coefficients
\beqa
&& \! \! \! \! \! \! \! \! \! \! \! \! \! \! \! \! \! \! \! \!
d^L_M (\kappa_1 m_1; \kappa_2 m_2) \equiv
\langle \Omega_{\kappa_1m_1} | C_{LM} | \Omega_{\kappa_2m_2}
\rangle
\nonumber \\ [2mm]
&=&
\frac{1}{\sqrt{2j_1+1}} \,
\langle j_2 L m_2 M| j_1 m_1 \rangle \;
\left<\ell_1 \1o2 j_1|| {\bf C}^{(L)} ||
\ell_2 \1o2 j_2\right>
\, .
\label{C.9}\eeqa
To simplify the formulas, here we use the quantum number $\kappa$
instead of $\ell$ and $j$ [see the Eqs.\ \req{B.14} and \req{B.15}].
Note that $d^L_M (\kappa_1,m_1;\kappa_2,m_2)=0$ unless $M=m_1-m_2$. The
non-vanishing coefficients are usually indicated by $d^L
(\kappa_1,m_1;\kappa_2,m_2)$, omitting the $M$ label. These
$d$-coefficients are the same as those introduced by \citep{Grant1961},
except for the fact that ours include the parity factor
$\upsilon(L,\ell,\ell')$, and we use a different phase convention.
Table \ref{table.dcoefs} gives the numerical values of
non-vanishing $d$-coefficients with $|\kappa_1|,|\kappa_2| \le 2$.

The $d$-coefficients satisfy the following symmetry relations
\begin{subequations}
\label{C.10}
\beqa
d^L_M (\kappa_1, m_1; \kappa_2, m_2)
&=& (-1)^{m_1-m_2}\, d^L_{-M} (\kappa_2, m_2; \kappa_1, m_1)
\label{C.10a} \\ [2mm]
&=& (-1)^{L-|\kappa_1|+|\kappa_2|}\, d^L_{-M} (\kappa_1,
-m_1; \kappa_2, -m_2)
\label{C.10b}\\ [2mm]
&=& d^L_M (-\kappa_1, m_1; - \kappa_2, m_2) \, ,
\label{C.10c}\eeqa
which follow directly from the symmetry properties of the Clebsch--Gordan
coefficients.  The last equality results from observing that, reversing
the signs of $\kappa_1$ and $\kappa_2$, the values of $j_1$ and $j_2$
and of the parity factor $\upsilon(L,\ell_1,\ell_2)$ remain unaltered.
We also have
\beq
d^L_M (\kappa_1, m_1; \kappa_2, m_2) \;
d^L_{M'} (-\kappa_1, m_1'; \kappa_2, m_2')=0\, ,
\label{C.10d}\eeq
because the values of $\ell_1$ corresponding to $\kappa_1$ and to
$-\kappa_1$ differ by one unit and, therefore, the parity factor
$\upsilon(L,\ell_1,\ell_2)$ is null for one of these coefficients.

The orthonormality of the spherical spinors \req{B.12} implies that
\beq
d^0_0 (\kappa_1 m_1; \kappa_2 m_2)
= \delta_{\kappa_1 \kappa_2} \, \delta_{m_1 m_2}\, .
\label{C.10e}\eeq
The $d^L_M$ coefficients satisfy the following sum rules
\beq
\sum_{m_1} d^L_M (\kappa_1 m_1; \kappa_1 m_1)
= (2j_1+1) \, \delta_{L 0} \, \delta_{M 0}\, ,
\label{C.10f}\eeq
\beqa
\sum_M \sum_{m_2} \left[ d^L_M (\kappa_1 m_1; \kappa_2
m_2)\right]^2 &=&
v(L,\ell_1,\ell_2) \,
\frac{2j_2+1}{2j_1+1} \, \langle L j_2 0
\1o2 | j_1 \1o2 \rangle^2
\nonumber \\ [2mm]
&=& \frac{1}{2j_1+1} \,
\left<\ell_1 \1o2 j_1|| {\bf C}^{(L)} ||
\ell_2 \1o2 j_2\right>^2\, ,
\label{C.10g}\eeqa
\beqa
\sum_M \sum_{m_1,m_2}
\left[ d^L_M (\kappa_1 m_1; \kappa_2 m_2) \right]^2 &=&
v(L,\ell_1,\ell_2) \, (2j_2+1) \, \langle L j_2 0 \1o2 |
j_1 \1o2 \rangle^2
\nonumber \\ [2mm]
&=& \left<\ell_1 \1o2 j_1|| {\bf C}^{(L)} || \ell_2
\1o2 j_2\right>^2\, ,
\label{C.10h}\eeqa
\beqa
&&\sum_{m_1,m_2} d^L_M(\kappa_1,m_1;\kappa_2,m_2) \,
d^{L'}_{M'}(\kappa_1,m_1;\kappa_2,m_2)
\nonumber \\ [2mm]
&& \rule{25mm}{0mm} \mbox{} =
\upsilon(L,\ell_1,\ell_2) \,
(2j_2+1) \, \langle L j_2 0 \1o2 |
j_1 \1o2 \rangle^2
\, \displaystyle{\frac{\delta_{LL'}\delta_{MM'}}{2L+1}}
\nonumber \\ [2mm]
&& \rule{25mm}{0mm} \mbox{} =
\left<\ell_1 \1o2 j_1|| {\bf C}^{(L)} ||
\ell_2 \1o2
j_2\right>^2 \, \frac{\delta_{LL'}\delta_{MM'}}{2L+1}\, ,
\label{C.10i}\eeqa
and
\beqa
&&\sum_{m_1,m_2} d^L_M(\kappa_1,m_1;\kappa_2,m_2) \,
d^{L'}_{M'}(-\kappa_1,m_1;\kappa_2,m_2) = 0 \, .
\label{C.10j}\eeqa
\end{subequations}

Matrix elements of spin operators can be evaluated by using the
following identity,
\beqa
&& \! \! \! \! \! \! \! \! \! \! \! \! \! \!
\Omega^\dagger_{\kappa_1m_1}
(\hat{\bf r}) \, \sigmab \, \Omega_{\kappa_2 m_2} (\hat{\bf r})
= \sum_{J,M} (-1)^M
\nonumber \\ [2mm]
&& \times \left\{ \sqrt{\frac{J}{4\pi}} \, \left(
\frac{\kappa_1+\kappa_2}{J}
- 1 \right) d^J_M(\kappa_1, m_1;-\kappa_2,m_2) \,
{\bf Y}_{J,-M}^{J-1}(\hat{\bf r}) \right.
\nonumber \\ [2mm]
&& \rule{7.5mm}{0mm}+ \sqrt{\frac{2J+1}{4\pi \, J(J+1)}}\,
\; (\kappa_1-\kappa_2)
\, d^J_M(\kappa_1, m_1;\kappa_2,m_2) \,
{\bf Y}_{J,-M}^{J}(\hat{\bf r})
\nonumber \\ [2mm]
&& \rule{7.5mm}{0mm} +\sqrt{\frac{J+1}{4\pi}} \, \left. \left(
\frac{\kappa_1+\kappa_2}{J+1} + 1 \right)
d^J_M(\kappa_1, m_1;-\kappa_2,m_2) \,
{\bf Y}_{J,-M}^{J+1}(\hat{\bf r})\rule{0mm}{7mm}\right\}\, ,
\label{C.11} \eeqa
where the functions
${\bf Y}_{JM}^{L}(\hat{\bf r})$ are the vector spherical
harmonics, defined by \citep[see, \eg,][]{AkhiezerBerestetskii1965}
\beq
{\bf Y}^L_{JM} (\hat{\bf r}) = \sum_{\nu=-1}^1
\langle L, 1, M+\nu, -\nu | J M \rangle \,
Y_{L, M+\nu} (\hat{\bf r}) \, \xib_{-\nu}\, ,
\label{C.12}\eeq
and $\xib_\nu$ are the spherical unit vectors,
\beq
\xib_{+1} = \frac{1}{\sqrt{2}}
\left( \begin{array}{c}
-1 \\ [1mm]
-{\rm i} \\ [1mm] 0 \end{array}\right),
\qquad
\xib_{0} = \left( \begin{array}{c} 0 \\ [1mm] 0 \\
[1mm] 1
\end{array}\right), \qquad
\xib_{-1} = \frac{1}{\sqrt{2}} \left( \begin{array}{c}
1 \\ [1mm]
-{\rm i} \\ [1mm] 0 \end{array}\right)\, .
\label{C.13}\eeq
The important relation \req{C.11} can be derived by expanding the
left-hand side in vector spherical harmonics,
\beq
\Omega^\dagger_{\kappa_1m_1}
(\hat{\bf r}) \, \sigmab \, \Omega_{\kappa_2 m_2} (\hat{\bf r}) =
\sum_{LJM} A_{JM}^{L}
{\bf Y}_{JM}^{L}(\hat{\bf r})\, ,
\label{C.14}\eeq
where
\beqa
A_{JM}^L &\equiv& \int \left[ {\bf
Y}_{JM}^{L}(\hat{\bf r})
\right]^\ast \, \dotprod \, \left[
\Omega^\dagger_{\kappa_1 m_1}
(\hat{\bf r}) \, \sigmab \, \Omega_{\kappa_2 m_2} (\hat{\bf r}) \right]
\, \d \hat{\bf r}
\nonumber \\ [2mm]
&=& (-1)^{L+1+J+M} \int
\Omega^\dagger_{\kappa_1 m_1}
(\hat{\bf r}) \, \sigmab \dotprod
{\bf Y}_{J,-M}^{L}(\hat{\bf r})
\, \Omega_{\kappa_2 m_2} (\hat{\bf r})
\, \d \hat{\bf r}
\nonumber \\ [2mm]
&=& (-1)^{L+1+J+M} \left< \Omega_{\kappa_1 m_1} \left|
\sigmab \dotprod {\bf Y}_{J,-M}^{L}\rule{0mm}{4mm}
\right| \Omega_{\kappa_2 m_2} \right>\, .
\label{C.15}\eeqa
The matrix elements $\langle \Omega_{\kappa_1 m_1} | \sigmab \dotprod {\bf
Y}_{J,M}^{L}| \Omega_{\kappa_2 m_2} \rangle$ can be calculated in a routine
fashion, by using the orthogonality of the spherical harmonics and the
properties of the vector-coupling coefficients. It is found that the
non-vanishing ones are
\begin{subequations}
\label{C.16}
\beqa
\left< \Omega_{\kappa_1 m_1} \left| \sigmab \dotprod {\bf
Y}_{J,M}^{J}\rule{0mm}{4mm}
\right| \Omega_{\kappa_2 m_2} \right> &=&
\sqrt{\frac{2J+1}{4\pi \, J(J+1)}}\, (\kappa_2-\kappa_1) \,
d^J_M(\kappa_1,m_1;\kappa_2, m_2)\, , \rule{15mm}{0mm}
\label{C.16a}\eeqa
\beqa
\left< \Omega_{\kappa_1 m_1} \left| \sigmab \dotprod
{\bf Y}_{J,M}^{J-1}
\right| \Omega_{\kappa_2 m_2} \right> &=&
\sqrt{\frac{J}{4\pi}} \,
\left( \frac{\kappa_2+\kappa_1}{J} - 1 \right)\,
d^J_M(\kappa_1, m_1;-\kappa_2, m_2)\, , \rule{15mm}{0mm}
\label{C.16b}\eeqa
and
\beqa
\left< \Omega_{\kappa_1 m_1} \left| \sigmab \dotprod
{\bf Y}_{J,M}^{J+1}
\right| \Omega_{\kappa_2 m_2} \right> &=&
\sqrt{\frac{J+1}{4\pi}} \,
\left( \frac{\kappa_2+\kappa_1}{J+1} + 1 \right) \,
d^J_M(\kappa_1, m_1;-\kappa_2, m_2)\, . \rule{15mm}{0mm}
\label{C.16c}\eeqa
\end{subequations}
Insertion of these matrix elements into the expansion
\req{C.14}-\req{C.15} leads to the result \req{C.11}.

PWBA calculations involve matrix elements of the type $\langle
\Omega_{\kappa_1 m_1} | C_{\lambda \mu} \, \sigmab | \Omega_{\kappa_2
m_2} \rangle$, which can be calculated as follows. From the definition
\req{C.12} we have
\beqa
\lefteqn{\int \d \hat{\bf r} \,
C_{\lambda \mu}(\hat{\bf r}) \, {\bf Y}_{J,-M}^{L}(\hat{\bf r})
= \sqrt{\frac{4\pi}{(2\lambda+1)}}
\sum_{\nu}
\langle L, 1, -M+\nu, -\nu | J,-M \rangle \,
}
\nonumber \\ [2mm]
&& \mbox{} \times \xib_{-\nu}
(-1)^\mu \int \d \hat{\bf r} \,
Y_{\lambda, -\mu}^\ast(\hat{\bf r}) \, Y_{L,-M+\nu}(\hat{\bf r})
\nonumber \\ [2mm]
&=&(-1)^\mu \sqrt{\frac{4\pi}{(2\lambda+1)}} \sum_{\nu}
\left< L, 1, -M+\nu, -\nu \left| J,- M \right> \right. \,
\xib_{-\nu}
\; \delta_{\lambda L} \, \delta_{-\mu,-M+\nu}.
\label{C.17}\eeqa
By using the equality \req{C.11}, we can write
\beqa
\left< \Omega_{\kappa_1 m_1} \left| C_{\lambda \mu} \, \sigmab
\rule{0mm}{4mm}\right| \Omega_{\kappa_2 m_2} \right> &=&
(-1)^\mu \sqrt{\frac{4\pi}{(2\lambda+1)}} \sum_{\nu}
\left\{ (-1)^{\mu+\nu}
\sum_{J} \right.
\left< \lambda, 1, -\mu, -\nu \left| J,-\mu-\nu \right> \right.
\nonumber \\ [2mm]
&& \! \! \! \! \! \! \! \! \! \! \! \! \! \! \! \! \! \! \!
\mbox{} \times \left[
\sqrt{\frac{J}{4\pi}} \, \left( \frac{\kappa_1+\kappa_2}{J}
- 1 \right) d^J_{\nu+\mu} \left( \kappa_1,m_1;-\kappa_2, m_2 \right) \,
\delta_{J,\lambda+1} \right.
\nonumber \\ [2mm]
&& \! \! \! \! \! \! \! \! \! \! \! \! \! \! \! \! \! \! \!
\mbox{} + \sqrt{\frac{2J+1}{4\pi \, J(J+1)}}\;
(\kappa_1-\kappa_2)
\, d^J_{\nu+\mu}(\kappa_1,m_1;\kappa_2, m_2) \,
\delta_{J,\lambda}
\nonumber \\ [2mm]
&& \! \! \! \! \! \! \! \! \! \! \! \! \! \! \! \! \! \! \!
\mbox{} + \left. \left. \sqrt{\frac{J+1}{4\pi}} \, \left(
\frac{\kappa_1+\kappa_2}{J+1} + 1 \right)
d^J_{\nu+\mu}(\kappa_1,m_1;-\kappa_2, m_2) \,
\delta_{J,\lambda-1} \right] \right\} \xib_{-\nu}\, .  \rule{10mm}{0mm}
\nonumber\eeqa
Hence,
\beq
\left< \Omega_{\kappa_1 m_1} \left| C_{\lambda \mu} \, \sigmab
\rule{0mm}{4mm}\right| \Omega_{\kappa_2 m_2} \right> =
\sum_{\nu} (-1)^{\nu} \,
c_{\lambda \mu}^\nu(\kappa_1 m_1; \kappa_2 m_2) \,
\xib_{-\nu}\, ,
\label{C.18}\eeq
with the coefficients
\beqa
c_{\lambda \mu}^\nu(\kappa_1 m_1; \kappa_2 m_2)
&=& \sqrt{\frac{1}{2\lambda+1}} \sum_{J}
\left< \lambda, 1, \mu, \nu \left| J, \mu+\nu \right>  \right.
\nonumber \\ [2mm]
\nonumber \\ [2mm]
&& \! \! \! \! \! \! \! \! \! \! \! \! \! \! \! \! \! \! \!
\mbox{} \times \left[
\sqrt{J+1} \, \left( \frac{\kappa_1+\kappa_2}{J+1} + 1 \right)
d^J_{\nu+\mu}(\kappa_1,m_1;-\kappa_2, m_2) \,
\delta_{J,\lambda-1}
\right.
\nonumber \\ [2mm]
&& \! \! \! \! \! \! \! \! \! \! \! \! \! \! \! \! \! \! \!
\mbox{} - \sqrt{\frac{2J+1}{J(J+1)}}\;
(\kappa_1-\kappa_2)
\, d^J_{\nu+\mu}(\kappa_1,m_1;\kappa_2, m_2) \,
\delta_{J,\lambda}
\nonumber \\ [2mm]
&& \! \! \! \! \! \! \! \! \! \! \! \! \! \! \! \! \! \! \!
\mbox{} + \left.
\sqrt{J} \, \left( \frac{\kappa_1+\kappa_2}{J}
- 1 \right) d^J_{\nu+\mu} \left( \kappa_1,m_1;-\kappa_2, m_2 \right) \,
\delta_{J,\lambda+1}
\right],
\label{C.19}\eeqa
where we have used the symmetry properties of the Clebsch--Gordan
coefficients.


\section{Collisions with free electrons at rest \label{secD}}
\setcounter{equation}{0}

For large values of the energy loss $W$, the longitudinal GOS, $\d f
(Q,W)/\d W$, considered as a function of $Q$, has a prominent peak near
$Q=W$, the Bethe ridge. As suggested by the impulse approximation
\citep[see,\eg,][]{Segui2002}, the structure of the Bethe ridge reflects the
momentum distribution of the bound target electron. The key point for
the calculation of the energy-loss DCS and its integrals is that, in the
limit of high energy losses, the finite width of this peak does not
affect the energy-loss DCS. To see this, we consider the case of
hydrogenic ions in their ground state, for which the non-relativistic
GOS is given by a simple analytical expression \citep{Inokuti1971}. Using
this expression, Inokuti has shown that the GOS takes appreciable values
if [Eq.\ (3.11) in \citeauthor{Inokuti1971}, \citeyear{Inokuti1971}]
\beq
| Q - W | \lesssim \xi \sqrt{W E_a}\; ,
\label{D.1}\eeq
where $\xi$ is a constant of the order of unity. Apparently, the width
of the Bethe ridge increases with the energy loss. For computing the
energy-loss DCS, however, it is more natural to consider the GOS as a
function of $\ln Q$ instead of $Q$, because this change of variable in
the integral \req{6.29}, $\d Q = Q \, \d (\ln Q)$, cancels out the
factor $Q^{-1}$ in the DDCS for longitudinal interactions [see Eq.\
\req{6.9}]. The above inequality implies that the GOS is appreciable
only when
\beq
| \ln (Q/E_a) - \ln (W/E_a) |
\lesssim \xi \sqrt{E_a/W}\, .
\label{D.2}\eeq
That is, the width of the Bethe ridge in the $\ln Q$ scale is $\sim \xi
(E_a/W)^{1/2}$. We thus see that the ``effective'' width of the ridge
does decrease when the energy loss increases. It is expected that this
conclusion also holds for the GOS and the TGOS of atoms and ions with
more than one electron. Indeed, for collisions with $Q$ and $W$ much
larger than the ionization energy $E_a$ of the active subshell, the target
electrons behave as if they were free and at rest. We will show below
that the GOS and the TGOS for collisions with a free electron at rest
are both equal to $\delta(W-Q)$. Accordingly, for sufficiently large
energy losses, the finite width of the Bethe ridge can be disregarded,
\ie, the GOS and the TGOS can be replaced with the delta distribution, $Z
\delta(W-Q)$, where $Z$ is the number of bound electrons in the target.

Here we consider collisions of a charged particle with free electrons at
rest; the calculation is adapted from \citet{Fano1963}.
We assume that the projectile has mass $M$ and charge $Z_0$, and
that it is distinguishable from the target electron. As usually, we will
calculate the DCS considering the interaction between the colliding
particles as a first-order perturbation. Hence, the unperturbed states
of both the projectile and the target electron are free states ($V=0$)
which we will represent as plane waves. Before the collision, the
projectile has kinetic energy $E$ and momentum $\hbar {\bf k}$; the
corresponding quantities after the collision are $E'$ and $\hbar {\bf
k}'$. The initial and final states of the projectiles will be described
as positive-energy plane waves of the form \req{B.5},
\beq
\phi_{{\bf k}, m_{{\rm S}},+1}({\bf r}) = (2\pi)^{-3/2}
\exp({\rm i} {\bf k} \cdot {\bf r}) \, U_{{\bf k}, m_{\rm S},+1}\, .
\label{D.3}\eeq
To keep the analogy with our calculation of collisions with bound
electrons (see Section \ref{sec4}), with initial states normalized to unity,
the states of the target electron need to be described by
positive-energy plane waves satisfying periodic boundary conditions on a
cubic box of side $L$,
\beq
\varphi_{{\bf k}, m_{{\rm S}},+1}({\bf r}) = L^{-3/2}
\exp({\rm i} {\bf k} \cdot {\bf r}) \, U_{{\bf k}, m_{\rm S},+1}\, ,
\label{D.4}\eeq
which represent one electron in the normalization box. The initial and
final states of the target electron are plane waves with respective
kinetic energies $\epsilon_a=0$ and $\epsilon_b = W=Q$ and wave numbers
${\bf k}_a =0$ and ${\bf k}_b$. The differential cross section for
transitions between states with defined spin quantum numbers is given by
Fermi's golden rule,
\beqa
\d \sigma^{\rm free} &=& \sum_{{\bf k}_b} \frac{(2\pi)^4}{\hbar v} \,
|T^{\rm free}_{fi}|^2 \, \delta(E - E' -\epsilon_b) \,
\d {\bf k}'
\label{D.5}\eeqa
with the following transition matrix elements [cf.\ Eq.\ \req{4.4}]
\beqa
T^{\rm free}_{fi}=&-& \frac{Z_0 e^2}{2\pi^2}
\int \d {\bf q} \,
\int \d {\bf r}_0 \,
\int_{L^3} \d {\bf r} \,
\phi^\dagger_{{\bf k}', m'_{\rm S},+1} ({\bf r}_0) \,
\varphi_{{\bf k}_b, m_{{\rm S}b},+1}^\dagger ({\bf r})
\nonumber \\ [2mm]
&\times&
\left( \frac{1}{q^2} - \frac{\widetilde{\alphab}_0
- (\widetilde{\alphab}_0 \dotprod \hat{\bf q}) \dotprod \hat{\bf q}
}{q^2 - (W/\hbar c)^2} \dotprod \widetilde{\alphab} \right)
\nonumber \\ [2mm]
 &\times&
\exp\left[ {\rm i} {\bf q} \dotprod \left({\bf r} -
{\bf r}_0 \right) \right] \,
\phi_{{\bf k}, m_{\rm S},+1} ({\bf r}_0)
\, \varphi_{{\bf 0}, m_{{\rm S}a},+1}({\bf r})\, .
\label{D.6}\eeqa
The integral over ${\bf r}_0$ gives a factor $\delta({\bf q} - {\bf
k} + {\bf k}')$ and, after integration over ${\bf q}$, we obtain
\beqa
\lefteqn{
T_{fi}^{\rm PW} =
- \frac{Z_0 e^2}{2\pi^2} \,
\left\{
U_{{\bf k}', m'_{\rm S},+1}^\dagger \, U_{{\bf k}, m_{\rm S},+1} \;
\frac{1}{q^2}  \left< \varphi_{{\bf k}_b, m_{{\rm S}b},+1} \left|
\exp ({\rm i} {\bf q}\dotprod {\bf r}) \rule{0mm}{4mm}
\right| \varphi_{{\bf 0}, m_{{\rm S}a},+1} \right> \right.
}
\nonumber \\ [2mm]
&& \! \! \!
\mbox{} - U_{{\bf k}', m'_{\rm S},+1}^\dagger \,
\left(  \frac{\widetilde{\alphab}_0
- (\widetilde{\alphab}_0 \dotprod \hat{\bf q}) \dotprod \hat{\bf q}
}{q^2 - (W/\hbar c)^2} \right)
U_{{\bf k}, m_{\rm S},+1} \dotprod  \,
\left. \left< \varphi_{{\bf k}_b, m_{{\rm S}b},+1}
 \left| \widetilde{\alphab} \,
\exp ({\rm i} {\bf q}\dotprod {\bf r}) \rule{0mm}{4mm}
\right| \varphi_{{\bf 0}, m_{{\rm S}a},+1}
\right> \right\}, \rule{9mm}{0mm}
\label{D.7}\eeqa
with ${\bf q} = {\bf k} - {\bf k}'$. To complete the evaluation of the
matrix elements we consider a reference frame with the $z$ axis in the
direction of ${\bf q}$ and write
\beq
T_{fi}^{\rm PW} =
- \frac{Z_0 e^2}{2\pi^2} \,
\left\{
U_{{\bf k}', m'_{\rm S},+1}^\dagger \, U_{{\bf k}, m_{\rm S},+1} \;
\frac{A}{q^2}
- U_{{\bf k}', m'_{\rm S},+1}^\dagger \,
\left(  \frac{\widetilde{\alpha}_{0x} D_x + \widetilde{\alpha}_{0y} D_y
}{q^2 - (W/\hbar c)^2} \right)
U_{{\bf k}, m_{\rm S},+1} \right\}
\label{D.8}\eeq
with
\beq
A = \left< \varphi_{{\bf k}_b, m_{{\rm S}b},+1}({\bf r})
\left| \exp ({\rm i} {\bf q}\dotprod {\bf r} ) \rule{0mm}{4mm}\right|
\varphi_{{\bf 0}, m_{{\rm S}a},+1}({\bf r}) \right>
\label{D.9}\eeq
and
\beq
{\bf G} =
\left< \varphi_{{\bf k}_b, m_{{\rm S}b},+1}({\bf r})
\left| \widetilde{\alphab}
\exp ({\rm i} {\bf q}\dotprod {\bf r} ) \rule{0mm}{4mm}\right|
\varphi_{{\bf 0}, m_{{\rm S}a},+1}({\bf r}) \right>.
\label{D.10}\eeq
We have
\beqa
A &=&
\sqrt{\frac{\epsilon_b+2\me c^2}{2\epsilon_b +2\me c^2}}
\left( \begin{array}{c}  \chi_{m_{{\rm S}b}} \\ [2mm]
\displaystyle{\frac{c \hbar (\sigmab \dotprod {\bf k}_b)
}{\epsilon_b+2\me c^2}} \, \chi_{m_{{\rm S}b}}
\end{array} \right)^\dagger
\left( \begin{array}{c}  \chi_{m_{{\rm S}a}} \\ [2mm]
0
\end{array} \right) \frac{1}{L^3} \int_{L^3} \exp[{\rm i} ({\bf q} -{\bf
k}_b)\dotprod {\bf r}] \, \d {\bf r}
\nonumber \\ [2mm]
&=&
\sqrt{\frac{\epsilon_b+2\me c^2}{2\epsilon_b +2\me c^2}}
\delta_{m_{{\rm S}b},m_{{\rm S}a}} \, \delta_{{\bf q},{\bf k}_b}
\label{D.11}\eeqa
plus terms that vanish in the limit $L\rightarrow \infty$. Similarly,
\beqa
{\bf G} &=&
\sqrt{\frac{\epsilon_b+2\me c^2}{2\epsilon_b +2\me c^2}}
\left( \begin{array}{c}  \chi_{m_{{\rm S}b}} \\ [2mm]
\displaystyle{\frac{c \hbar (\sigmab \dotprod {\bf k}_b)
}{\epsilon_b+2\me c^2}} \, \chi_{m_{{\rm S}b}}
\end{array} \right)^\dagger \widetilde{\alphab}
\left( \begin{array}{c}  \chi_{m_{{\rm S}a}} \\ [2mm]
0
\end{array} \right) \frac{1}{L^3} \int \exp[{\rm i} ({\bf q} -{\bf
k}_b)\dotprod {\bf r}] \, \d {\bf r}
\nonumber \\ [2mm]
&=&
\sqrt{\frac{\epsilon_b+2\me c^2}{2\epsilon_b +2\me c^2}}
\left( \begin{array}{c}  \chi_{m_{{\rm S}b}} \\ [2mm]
\displaystyle{\frac{c \hbar (\sigmab \dotprod {\bf k}_b)
}{\epsilon_b+2\me c^2}} \, \chi_{m_{{\rm S}b}}
\end{array} \right)^\dagger \left( \begin{array}{c}  0 \\ [2mm]
 \sigmab \chi_{m_{{\rm S}a}}
\end{array} \right)
 \, \delta_{{\bf q},{\bf k}_b}
\nonumber \\ [2mm]
&=&
\sqrt{\frac{\epsilon_b+2\me c^2}{2\epsilon_b +2\me c^2}} \;
\chi_{m_{{\rm S}b}}^\dagger
\displaystyle{\frac{c \hbar (\sigmab \dotprod {\bf k}_b)
}{\epsilon_b+2\me c^2}} \sigmab \chi_{m_{{\rm S}a}} \;
\delta_{{\bf q},{\bf k}_b}
\nonumber \\ [2mm]
&=&
\sqrt{\frac{\epsilon_b}{2\epsilon_b +2\me c^2}} \;
\chi_{m_{{\rm S}b}}^\dagger
\sigma_z \sigmab \chi_{m_{{\rm S}a}}
\; \delta_{{\bf q},{\bf k}_b}\, ,
\label{D.12}\eeqa
where we have used that $\sigmab \dotprod \hat{\bf k}_b = \sigma_z$,
because the vector $\hat{\bf k}_b = \hat{\bf q}$ is
parallel to the $z$ axis. Considering the expressions of the Pauli
matrices, a routine calculation shows that the components of the vector
${\bf G}$ are
\begin{subequations}
\label{D.13}
\beqa
G_x &=& \sqrt{\frac{\epsilon_b}{2\epsilon_b +2\me c^2}} \;
2 m_{{\rm S}b}\,
(1-\delta_{ m_{{\rm S}b}, m_{{\rm S}a}})\; \delta_{{\bf q},{\bf k}_b},
\label{D.13a} \\ [2mm]
G_y &=& \sqrt{\frac{\epsilon_b}{2\epsilon_b +2\me c^2}} \;
2 m_{{\rm S}b}\; {\rm i}
2 m_{{\rm S}a}
(1-\delta_{ m_{{\rm S}b}, m_{{\rm S}a}})\; \delta_{{\bf q},{\bf k}_b},
\label{D.13b} \\ [2mm]
G_z &=& \sqrt{\frac{\epsilon_b}{2\epsilon_b +2\me c^2}} \;
2 m_{{\rm S}b}\;
2 m_{{\rm S}a}
\delta_{ m_{{\rm S}b}, m_{{\rm S}a}}\; \delta_{{\bf q},{\bf k}_b}\, .
\label{D.13c}\eeqa
\end{subequations}
We thus see that the longitudinal and transverse interactions cause
transitions to states with different spins and do not interfere. This
property implies that the DCS can be evaluated by following the same
steps as in the case of bound electrons (see Section \ref{sec4.1}).

Using the identity \req{2.24}, we can write the DCS in the form
\beqa
\d \sigma^{\rm free} &=& \sum_{{\bf k}_b} \frac{(2\pi)^4}{\hbar v} \,
|T^{\rm free}_{fi}|^2 \, \delta(W -\epsilon_b) \,
k' \, \frac{E-W+M c^2}{c^2\hbar^2} \, \d W \, \d \hat {\bf k}'\, ,
\label{D.14}\eeqa
where we have introduced the energy loss $W=E-E'$.
In the common situation where both the projectile and the
target electron have their initial spins unpolarized and the final spins
are not observed, the DCS is obtained by summing over final spin states
and averaging over initial spin states,
\beqa
\d \sigma^{\rm free} &=& \sum_{{\bf k}_b} \frac{(2\pi)^4}{\hbar v} \,
{\cal T}^{\rm free}_{fi} \, \delta(W -\epsilon_b) \,
k' \, \frac{E-W+M c^2}{c^2\hbar^2} \, \d W \, \d \hat {\bf k}'\, ,
\label{D.15}\eeqa
where
\beqa
{\cal T}^{\rm free}_{fi}
&\equiv& \frac{1}{4} \sum_{m_{{\rm S}a}, m_{{\rm S}b}}
\sum_{m_{\rm S}, m'_{\rm S}} | T_{fi}^{\rm free} |^2\, .
\label{D.16}\eeqa
A calculation analogous to the derivation of Eq.\ \req{4.40} [see
Section \ref{sec4.1}] yields the result
\beqa
{\cal T}^{\rm free}_{fi}
&=&
\frac{Z_0^2 e^4}{4\pi^4} \, \frac{1}{q^4} \,
\frac{(2E-W + 2 M c^2)^2 - (c\hbar q)^2 }{4\,
(E+M c^2)(E-W+M c^2)}
\nonumber \\ [2mm]
&& \rule{5mm}{0mm}\times
\sum_{m_{{\rm S}a}, m_{{\rm S}b}} \left|
\left< \varphi_{{\bf k}_b, m_{{\rm S}b},+1} \left|
\exp ({\rm i} {\bf q}\dotprod {\bf r}) \rule{0mm}{4mm}
\right| \varphi_{{\bf 0}, m_{{\rm S}a},+1} \right> \right|^2
\nonumber \\ [2mm]
&+& \frac{Z_0^2 e^4}{4\pi^4} \, \frac{1}{[q^2-(W/c\hbar)^2]^2} \,
\frac{E+M c^2}{E-W+M c^2}
\nonumber \\ [2mm]
&& \rule{5mm}{0mm}\times
\left( \beta^2 \sin^2\theta_{\rm r}
+ \frac{(c\hbar q)^2 - W^2 }{2(E+ M c^2)^2}\,
\right) \sum_{m_{{\rm S}a}, m_{{\rm S}b}}  \,
\left| G_x \right|^2\, . \rule{10mm}{0mm}
\label{D.17}\eeqa
where the recoil angle $\theta_{\rm r}$, the angle between the vectors
${\bf p} = \hbar {\bf k}$ and ${\bf q}$ (see Fig.\ \ref{fig35}), is given by
\beqa
\cos\theta_{\rm r} =
\frac{\epsilon_b}{\beta (c\hbar q)}
\left( 1 + \frac{ (c\hbar q)^2-\epsilon_b^2}
{2 \epsilon_b (E+M c^2)} \right)\, .
\label{D.18}\eeqa

As in the case of collisions with bound electrons, we introduce the
recoil energy $Q$, defined by Eq.\ \req{A.18}. Of course, for collisions
with free electrons at rest, $Q=W=\epsilon_b$, because the energy lost
by the projectile is equal to the kinetic energy of the recoiling target
electron. Since
\beq
\d \hat{\bf k}' = 2 \pi \d (\cos\theta) = 2\pi \,
\frac{Q+\me c^2}{c^2 \hbar^2 kk'} \, \d Q\, ,
\label{D.19}\eeq
the DCS is
\beq
\frac{\d^2 \sigma^{\rm free}}{\d W\, \d Q}
= \frac{(2\pi)^4}{\hbar v}
\, k' \, \frac{E-W + M c^2}{c^2\hbar^2} \, 2\pi \,
\frac{Q+\me c^2}{c^2 \hbar^2 kk'} \, \delta(W-Q)
\sum_{{\bf k}_b} {\cal T}^{\rm free}_{fi} \, .
\label{D.20}\eeq
with $\epsilon_b = W$. We can now express the DCS in the form of Eq.\
\req{4.46},
\beqa
\lefteqn{
\frac{\d \sigma^{\rm free}}{\d W\d Q} =
\frac{2\pi Z_0^2 e^4}{\me v^2} \left[
\frac{2\me c^2}{WQ(Q+2\me c^2)} \right.
}
\nonumber \\ [2mm]
&& \mbox{} \times
\left\{ \frac{(2 E -W  + 2 M c^2)^2 - Q(Q+2\me c^2)}{4\,
(E+M c^2)^2} \right\}
\frac{\d f^{\rm free}(Q,W)}{\d W}
+ \frac{2\me c^2 W}{[Q(Q+2\me c^2)-W^2]^2}
\nonumber \\ [2mm]
&& \mbox{} \times \left.
\left( \beta^2 \sin^2\theta_{\rm r}
+ \left\{ \frac{Q(Q+2\me c^2)- W^2 }{2(E+ M c^2)^2}\,
\right\} \right) \frac{\d g^{\rm free}(Q,W)}{\d W} \right],
\rule{10mm}{0mm}
\label{D.21}\eeqa
where we have introduced the longitudinal and transverse
generalized oscillator strengths, defined by
\beqa
&& \!\!\!\!\!\!\!\!\!\!\!\!\!\!\!
\frac{\d f^{\rm free}(Q,W)}{\d W} \equiv \frac{W 2(Q+\me
c^2)}{Q(Q+2\me c^2)} \, \delta(W-Q)
\nonumber \\ [2mm]
&\times&
\frac{1}{2} \sum_{{\bf k}_b} \sum_{m_{{\rm S}a}, m_{{\rm S}b}}
\left| \left< \varphi_{{\bf k}_b, m_{{\rm S}b},+1}({\bf r})
\left| \exp ({\rm i} {\bf q}\dotprod {\bf r} ) \rule{0mm}{4mm}\right|
\varphi_{{\bf 0}, m_{{\rm S}a},+1}({\bf r}) \right> \right|^2\, ,
\label{D.22}\eeqa
and
\beqa
&& \!\!\!\!\!\!\!\!\!\!\!\!\!\!\!
\frac{\d^{\rm free} g(Q,W)}{\d W} \equiv \frac{2(Q+\me
c^2)}{W}\,\delta(W-Q)
\nonumber \\ [2mm]
&\times&
\frac{1}{2} \sum_{{\bf k}_b} \sum_{m_{{\rm S}a}, m_{{\rm S}b}}
\left| \left< \varphi_{{\bf k}_b, m_{{\rm S}b},+1}({\bf r})
\left| \rule{0mm}{4mm} \widetilde {\alpha}_x \exp ({\rm i} {\bf
q}\dotprod {\bf r} ) \right|
\varphi_{{\bf 0}, m_{{\rm S}a},+1}({\bf r}) \right> \right|^2\, ,
\label{D.23}\eeqa
respectively. Inserting the expressions \req{D.11} and \req{D.13} of the
matrix elements, and recalling that $\epsilon_b =W =Q$, we obtain
\beq
\frac{\d f^{\rm free}(Q,W)}{\d W} = \frac{W 2(Q+\me
c^2)}{Q(Q+2\me c^2)}\,
\frac{W+2\me c^2}{2W +2\me c^2}
\,\delta(W-Q) \, \sum_{{\bf k}_b} \, \delta_{{\bf q},{\bf
k}_b}=\delta(W-Q)\, ,
\label{D.24}\eeq
and
\beq
\frac{\d g^{\rm free}(Q,W)}{\d W} = \frac{2(Q+\me
c^2)}{W}\,
\frac{W}{2W +2\me c^2} \, \delta(W-Q)
 \, \sum_{{\bf k}_b} \delta_{{\bf q},{\bf k}_b}= \delta(W-Q) \, .
\label{D.25}\eeq

Finally, the DDCS \req{D.21} for collisions with a free electron at
rest can be expressed as
{\allowdisplaybreaks
\beqa
\frac{\d^2 \sigma^{\rm free}}{\d Q \d W}
&=& \frac{2\pi Z_0^2 e^4}{\me v^2} \,
\left[\rule{0mm}{8mm}
\frac{2\me c^2}{WQ(Q+2\me c^2)} \,
\left\{ \frac{(2 E -W  + 2 M c^2)^2 - Q(Q+2\me c^2)}{4\,
(E+ M c^2)^2} \right\}
 \right.
\nonumber \\ [2mm]
&& \mbox{} + \, \frac{2\me c^2 W}{[Q(Q+2\me c^2)-W^2]^2} \,
\nonumber \\ [2mm]
&& \rule{5mm}{0mm} \left.
\times \left( \beta^2 \sin^2\theta_{\rm r}
+ \left\{ \frac{Q(Q+2\me c^2)- W^2 }{2(E+ M c^2)^2}\,
\right\} \right) \rule{0mm}{8mm}\right] \, \delta(Q-W)\, .
\label{D.26}\eeqa
}
Setting $Q=W$, we obtain
\beqa
\frac{\d^2 \sigma^{\rm free}}{\d Q \d W}
&=& \left( \frac{2\pi Z_0^2 e^4}{\me
v^2} \, \frac{1}{W^2} \right) \, F_{\rm rel}(W)
\, \delta(Q-W)
\label{D.27}\eeqa
with
\beq
F_{\rm rel}(W) = 1 - \frac{\left[(2E-W+2 M c^2)\me c^2 +M^2c^4+\me^2
c^4\right]W} {2\me c^2(E+M c^2)^2}\, .
\label{D.28}\eeq
The DDCS \req{D.27} coincides with the result obtained from elementary
quantum electrodynamics for collisions of two distinguishable electrons
\citep[see, \eg,][]{GreinerReinhardt1994}. The quantity in parenthesis is the
non-relativistic (Thomson) energy-loss DCS, but with the relativistic
speed $v=\beta c$. Hence, the factor $F_{\rm rel}(W)$ accounts for the
remaining relativistic corrections.


\section{Electron energy-loss spectra \label{secE}}
\setcounter{equation}{0}

Here we provide a brief presentation of the basic theory of electron
energy-loss spectroscopy (EELS). This technique utilizes a parallel
electron beam with a narrow energy spectrum, usually in the range from
100 keV to 300 keV, that impinges of a thin solid target; the
energy-spectrum of transmitted electrons is measured with a detector
that covers a small solid angle around the direction of the incident
beam \citep[see, \eg,][]{Egerton2011}. The measured EEL spectrum
corresponds to electrons that have undergone inelastic interactions with
scattering angle less than the effective collection semi-angle
$\vartheta$ of the experimental setup. Typically, $\vartheta$ is small
(a few degrees) and the initial kinetic energy is much larger than the
observed energy losses. Under these circumstances ($\vartheta \ll 1$, $W
\ll E$), the relevant interactions involve small recoil energies ($Q \ll
2 \me c^2$) in the interval between the kinematical minimum $Q_-$ (for
$\theta = 0$) given by
\beq
Q_-(Q_-+2\me c^2) = c^2(p-p')^2
\label{E.1}\eeq
and the value $Q(\vartheta)$ corresponding to the maximum
deflection of the detected electrons,
\beqa
Q(\vartheta) \left[ Q(\vartheta) + 2 \me c^2 \right]
& = &  c^2(p-p')^2 + 2\, c p \, cp' \, (1-\cos\vartheta)
\nonumber \\ [2mm]
& = &  Q_-(Q_-+2\me c^2) + 4 \, c p \, cp' \, \sin^2(\vartheta/2) .
\label{E.2}\eeqa
The measured energy-loss spectrum is given by
\beq
I(W) = I_0 \, {\cal E}_{\rm d} \; t \, {\cal N}
\left[ \frac{\d\sigma}{\d W} \right]_{\theta<\vartheta},
\label{E.3}\eeq
where $I_0$ is the number of incident electrons, ${\cal E}_{\rm d}$ is
the efficiency of the detector, $t$ is the thickness of the specimen,
${\cal N}$ is the number of atoms per unit volume, and the last factor
is the restricted energy-loss DCS,
\beq
\left[ \frac{\d\sigma}{\d W} \right]_{\theta<\vartheta} =
\int_{Q_-}^{Q(\vartheta)} \frac{\d^{2}\sigma}{\d Q\d W} \, \d Q.
\label{E.4}\eeq
The integrand is the atomic DDCS given by Eqs.\ \req{6.8} to \req{6.11}.
When only small-angle collisions are of interest, the DDCS
can be simplified by replacing the GOSs with the OOS. This replacement
amounts to assuming that the complete angular dependence of the DCS
comes from kinematical factors and, consequently, angular integrals can
be evaluated irrespective of the characteristics of the target (atom,
molecule, or solid). Ignoring the terms within curly braces in Eqs.\
\req{6.9} and \req{6.10}, the DDCS at small angles takes the form
\beqa
\frac{\d^2 \sigma}{\d W \, \d Q} &=&
\frac{2\pi Z_0^2 e^4}{\me v^2} \,
\frac{2\me c^2}{WQ(Q+2\me c^2)} \, \frac{\d f (W)}{\d W}
\nonumber \\ [2mm]
&& \mbox{} +
\frac{2\pi Z_0^2 e^4}{\me v^2} \,
\frac{2\me c^2 W}{[Q(Q+2\me c^2)-W^2]^2} \,
\nonumber \\ [2mm]
&& \mbox{} \times
\left[
\beta^2
- \frac{W^2}{Q(Q+2\me c^2)} \left( 1 + \frac{ Q(Q+2\me c^2)-W^2}
{2 W (E+ \me c^2)} \right)^2
\right] \frac{\d f (W)}{\d W}\, .
\label{E.5}\eeqa

In order to obtain handy analytical expressions, we limit our
considerations to situations where $W \ll E$, $W \ll \me c^2$, and $Q
\ll W$, that is, interactions of high-energy projectiles with small energy
losses and small scattering angles. Notice that the last condition may
be violated in interactions with very small energy transfers, \ie, in
low-loss EEL spectra. Under the assumed circumstances,
\beq
Q_-(Q_-+2\me c^2) \simeq \frac{W^2}{\beta^2}\, ,
\label{E.6}\eeq
and the value $Q(\vartheta)$ corresponding to the maximum
deflection of the detected electrons can be approximated as
\beq
Q(\vartheta) \left[ Q(\vartheta) + 2 \me c^2 \right]
 \simeq  \frac{W^2}{\beta^2} + cp\, cp' \, 4 \sin^2(\vartheta/2).
\label{E.7}\eeq
In addition, the DDCS \req{E.5} simplifies to
\beqa
\frac{\d^2 \sigma}{\d W \, \d Q} &\simeq&
\frac{2\pi Z_0^2 e^4}{\me v^2} \left[
\frac{1}{WQ}
+ \frac{2\me c^2 W}{[Q \, 2\me c^2-W^2]^2} \,
\left(
\beta^2
- \frac{W^2}{Q \, 2\me c^2}
\right) \right] \frac{\d f (W)}{\d W}\,
\nonumber \\ [2mm]
&=&
\frac{2\pi Z_0^2 e^4}{\me v^2} \, \frac{1}{W} \,  \frac{\d f (W)}{\d W}
\left[ \frac{1}{Q}
+ \frac{2\me c^2}{W^2 [Q \, 2\me c^2/W^2 -1 ]^2} \,
\left( \beta^2 - \frac{W^2}{Q \, 2\me c^2}
\right) \right] . \rule{10mm}{0mm}
\label{E.8}\eeqa
The restricted energy-loss DCS can now be evaluated analytically,
\beqa
\left[ \frac{\d\sigma}{\d W} \right]_{\theta<\vartheta} &=&
\frac{2\pi Z_0^2 e^4}{\me v^2} \, \frac{1}{W} \,
\frac{\d f (W)}{\d W}
\nonumber \\ [2mm]
&& \mbox{} \times
\int_{Q_-}^{Q(\vartheta)} \d Q
\left[ \frac{1}{Q}
+ \frac{2\me c^2}{W^2 [Q \, 2\me c^2/W^2 -1 ]^2} \,
\left( \beta^2 - \frac{W^2}{Q \, 2\me c^2}
\right) \right]
\nonumber\eeqa
with
\beq
Q_- \, 2\me c^2 \simeq \frac{W^2}{\beta^2}
\qquad \mbox{and} \qquad
Q(\vartheta) \, 2 \me c^2
\simeq \frac{W^2}{\beta^2} + cp \, cp'\, 4 \sin^2(\vartheta/2).
\label{E.9}\eeq
The integral is
\beqa
&& \! \! \! \! \! \! \! \! \! \! \! \! \! \! \! \! \! \! \!
\! \! \! \! \! \! \! \! \!
\int_{Q_-}^{Q(\vartheta)} \d Q
\left[ \frac{1}{Q}
+ \frac{2\me c^2}{W^2 [Q \, 2\me c^2/W^2 -1 ]^2} \,
\left( \beta^2 - \frac{W^2}{Q \, 2\me c^2}
\right) \right]
\nonumber \\ [2mm]
&=& \ln \left[ 1 + \beta^2 W^{-2}
cp\, cp' \, 4 \sin^2(\vartheta/2) \rule{0mm}{4mm}\right]
\nonumber \\ [2mm]
&& \mbox{} + (1-\beta^2) \left\{
\frac{\beta^2}{
1-\beta^2 + \beta^2 W^{-2} cp\, cp' \, 4 \sin^2(\vartheta/2)
}
- \frac{\beta^2}{1 - \beta^2} \right\}
\nonumber \\ [2mm]
&& \mbox{} - \ln \left[
(1-\beta^2) \,
\frac{ 1 + \beta^2 W^{-2} cp\, cp' \, 4 \sin^2(\vartheta/2)}
{1 - \beta^2 + \beta^2 W^{-2} cp\, cp' \, 4 \sin^2(\vartheta/2)
} \right].
\nonumber\eeqa
Hence
\beqa
\left[ \frac{\d\sigma}{\d W} \right]_{\theta<\vartheta} &=&
\frac{2\pi Z_0^2 e^4}{\me v^2} \, \frac{1}{W} \,
\frac{\d f (W)}{\d W}
\left\{
\ln \left( 1 + {\cal X} \right) \rule{0mm}{6    mm}\right.
\nonumber \\ [2mm]
&& \mbox{} + \beta^2 \left[
\frac{1 - \beta^2}{1-\beta^2 + {\cal X}} - 1 \right]
\left. - \ln \left[ (1-\beta^2) \, \frac{ 1 + {\cal X}}
{1 - \beta^2 + {\cal X}} \right] \right\}
\label{E.10}\eeqa
with
\beq
{\cal X} = \beta^2 \, \frac{cp\, cp'}{W^2} \, 4 \sin^2(\vartheta/2) .
\label{E.11}\eeq
Reorganizing terms we obtain the following compact formula
\beqa
\left[ \frac{\d\sigma}{\d W} \right]_{\theta<\vartheta} &=&
\frac{2\pi Z_0^2 e^4}{\me v^2} \, \frac{1}{W} \,
\frac{\d f (W)}{\d W}
\left[
\ln {\cal Y}
- \beta^2 \left( 1- \frac{1}{\cal Y}
\right)
\right]
\label{E.12}\eeqa
with
\beqa
{\cal Y} &=& 1 + \frac{{\cal X}}{1-\beta^2}
= 1 + \frac{\beta^2}{1-\beta^2}
\, \frac{cp\, cp'}{W^2} \, 4 \sin^2(\vartheta/2)
\nonumber \\ [2mm]
&=& 1 + \frac{
\sqrt{[E(E+2\me c^2)]^3 \; (E-W) (E-W+2\me c^2)}}{\me^2 c^4 \, W^2} \,
4 \sin^2(\vartheta/2)\, .
\label{E.13}\eeqa
Considering a beam energy of 300 keV and an aperture $\vartheta$ of 1
degree (17.5 mrad), the percentage differences between the results from
expression \req{E.12} and from the exact formula \req{E.4}
with the DDCS given by Eq.\ \req{E.5} for $W=1$, 5 and 10 keV are 0.04,
0.5, and 1.4, respectively


\subsection{DDCS in terms of the scattering angle}

We now show that our DDCS are consistent with those considered by
\citet{Fano1956}, which are the basis of the formulas given in Egerton's
(\citeyear{Egerton2011}) book.

The definition $Q(Q+2\me c^2)=(cq)^2$ of the recoil energy $Q$ implies, that
\beq
2 \left( Q + \me c^2 \right) \, \d Q = 2 \, cp \, cp'
\, \sin \vartheta \d \vartheta
\label{E.14}\eeq
and
\beq
\d \Omega \equiv 2\pi \, \sin \vartheta \, \d \vartheta =
\frac{2\pi \left( Q + \me c^2 \right)}{cp\, cp'}\, \d Q .
\label{E.15}\eeq
We can then express the DDCS in terms of the energy loss
and the angular deflection,
\beq
\frac{\d^2 \sigma}{\d W \, \d \Omega} =
\frac{\d^2 \sigma}{\d W \, \d Q} \, \frac{\d Q}{\d \Omega}
= \frac{\d^2 \sigma}{\d W \, \d Q} \,
\frac{cp\, cp'}{2\pi \left( Q + \me c^2 \right) } \, .
\label{E.16}\eeq
Equation \req{E.8} leads to the small-angle formula
\beqa
\frac{\d^2 \sigma}{\d W \, \d \Omega} &=&
\frac{Z_0^2 e^4}{\me v^2} \,
\frac{cp\, cp'}{Q + \me c^2} \, \frac{1}{W} \,
\frac{\d f (W)}{\d W}
\nonumber \\ [2mm]
&& \mbox{} \times \left[ \frac{1}{Q}
+ \frac{2\me c^2}{W^2 [Q \, 2\me c^2/W^2 -1 ]^2} \,
\left( \beta^2 - \frac{W^2}{Q \, 2\me c^2}
\right) \right] \, ,
\label{E.17}\eeqa
which is valid when
\beq
Q \simeq \frac{(c \hbar q)^2}{2\me c^2}, \qquad Q < W \ll \min\{E,
\me c^2\}.
\label{E.18}\eeq
That is,
\beqa
\frac{\d^2 \sigma}{\d W \, \d \Omega} &=&
\frac{Z_0^2 e^4}{\me v^2} \,
\frac{cp\, cp'}{\me c^2} \, \frac{1}{W} \,
\frac{\d f (W)}{\d W}
\nonumber \\ [2mm]
&& \mbox{} \times \left[ \frac{2\me c^2}{(c\hbar q)^2}
+ \frac{2\me c^2}{W^2 [(c\hbar q)^2/W^2 -1 ]^2} \,
\left( \beta^2 - \frac{W^2}{(c\hbar q)^2}
\right) \right]
\nonumber \\ [2mm]
& \simeq &
\frac{Z_0^2 e^4}{\me v^2} \,
2 \, cp\, cp' \, \frac{1}{W} \,
\frac{\d f (W)}{\d W} \, \frac{1}{(c\hbar q)^2}
\nonumber \\ [2mm]
&& \mbox{} \times \left[ 1
+ \frac{(c\hbar q)^2}{W^2 [(c\hbar q)^2/W^2 -1 ]^2} \,
\left( \beta^2 - \frac{W^2}{(c\hbar q)^2}
\right) \right]
\nonumber \\ [2mm]
& \simeq &
\frac{2 Z_0^2 e^4}{\me v^2} \,
\frac{1}{W} \, \frac{\d f (W)}{\d W} \, \frac{(cp^2)}{(c\hbar q)^2}
\left[ 1 + \frac{W^2}{[ (c\hbar q)^2 - W^2]^2} \,
\left[ \beta^2 (c\hbar q)^2- W^2 \right] \right] . \rule{10mm}{0mm}
\label{E.19}\eeqa
It is appropriate to recall that the first and second terms in this
expression account for the contribution of longitudinal interactions
(through the instantaneous Coulomb potential) and transverse
interactions (exchange of virtual photons), respectively.

In the assumed small-$W$ and small-angle regime ($W\ll E$, $\sin\theta \simeq
\theta$), the relation
\beq
(c\hbar q)^2 = (c\hbar q_-)^2 + 4 \, cp \, cp' \sin^2(\theta/2)
\nonumber\eeq
reduces to
\beq
(c\hbar q)^2 \simeq \frac{W^2}{\beta^2} + (cp)^2 \theta^2.
\label{E.20}\eeq
Following \citet{Egerton2011} and others, we introduce the characteristic angle
\beq
\theta_{\rm W} \equiv \frac{W}{\beta cp}
\label{E.21}\eeq
and write
\beq
(c\hbar q)^2 = (cp)^2 \left[ \theta_W^2 + \theta^2 \right]
\label{E.22}\eeq
Thus, Eq.\ \req{E.19} becomes
\beqa
\frac{\d^2 \sigma}{\d W \, \d \Omega} &=&
\frac{2 Z_0^2 e^4}{\me v^2} \,
\frac{1}{W} \, \frac{\d f (W)}{\d W} \, \frac{1}{\theta_W^2 + \theta^2}
\left[ 1 + W^2 \frac{
\beta^2 (cp)^2 \left[ \theta_W^2 + \theta^2 \right]- W^2}
{[ (cp)^2 \left[ \theta_W^2 + \theta^2 \right] - W^2]^2} \right]
\nonumber \\ [2mm]
&=&
\frac{2 Z_0^2 e^4}{\me v^2} \,
\frac{1}{W} \, \frac{\d f (W)}{\d W} \, \frac{1}{\theta_W^2 + \theta^2}
\left[ 1 + \frac{W^2}{(cp)^2} \,  \frac{\beta^2 \theta^2 }
{\left[ \theta_W^2 + \theta^2 - W^2 (cp)^{-2} \right]^2} \right]
\nonumber \\ [2mm]
&=&
\frac{2 Z_0^2 e^4}{\me v^2} \,
\frac{1}{W} \, \frac{\d f (W)}{\d W} \, \frac{1}{\theta_W^2 + \theta^2}
\left[ 1 + \frac{\beta^4 \theta_W^2 \theta^2 }
{\left[ (1-\beta^{2}) \theta_W^2 + \theta^2 \right]^2} \right].
\nonumber \eeqa
That is,
\beq
\frac{\d^2 \sigma}{\d W \, \d \Omega} =
\frac{2 Z_0^2 e^4}{\me v^2} \,
\frac{1}{W} \, \frac{\d f (W)}{\d W} \, \frac{1}{\theta_W^2 + \theta^2}
\left[ 1 + \left( \frac{\beta^2 \theta_W \theta }
{\gamma^{-2} \theta_W^2 + \theta^2} \right)^2 \right] \, .
\label{E.23}\eeq
This approximate formula was first derived by \citet{Fano1956}, and is given
in Eq.\ (A.4) of Egerton's (\citeyear{Egerton2011}) book, with a typo.

\newpage




  \addcontentsline{toc}{section}{References}



\begin{center}
\rule[-0.125mm]{2cm}{0.25mm}\rule[-0.3mm]{8cm}{.6mm}\rule[-0.125mm]{2cm}{0.25mm}
\end{center}

\end{document}